\documentclass[12pt,oneside]{book}
\usepackage{latexsym}
\usepackage[dvips]{graphicx}
\usepackage{fancyhdr}
\usepackage[italian]{babel}
\usepackage{feynmf}

\begin{document}

\input epsf
\def\pct#1{\centerline{ \epsfbox{#1.eps}}}

\newcommand{\dsl}{{\not{\! \partial}}}
\newcommand{\Dsl}{{\not{\! D}}}
\newcommand{\psisl}{{\not{\! \psi}}}
\newcommand{\epsilonsl}{{\not{\! \epsilon}}}
\newcommand{\Ssl}{{\not{\! {\cal S}}}}
\newcommand{\nablasl}{{\not{\! \nabla}}}
\newcommand{\psl}{{\not{\! p}}}
\newcommand{\qsl}{{\not{\! q}}}
\newcommand{\ksl}{{\not{\! k}}}
\newcommand{\lsl}{{\not{\! \ell}}}

\newtheorem{CM}{Teorema}


\pagenumbering{roman}

\thispagestyle{empty}

\vspace*{-1.5truecm}

\begin{center}
{\LARGE UNIVERSIT\`A DEGLI STUDI DI ROMA \\
       ``TOR \rule{0pt}{24pt}VERGATA''} \\
\end{center}

\vspace{0.8cm}

\begin{center}
FACOLT\`A DI SCIENZE MATEMATICHE, FISICHE E NATURALI \\
Dipartimento di Fisica \\
\end{center}

\vspace{1.5cm}

\begin{center}
{\LARGE {\bf Sulle interazioni dei campi di gauge di spin
arbitrario}}
\end{center}

\vspace{1.0cm}

\begin{center}
{Tesi di laurea in Fisica} \\
\end{center}

\vspace{2.5cm} \noindent
Relatore \hspace{9.5cm} Laureando \\
\rule{0pt}{20pt}{\large{Prof. {\em Augusto Sagnotti
\hspace{5.8cm} Carlo Iazeolla}}} \\

\vspace{0.5cm} \noindent
Relatore esterno \\
\rule{0pt}{20pt}{\large{Prof. {\em Damiano Anselmi}}} \\
(Dipartimento di Fisica ``E. Fermi'', Universit\`{a} di Pisa) \\

\vspace{2cm}
\begin{center}
Anno Accademico 2003-2004
\end{center}

\newpage \vspace*{16cm}
\begin{flushright}

\vspace*{6cm}

{\large {\em Ai miei genitori}}
\end{flushright}

 \vspace*{16cm}
\newpage
{\LARGE {\textbf{Prefazione}}}

\vspace{1.5cm}

Questo lavoro di Tesi \`{e} il risultato della formazione
impartitami dal Corso di Laurea in Fisica e della mia prima
attivit\`{a} di ricerca, condotta presso il Dipartitimento di
Fisica dell'Universit\`{a} di Roma ``Tor Vergata'', i cui membri
desidero ringraziare per la disponibilit\`{a} dimostratami durante
il corso dei miei studi. Il lavoro \`{e} stato in parte svolto
anche presso il Dipartimento di Fisica ``E. Fermi''
dell'Universit\`{a} di Pisa, che pure ringrazio per
l'ospitalit\`{a} e le risorse offerte.

Un particolare ringraziamento va al mio relatore Prof. Augusto
Sagnotti, al cui insegnamento \`{e} legata una grande parte della
mia formazione, segnata dal suo straordinario entusiasmo per la
fisica. Egli mi ha inoltre introdotto alla ricerca scientifica,
offrendomi la sua guida ed il suo essenziale aiuto nelle
inevitabili difficolt\`{a} di approccio ad un campo affascinante,
dandomi anche l'occasione di conoscerne da vicino altri
protagonisti.

Tra questi, il Prof. Damiano Anselmi, dell'Universit\`{a} di Pisa,
cui sono grato per l'importante e istruttiva collaborazione su
tutti i risultati riportati nell'ultimo capitolo della Tesi, ed il
Prof. Per Sundell, dell'Universit\`{a} di Uppsala, che ringrazio
per molte utili discussioni che hanno costituito la base per il
mio avvicinamento alla teoria interagente dei campi di spin
arbitrario.

Desidero inoltre rivolgere un ringraziamento agli altri membri del
gruppo di Teoria delle Stringhe, il Prof. Massimo Bianchi, il Dr.
Gianfranco Pradisi e il Dr. Yassen Stanev per il contributo
formativo da essi ricevuto riguardo a questo vasto campo di
ricerca, e a tutti i dottorandi del gruppo per i loro utili e
graditi suggerimenti: in particolare desidero menzionare Marco
D'Alessandro e Gigi Genovese (anche per il loro aiuto nella
composizione del testo), e soprattutto Dario Francia per le
innumerevoli e sempre proficue discussioni sugli argomenti pi\`{u}
disparati, e per la stimolante collaborazione su alcune delle
questioni affrontate nel capitolo 7.

Non posso non ringraziare infine Alessandra, Lorenzo e Anna per la
loro amicizia ed il

\newpage

\noindent{sostegno della loro compagnia, ed i miei genitori, cui
questo lavoro \`{e} dedicato, per l'affetto e l'incoraggiamento
che mi hanno sempre dato}.

\vspace{0,5cm}

Roma, Luglio 2004 \hspace{8cm} Carlo Iazeolla

\pagebreak

\tableofcontents

\chapter{Introduzione}

\pagenumbering{arabic} \setcounter{page}{1}

Una teoria di gauge di spin elevato (\emph{higher spin} o HS)
\`{e} una teoria contenente campi di gauge di spin superiore a
due. Non \`{e} dunque sorprendente che la costruzione di una
teoria di questo tipo abbia stimolato l'interesse dei fisici e dei
matematici sin dagli anni '30, con importanti contributi di Dirac
\cite{Dirac:1936tg}, Fierz e Pauli \cite{Fierz:1939ix}, Rarita e
Schwinger \cite{Rarita:mf} e molti altri. Si tratta oggi di
cercare generalizzazioni delle teorie di gauge note basate su
campi di gauge di spin 1 (la teoria di Maxwell dell'
elettromagnetismo e le teorie di Yang-Mills), di spin 2 (la
Relativit\`{a} Generale di Einstein), e di spin 3/2 (la
supergravit\`{a}).

La peculiarit\`{a} di una teoria di gauge \`{e} la presenza di
simmetrie locali, ovvero l'invarianza sotto trasformazioni i cui
parametri sono funzioni arbitrarie delle coordinate
spazio-temporali. La costruzione di teorie di gauge di HS
consistenti corrisponde quindi alla scoperta di principi di
simmetria pi\`{u} generali associati ai rispettivi campi di massa
nulla.

\vspace{0.5cm}

L'interesse per queste teorie ha avuto, nel corso del tempo,
molteplici motivazioni. Anzitutto, esistono rappresentazioni
unitarie irriducibili (UIRs) di HS del gruppo di Poincar\'{e} , ed
\`{e} perci\`{o} naturale lo studio di teorie di campo che
descrivano particelle corrispondenti a tali rappresentazioni.
Ci\`{o} spinse Fierz e Pauli a studiare equazioni libere di campi
massivi di HS nello spaziotempo piatto gi\`{a} nel 1939; i loro
risultati furono poi ulteriormente sviluppati da Singh e Hagen nel
1974 \cite{Singh:qz, Singh:rc}, mentre il limite di massa nulla di
tali equazioni venne studiato inizialmente da Fronsdal e da Fang e
Fronsdal nel 1978 \cite{Fronsdal:1978rb, Fang:1978wz,
Fronsdal:1978vb, Fang:1979hq}. Il risultato fu una formulazione
della dinamica libera dei campi di spin elevato, che generalizzava
quelle dell'elettromagnetismo e della gravit\`{a} linearizzata, in
termini di tensori totalmente simmetrici ma soggetti al vincolo di
doppia traccia nulla, con parametri di gauge simmetrici e a
singola traccia nulla. \`{E} stata poi recentemente scoperta da
Francia e Sagnotti \cite{Francia:2002aa, Francia:2002pt} una
formulazione equivalente ma pi\`{u} generale, che elimina i
vincoli sui campi e sui corrispondenti parametri di gauge e,
utilizzando una costruzione dovuta a de Wit e Freedman
\cite{deWit:pe}, esibisce la geometria che governa tali equazioni,
in modo simile ai casi di Maxwell e Einstein.

\vspace{0.5cm}

Le difficolt\`{a} nel costruire una teoria interagente, evidenti
gi\`{a} dai primi lavori, vennero formalizzate negli anni '60 in
teoremi ``no-go'' come quello di Coleman-Mandula \cite{Coleman:ad}
ed alcune sue generalizzazioni \cite{Haag:1974qh}, come vedremo
meglio in seguito. Fu la sorprendente scoperta, nel 1976, della
supergravit\`{a} \cite{Freedman:1976xh,Deser:1976eh}, una teoria
interagente di un campo di massa nulla e spin 3/2 , il gravitino,
col campo gravitazionale di spin 2, a risvegliare l'interesse per
la costruzione di interazioni consistenti fra campi di spin
arbitrario. Come nelle altre teorie di gauge, nella
supergravit\`{a} la forma delle interazioni \`{e} fissata dal
principio di gauge, ovvero dalla richiesta di invarianza sotto
diffeomorfismi e trasformazioni di supersimmetria locali.

Nuove motivazioni vennero proprio dalle teorie di
supergravit\`{a}: la limitazione $ s\leq2$ sullo spin delle
particelle nei supermultipletti di massa nulla delle
supergravit\`{a} estese induce infatti la restrizione $ {\cal
N}\leq 8 $ sul numero delle cariche di supersimmetria, ovvero sul
numero dei gravitini, un limite sui tipi di possibili candidati
per teorie di grande unificazione. Inoltre, non era escluso che
l'inclusione di campi di HS potesse migliorare il comportamento
quantistico della teoria.

\vspace{0.5cm}

A partire dai lavori di Scherk e Schwarz \cite{Scherk:1974ca,
Scherk:jh} e Yoneya \cite{Yoneya:1973ca, Yoneya:jg} negli anni
'70, la Teoria delle Superstringhe \`{e} emersa quale miglior
candidato per una teoria unificata delle interazioni fondamentali,
il solo attualmente in grado di fondere, almeno potenzialmente, la
gravit\`{a} con il Modello Standard. In questa teoria l'oggetto
fondamentale non \`{e} una particella puntiforme, ma una stringa
unidimensionale chiusa o aperta di lunghezza tipica estremamente
piccola ($ l_{s}\sim10^{-33} cm $), i cui vari stati vibrazionali
si manifestano come un'infinit\`{a} di particelle. Le
propriet\`{a} fondamentali di queste ultime, massa e spin, sono
legate tra loro, alla frequenza dei modi vibrazionali e alla
tensione della stringa da relazioni ben precise. In particolare,
in unit\`{a} $\hbar=c=1$,
\begin{equation}
l_{s}=\sqrt{2\alpha'}=\frac{1}{\sqrt{\pi T}}\ ,
\end{equation}
dove $\alpha'$ \`{e} una costante detta \emph{pendenza di Regge} e
$T$ la tensione, e, qualitativamente
\begin{equation}
m^{2}\sim \frac{1}{\alpha'}(N-a)\ , \qquad m^{2}\sim s\ ,
\end{equation}
dove $m^{2}$ \`{e} la massa degli stati vibrazionali di stringa,
classificati dall'operatore numero $N$, ed $s$ il loro spin. La
Teoria delle Stringhe dunque descrive naturalmente una torre di
infiniti stati eccitati \emph{massivi}, con massa e spin via via
crescenti. La tensione di stringa viene scelta dell'ordine della
massa di Planck, $ M_{Pl}\sim10^{19} GeV $, sicch\'{e} alle
energie ordinarie soltanto i modi a frequenza pi\`{u} bassa
possono essere eccitati, quelli corrispondenti a particelle di
massa nulla di spin $ s\leq2 $ \footnote {Pi\`{u} precisamente, la
stringa bosonica vive in $26$ dimensioni, e i suoi modi di massa
nulla contengono sempre un bosone di spin 2 nello spettro di massa
della stringa bosonica chiusa e un bosone vettore in quello della
stringa bosonica aperta. Questo mostra come la Teoria delle
Stringhe abbia le potenzialit\`{a} di unificare gravit\`{a} e
Modello Standard, includendone naturalmente i campi di gauge fra
le proprie eccitazioni, in quanto \`{e} possibile dimostrare che
teorie di sole stringhe aperte sono inconsistenti, e
l'introduzione delle stringhe chiuse \`{e} inevitabile. Il settore
di massa nulla dello spettro delle superstringhe, che vivono in
$10$ dimensioni, include anche uno spinore di spin $3/2$,
sicch\'{e} le teorie corrispondenti ammettono le supergravit\`{a}
come limiti di bassa energia.}, mentre tutte le eccitazioni di HS
hanno masse enormemente elevate e non possono essere osservate
alle usuali basse energie. Tuttavia, esse sono essenziali per la
consistenza della teoria.

Questa situazione cambia per\`{o} drasticamente se ci mettiamo nel
limite opposto, quello di altissime energie, rispetto alle quali
anche le masse dei modi di HS risultano trascurabili, e ne risulta
quindi una teoria di campi di massa nulla e spin arbitrariamente
elevato, vale a dire una teoria di gauge di HS. Le diverse teorie
di stringa, legate l'una all'altra da dualit\`{a}, sembrano
ammettere come comune origine una teoria pi\`{u} fondamentale,
detta M-teoria, della quale attualmente conosciamo solo una serie
di manifestazioni indirette. Le teorie di gauge di HS possono in
questo senso offrire un principio guida per esplorare la fisica di
questo regime ancora sconosciuto delle teorie di stringa. Secondo
tale punto di vista, queste ultime potrebbero essere ottenute
attraverso un meccanismo di rottura spontanea della simmetria a
partire da una fase pi\`{u} simmetrica, con infinite simmetrie di
gauge di spin arbitrario: i generatori delle simmetrie rotte a
bassa energia corrisponderebbero alle rappresentazioni di HS del
gruppo di Lorentz ed i campi di gauge ad essi legati apparirebbero
massivi, mentre la simmetria residua, una sottoalgebra finita
dell'algebra infinito-dimensionale di partenza, comprenderebbe le
invarianze di gauge ordinarie, alle quali corrispondono campi di
massa nulla e spin $ s\leq2 $.

\vspace{0.5cm}

Le teorie di gauge di HS hanno recentemente suscitato ulteriore
interesse nel contesto degli studi sulla cosiddetta corrispondenza
AdS/CFT \cite{Maldacena:1997re}. Si tratta di una congettura
riguardante l'equivalenza, a livello delle rispettive funzioni di
partizione, tra la teoria di superstringa di tipo IIB sullo
spazio-tempo $ AdS_{5}\times S^{5} $ (dove $AdS_{5}$ \`{e} lo
spazio-tempo di Anti-de Sitter pentadimensionale) e la teoria di
Yang-Mills supersimmetrica (SYM) $ {\cal N}=4 $ che vive sul bordo
quadridimensionale di $AdS_{5}$. Per poter esplorare il contenuto
di questa congettura, \`{e} tuttavia necessario identificare
opportuni limiti in cui essa diventi pi\`{u} trattabile (la
quantizzazione di una teoria di superstringa su una variet\`{a}
non piatta non \`{e} ancora compresa, se non in casi particolari).
Nel suo senso pi\`{u} debole, la dualit\`{a} riguarda il limite
classico della supergravit\`{a} di tipo IIB sullo spazio-tempo $
AdS_{5} $, legato al comportamento a bassa energia della teoria di
superstringa di tipo IIB, e il limite di forte accoppiamento della
$ {\cal N}=4 $ SYM sul bordo quadridimensionale di $ AdS_{5} $. Si
pu\`{o} infatti mostrare che l'espansione in potenze di $\alpha'$
dell'azione effettiva di stringa su $AdS$ corrisponde
all'espansione in potenze di $\lambda^{-1/2}$ della teoria di
bordo, dove $\lambda$ \`{e} la cosiddetta costante di 't Hooft
della SYM, proporzionale al quadrato della costante di
accoppiamento di gauge $g_{YM}$.

Tale corrispondenza (detta olografica, dal momento che riguarda
due teorie che vivono in spazi-tempo con diversi numeri di
dimensioni) si \`{e} gi\`{a} dimostrata feconda di risultati
interessanti, ma ben poco si sa invece sulla fisica in $AdS$ nel
regime opposto, in cui la SYM di bordo \`{e} debolmente accoppiata
o libera. Tuttavia si sa che in questo limite accadono due cose:
la massa degli stati di stringa ($ \sim\alpha'\,^{-1} $) di HS
diventa piccola, sicch\'{e} la supergravit\`{a} non \`{e} pi\`{u}
una buona approssimazione della teoria che vive nel volume
(\emph{bulk}) di $AdS$  e, d'altra parte, la teoria di YM di bordo
libera ammette un numero infinito di correnti conservate di spin
arbitrariamente alto. Ci\`{o} ha dunque condotto all'idea che la
SYM libera sul bordo di $ AdS_{5} $ sia, nello spirito della
corrispondenza, duale olografico di una teoria di gauge
interagente di HS nel \emph{bulk}, basata su una estensione
infinito-dimensionale della superalgebra di $AdS$
\cite{Sezgin:2002rt}. Un ulteriore progresso in questa direzione
\`{e} stato possibile notando che una costante di accoppiamento
non nulla della SYM d\`{a} effettivamente origine ad un meccanismo
di Higgs nel \emph{bulk} in cui le simmetrie di HS sono
spontaneamente rotte a quelle della supergravit\`{a} ordinaria,
ovvero alla suddetta superalgebra di $AdS$, di dimensione finita.

\vspace{0.5cm}

Alla luce di quanto detto appare dunque chiara l'importanza sempre
maggiore che le teorie di gauge di HS hanno assunto nel corso
degli anni, e come attualmente uno studio dettagliato delle loro
propriet\`{a} sembri promettere risultati estremamente
interessanti e la possibilit\`{a} di far luce sulla ancora
sconosciuta M-teoria. Come gi\`{a} accennato in precedenza,
tuttavia, costruire una teoria interagente invariante sotto
trasformazioni di gauge di HS \`{e} tutt'altro che banale, come
testimoniato dal fatto che, storicamente, le ricerche sulla teoria
libera hanno preceduto di molti anni i primi significativi
progressi nel costruire interazioni tra campi di HS consistenti
con le simmetrie suddette.

Una delle ragioni di questa difficolt\`{a} \`{e} contenuta nel
gi\`{a} citato teorema no-go di Coleman-Mandula. Esso afferma che
la matrice S unitaria di una teoria di campo non pu\`{o} essere
invariante sotto trasformazioni i cui generatori abbiano spin $
s\geq\frac{3}{2} $ (i quali si accoppiano, nel costruire una
derivata covariante, a campi di spin $ \frac{5}{2} $). Poich\`{e}
la matrice S contiene informazioni sulle interazioni tra i campi
della teoria, ci\`{o} implica che non \`{e} possibile costruire
interazioni con campi di spin arbitrario. Bisogna precisare che
tale teorema assume come ipotesi l'invarianza di Lorentz della
teoria in questione e prende in esame soltanto un numero finito di
campi, assunzioni comunque piuttosto naturali nel contesto delle
teorie di gauge ordinarie.

Tuttavia, rilassando la seconda ipotesi ed includendo
simultaneamente tutti gli spin, interazioni cubiche consistenti
tra campi di HS (ma non tra campi di HS ed il campo
gravitazionale) nello spazio piatto vennero effettivamente
costruite verso la met\`{a} degli anni '80
\cite{Bengtsson:1983pd,Bengtsson:1983pg,Berends:wp,Berends:1984rq}.
Ogni tentativo di includere la gravit\`{a} fu per\`{o} vano, per
una ragione semplice: per introdurre l'interazione con il campo
gravitazionale rispettando l'invarianza per trasformazioni
generali di coordinate, le derivate devono essere covariantizzate
includendo un termine di accoppiamento con la connessione di spin
2: $
\partial\longrightarrow D=\partial+\Gamma $. Ma questo rompe l'invarianza
sotto le trasformazioni di gauge di HS! La lagrangiana libera
covariantizzata contiene infatti termini del tipo $
\frac{1}{2}(D\varphi)^{2}-\frac{s}{2}(D\cdot\varphi)^{2} $, ove $
\varphi $ \`{e} il campo di spin $s$, e la legge di trasformazione
di quest'ultimo, anch'essa covariantizzata, \`{e}:
$\delta\varphi_{\mu_{1}\mu_{2}...\mu_{s}}=D_{(\mu_{1}}\epsilon_{\mu_{2}...\mu_{s})}$.
Si pu\`{o} quindi mostrare che la variazione dovuta ad una
trasformazione di gauge della lagrangiana in presenza della
gravit\`{a} si riduce ad un commutatore di derivate covarianti,
che sappiamo essere proporzionale (in assenza di torsione) al
tensore di Riemann, $ [D...,D...]=\Re... $, sicch\'{e}
\begin{equation}
\delta{\cal L} =\Re...(\epsilon...D\varphi...)\neq 0 \ .
\end{equation}
Tale termine \`{e} diverso da zero e, per $ s>2 $, contiene anche
la parte a traccia nulla del tensore di Riemann, il tensore di
Weyl \footnote {\`{E} interessante notare come il caso di spin
$3/2$ sia l'ultimo, a parte $s=2$, per il quale la variazione di
gauge della lagrangiana covariantizzata d\`{a} origine soltanto al
tensore di Ricci \cite{deWit:pe} e non all'intero tensore di
Riemann, sicch\'{e} l'equazione di Rarita-Schwinger per il
gravitino ammette un'interazione consistente col campo
gravitazionale, che soddisfa le equazioni di Einstein nel vuoto
$\Re_{\mu\nu}=0$, e d\`{a} luogo alla supergravit\`{a}.}, che non
\`{e} possibile compensare con una modifica della lagrangiana
stessa, poich\'{e}, almeno nello spazio piatto, non esiste alcun
termine che sotto una variazione della metrica dia il tensore di
Riemann (dal momento che questo porterebbe ad una equazione del
moto tipo $\Re^{\alpha}_{\mu\nu\rho}=0$, che implicherebbe
l'assenza di gravit\`{a}).

Queste difficolt\`{a} vennero superate da Fradkin e Vasiliev
\cite{Fradkin:ks} riconsiderando il problema dell'interazione
gravitazionale in un background non piatto, ma con curvatura
costante, vale a dire negli spazi-tempo di de Sitter ($dS$) e
Anti-de Sitter ($AdS$). Anzitutto, in questo modo il teorema di
Coleman-Mandula non poneva ostacoli, poich\'{e} ne erano evitate
le ipotesi: la simmetria della teoria era $AdS$ e non Lorentz (e
non esiste una matrice S su $AdS$). In secondo luogo, lavorare con
una costante cosmologica $ \Lambda\neq0 $ consentiva di modificare
la lagrangiana con termini cubici alto-derivativi come
\begin{equation}
{\cal
L}^{int}=\sum_{A,B}\alpha(A,B)\Lambda^{-\left[\frac{(A+B)}{2}\right]}D^{A}\varphi
D^{B}\varphi R \ ,
\end{equation}
una combinazione appropriata di derivate di ordine $A$ e $B$ del
campo di spin $s$ con certi coefficienti $\alpha(A,B)$, dove $A$ e
$B$ sono limitati dalla condizione $A+B\leq s$, ed $ R $
rappresenta le fluttuazioni del tensore di Riemann rispetto alla
curvatura $ R_{0}=\Lambda gg $ del background $AdS$, nell'ambito
di una espansione attorno ad $ R_{0} $. A differenza di
un'espansione attorno ad un background piatto, infatti, in questo
caso il tensore di Riemann non \`{e} piccolo, ma \`{e} $
\Re=R+\Lambda gg $, e l'espansione \`{e} in potenze di $ R $.
Ci\`{o} implica che, nella variazione di gauge di HS di $ {\cal
L}^{int} $, il commutatore di due derivate covarianti produce,
all'ordine pi\`{u} basso, $ \Lambda gg $. Questo, in ultima
analisi, porta a termini del tipo $ \delta {\cal L}^{int}\sim R\;
D\varphi\;\epsilon $ con coefficienti indipendenti da $\Lambda$,
che, con una scelta appropriata degli $\alpha(A,B)$, cancellano
esattamente la variazione della lagrangiana libera covariantizzata
del campo di spin $s$.

Risulta perci\`{o} chiaro che:
\begin{enumerate}
\item Come gi\`{a} evidenziato dall'analisi nello spazio-tempo
piatto \cite{Bengtsson:1983pd,Bengtsson:1983pg}, una teoria di
gauge di HS interagente richiede, per essere consistente,
l'introduzione simultanea di infiniti campi di HS di massa nulla.

\item Interazioni di HS consistenti richiedono derivate di ordine
pi\`{u} alto del secondo dei campi fondamentali, dinamici. Questa
propriet\`{a} di non localit\`{a} delle interazioni fra campi di
HS \`{e} quanto mai interessante, specialmente perch\'{e}, alla
luce della connessione con la Teoria delle Stringhe, pu\`{o}
essere un indizio del fatto che tali campi esistono come modi di
vibrazione di oggetti fondamentali estesi, unidimensionali
(stringhe) o $p$-dimensionali ($p$-brane)
\cite{Bergshoeff:1988jm}.

\item Se si vuole includere la gravit\`{a},la richiesta di
simmetrie di HS non rotte ha come conseguenza la non
analiticit\`{a} delle interazioni di HS nella costante cosmologica
$ \Lambda $, ovvero la necessit\`{a} che la costante cosmologica
sia diversa da zero e che il limite piatto non sia ben definito.
\end{enumerate}
Ovviamente le ultime due osservazioni sono strettamente collegate,
dato che termini alto-derivativi nell'azione sono possibili solo
se dimensionalmente riscalati con potenze negative della costante
cosmologica.

\vspace{0.5cm}

Importanti indicazioni dell'importanza di un background non piatto
sono state fornite anche da Flato e Fronsdal \cite{Flato:1978qz}
partendo da un'analisi delle propriet\`{a} delle rappresentazioni
del gruppo di simmetria di $AdS_{4}$, $SO(3,2)$: il prodotto
diretto di due rappresentazioni dette singletoni si decompone,
sotto l'azione di $SO(3,2)$, in una somma diretta di
rappresentazioni unitarie irriducibili corrispondenti a campi di
massa nulla e spin arbitrario, caratteristiche preservate dalla
loro restrizione al sottogruppo di Poincar\'{e}. Ma prendere il
limite piatto ($\Lambda\rightarrow 0$) della decomposizione del
prodotto tensoriale di due singletoni non \`{e} equivalente a
calcolare il prodotto tensoriale del limite piatto di due
singletoni, che produce infatti stati con autovalore nullo di
energia ed impulso, rappresentazioni banali del gruppo di
Poincar\'{e}.

\vspace{0.5cm}

La conclusione fondamentale quindi \`{e} che $ \Lambda $
dev'essere diversa da zero nella fase con simmetrie di HS non
rotte. Tuttavia, come gi\`{a} notato, ci si aspetta che tali
simmetrie siano rotte da qualche meccanismo di rottura sponatanea.
Tale meccanismo potrebbe allora ridefinire anche il valore della
costante cosmologica, attraverso i valori di vuoto che alcuni
campi tipo-Higgs acquisterebbero. La non analiticit\`{a} nella
costante cosmologica non impedisce quindi a priori di ottenere, a
partire dalla fase altamente simmetrica delle teorie di gauge di
HS, una fase in cui i campi di HS abbiano masse non nulle e
$\Lambda=0$ o piccola.

In una serie di lavori \cite{Fradkin:ks, Fradkin:1986qy,
Fradkin:ah}, Fradkin e Vasiliev hanno costruito interazioni
cubiche consistenti a livello di azione in $AdS$, e
successivamente Vasiliev \cite{Vasiliev:sa, Vasiliev:1989yr,
Vasiliev:en, Vasiliev:vu, Vasiliev:1990bu, Vasiliev:1992av} (per
articoli di rassegna cfr. \cite{Vasiliev:1995dn, Vasiliev:1999ba,
Vasiliev:2001ur}) ha realizzato una teoria consistente a tutti gli
ordini nelle interazioni a livello delle equazioni del moto,
utilizzando un formalismo basato su un'estensione della
formulazione del vielbein della gravit\`{a}, che include campi di
gauge di spin arbitrario in alcuni \emph{master fields} a valori
in un'algebra di HS ottenuta come un'appropriata estensione
infinito-dimensionale della (super)algebra di simmetria del
background $AdS$ \footnote {A grandi linee tale estensione
corrisponde, a parte alcune importanti sottigliezze che
esamineremo, alla cosiddetta \emph{enveloping algebra}
dell'algebra di simmetria di $AdS$, i cui generatori sono dati da
potenze arbitrarie dei generatori di quest'ultima.}. La simmetria
di HS determina in gran parte la forma delle interazioni tra i
campi di gauge di spin arbitrario e tra questi e i campi di
materia di spin $0$ e $1/2$, includendole in alcune eleganti
equazioni differenziali di primo ordine nei \emph{master fields},
secondo lo schema dei cosiddetti \emph{sistemi integrabili di
Cartan} (o \emph{free differential algebras}
\cite{D'Auria:1982pm}), mentre una formulazione \emph{off-shell}
\`{e} tuttora oggetto di studio. La simmetria locale sotto
l'algebra di HS richiede l'introduzione di infiniti campi
ausiliari, equivalenti di HS della connessione di Lorentz della
gravit\`{a} di Einstein, eliminabili in termini di derivate dei
campi fisici attraverso vincoli di torsione generalizzati.
L'azione gauge-invariante che d\`{a} luogo alla corretta dinamica
all'ordine cubico, la cui lagrangiana ha la forma $R\wedge R$,
dove $R$ \`{e} la curvatura di spin $s$, non contiene tali
vincoli, che devono quindi essere imposti dall'esterno, mentre
essi sono automaticamente inclusi nel sistema integrabile di
equazioni sopra citato, che in tal senso sembra costituire un
formalismo pi\`{u} generale per l'analisi di un sistema fisico
complicato come quello delle teorie di gauge di HS. Tuttavia,
sebbene le equazioni di Vasiliev contengano tutta l'informazione
sulla dinamica dei campi di massa nulla e spin arbitrario, la
proliferazione di campi ausiliari le rende estremamente difficili
da trattare, e particolarmente complicato \`{e}, per esempio,
estrarre i vertici di interazione tra soli campi fisici.

Per questo ed altri motivi si sta inoltre tentando di esplorare la
fisica dei campi di gauge di spin elevato studiando la Teoria
delle Stringhe nel limite di tensione nulla, in cui le eccitazioni
di spin $s>2$ hanno massa nulla, sia su un background piatto che
su $(A)dS$. Tale linea di ricerca ha gi\`{a} dato luogo a
risultati interessanti, come il fatto che le equazioni libere del
campo di stringa descrivono, nel limite di tensione nulla
$\alpha'\rightarrow\infty$, la propagazione di campi di spin
arbitrario attraverso le equazioni per ``tripletti'' di campi
(prendendo in esame i soli tensori totalmente simmetrici), che si
riducono infatti alle equazioni libere di Francia e Sagnotti (cfr.
\cite{Sagnotti:2003qa}, anche per generalizzazioni di questo
risultato al caso di tensori a simmetria mista e al background
$(A)dS$, oltre che per ulteriori referenze) piuttosto che alle
equazioni di Fronsdal.

La struttura di questo lavoro \`{e} la seguente. Nel capitolo 2
vengono introdotti gli spazi-tempo di de Sitter e di Anti-de
Sitter, importanti per le teorie di gauge di HS, trattando in
dettaglio alcune conseguenze della presenza di una costante
cosmologica non nulla. Il capitolo $3$ \`{e} invece dedicato
all'esposizione di una formulazione della teoria della gravit\`{a}
equivalente a quella di Einstein, che ha fornito il punto di
partenza per l'approccio di Vasiliev. Nel capitolo $4$ viene
esaminata la dinamica dei campi di gauge di spin arbitrario
partendo dalla teoria libera, esaminandone sia la formulazione
``tradizionale'' di Fronsdal che quella ``geometrica'' e
completamente covariante dovuta a Francia e Sagnotti, per poi
passare ad una rassegna dei risultati di Vasiliev che costituisce
l'oggetto dei capitoli $5$ e $6$. In particolare, tutte le
conclusioni raggiunte in questa Introduzione riemergono
nell'ambito della teoria non lineare di Vasiliev. Vedremo anche
come, da argomenti puramente algebrici ed indipendenti da quanto
detto sulla Teoria delle Stringhe, risulter\`{a} che teorie di
gauge di HS consistenti contengono necessariamente infiniti campi
di spin $ 0\leq s<\infty $: l'algebra di simmetria di HS \`{e}
quindi infinito-dimensionale, come il corrispondente insieme di
campi in mutua interazione. Completano la trattazione tre
appendici: la prima contiene l'enunciato e la dimostrazione del
teorema di Coleman e Mandula; nella seconda si richiama invece il
linguaggio degli spinori ed il formalismo a due componenti,
fondamentale per l'approccio di Vasiliev in $d=4$; la terza,
infine, \`{e} strettamente legata ai capitoli $5$ e $6$ e riporta
alcune formule utili ivi utilizzate.

Il contenuto dell'ultimo capitolo della tesi, legato a
propriet\`{a} delle correnti di spin arbitrario, viene omesso e
verr\`{a} incluso in una pubblicazione in preparazione con D.
Anselmi, D. Francia e A. Sagnotti.

\chapter{Spazi-tempo di (Anti-)de Sitter}

\section{Introduzione}
\label{2.1}

Gli spazi-tempo di de Sitter ($dS$) e Anti-de Sitter ($AdS$) sono
due esempi notevoli di spazi massimamente simmetrici, spazi dotati
del massimo numero possibile di isometrie della metrica, pari a $
\frac{d(d+1)}{2} $ per una metrica in $ d $ dimensioni.

Assegnata la segnatura della metrica, che prendiamo
minkowskiana, $ \eta_{\mu\nu}=diag(-+...+) $ , 
gli spazi massimamente simmetrici sono generalizzazioni  dei tre
spazi euclidei massimamente simmetrici a curvatura costante: il
piano (curvatura costante nulla), la sfera $ S^{d} $ (curvatura
costante positiva) e lo spazio iperbolico $ H^{d} $ (curvatura
costante negativa).

\`{E} conveniente descrivere uno spazio massimamente simmetrico in
$d$ dimensioni in coordinate cartesiane, immergendolo in uno
spazio piatto $(d+1)$-dimensionale attraverso la condizione
\begin{equation}\label{gensph}
k\eta_{\mu\nu}x^{\mu}x^{\nu}+z^{2}=L^{2}\qquad\mu,\nu=1,...,d\ ,
\end{equation}
un vincolo che restringe le variabili su una superficie non
euclidea, e
\begin{equation}\label{ambsp}
ds^{2}=\eta_{\mu\nu}dx^{\mu}dx^{\nu}+\frac{1}{k}dz^{2}\ ,
\end{equation}
la definizione della metrica dello spazio ambiente piatto a $ d+1
$ dimensioni.

Della costante k, detta costante di curvatura, conter\`{a}
evidentemente solo il segno, poich\'{e} ogni riscalamento per un
fattore positivo pu\`{o} esser riassorbito nelle variabili $
x^{\mu} $.

Ricavando z dalla (\ref{gensph}), differenziando e sostituendo per
$ dz^{2} $ in (\ref{ambsp}) si ottiene
\begin{equation}
ds^{2}=\eta_{\mu\nu}dx^{\mu}dx^{\nu}+k\frac{\eta_{\mu\alpha}\eta_{\nu\beta}x^{\alpha}x^{\beta}}{L^{2}-k\eta_{\mu\nu}x^{\mu}x^{\nu}}dx^{\mu}dx^{\nu}
\ ,
\end{equation}
e dunque la metrica di uno spazio massimamente simmetrico \`{e}
\begin{equation}\label{max.sym}
g_{\mu\nu}=\eta_{\mu\nu}+k\frac{\eta_{\mu\alpha}\eta_{\nu\beta}x^{\alpha}x^{\beta}}{L^{2}-k\eta_{\mu\nu}x^{\mu}x^{\nu}}
\ ,
\end{equation}
che ha per inversa
\begin{equation}
g^{\mu\nu}=\eta^{\mu\nu}-k\frac{x^{\mu}x^{\nu}}{L^{2}}\ .
\end{equation}
A questo punto \`{e} semplice calcolare i simboli di Christoffel,
\begin{equation}
\Gamma^{\alpha}_{\mu\nu}=\frac{1}{2}g^{\alpha\beta}(\partial_{\nu}g_{\beta\mu}+\partial_{\mu}g_{\beta\nu}-\partial_{\beta}g_{\mu\nu})=\frac{k}{L^{2}}x^{\alpha}g_{\mu\nu}
\ ,
\end{equation}
ed il tensore di Riemann,
\begin{equation}
R^{\alpha}_{\beta\mu\nu}=\partial_{\mu}\Gamma^{\alpha}_{\beta\nu}-\partial_{\nu}\Gamma^{\alpha}_{\beta\mu}+\Gamma^{\alpha}_{\rho\mu}\Gamma^{\rho}_{\beta\nu}-\Gamma^{\alpha}_{\rho\nu}\Gamma^{\rho}_{\beta\mu}=\frac{k}{L^{2}}(\delta^{\alpha}_{\mu}g_{\beta\nu}-\delta^{\alpha}_{\nu}g_{\beta\mu})
\ ,
\end{equation}
sicch\'{e}
\begin{equation}
R_{\alpha\beta\mu\nu}=\frac{k}{L^{2}}(g_{\alpha\mu}g_{\beta\nu}-g_{\alpha\nu}g_{\beta\mu})
\ .
\end{equation}
Contraendo il primo e il terzo indice del tensore di Riemann si
ottiene il tensore di Ricci,
\begin{equation}\label{ricci}
R_{\mu\nu}\equiv
R^{\alpha}_{\mu\alpha\nu}=\frac{k}{L^{2}}(d-1)g_{\mu\nu}\ .
\end{equation}
Contraendo ulteriormente i due indici del tensore di Ricci si
ottiene infine la curvatura scalare,
\begin{equation}\label{scal}
R\equiv R^{\mu}_{\mu}=\frac{k}{L^{2}}d(d-1)\ .
\end{equation}
Si vede quindi che, per uno spazio a curvatura costante, il
tensore di Riemann \`{e} completamente deteminato dalla curvatura
scalare $ R $, e
\begin{equation}
R_{\alpha\beta\mu\nu}=\frac{1}{d(d-1)}R(g_{\alpha\mu}g_{\beta\nu}-g_{\alpha\nu}g_{\beta\mu})
\ ,
\end{equation}
il che implica che la curvatura di Weyl per questi spazi sia
identicamente nulla:
\begin{eqnarray}\label{Weyl}
C_{\alpha\beta\mu\nu} & = &
R_{\alpha\beta\mu\nu}+\frac{1}{d-2}(g_{\alpha\nu}R_{\mu\beta}-g_{\alpha\mu}R_{\nu\beta}+g_{\beta\mu}R_{\nu\alpha}-g_{\beta\nu}R_{\mu\alpha}){}\nonumber\\
& & {}
+\frac{1}{(d-1)(d-2)}R(g_{\alpha\mu}g_{\beta\nu}-g_{\alpha\nu}g_{\beta\mu})=0
\ .
\end{eqnarray}
Inoltre, la curvatura scalare $R$, che abbiamo detto essere
costante, risulta $ \sim k $, e il segno di $ k $ individua
univocamente tre diversi tipi di spazio-tempo massimamente
simmetrico. Per $ k=0 $ riotteniamo lo spazio-tempo piatto di
Minkowski $ M_{d} $, a curvatura nulla. Il valore $ k=1 $
definisce lo spazio-tempo di de Sitter (dS), una generalizzazione
nel minkowskiano della sfera euclidea (curvatura costante
positiva). Infine,  $ k=-1 $ definisce lo spazio-tempo di Anti-de
Sitter(AdS), una generalizzazione nel minkowskiano dello spazio
iperbolico euclideo.

Tali spazi sono soluzioni delle equazioni di Einstein nel vuoto in
presenza di una costante cosmologica,
\begin{equation}\label{einstein}
R_{\mu\nu}-\frac{1}{2}g_{\mu\nu}R=-\Lambda g_{\mu\nu}\ ,
\end{equation}
ottenibili estremizzando rispetto alla metrica l'azione di
Einstein-Hilbert con termine cosmologico
\begin{equation}\label{e-h}
S=\frac{1}{16\pi G_{d}}\int d^{d}x\sqrt{-g}(R-2\Lambda)\ .
\end{equation}
Infatti, per le (\ref{ricci}) e (\ref{scal})
\begin{equation}
R_{\mu\nu}-\frac{1}{2}g_{\mu\nu}R=-k\frac{(d-1)(d-2)}{2L^{2}}g_{\mu\nu}
\ ,
\end{equation}
sicch\'{e}, per confronto con la (\ref{einstein}), si deduce che
deve essere
\begin{equation}
\Lambda=k\frac{(d-1)(d-2)}{2L^{2}}\ ,
\end{equation}
da cui risulta evidente la relazione tra $ \Lambda $ e $ k $: il
segno della costante cosmologica segue quello di k, ovvero della
curvatura, in $ d>2 $ \footnote {I casi $ d=1,2 $ sono banali dal
punto di vista della gravit\`{a}, poich\'{e} in $ d=1 $ non esiste
curvatura, mentre in $ d=2 $, sebbene sia possibile definire una
curvatura, l'azione di Einstein-Hilbert, che codifica la dinamica
del campo gravitazionale, definisce un invariante topologico, la
caratteristica di Eulero.}, sicch\'{e} allo spazio-tempo dS resta
associata una costante cosmologica positiva, e allo spazio AdS una
costante cosmologica negativa.

\vspace{0.5cm}

\`{E} interessante notare che questi spazi possono essere visti
come soluzioni delle equazioni di Einstein per un fluido perfetto
con densit\`{a} costante $\rho=\frac{\Lambda}{8\pi G_{d}}$ e
pressione costante $p=-\frac{\Lambda}{8\pi G_{d}}$. Infatti il
tensore di energia-impulso per un fluido perfetto ha la forma
$T_{\mu\nu}=(\rho+p)u_{\mu} u_{\nu}-p g_{\mu\nu}$ (ove $u_{\mu}$
rappresenta la quadrivelocit\`{a}, $\rho$ la densit\`{a} di
energia e $p$ la pressione) e compare nelle equazioni di Einstein
in presenza di materia come
\begin{equation}
R_{\mu\nu}-\frac{1}{2}g_{\mu\nu}R=-8\pi G_{d}T_{\mu\nu}\ .
\end{equation}
Per confronto con la (\ref{einstein}) si comprende l'analogia
suddetta. Essa suggerisce l'importanza che la costante cosmologica
assume nel determinare l'evoluzione dell'Universo.

\`{E} ben noto infatti che l'Universo, pensato come omogeneo ed
isotropo su grande scala (come indicato dalla legge di Hubble),
pu\`{o} essere descritto come uno spazio-tempo di
Friedmann-Robertson-Walker (FRW). Per un siffatto spazio-tempo
\`{e} possibile scegliere un sistema di coordinate in cui la
metrica appaia nella forma
\begin{equation}
ds^{2}=-dt^{2}+a^{2}(t/L)d\sigma^{2}\ ,
\end{equation}
dove $d\sigma^{2}$ \`{e} la metrica di uno spazio massimamente
simmetrico tridimensionale ed \`{e} indipendente dal tempo. La
metrica di FRW corrisponde ad una ipersuperficie spaziale a
curvatura costante che si espande o si contrae a seconda
dell'andamento nel tempo del fattore di scala cosmologico
$a^{2}(t/L)$. Tale curvatura costante, positiva, negativa o nulla
\`{e} parametrizzata dal segno della costante $k$ gi\`{a}
introdotta, essendo sempre possibile assorbire fattori di
normalizzazione in $a^{2}(t/L)$. La formula (\ref{max.sym})
fornisce la metrica di questa ipersuperficie spaziale,
\begin{equation}
d\sigma^{2}=d\vec{x}\cdot d\vec{x}+k\frac{(\vec{x}\cdot
d\vec{x})^{2}}{L^{2}-k\vec{x}\cdot\vec{x}}\ .
\end{equation}
In coordinate polari
\begin{equation}
d\sigma^{2}=\frac{dr^{2}}{1-k\frac{r^{2}}{L^{2}}}+r^{2}d\Omega^{2}
\ ,
\end{equation}
(ove $d\Omega^{2}=d\theta^{2}+\sin^{2}\theta d\phi^{2}$) ed in
termini della variabile adimensionale $r'\equiv \frac{r}{L}$
\begin{equation}
d\sigma^{2}=L^{2}\left\{\frac{dr'^{2}}{1-kr'^{2}}+r'^{2}d\Omega^{2}
\right\} \ .
\end{equation}
\`{E} conveniente passare infine alla forma trigonometrica,
\begin{equation}
d\sigma^{2}=L^{2}\{d\chi^{2}+f^{2}(\chi)d\Omega^{2}\} \ ,
\end{equation}
dove la variabile $\chi$ \`{e} definita in modo tale che
$\frac{dr'^{2}}{1-kr'^{2}}=d\chi^{2},\quad k=0,\pm 1$, ovvero
\begin{eqnarray}
f(\chi)= \left \{ \begin{array}{ll} \sin\chi & \textrm{se $k=+1$}\ ,\\
\chi & \textrm{se $k=0$}\ ,\\
\sinh\chi & \textrm{se $k=-1$} \ ,\end{array}\right.\
\end{eqnarray}
con $0\leq\chi<\infty$ per $k=0,-1$, mentre $0\leq\chi<2\pi$ per
$k=+1$.

La metrica di FRW in coordinate trigonometriche \`{e} dunque
\begin{equation}
ds^{2}=-dt^{2}+R^{2}(t/L)\{d\chi^{2}+f^{2}(\chi)d\Omega^{2}\}\ ,
\end{equation}
dove $R^{2}(t/L)\equiv L^{2}a^{2}(t/L)$. Le equazioni di Einstein
per uno spazio-tempo di FRW danno luogo a due equazioni
differenziali ordinarie per il fattore di scala $R^{2}(t/L)$ che
ne governa l'evoluzione,
\begin{equation}\label{friedmann}
\left(\frac{\dot{R}}{R}\right)^{2}+\frac{k}{R^{2}}=\frac{8\pi
G}{3}\rho+\frac{1}{3}\Lambda
\end{equation}
(ove stiamo specializzando a $d=4$ e abbiamo posto $G_{4}\equiv G
$), detta \emph{equazione di Friedmann}, e
\begin{equation}\label{accel}
\frac{\ddot{R}}{R}=-\frac{4\pi G}{3}(\rho+3p)+\frac{1}{3}\Lambda \
.
\end{equation}
\`{E} naturale assumere, per il fluido perfetto col quale
schematizziamo la distribuzione di materia nell'Universo su grande
scala, che $\rho$ sia positiva e $p$ non negativa. La fondamentale
conseguenza di queste equazioni \`{e} dunque che, se $\Lambda=0$,
$R(t/L)$ non pu\`{o} essere costante, ovvero che le equazioni di
Einstein non ammettono soluzioni statiche, ma descrivono universi
in espansione o contrazione.

Questo fu precisamente il motivo che indusse Einstein ad
introdurre il termine di costante cosmologica nelle sue equazioni,
poich\'{e} solo per un valore critico di $\Lambda$ esiste una
soluzione statica, che chiameremo nel seguito \emph{universo
statico di Einstein}, descritto dalla metrica
\begin{equation}\label{static}
d\bar{s}^{2}=-dt^{2}+d\chi^{2}+\sin^{2}\chi \,d\Omega^{2} \ ,
\end{equation}
Una discussione dettagliata delle diverse soluzioni delle
equazioni di Einstein ci porterebbe troppo lontano. Qui ci
limitiamo a far notare come l'introduzione di $\Lambda\neq 0$,
caratteristica degli spazi $dS$ ed $AdS$ che stiamo trattando,
abbia notevoli conseguenze sull'evoluzione cosmologica. Ad
esempio, \`{e} evidente da (\ref{accel}) che una $\Lambda>0$
agisce come una repulsione che contrasta l'attrazione
gravitazionale, rendendo possibile, tra l'altro, un'espansione
accelerata e l'assenza della singolarit\`{a} iniziale tipica dei
modelli FRW senza costante cosmologica.

$dS$ ed $AdS$ non sono che casi particolari di spazi FRW. Nel
vuoto, vale a dire con $\rho=p=0$, una costante cosmologica non
nulla comporta un andamento esponenziale del fattore di scala
(qualitativamente $R(t)\sim \exp [a\sqrt{\Lambda}t]$) che si
traduce evidentemente in un'espansione accelerata e infinita nel
caso $\Lambda>0$ e in una soluzione oscillante per $\Lambda<0$.

Esaminiamo ora $dS_{4}$ e $AdS_{4}$ in maggior dettaglio,
mettendone in evidenza alcune propriet\`{a} globali, dopo aver
introdotto alcuni concetti e strumenti necessari facendo
riferimento al caso pi\`{u} semplice dello spazio-tempo
minkowskiano.

\section{ Spazio-tempo piatto: propriet\`a globali}
\label{2.2}

\begin{figure}
\begin{center}
\includegraphics[width=10cm]{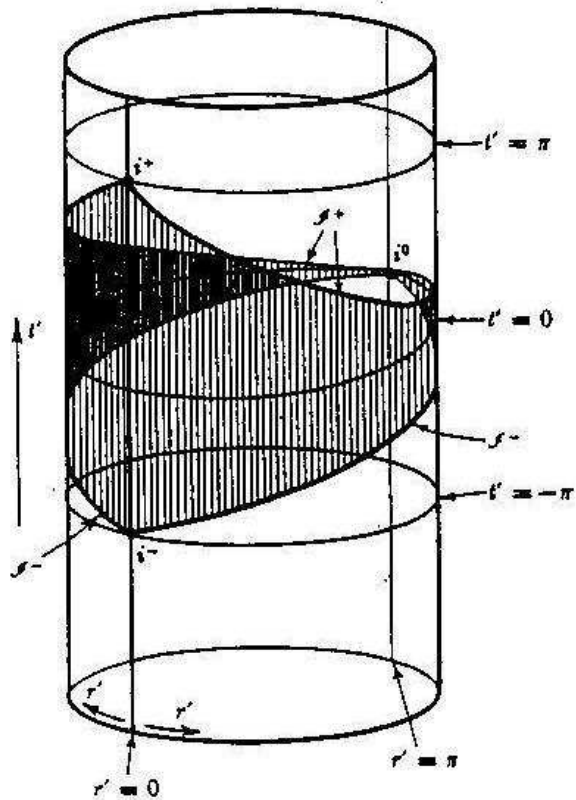}
\caption{L'universo statico di Einstein rappresentato da un
cilindro immerso in uno spazio piatto con una dimensione in
pi\`{u}. Le coordinate $\theta$ e $\phi$ sono state soppresse.
Ogni punto sulla superficie corrisponde a met\`{a} di una
$2$-sfera di area $4\pi\sin^{2}r'$. La regione ombreggiata \`{e}
conforme all'intero spazio di Minkowski, e sul suo bordo sono
mappati i punti all'infinito di quest'ultimo \cite{HE}.}
\label{fig1}
\end{center}
\end{figure}

Riprendiamo in considerazione l'universo statico di Einstein, che
pu\`{o} esser rappresentato come il cilindro
$x^{2}+y^{2}+z^{2}+w^{2}=1$ immerso nello spazio di Minkowski
pentadimensionale con metrica
$ds^{2}=-dt^{2}+dx^{2}+dy^{2}+dz^{2}+dw^{2}$. Sopprimendo due
dimensioni, lo mostriamo in fig. \ref{fig1}, dove a ciascun punto
sulla superficie cilindrica corrisponde met\`{a}\footnote {Questo
perch\'{e} in fig. \ref{fig1} ciascun valore di $\chi$ \`{e} rappresentato 
due volte.} di una $2$-sfera di area $4\pi\sin^{2}\chi$ ($\chi$
\`{e} sostituito da $r'$ in figura, come spiegato pi\`{u} avanti).

Un cilindro \`{e} una superficie conformemente piatta, ovvero
esiste una trasformazione conforme che porta la metrica piatta
minkowskiana $ds^{2}=-dt^{2}+dr^{2}+r^{2}d\Omega^{2}$ nella
(\ref{static}) a meno di un fattore conforme. Ricordiamo che, in
generale, una trasformazione conforme \`{e} un diffeomorfismo
$x\rightarrow x'(x)$ che lascia immutata la forma funzionale della
metrica a meno di un riscalamento locale (trasformazione di Weyl):
\begin{equation}
g'_{\mu\nu}(x')=\frac{\partial x^{\alpha}}{\partial
x'^{\mu}}(x')\frac{\partial x^{\beta}}{\partial
x'^{\nu}}(x')g_{\alpha\beta}(x(x'))=\lambda^{2}(x')g_{\mu\nu}(x')
\ .
\end{equation}
Il cambio di coordinate in questione \`{e}
\begin{equation}\label{map}
t'=\arctan (t+r)+\arctan (t-r)\ ,\quad r'=\arctan (t+r)-\arctan
(t-r)\ ,
\end{equation}
con $t'$ ed $r'$ che variano sul dominio
\begin{equation}\label{dom}
-\pi<t'+r'<\pi\ ,\quad -\pi<t'-r'<\pi\ ,\quad r'>0 \ .
\end{equation}
La metrica piatta diventa quindi
\begin{equation}
ds'^{2}=\frac{1}{4}\sec^{2}\left(\frac{1}{2}(t'+r')\right)\sec^{2}\left(\frac{1}{2}(t'-r')\right)d\bar{s}^{2}
\ ,
\end{equation}
dove $d\bar{s}^{2}$ \`{e} dato dalla (\ref{static}) con le
sostituzioni $t\rightarrow t'$ e $\chi\rightarrow r'$. Questo
significa che l'intero spazio-tempo di Minkowski \`{e} conforme
alla regione (\ref{dom}) dell'universo statico di Einstein (la
regione ombreggiata in fig. \ref{fig1}). Attraverso la
trasformazione di Weyl
\begin{equation}
d\tilde{s}^{2}=4\left[\sec^{2}\left(\frac{1}{2}(t'+r')\right)\sec^{2}\left(\frac{1}{2}(t'-r')\right)\right]^{-1}ds'^{2}
\end{equation}
\`{e} inoltre possibile estendere in modo massimale il sistema di
coordinate a coprire l'intero universo statico di Einstein, con
$-\infty<t'<+\infty$ e compattificando la parte spaziale con
l'aggiunta del punto all'infinito $r'=\pi$. La conclusione \`{e}
dunque che l'universo statico di Einstein \`{e} la
compattificazione conforme e massimamente estesa dello
spazio-tempo piatto di Minkowski.

Ricordiamo, a questo punto, alcune definizioni. Una metrica con
segnatura minkowskiana, $g_{\mu\nu}=diag(-++...+)$ separa i
vettori $V^{\mu}$ diversi dal vettore nullo in tre classi, a
seconda che $V^{\mu}g_{\mu\nu}V^{\nu}$ sia positivo, negativo o
nullo: nel primo caso il vettore si dir\`{a} \emph{di tipo
spazio}, nel secondo \emph{di tipo tempo} e nel terzo \emph{di
tipo luce}. Una curva viene dunque detta \emph{di tipo spazio},
\emph{di tipo tempo} o \emph{di tipo luce} se il suo versore
tangente $v^{\mu}$ \`{e} tale che
$v^{\mu}g_{\mu\nu}v^{\nu}=1,0,-1$, rispettivamente. La
classificazione delle ipersuperfici procede in modo analogo, ove
naturalmente il vettore $n^{\mu}$ che si prende in considerazione
\`{e} normale alla ipersuperficie stessa: se
$n^{\mu}g_{\mu\nu}n^{\nu}=1,0,-1$ la superficie si dir\`{a},
rispettivamente,  \emph{di tipo tempo}, \emph{di tipo luce} e
\emph{di tipo spazio}.

Si noti che una trasformazione conforme non altera la struttura
causale di uno spazio-tempo, poich\'{e} il riscalamento per un
fattore positivo $g'=\lambda^{2} g$ non altera gli angoli n\'{e}
il tipo (spazio, tempo o luce) di un vettore \footnote {\`{E}
chiaro infatti che $V^{\mu}g_{\mu\nu}V^{\nu}>0,=0,<0\Rightarrow
V^{\mu}g'_{\mu\nu}V^{\nu}>0,=0,<0$. }.

In Relativit\`{a} Generale si assume la validit\`{a} del postulato
di causalit\`{a} locale, secondo cui, dal momento che nessun
segnale pu\`{o} viaggiare con velocit\`{a} superiore a quella
della luce, i valori dei campi in un punto $p$ dello spazio-tempo
possono essere influenzati da quelli in un punto $q$ (e dalle loro
derivate, sino ad un opportuno ordine finito) soltanto se esiste
una curva di tipo tempo o luce che congiunge $p$ e $q$, ovvero
soltanto se $p$ non giace esternamente al cono di luce che ha
vertice in $q$. Data allora una ipersuperficie $\textsl{S}$ di
tipo spazio, il suo sviluppo di Cauchy nel futuro (passato)
$D^{+}(\textsl{S})$ ($D^{-}(\textsl{S})$) \`{e} definito come
l'insieme di tutti i punti $q$ della variet\`{a} spaziotemporale
$\textsl{M}$ tali che ogni curva di tipo tempo o luce che li
attraversa diretta verso il passato (futuro) interseca
$\textsl{S}$. In altri termini, $D^{+}(\textsl{S})$ costituisce
l'evoluzione causale di $\textsl{S}$. Se $D^{+}(\textsl{S})\cup
D^{-}(\textsl{S})=\textsl{M}$, ovvero se ogni curva di tipo tempo
o luce in $\textsl{M}$ interseca $\textsl{S}$, $\textsl{S}$ viene
detta \emph{superficie di Cauchy }. Ad esempio, nello spazio-tempo
di Minkowski le superfici di tipo spazio \{$x_{0}=cost$ \}
costituiscono una famiglia di superfici di Cauchy che copre
l'intero spazio-tempo.

Il confine della regione (\ref{dom}) rappresenta la struttura
conforme dell'infinito nello spazio di Minkowski. Esso \`{e}
costituito dalle superfici di tipo luce
$p\equiv\frac{1}{2}(t'+r')=\frac{1}{2}\pi$ ($I^{+}$) e
$q\equiv\frac{1}{2}(t'-r')=-\frac{1}{2}\pi$ ($I^{-}$), insieme con
i punti $p=q=\frac{1}{2}\pi$ ($i^{+}$),
$p=\frac{1}{2}\pi,q=-\frac{1}{2}\pi$ ($i^{0}$) e
$p=-\frac{1}{2}\pi,q=-\frac{1}{2}\pi$ ($i^{-}$).

Quello di Minkowski \`{e} inoltre un esempio di spazio-tempo
\emph{geodeticamente completo}, il che significa che qualsiasi
geodetica \`{e} completa, ovvero \textit{esiste per valori
arbitrariamente grandi del parametro affine}. Secondo la mappa
(\ref{map}) ogni geodetica di tipo tempo diretta verso il futuro
si avvicina ad $i^{+}$ ($i^{-}$) per valori arbitrariamente grandi
e positivi (negativi) del suo parametro affine. Si pu\`{o} dunque
pensare a questi due punti come all'immagine, tramite la suddetta
mappa, dell'infinito di tipo tempo nel futuro e nel passato. Le
geodetiche di tipo luce nel piano, invece, partono da $I^{-}$ e
terminano su $I^{+}$, superfici che rappresentano quindi
l'infinito di tipo luce sul cilindro. Infine, le geodetiche di
tipo spazio iniziano e finiscono in $i^{0}$, immagine sul cilindro
dell'infinito di tipo spazio sul piano.

\begin{figure}[!t]
\begin{center}
\includegraphics[width=17cm]{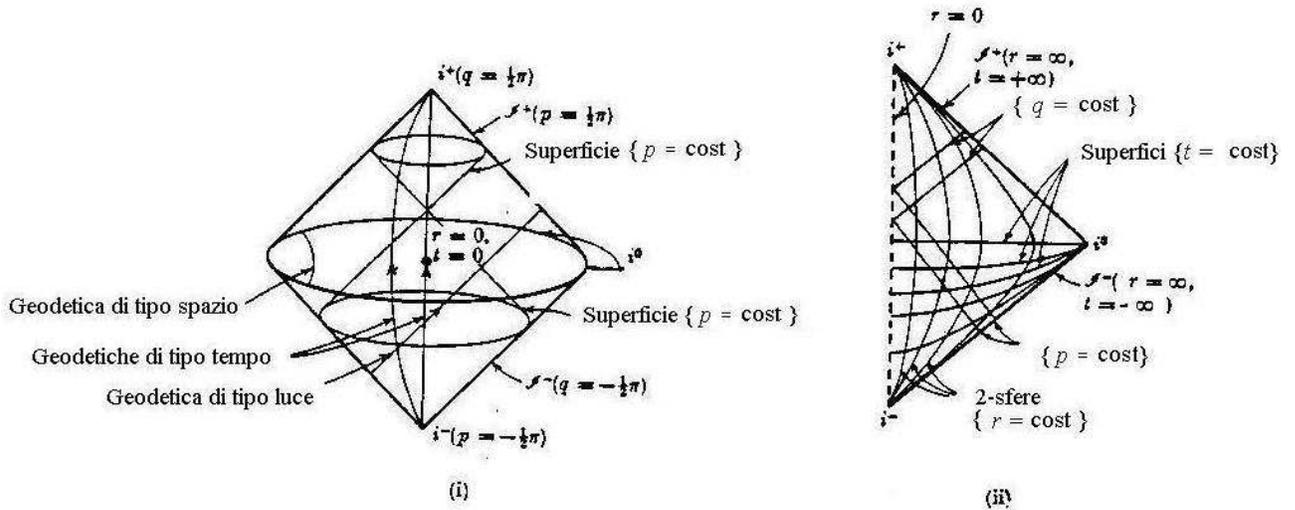}
\caption{ (i) La regione ombreggiata di fig. \ref{fig1}, con una
sola coordinata soppressa. (ii) Il diagramma di Penrose per lo
spazio-tempo di Minkowski; ogni punto rappresenta una $2$-sfera,
ad eccezione di $i^+$, $i^-$ ed $i^0$, ciascuno dei quali
rappresenta un singolo punto, e dei punti sulla linea $r=0$ (dove
le coordinate polari sono singolari) \cite{HE}.} \label{fig2}
\end{center}
\end{figure}

Aprendo il cilindro con un taglio lungo le generatrici $r'=0,\pi$,
si pu\`{o} inoltre stendere tale regione su un piano di coordinate
$(t',r')$ (fig. \ref{fig2} (ii)), ogni punto del quale corrisponde
ad una $2$-sfera, con l'eccezione di $i^{+}$, $i^{-}$ ed $i^{0}$,
e sul quale geodetiche di tipo luce corrispondono a rette a $\pm
45^{\circ}$. Tale diagramma viene detto \emph{diagramma di
Penrose}. Il fatto che una trasformazione conforme non alteri i
coni di luce di uno spazio-tempo fa s\`{i} che la sua
rappresentazione in termini di una regione dell'universo statico
di Einstein o il suo diagramma di Penrose contengano in effetti
tutta l'informazione sulla struttura causale, anche se le distanze
sono altamente distorte.

Possiamo a questo punto rivolgere la nostra attenzione agli spazi
$dS$ e $AdS$, sottoponendoli ad un'analisi analoga e mettendo in
luce le novit\`{a} introdotte dalla presenza di una costante
cosmologica .

\section{Spazio-tempo dS: propriet\`a globali}
\label{2.3}

\begin{figure}[!t]
\begin{center}
\includegraphics[width=13cm]{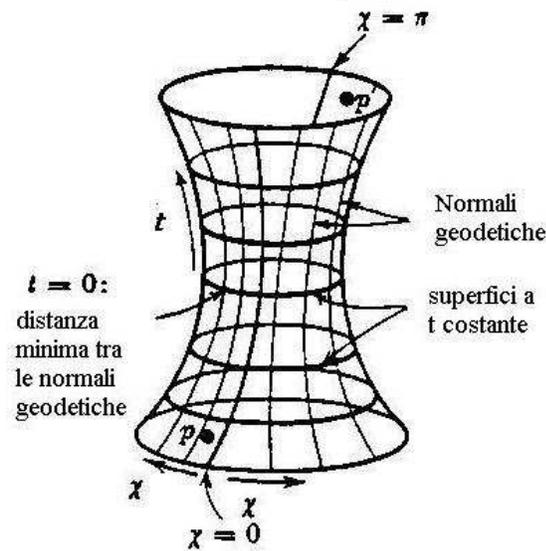}
\caption{ Lo spazio-tempo di de Sitter rappresentato da un
iperboloide immerso in uno spazio-tempo piatto a cinque dimensioni
(nella figura due dimensioni sono soppresse). Le sezioni
$\{t=cost\}$ sono superfici con curvatura $k=1$ \cite{HE}.}
\label{fig3}
\end{center}
\end{figure}

Lo spazio-tempo quadridimensionale di de Sitter, $dS_{4}$, pu\`{o}
esser facilmente visualizzato come l'iperboloide
\begin{equation}
-v^{2}+w^{2}+x^{2}+y^{2}+z^{2}=L^{2}
\end{equation}
immerso in uno spazio-tempo minkowskiano pentadimensionale con
metrica
\begin{equation}
-dv^{2}+dw^{2}+dx^{2}+dy^{2}+dz^{2}=ds^{2}
\end{equation}
(fig. \ref{fig3}). L'iperboloide pu\`{o} essere parametrizzato con
le coordinate $(t,\chi,\theta,\phi)$, definite dalle relazioni
\begin{eqnarray}\label{dScoord}
v & = & L\sinh(L^{-1}t) \ , \nonumber \\
w & = & L\cosh(L^{-1}t)\cos\chi  \ , \nonumber \\
x & = & L\cosh(L^{-1}t)\sin\chi\cos\theta  \ ,\\
y & = & L\cosh(L^{-1}t)\sin\chi\sin\theta\cos\phi \ , \nonumber  \\
z & = & L\cosh(L^{-1}t)\sin\chi\sin\theta\sin\phi \ . \nonumber
\end{eqnarray}
In questo sistema di coordinate la metrica assume la forma
\begin{equation}\label{dSmetric}
ds^{2}=-dt^{2}+L^{2}\cosh^{2}(L^{-1}t)\{d\chi^{2}+\sin^{2}\chi\,
d\Omega^{2}\}\ ,
\end{equation}
ed ha apparenti singolarit\`{a} per $\chi=0,\pi$ e $\theta=0,\pi$;
esse tuttavia non sono singolarit\`{a} vere dello spazio-tempo
(come si pu\`{o} controllare dal fatto che lo scalare di curvatura
$R$ non diverge in questi punti), ma sono legate alla scelta del
sistema di coordinate polari \footnote {Lo stesso accade nella
descrizione dello spazio-tempo piatto minkowskiano con coordinate
polari: la metrica assume la forma
$ds^{2}=-dt^{2}+dr^{2}+r^{2}d\Omega^{2}$ ed ha singolarit\`{a}
apparenti per $r=0$ e $\theta=0,\pi$. Per ottenere una descrizione
regolare in coordinate polari si deve restringere le variabili a
$0<r<\infty$, $0<\theta<\pi$, $0<\phi<2\pi$, utilizzando almeno
due parametrizzazioni locali di questo tipo per coprire
interamente la spazio di Minkowski.}. A parte queste
singolarit\`{a}, le coordinate $(t,\chi,\theta,\phi)$ con
$-\infty<t<\infty$, $0\leq\chi\leq\pi$, $0\leq\theta\leq\pi$ e
$0\leq\phi\leq 2\pi$ coprono tutto lo spazio.

\`{E} evidente che $dS_{4}$ ha la topologia $R^{1}\times S^{3}$.
Le ipersuperfici spaziali a $t=cost$ sono $3$-sfere di curvatura
costante $k=+1$ e la (\ref{dSmetric}) descrive tale spazio-tempo
come una $3$-sfera di raggio $L\cosh(L^{-1}t)$ variabile nel tempo
: esso si contrae da $t=-\infty$ fino ad un valore minimo, pari ad
$L$, raggiunto a $t=0$, per poi espandersi indefinitamente per
$t\rightarrow\infty$.

Tali $3$-sfere, sezioni spaziali a $t$ costante di $dS_{4}$, sono
superfici di Cauchy e la distanza relativa tra le loro normali
geodetiche si contrae monotonamente per $t$ crescente fino ad un
minimo, raggiunto in corrispondenza di $t=0$, e poi torna a
crescere all'infinito (fig. \ref{fig3}).

Come lo spazio-tempo di Minkowski, anche quello di de Sitter \`{e}
geodeticamente completo.

Per studiare la struttura globale di $dS$ \`{e} utile definire una
nuova coordinata temporale $t'$, detta \emph{tempo conforme}, per
mezzo del cambio di variabili
\begin{equation}\label{domdS}
t'=2\arctan(\exp L^{-1}t)-\frac{1}{2}\pi \ , \quad
-\frac{1}{2}\pi<t'<\frac{1}{2}\pi \ ,
\end{equation}
che trasforma i punti all'infinito in valori finiti di $t'$.
Inoltre, in termini delle variabili $(t',\chi,\theta,\phi)$, la
metrica prende la forma
\begin{equation}\label{dS/static}
ds^{2}=L^{2}\cosh^{2}(L^{-1}t')\,d\bar{s}^{2}\ ,
\end{equation}
dove $d\bar{s}^{2}$ \`{e} dato dalla (\ref{static}).

\begin{figure}[!t]
\begin{center}
\includegraphics[width=11cm]{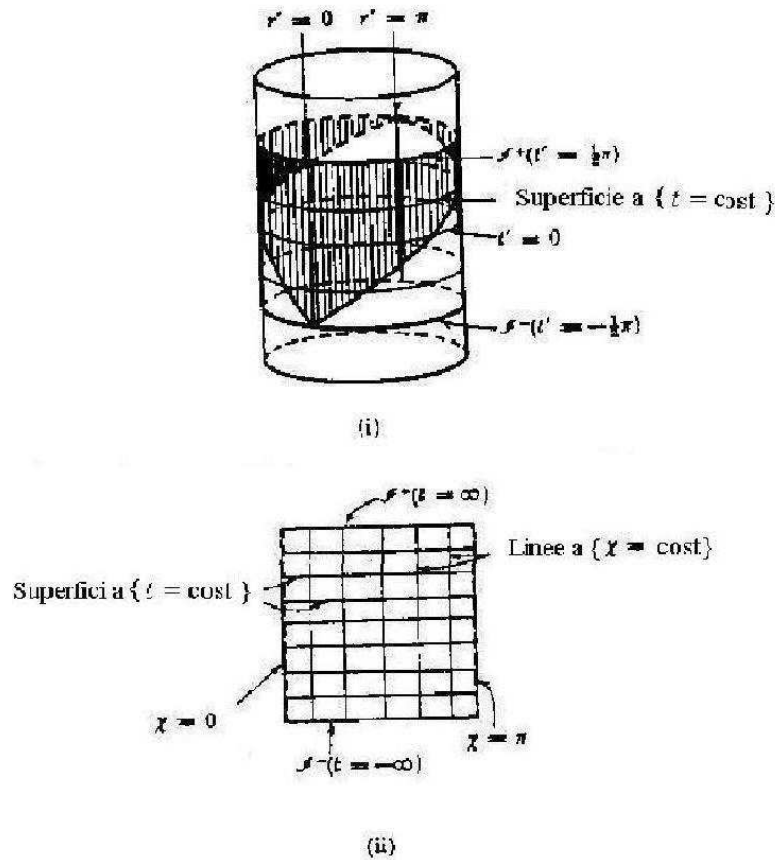}
\caption{ (i) Lo spazio-tempo di de Sitter \`{e} conforme alla
regione $-\frac{1}{2}\pi<t'<\frac{1}{2}\pi$ dell'universo statico
di Einstein. (ii) Il diagramma di Penrose dello spazio-tempo di de
Sitter \cite{HE}.} \label{fig4}
\end{center}
\end{figure}

Questo mostra che lo spazio-tempo di de Sitter \`{e} conforme alla
regione definita da (\ref{domdS}) dell'universo statico di
Einstein (fig. \ref{fig4} (i)) \footnote {Entrambi sono inoltre
conformemente piatti, come era evidente per $dS$ dal fatto che la
curvatura di Weyl, invariante sotto trasformazioni conformi, \`{e}
nulla (vedi (\ref{Weyl})).}. Il bordo della suddetta regione
rappresenta dunque la struttura conforme dell'infinito nello
spazio-tempo di de Sitter, e consiste delle superfici di tipo
spazio $t'=\frac{1}{2}\pi$ ($I^{+}$) e $t'=-\frac{1}{2}\pi$
($I^{-}$). Il corrispondente diagramma di Penrose \`{e} mostrato
in fig. \ref{fig4} (ii).

La principale differenza rispetto al caso minkowskiano sta nel
fatto che in $dS$ l'infinito di tipo tempo e luce, sia nel futuro
che nel passato, corrisponde sul cilindro alle superfici di tipo
spazio $I^{+}$ ed $I^{-}$, ove hanno origine e termine tutte le
geodetiche di tipo tempo o luce. Per lo spazio-tempo piatto si era
invece visto che l'infinito di tipo tempo e luce corrisponde alle
superfici $I^{+}$ ed $I^{-}$ di tipo luce insieme con i punti
$i^{+}$ ed $i^{-}$. Questo fatto \`{e} all'origine di un fenomeno
del tutto peculiare dello spazio-tempo di de Sitter: la presenza
di un \emph{orizzonte delle particelle} e di un \emph{orizzonte
degli eventi} per famiglie di osservatori geodetici.

\begin{figure}
\begin{center}
\includegraphics[width=15cm]{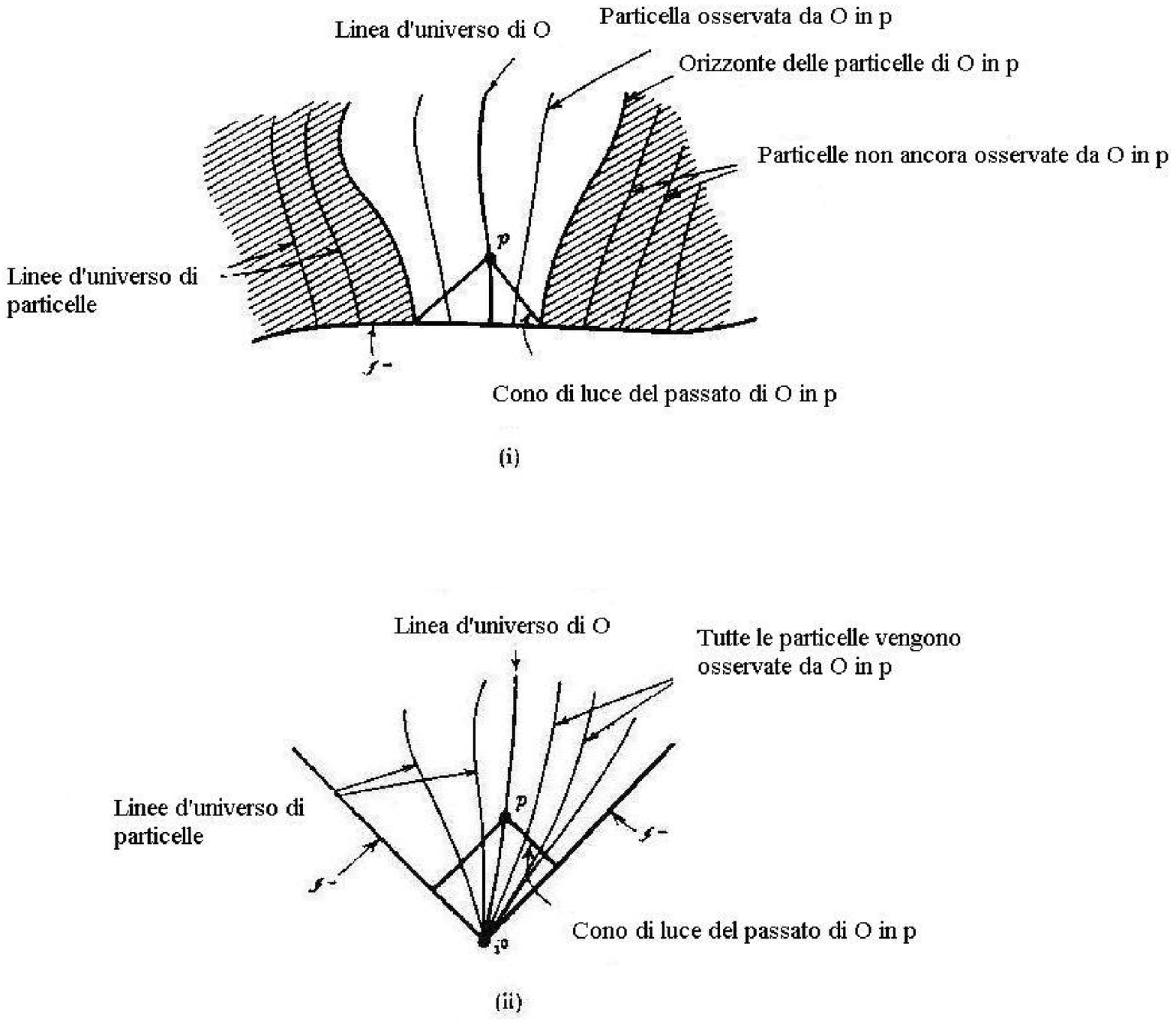}
\caption{(i) L'orizzonte delle particelle per l'osservatore
geodetico $O$ in $p$, come conseguenza del fatto che l'infinito
nella direzione temporale nel passato corrisponde ad una
superficie di tipo spazio. (ii) Assenza dell'orizzonte delle
particelle se tale superficie \`{e} di tipo luce \cite{HE}.}
\label{fig5}
\end{center}
\end{figure}

Data infatti in $dS$ una famiglia di particelle, le cui storie
sono rappresentate da geodetiche di tipo tempo, che hanno inizio
su $I^{-}$ e fine su $I^{+}$, prendiamo un evento $p$ sulla linea
d'universo di una particella $O$ di questa famiglia (sia $p$ un
dato istante di tempo proprio misurato lungo tale linea
d'universo). In accordo con il postulato di causalit\`{a} locale,
gli eventi nello spazio-tempo che $O$ pu\`{o} osservare, al tempo
$p$, sono tutti e soli quelli compresi nel cono di luce del
passato con vertice in $p$. Ovvero, solo le particelle le cui
linee d'universo intersecano tale cono di luce sono visibili ad
$O$. Ma possono evidentemente esistere particelle le cui linee
d'universo non lo intersecano: esse non sono quindi visibili ad
$O$ in $p$, ma possono naturalmente diventarlo in qualche istante
successivo $q$. Tuttavia, ci saranno ancora, in $q$, particelle
non visibili ad $O$. Definiamo allora \emph{orizzonte delle
particelle} per l'osservatore $O$ in $p$ la separazione tra
particelle visibili e non visibili ad $O$ in $p$, vale a dire la
geodetica che rappresenta la storia delle particelle che giacciono
ai limiti del cono di luce del passato di $O$ in $p$,
all'intersezione di questo con $I^{-}$ (vedi fig. \ref{fig5} (i)).
Per contrasto, notiamo che non esiste alcun orizzonte delle
particelle per un osservatore geodetico nello spazio-tempo di
Minkowski e tutte le particelle sono per lui visibili. Questo
\`{e} conseguenza del fatto che, in questo caso, $I^{-}$ \`{e} una
superficie di tipo luce e tutte le geodetiche di tipo tempo hanno
origine in un punto, $i^{-}$ (vedi fig. \ref{fig5} (ii)).

\begin{figure}
\begin{center}
\includegraphics[width=15cm]{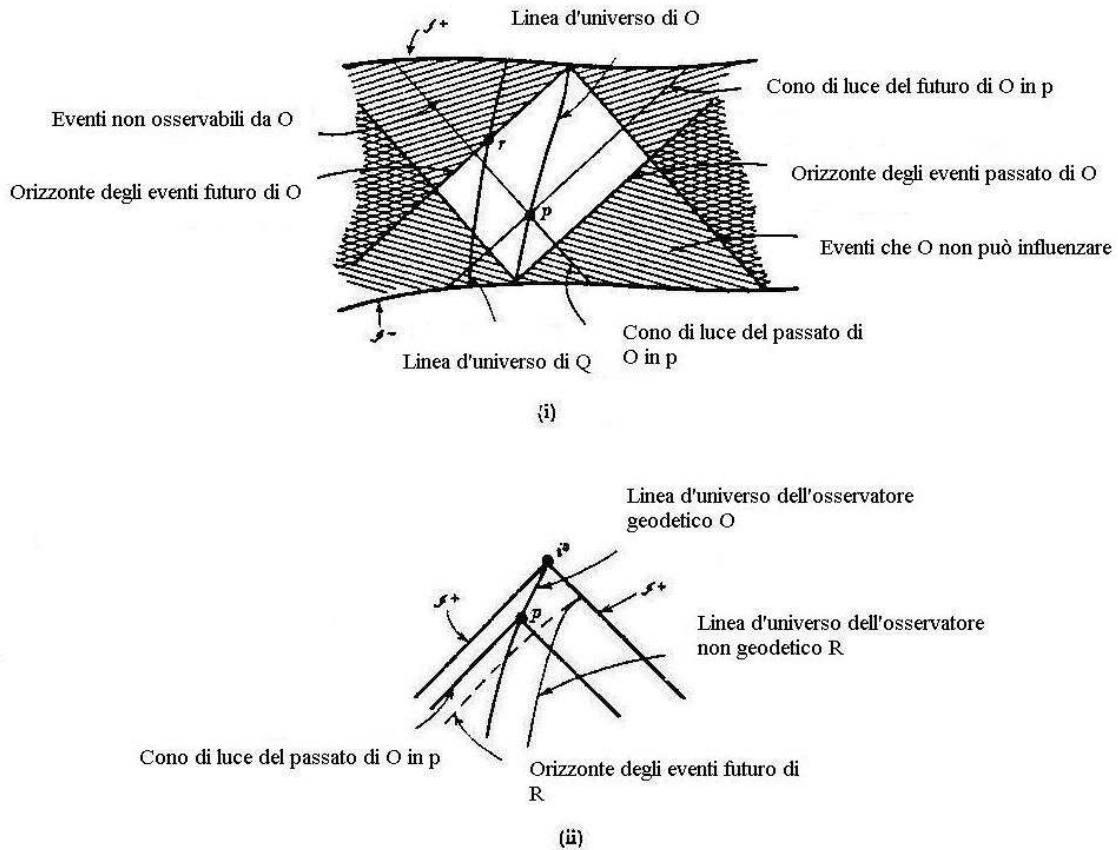}
\caption{(i) L'orizzonte degli eventi futuro per l'osservatore
geodetico O esiste in quanto l'infinito temporale futuro
corrisponde, in $dS$, ad una superficie di tipo spazio; cos\`{i}
anche l'orizzonte degli eventi passato \`{e} conseguenza del fatto
che $I^{-}$ \`{e} una superficie di tipo spazio. (ii) Assenza
dell'orizzonte degli eventi futuro per un osservatore geodetico
nello spazio-tempo piatto. Anche in questo caso, tuttavia, un
osservatore accelerato pu\`{o} avere un orizzonte degli eventi
futuro \cite{HE}.} \label{fig6}
\end{center}
\end{figure}

Inoltre, in $dS$ esiste un limite alla linea d'universo di $O$
sulla superficie $I^{+}$. Questo significa che il cono luce del
passato con vertice nel punto in cui tale linea d'universo
interseca $I^{+}$ costituisce una superficie di separazione tra
gli eventi che, per qualche $p<\infty$, saranno osservabili da
parte di $O$ e quelli che non lo saranno mai. Tale superficie
viene chiamata \emph{orizzonte degli eventi futuro}. Ancora,
notiamo che nello spazio-tempo di Minkowski la linea d'universo
dell'osservatore geodetico $O$ finisce in $i^{+}$ e la superficie
$I^{+}$ \`{e} di tipo luce: la conseguenza \`{e} che il cono di
luce con vertice in $i^{+}$ include l'intero spazio-tempo, e
nessun evento ne \`{e} al di fuori. Ovvero, non esiste alcun
orizzonte degli eventi futuro, nello spaziotempo piatto. Questo
per\`{o} non \`{e} vero per un osservatore che si muova con
un'accelerazione uniforme: la sua linea d'universo allora non
\`{e} pi\`{u} una geodetica e pu\`{o}, ad esempio, terminare su
$I^{+}$ anzich\'{e} in $i^{+}$, dando luogo ad un orizzonte degli
eventi futuro anche nello spazio-tempo piatto (vedi fig.
\ref{fig6} (ii)).

Questa considerazione ci suggerisce la ragione fisica
dell'esistenza dell'orizzonte. Rappresentando $dS$ come una ben
determinata regione dell'universo statico di Einstein abbiamo
osservato che, almeno per quanto riguarda un osservatore
geodetico, l'orizzonte delle particelle e quello degli eventi
futuro hanno origine dal fatto che $I^{-}$ ed $I^{+}$ sono
superfici di tipo spazio. Tuttavia, abbiamo visto in precedenza
che una costante cosmologica positiva conduce ad un'espansione
esponenziale dello spazio-tempo. \`{E} chiaro quindi che
l'esistenza dell'orizzonte \`{e} dovuta al fatto che lo
spazio-tempo si espande con velocit\`{a} superiori a quella della
luce, sicch\'{e} esisteranno alcuni eventi, quelli all'esterno
dell'orizzonte, la cui luce non raggiunge mai l'osservatore. In
altri termini, una data regione di spazio-tempo si espande pi\`{u}
rapidamente di quanto non faccia il cono di luce di un
osservatore, le cui falde non potranno quindi mai contenere al
loro interno tutti i punti della suddetta regione: anche per tempi
arbitrariamente grandi, resteranno sempre eventi non causalmente
connessi all'osservatore.

Diamo un'ulteriore caratterizzazione dell'orizzonte degli eventi
futuro. Consideriamo la linea d'universo della particella $Q$ ed
il punto $r$ in cui essa interseca l'orizzonte degli eventi futuro
di $O$. Per definizione di quest'ultimo, $O$ vede l'evento $r$
solo dopo un tempo proprio infinito (cio\`{e} una volta giunto su
$I^{+}$), mentre solo un intervallo finito di tempo proprio
trascorre lungo la linea d'universo di $Q$ tra un qualsiasi altro
evento ed $r$. $O$ vede in un tempo infinito una parte finita
della storia di $Q$, ovvero la luce degli eventi lungo la linea
d'universo di Q gli giunge con un redshift che tende ad infinito
per eventi arbitrariamente vicini ad $r$.

Dal momento che in $dS$ anche $I^{-}$ \`{e} una superficie di tipo
spazio, si comprende con ragionamenti analoghi l'esistenza di un
\emph{orizzonte degli eventi passato} di $O$: esso consiste della
superficie del cono di luce del futuro con vertice nel punto in
cui la linea d'universo di $O$ interseca $I^{-}$. I punti interni
costituiscono l'insieme massimale di eventi che $O$ pu\`{o}
influenzare, secondo il postulato di causalit\`{a} locale.
L'interno della regione ottenuta dall'intersezione dell'orizzonte
degli eventi futuro con l'orizzonte degli eventi passato contiene
dunque l'insieme massimale degli eventi causalmente connessi ad
$O$ (vedi fig. \ref{fig6} (i)).

Nessuna di queste osservazioni dipende dalla dimensionalit\`{a}
$d+1$ dello spazio-tempo: tutte le propriet\`{a} di $dS_{4}$ sin
qui esposte si generalizzano in modo analogo al caso $dS_{d+1}$,
con topologia $R\times S^{d}$, avendo cura soltanto di sostituire,
nella metrica, all'elemento di angolo solido bidimensionale
$d\Omega^{2}$ quello $(d-1)$-dimensionale $d\Omega^{2}_{d-1}$. Le
(\ref{dScoord}) andranno inoltre conseguentemente modificate nella
parte angolare:
\begin{eqnarray}\label{dScoord2}
x_{0} & = & L\sinh(L^{-1}t) \ , \nonumber \\
x_{i} & = & L\cosh(L^{-1}t)\Omega_{i}\ ,\quad i=1,2,...,d+1 \ ,
\end{eqnarray}
con le variabili angolari $\Omega_{i}$ tali che
$\sum_{i}\Omega_{i}^{2}=1$.

\section{Spazio-tempo AdS: propriet\`a globali}
\label{2.4}

\begin{figure}[!t]
\begin{center}
\includegraphics[width=12cm]{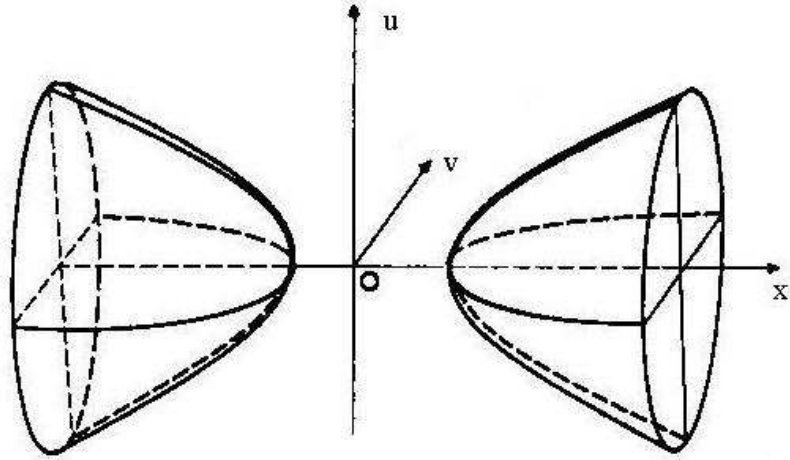}
\caption{Lo spazio-tempo di Anti-de Sitter rappresentato come un
iperboloide a due falde immerso in uno spazio-tempo piatto a
cinque dimensioni (due dimensioni spaziali sono soppresse nella
figura).} \label{fig7}
\end{center}
\end{figure}

Lo spazio-tempo quadridimensionale di Anti-de Sitter, $AdS_{4}$,
pu\`{o} convenientemente esser descritto come l'iperboloide
\begin{equation}\label{AdShyperb}
-u^{2}-v^{2}+x^{2}+y^{2}+z^{2}=1
\end{equation}
(dove prendiamo $L=1$ per semplicit\`{a}) immerso nello
spazio-tempo piatto pentadimensionale con metrica
\begin{equation}
-du^{2}-dv^{2}+dx^{2}+dy^{2}+dz^{2}=ds^{2} \ .
\end{equation}
Evidentemente la (\ref{AdShyperb}) descrive una superficie non
semplicemente connessa, dal momento che non esistono valori di $u$
e $v$ reali che la soddisfino per $(x^{2}+y^{2}+z^{2})^{1/2}<1$.
Sopprimendo due dimensioni spaziali, \`{e} possibile rappresentare
$AdS_{4}$ come l'iperboloide a due falde in fig. \ref{fig7}
\footnote {Ovviamente sopprimendo due dimensioni spaziali $AdS$ si
riduce alla superficie $-u^{2}-v^{2}+x^{2}=1$, l'iperboloide a due
falde disegnato, una superficie non connessa. }.

Trattandosi di uno spazio-tempo con ipersuperfici spaziali a
$t=cost$ a curvatura negativa $k=-1$ e con costante cosmologica
$\Lambda<0$, per quanto detto nella sezione $2.1$ \`{e} chiaro che
esso \`{e} descritto dalla metrica
\begin{equation}\label{AdSmetric}
ds^{2}=-dt^{2}+\cos^{2}t\{d\chi^{2}+\sinh^{2}\chi\, d\Omega^{2}\}
\ ,
\end{equation}
corrispondente ad un iperboloide tridimensionale (dunque una
ipersuperficie non compatta, a differenza di quanto accadeva per
$dS$) il cui raggio di curvatura varia nel tempo come $\cos t$ per
$-\frac{1}{2}\pi<t<\frac{1}{2}\pi$. Tuttavia, per
$t=\pm\frac{1}{2}\pi$ la metrica ha singolarit\`{a} soltanto
apparenti. Un sistema di coordinate globale per $AdS_{4}$ \`{e}
definito dalle relazioni
\begin{eqnarray}\label{newcoord}
u & = & \sinh r\cos t' \ , \nonumber  \\
v & = & \sinh r\sin t' \ , \nonumber  \\
x & = & \cosh r\cos\theta  \ , \\
y & = & \cosh r\sin\theta\cos\phi \ , \nonumber  \\
z & = & \cosh r\sin\theta\sin\phi  \ , \nonumber
\end{eqnarray}
con $r\geq 0$ e $0\leq t'<2\pi$, e nelle coordinate
$(t',r,\theta,\phi )$ la metrica si scrive
\begin{equation}\label{AdSmetric2}
ds^{2}=-\cosh^{2}r\, dt'^{2}+dr^{2}+\sinh^{2}r\, d\Omega^{2}\ .
\end{equation}
Nel limite $r\rightarrow 0$, la metrica (\ref{AdSmetric2}) tende a
$ds^{2}=-dt'^{2}+dr^{2}+r^{2}\, d\Omega^{2}$, sicch\'{e} questo
spazio-tempo ha la topologia $S^{1}\times R^{3}$ e presenta curve
chiuse di tipo tempo. Tuttavia, la causalit\`{a} \`{e} recuperata
se si apre il cerchio $S^{1}$ e si prende il suo ricoprimento
$R^{1}$, ovvero se si ammette $-\infty<t'<+\infty$, ottenendo
cos\`{i} il ricoprimento universale dello spazio-tempo di Anti-de
Sitter, con topologia $R^{4}$. D'ora in poi con $AdS_{4}$ si
intender\`{a} sempre questo ricoprimento.

\begin{figure}
\begin{center}
\includegraphics[width=10cm]{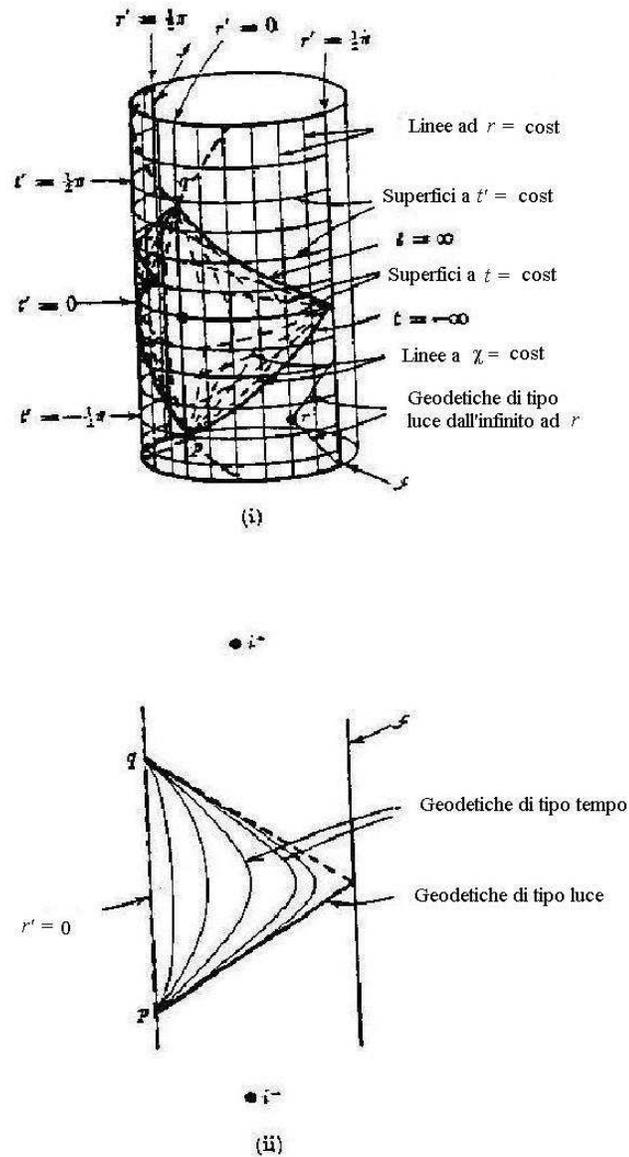}
\caption{(i) Il ricoprimento universale dello spazio-tempo di
Anti-de Sitter \`{e} conforme a met\`{a} dell'universo statico di
Einstein. Mentre le coordinate $(t',r,\theta,\phi)$ coprono
l'intero spazio, le coordinate $(t,\chi,\theta,\phi)$ coprono
soltanto la regione a forma di diamante mostrata. (ii) Il
diagramma di Penrose del ricoprimento universale dello spazio di
Anti-de Sitter. L'infinito \`{e} rappresentato dalla superficie di
tipo tempo $I$ e dai punti disgiunti $i^+$ e $i^-$ \cite{HE}.}
\label{fig8}
\end{center}
\end{figure}

Quindi in $AdS$ sia la coordinata $r$ che la coordinata $t'$
variano su intervalli infiniti. Il cambio di coordinate che porta
l'infinito nella direzione $r$ di $AdS$ in punti al finito \`{e}
dato da
\begin{equation}\label{domAdS}
r'=2\arctan(\exp r)-\frac{1}{2}\pi \ , \quad 0\leq
r'<\frac{1}{2}\pi \ ,
\end{equation}
e riduce la metrica a
\begin{equation}
ds^{2}=\cosh^{2}r\, d\bar{s}^{2} \ .
\end{equation}
A meno di un riscalamento di Weyl, abbiamo riscritto la metrica di
$AdS$ come la (\ref{static}): tuttavia, la variabile $r'$ varia
nell'intervallo $0\leq r'<\frac{1}{2}\pi$, e quindi $AdS$ \`{e}
conforme a \emph{met\`{a}} dell'universo statico di Einstein. In
fig. \ref{fig8} viene mostrata tale regione, coperta dalle
variabili $(t',r',\theta,\phi )$ e pari a met\`{a} del cilindro,
insieme con quella, a forma di diamante, coperta dalle coordinate
$(t,\chi,\theta,\phi )$; viene inoltre mostrato il corrispondente
diagramma di Penrose.

Si vede come $AdS$ risulti avere propriet\`{a} in qualche modo
opposte a quelle di $dS$: intuitivamente (e trascurando le
differenze introdotte dal prendere  per $AdS$ il ricoprimento
universale) possiamo dire che le caratteristiche di $dS$ nelle
direzioni $t'$ e $\chi\equiv r'$ diventano, rispettivamente,
quelle delle direzioni $r'$ e $t'$ di $AdS$. Per esempio, in
questo caso l'infinito di tipo luce e quello di tipo spazio
corrispondono ad una superficie di tipo tempo ($I$) sul cilindro:
essa corrisponde al bordo di $AdS$, che si raggiunge nel limite
$r\rightarrow\infty$, ovvero per $r'=\frac{1}{2}\pi$. Inoltre la
regione (\ref{domAdS}) sul cilindro \`{e} infinitamente sviluppata
nella direzione tempo. Rappresenteremo l'infinito di tipo tempo
nel diagramma di Penrose attraverso i due punti disgiunti $i^{+}$
ed $i^{-}$.

Notiamo inoltre che ogni ipersuperficie di $AdS_{4}$ a $t'=cost$
viene mappata in una semisfera  tridimensionale\footnote
{Coerentemente col fatto che le sezioni spaziali a tempo costante
sono superfici non compatte, con bordo, la trasformazione conforme
le manda in semisfere, anch'esse dotate di bordo. In $dS$ si
avevano invece sezioni a tempo costante compatte, $3$-sfere, che
rimanevano tali dopo la trasformazione.}: per ogni $t'$, quindi,
ciascun punto appartenente ad $I$ corrisponder\`{a} al bordo
bidimensionale $S^{2}$ di ciascuna semisfera, il suo equatore.
Nella sua evoluzione temporale, tale $2$-sfera descrive, in
effetti, il bordo di $AdS_{4}$, ovvero lo spazio-tempo
tridimensionale $R\times S^{2}$. Questo significa che il bordo di
$AdS_{4}$, o meglio della sua compattificazione conforme, coincide
con la compattificazione conforme massimamente estesa dello
spazio-tempo piatto di Minkowski tridimensionale. Questa
osservazione, estendibile a dimensione $d$ arbitraria, gioca un
ruolo cruciale nella corrispondenza $AdS_{d+1}/CFT_{d}$.

Le curve ${\chi,\theta,\phi=cost}$, geodetiche ortogonali alle
superfici $t=cost$, convergono nei punti $q$, nel futuro, e $p$,
nel passato, vertici del diamante. Questo rende ragione della
singolarit\`{a} apparente del sistema di coordinate
$(t,\chi,\theta,\phi )$. Esso copre soltanto la regione a forma di
diamante compresa tra la superficie $t=0$ e le superfici di tipo
luce su cui tali normali geodetiche diventano degeneri. Tutte le
geodetiche di tipo tempo, in $AdS$, sono fatte in questo modo:
partono divergendo da un punto per poi riconvergere in un punto
immagine, sia nel passato che nel futuro, dal quale divergono
ancora per tornare a congiungersi in un secondo punto immagine, e
cos\`{i} via, disegnando una successione di regioni a diamante,
come in fig. \ref{fig8}. Le geodetiche di tipo tempo, dunque, non
raggiungono mai il bordo di $AdS$ $I$, al contrario delle
geodetiche di tipo luce, che delimitano il futuro di $p$. Questo
comporta, ad esempio, che in $AdS$ particelle massive non possono
mai arrivare sino al bordo, mentre un raggio di luce impiega un
tempo finito per andare da $p$ al bordo e dal bordo a $q$ secondo
ogni osservatore geodetico. Il comportamento peculiare delle
geodetiche in $AdS$ ha come conseguenza anche l'esistenza di
regioni, nel futuro di $p$ (ovvero, i cui punti possono essere
congiunti a $p$ da una generica curva di tipo tempo), che non
possono essere raggiunte da $p$ attraverso una geodetica di tipo
tempo. I punti nel futuro di $p$ raggiungibili lungo geodetiche di
tipo tempo costituiscono l'interno della catena di regioni a forma
di diamante gi\`{a} citate.

L'infinito sviluppo lungo la direzione tempo, insieme con
l'esistenza di un bordo, mappato a $r'=\frac{1}{2}\pi$, fa s\`{i}
che non esistano superfici di Cauchy in $AdS$. In particolare,
sebbene si possano trovare famiglie di superfici di tipo spazio
che coprono tutto $AdS$ (come le superfici $\{t'=cost\}$),
ciascuna delle quali seziona completamente lo spazio-tempo,
esistono sempre geodetiche di tipo luce che arrivano al bordo
senza intersecarle. Questo fatto non ha riscontro in nessuno dei
casi precedentemente esaminati: sia nello spazio-tempo di
Minkowski che in quello di de Sitter, infatti, superfici di tipo
spazio che siano sezioni dell'intero spazio-tempo sono sempre
anche superfici di Cauchy. Ci\`{o} ha conseguenze importanti. Data
una superficie di Cauchy, si pu\`{o} predire lo stato
dell'universo in ogni istante nel passato o nel futuro, noti i
dati iniziali sulla superficie stessa, poich\'{e} ogni curva di
tipo tempo e di tipo luce la interseca. In $AdS$, invece, forniti
i dati iniziali su una superficie a $t'=cost$, si pu\`{o} al
massimo predire lo sviluppo di Cauchy della superficie stessa: ad
esempio, partendo sulla superficie $t'=0$, la regione massimale in
cui \`{e} possibile fare predizioni, in quanto causalmente
determinata dai dati iniziali su di essa, corrisponde al diamante
coperto dalle coordinate $(t,\chi,\theta,\phi)$. Al di fuori di
questa regione, sull'intero spazio-tempo di Anti-de Sitter, il
problema di Cauchy \`{e} mal definito, viziato dal fatto che
ulteriore informazione viaggia lungo geodetiche di tipo luce
provenienti da $t'=\pm\infty$. \`{E} necessario quindi dare anche
condizioni al contorno sul bordo $R\times S^{2}$ a
$r'=\frac{1}{2}\pi$.

Anche in questo caso tutte le propriet\`{a} descritte per
$AdS_{4}$ si estendono naturalmente ad $AdS_{d+1}$, con le
modifiche alla parte angolare di cui si \`{e} gi\`{a} parlato
nella sezione precedente.

\chapter{Formulazione geometrica della gravit\`{a}}

\section{Formulazione di MacDowell-Mansouri}
\label{3.1}

Nel seguito ci occuperemo soprattutto di AdS, lo spazio-tempo
pi\`{u} interessante dal punto di vista delle teorie di gauge di
spin elevato, poich\'{e} consente estensioni supersimmetriche
delle algebre di simmetria.

Come osservato in precedenza, uno spazio-tempo $AdS_{d}$
$d$-dimensionale pu\`{o} essere descritto semplicemente, in
termini delle coordinate cartesiane di immersione in uno
spazio-tempo minkowskiano $ M_{d+1} $ $(d+1)$-dimensionale, come
l'iperboloide
\begin{equation}
x_{\alpha}x^{\alpha}=-x_{0}^{2}+x_{1}^{2}+x_{2}^{2}+...+x_{d-1}^{2}-x_{d}^{2}=L^{2}
\ .
\end{equation}
La componente connessa del gruppo di trasformazioni lineari che
preserva la forma $x_{\alpha}x^{\alpha}$ \`{e} $ SO(d-1,2) $, la
cui algebra $ so(d-1,2) $ \`{e} dunque l'algebra di isometria
dello spazio-tempo $AdS_{d}$. Ad essa faremo dunque riferimento,
d'ora in poi, col nome di algebra $AdS_{d}$.

Specializzando al caso $d=4$, l'algebra di $ AdS_{4} $, $ so(3,2)
$, \`{e}
\begin{equation}
[J_{AB},J_{CD}]=i(\eta_{AC}J_{BD}+\eta_{BD}J_{AC}-\eta_{AD}J_{BC}-\eta_{BC}J_{AD})
\ ,
\end{equation}
dove
\begin{eqnarray*}
J_{AB}=-J_{BA}\ ,\qquad\eta_{AB}=diag(-,+,+,+,-)\ ,\qquad
A,B=0,1,...,4 \ .
\end{eqnarray*}
La sottoalgebra generata dai $ J_{ab} $ con $ a,b=0,1,2,3 $
pu\`{o} evidentemente essere identificata con l'algebra di
Lorentz, mentre i $ J_{a4} $ con i generatori delle traslazioni $
P_{a} $, a meno di un riscalamento con l'inverso del raggio di
curvatura di $ AdS_{4} $, necessario per dar loro le giuste
dimensioni: $ P_{a}=L^{-1}J_{a4} $. L'algebra di $ AdS_{4} $ si
riscrive dunque, in termini dei $ J_{ab}\equiv M_{ab} $ e $
J_{a4}\equiv P_{a} $, come
\begin{eqnarray}
\left[M_{ab},M_{cd}\right] & = &
i(\eta_{ac}M_{bd}+\eta_{bd}M_{ac}-\eta_{ad}M_{bc}-\eta_{bc}M_{ad})
\ , \nonumber \\
\left[ P_{a},M_{bc} \right]  & = &
i(\eta_{ac}P_{b}-\eta_{ab}P_{c}) \ ,
\\
\left[ P_{a},P_{b} \right] & = & -iL^{-2}M_{ab}\sim i\Lambda
M_{ab} \ . \nonumber
\end{eqnarray}
L'ultima equazione mostra come, su uno spazio a curvatura costante
non nulla, due traslazioni non commutino, ma il loro commutatore
sia equivalente ad una trasformazione di Lorentz. Allo stesso
tempo, tuttavia, risulta chiaro come sia possibile recuperare
l'algebra di Poincar\'{e} da $ so(3,2) $ attraverso una
contrazione di Inon\"{u}-Wigner, vale a dire nel limite piatto $
L\longrightarrow\infty $ che corrisponde a $
\Lambda\longrightarrow 0 $.

Tale procedimento consente di collegare un' algebra di Lie non
semisemplice, quale l'algebra di Poincar\'{e}, ad una famiglia a
un parametro (il raggio di $ AdS $, $L$) di algebre di Lie
semplici.

Teorie di gauge di spin ``basso'', come la teoria di Yang-Mills
(YM), possono essere costruite rendendo locali (\emph{gauging})
algebre di simmetria semplici.

Similmente, 
la gravit\`{a} einsteiniana pu\`{o} essere interpretata come una
teoria invariante sotto diffeomorfismi (a livello infinitesimo
traslazioni dipendenti dal punto) e trasformazioni di Lorentz
locali, 
ovvero come una teoria di gauge corrispondente ad un'opportuna
algebra di simmetria, di solito identificata con l'algebra di
Poincar\'{e} $ iso(3,1) $.

Tuttavia, per quanto detto sopra, tale algebra pu\`{o} essere
ottenuta a partire da un'algebra di Lie semplice, con il vantaggio
che il \emph{gauging} di quest'ultima rende manifeste, a livello
della costruzione della teoria e dell'azione, alcune somiglianze
con la teoria di YM che sono invece meno evidenti nella
formulazione einsteiniana.

Questa formulazione ``geometrica'' della gravit\`{a}, dovuta a
MacDowell e Mansouri \cite{MacDowell:1977jt}, \`{e} basata
sull'algebra di simmetria spazio-temporale $ so(3,2) $ e conduce
ad una azione, scritta in termini del prodotto esterno di
curvature, simile a quella di YM, che riproduce l'azione di
Einstein-Hilbert con termine cosmologico (\ref{e-h}).

Una generalizzazione appropriata di questa formulazione della
gravit\`{a} sar\`{a} la chiave dell'approccio ad una teoria di
gauge di campi di spin arbitrario. \`{E} dunque istruttivo
esaminarla in dettaglio.

Introduciamo la 1-forma di connessione a valori nell'algebra $
AdS_{4} $, $ A^{AB}=-A^{BA}=dx^{\mu}A_{\mu}^{AB} $, con $
\mu=0,1,2,3 $ indice di spazio-tempo (di base, nel linguaggio dei
fibrati) ed $ A,B=0,1,...,4 $, indice  dell'algebra di simmetria $
so(3,2) $ (di fibra). $A_{\mu}^{AB}$ \`{e} il campo di gauge
necessario per promuovere $ so(3,2) $ a simmetria locale, come
accade per YM. Possiamo espandere nei generatori, rispetto alla
base $(P_{a},M_{ab})$, come
\begin{equation}
A=A^{AB}J_{AB}=i(e^{a}P_{a}+\frac{1}{2}\omega^{ab}M_{ab}) \ ,
\end{equation}
dove la 1-forma $\omega^{ab}$ \`{e} la connessione di spin, campo
di gauge associato alla sottoalgabra di Lorentz $so(3,1)$, mentre
la 1-forma $e^{a}$ \`{e} il vielbein, associato alle traslazioni
$P_{a}$ su $AdS_{4}$ , vale a dire al quoziente $so(3,2)/so(3,1)$,
e legato alla connessione di $so(3,2)$ da $e^{a}=L A^{a4}$.

Con la connessione $A^{AB}$ possiamo costruire la 2-forma di
curvatura
\begin{equation}\label{covcurv}
\Re^{AB}=dA^{AB}+A^{AC}\wedge A_{C}\,^{B}\ ,
\end{equation}
dove $d=dx^{\mu}\frac{\partial}{\partial x^{\mu}}$ \`{e} la
derivata esterna e $\Re^{AB}=dx^{\mu}\wedge
dx^{\nu}\,\Re^{AB}_{\mu\nu}$. Osserviamone la decomposizione sui
generatori:
\begin{equation}\label{defRe}
\Re=\frac{i}{2}\ dx^{\mu}\wedge
dx^{\nu}\,[T^{a}_{\mu\nu}P_{a}+\frac{1}{2}(R^{ab}_{\mu\nu}+2L^{-2}e^{a}_{\mu}e^{b}_{\nu})M_{ab})]
\ ,
\end{equation}
dove
\begin{eqnarray}
T^{a}_{\mu\nu} & = & \partial_{\mu}
e^{a}_{\nu}+\omega_{\mu}\,^{a}\,_{c}\, e^{c}_{\nu} - (\mu\leftrightarrow\nu) \ , \\
R^{ab}_{\mu\nu} & = &
\partial_{\mu}\omega_{\nu}\,^{ab}+\omega_{\mu}\,^{a}\,_{c}\,\omega_{\nu}\,^{cb}-
(\mu\,\leftrightarrow\,\nu)\ .
\end{eqnarray}
$T^{a}_{\mu\nu}$ \`{e} il tensore di torsione, ed il vincolo
\begin{equation}\label{torsion}
T^{a}_{\mu\nu}=0
\end{equation}
consente di esprimere la connessione di spin (o connessione di
Lorentz) $\omega_{\mu}\,^{ab}$ in termini di derivate del vielbein
$e^{a}_{\mu}$. In virt\`{u} di questo vincolo si pu\`{o} dunque
distinguere tra un campo fondamentale, $e^{a}_{\mu}$, che contiene
i gradi di libert\`{a} fisici, ed un campo ausiliario,
$\omega_{\mu}\,^{ab}$, che non descrive gradi di libert\`{a}
indipendenti, poich\'{e} $\omega\sim e^{-1}\partial e$.

$R^{ab}_{\mu\nu}$ invece \`{e} l'usuale tensore di Riemann, che
compare in $\Re$ assieme al termine cosmologico $ L^{-2}ee\sim
\Lambda ee $. Quest'ultimo nasce dal fatto che il commutatore di
due traslazioni su uno spazio AdS \`{e} non nullo e scompare nel
limite piatto.

Come gi\`{a} accennato nell'introduzione, le equazioni:
\begin{eqnarray}\label{AdSvac}
T^{a}_{\mu\nu} & = & 0 \ , \\
\Re^{ab}_{\mu\nu} & = &
R^{ab}_{\mu\nu}+2L^{-2}e^{a}_{\mu}e^{b}_{\nu}=0 \ ,
\end{eqnarray}
descrivono uno spazio-tempo AdS di raggio $L^{-2}$, ovvero di
costante cosmologica $\Lambda$.

L'azione di MacDowell e Mansouri, scritta in termini delle
curvature in modo simile a quella delle YM e che riproduce
l'azione di Einstein-Hilbert col termine cosmologico, \`{e}
\begin{equation}\label{MM}
S_{MM} =
-\,\frac{1}{4\kappa^{2}\Lambda}\int_{M_{4}}\Re^{ab}\wedge\Re^{cd}\,\varepsilon_{abcd}=
-\,\frac{1}{4\kappa^{2}\Lambda}\int d^{4}x\,
\epsilon^{\mu\nu\rho\sigma}\varepsilon_{abcd}\,\Re_{\mu\nu}\,^{ab}\Re_{\rho\sigma}\,^{cd}
\ .
\end{equation}
Sostituendo nella (\ref{MM}) la (\ref{defRe}), tenendo conto della
(\ref{torsion}), e ricordando che $L^{-2}\sim \Lambda$ si ottiene
\begin{eqnarray}\label{svilMM}
S_{MM} & = & -\,\frac{1}{4\kappa^{2}\Lambda}\int d^{4}x\,
\epsilon^{\mu\nu\rho\sigma}\varepsilon_{abcd}(R_{\mu\nu}\,^{ab}R_{\rho\sigma}\,^{cd}+4L^{-2}\,e^{a}_{\mu}e^{b}_{\nu}R_{\rho\sigma}\,^{cd}
\nonumber \\
& & {} +4L^{-4}\,e^{a}_{\mu}e^{b}_{\nu}e^{c}_{\rho}e^{d}_{\sigma})
\ .
\end{eqnarray}
Esaminiamo uno ad uno i termini appena ricavati:
\begin{enumerate}
\item Il primo termine a secondo membro \`{e} proporzionale a
$\Lambda^{-1}$ e alto-derivativo (coinvolge cio\`{e} pi\`{u} di
due derivate del campo fondamentale, il vielbein). Tuttavia, \`{e}
proporzionale ad un termine topologico, la caratteristica di
Eulero della variet\`{a}
\begin{equation}
\chi=\frac{1}{32\pi^{2}}\int_{M_{4}}R^{ab}\wedge
R^{cd}\varepsilon_{abcd} \ ,
\end{equation}
e non contribuisce alle equazioni del moto.

\item Il secondo \`{e} un termine indipendente da $\Lambda$ e
produce la curvatura scalare ed il determinante del vielbein $e$.
Ricordando infatti che $
e\sim\varepsilon^{\mu\nu\rho\sigma}\varepsilon_{abcd}e^{a}_{\mu}e^{b}_{\nu}e^{c}_{\rho}e^{d}_{\sigma}$,
\`{e} semplice vedere che esso si riarrangia in
$e\,e^{\mu}_{a}\,e^{\nu}_{b}\,R^{ab}_{\mu\nu}$.

\item Infine compare un termine proporzionale ad $e\Lambda$,
ovvero un termine cosmologico.
\end{enumerate}

Come preannunciato, gli ultimi due termini ricostruiscono quindi
l'azione di Einstein-Hilbert gravitazionale in presenza di
costante cosmologica.

Il primo costituisce invece un termine ulteriore di interazione,
che qui non influisce poich\'{e} \`{e} una derivata totale.
Termini di questo tipo saranno tuttavia caratteristici delle
teorie di HS, e vale la pena di riassumerne le caratteristiche:
\begin{itemize}
\item \`{e} bilineare nelle curvature $R^{ab}$ di fluttuazione
rispetto al background $AdS$, sicch\'{e} nell'ottica di
un'espansione perturbativa attorno a quest'ultimo tali termini non
contribuiscono all'ordine lineare,

\item contiene pi\`{u} di due derivate del campo fondamentale, in
virt\`{u} del vincolo di torsione (\ref{torsion}),

\item ha un coefficiente proporzionale all'inverso della costante
cosmologica, e dunque il suo limite piatto non \`{e} ben definito.
Ci\`{o} significa che questo \`{e} un termine intrinsecamente
associato allo spazio-tempo curvo e che, specialmente in vista di
un'estensione di questo formalismo ai campi di HS, per i quali non
\`{e} un termine topologico, \`{e} necessario lavorare con
$\Lambda\neq 0$.

\end{itemize}

\vspace{0.5cm}

\section{Formulazione di Stelle-West}
\label{3.2}

\`{E} anche possibile costruire una versione $so(3,2)$-covariante
dell'azione di MacDowell-Mansouri, dovuta a Stelle e West
\cite{Stelle:aj}, al prezzo di introdurre un campo $V^{A}(x)$ che
non descrive gradi di libert\`{a} fisici, detto
\emph{compensatore}. Tale azione \`{e} simile, in forma, alla
(\ref{MM}), essendo quadratica nelle curvature $\Re^{AB}$, tensori
antisimmetrici di $so(3,2)$. Ha dunque bisogno del compensatore
per saturare l'ulteriore indice del tensore totalmente
antisimmetrico $\varepsilon_{ABCDE}$. Si introduce quindi il
vettore di tipo tempo di $so(3,2)$ $V^{A}$ (una $0$-forma dal
punto di vista spazio-temporale), con la normalizzazione
\begin{equation}\label{compnorm}
V^{A}V_{A}=-1 \ .
\end{equation}
L'algebra di Lorentz corrisponder\`{a} alla sottoalgebra di
stabilit\`{a} di $V^{A}$. Utilizzando il compensatore diamo
definizioni covarianti del vielbein \footnote {Naturalmente, il
vielbein $E_{\mu}^{A}$ deve essere una matrice non degenere per
dar luogo ad un tensore metrico non degenere, secondo la
$g_{\mu\nu}=E_{\mu}^{A}E_{\nu}^{B}\eta_{AB}$.} e della connessione
di Lorentz:
\begin{equation}\label{Vbcov}
L^{-1}E^{A}=DV^{A}\equiv dV^{A}+A^{AB}V_{B} \ ,
\end{equation}
\begin{equation}
\omega^{AB}=A^{AB}+L^{-1}(E^{A}V^{B}-E^{B}V^{A}) \ .
\end{equation}
Tenendo conto della (\ref{compnorm}), \`{e} facile vedere che
queste definizioni implicano
\begin{equation}
E^{A}V_{A}=0 \ ,
\end{equation}
\begin{equation}
\nabla V^{A}\equiv dV^{A}+\omega^{AB}V_{B}=0 \ .
\end{equation}
La $2$-forma di curvatura di $so(3,2)$, definita in
(\ref{covcurv}), ammette la decomposizione covariante
\begin{equation}
\Re^{AB}=R^{AB}-L^{-1}(T^{A}V^{B}-T^{B}V^{A})+2L^{-2}(E^{A}\wedge
E^{B}) \ ,
\end{equation}
dove abbiamo introdotto la $2$-forma di torsione $T^{A}$, definita
come
\begin{equation}\label{covtors}
T^{A}\equiv DE^{A}=L\,\Re^{AB}V_{B} \ .
\end{equation}
L'azione di Stelle-West per la gravit\`{a} \`{e} quindi
\begin{equation}\label{SW}
S_{SW}=-\,\frac{1}{4\kappa^{2}\Lambda}\int_{M_{4}}\varepsilon_{ABCDE}\,\Re^{AB}\wedge\Re^{CD}\,V^{E}
\ .
\end{equation}
Ricordando che
\begin{equation}
\delta\Re^{AB}=D\delta A^{AB} \quad, \quad V_{A}\delta V^{A}=0 \ ,
\end{equation}
e utilizzando l'identit\`{a}
\begin{equation}
\varepsilon_{ABCDE}=V^{A}V_{F}\,\varepsilon_{FBCDE}+...+V^{E}V_{F}\,\varepsilon_{ABCDF}
\ ,
\end{equation}
si pu\`{o} verificare che
\begin{eqnarray}
\delta S_{SW} & = &
-\,\frac{1}{4\kappa^{2}\Lambda}\int_{M_{4}}\varepsilon_{ABCDE}\,\Re^{AB}\wedge\delta
A^{CD}\wedge 2L^{-1}E^{E} \nonumber \\
& & {}
-\,\frac{1}{4\kappa^{2}\Lambda}\int_{M_{4}}\varepsilon_{ABCDE}\,\Re^{AB}\wedge
T^{C}\,L^{-1}\,4V^{D}\,\delta V^{E} \ .
\end{eqnarray}
Si noti che il contributo alla variazione dell'azione dovuto alla
variazione del compensatore contiene la torsione.

Come gi\`{a} sottolineato in precedenza, la condizione
\begin{equation}\label{curvnulla}
\Re^{AB}=0
\end{equation}
descrive lo spazio-tempo $AdS_{4}$. Consideriamo invece
$\Re^{AB}\neq 0$ ma piccolo, ovvero poniamoci nell'ambito di
un'espansione perturbativa attorno al background $AdS$. Tenendo
conto della (\ref{covtors}), possiamo allora concludere che,
almeno perturbativamente, esiste una variazione dei campi del tipo
\begin{equation}\label{furthsym}
\delta A^{AB}=\eta^{AB}(\Re,\epsilon)\ , \qquad \delta
V^{A}=\epsilon^{A} \ ,
\end{equation}
con $\epsilon^{A}V_{A}=0$ e dove $\eta^{AB}(\Re,\epsilon)$ \`{e}
una funzione bilineare in $\Re^{AB}$ ed $\epsilon^{A}$, sotto cui
l'azione \`{e} invariante. Questo significa che oltre alla
manifesta simmetria locale $so(3,2)$
\begin{equation}
\delta A^{AB}=D\epsilon^{AB}\ , \qquad \delta
V^{A}=\epsilon^{AB}V_{B} \ ,
\end{equation}
$S_{SW}$ possiede un'ulteriore simmetria di gauge sotto le
(\ref{furthsym}), che pu\`{o} essere utilizzata per fissare un
valore di $V^{A}$ che soddisfi la (\ref{compnorm}). Risulta
inoltre evidente dalla sua trasformazione (\ref{furthsym}),
proporzionale al parametro, che il compensatore \`{e} pura gauge,
un campo di Stueckelberg, e non trasporta gradi di libert\`{a}
fisici, come anticipato. Ovviamente, una volta fissato un valore
per $V^{A}$, la simmetria residua della teoria si riduce alle
combinazioni di trasformazioni di $so(3,2)$ e del tipo
(\ref{furthsym}) che non alterano tale scelta particolare, vale a
dire alle trasformazioni caratterizzate da parametri tali che
\begin{equation}
0=\delta V^{A}=\epsilon^{A}(x)+\epsilon^{AB}(x)V_{B} \ .
\end{equation}
L'origine della ulteriore simmetria (\ref{furthsym}) dell'azione
di Stelle e West pu\`{o} essere facilmente compresa tenendo conto
del fatto che quest'ultima \`{e} esplicitamente invariante, oltre
che sotto trasformazioni di gauge di $so(3,2)$, anche sotto
diffeomorfismi, essendo costruita in termini di forme
differenziali. A livello infinitesimo, l'azione dei diffeomorfismi
sulla $1$-forma di connessione, indotta dalla trasformazione di
coordinate $x^{\mu}\rightarrow x^{\mu}+\xi^{\mu}(x)$, \`{e} data
da
\begin{equation}\label{preFDA}
\delta
A_{\mu}^{AB}=-\xi^{\nu}\partial_{\nu}A_{\mu}^{AB}-(\partial_{\mu}\xi^{\nu})A_{\mu}^{AB}=-\xi^{\nu}\Re_{\nu\mu}^{AB}-D_{\mu}\epsilon^{AB}
\ ,
\end{equation}
mentre quella sulla $0$-forma $V^{A}$ \`{e}
\begin{equation}
\delta
V^{A}=-\xi^{\nu}\partial_{\nu}V^{A}=-\xi^{\nu}L^{-1}E^{A}_{\nu}+\epsilon^{AB}V_{B}
\ ,
\end{equation}
dove $\epsilon^{AB}=\xi^{\nu}A_{\nu}^{AB}$. Per confronto con la
(\ref{furthsym}) \`{e} ora chiaro che la simmetria di gauge
aggiuntiva, di parametro
$\epsilon^{A}(x)=-\xi^{\nu}L^{-1}E^{A}_{\nu}(x)$, corrisponde ad
una combinazione di diffeomorfismi e trasformazioni di $so(3,2)$
locali.

La scelta di gauge $V^{a}=0$ rompe $so(3,2)$ alla sottoalgebra di
stabilit\`{a} $so(3,1)$, ed in questa gauge l'azione di Stelle e
West diventa quella di MacDowell e Mansouri. C'\`{e} quindi una
completa equivalenza tra le due formulazioni. La prima rende
manifesta l'intera simmetria $so(3,2)$ per mezzo del compensatore,
un campo puramente ausiliario; un'opportuna scelta di gauge per
quest'ultimo consente poi di recuperare la seconda come fase
spontaneamente rotta dell'azione $S_{SW}$.

Nella formulazione di Stelle e West della gravit\`{a} compare
inoltre una caratteristica che ritroveremo nella trattazione delle
teorie di HS. Dalla (\ref{preFDA}) possiamo infatti osservare che,
per la soluzione dell'equazione di curvatura nulla
(\ref{curvnulla}), i diffeomorfismi si riducono ad una
trasformazione di gauge: in altri termini, l'equazione di
curvatura nulla incorpora i diffeomorfismi nel gruppo di gauge. A
posteriori, questo pu\`{o} esser visto come il motivo per cui (o,
se si preferisce, costituisce la miglior prova del fatto che) tale
formulazione geometrica della gravit\`{a} porta la simmetria
locale sotto riparametrizzazioni della gravit\`{a} einsteiniana
sullo stesso piano \footnote {Nel senso inteso all'inizio di
questo capitolo, cio\`{e} per quanto riguarda il tipo di algebra
di simmetria coinvolta, semplice, e la costruzione della teoria
attraverso la promozione dell'algebra stessa a simmetria locale.}
delle simmetrie locali interne, di tipo YM, attraverso il
\emph{gauging} dell'algebra di simmetria globale di un opportuno
background gravitazionale, $AdS$.

Quanto detto sulla formulazione di Stelle e West pu\`{o} essere
esteso a dimensione $d$ arbitraria, per l'algebra $so(d-1,2)$, con
le necessarie modifiche all'azione, introducendo un numero
opportuno di potenze di $E^{A}$.

Nel capitolo seguente esporremo la costruzione delle teorie di
gauge di HS come generalizzazioni dell'approccio geometrico alla
gravit\`{a}, sottolineando le caratteristiche da esso ereditate.


\chapter{Higher spins: teoria libera}

\section{Equazioni di Fronsdal nello spazio-tempo piatto}
\label{Fr}

In questo capitolo iniziamo lo studio delle propriet\`{a} dei
campi di gauge di spin arbitrario prendendo le mosse dalla teoria
libera, la cui struttura costituisce un'importante
generalizzazione di quella delle teorie di gauge di spin ``basso''
intero (la teoria di Maxwell dell'elettromagnetismo e la teoria di
Einstein della gravit\`{a} linearizzata) e semi-intero (la teoria
libera del campo di gauge di spin $3/2$, dovuta a Rarita e
Schwinger). Le equazioni di moto libere per campi di gauge di spin
arbitrario rappresentati da tensori e spinor-tensori totalmente
simmetrici furono ottenute per la prima volta da Fronsdal nel
1978, studiando il limite di massa nulla delle corrispondenti
equazioni massive formulate da Fierz e Pauli nel 1939 e
successivamente derivate da un principio di azione da Singh e
Hagen nel 1974. In uno spazio-tempo a quattro dimensioni i tensori
totalmente simmetrici esauriscono, a meno di dualit\`{a}, tutte le
rappresentazioni irriducibili di spin arbitrario del gruppo di
Lorentz, e nel seguito ci ridurremo quindi a questo caso, sebbene
la trattazione rimanga valida in generale, con la sola differenza
che in $d>4$ anche i tensori a simmetria mista giocano un ruolo e
devono essere tenuti in considerazione.

Si pu\`{o} mostrare che l'equazione
\begin{equation}\label{SHs}
(\Box-m^{2})\Phi^{(s)}=0 \ , \qquad \partial\cdot\Phi^{(s)}=0 \ ,
\end{equation}
dove il campo reale $\Phi^{(s)}\equiv\Phi_{\mu_{1}...\mu_{s}}$
(per semplicit\`{a} di notazione gli indici saranno sottintesi
dove ci\`{o} non crei confusione) \`{e} rappresentato da un
tensore di rango $s$ totalmente simmetrico e a traccia nulla,
\begin{equation}\label{tr0}
\Phi^{\nu}_{\phantom{\nu}\nu\mu_{3}...\mu_{s}}=0 \ ,
\end{equation}
descrive un campo bosonico libero di spin $s$ e massa $m$, ovvero
propaga esattamente il numero di gradi di libert\`{a}, $2s+1$,
corrispondenti alle polarizzazioni indipendenti di un campo
massivo di spin $s$ in quattro dimensioni. Infatti, un tensore
totalmente simmetrico con $s$ indici in $d$ dimensioni ha ${s+d-1
\choose d-1}$ componenti indipendenti, cui si devono sottrarre gli
${s+d-3 \choose d-1}$ vincoli indipendenti (\ref{tr0}), per
giungere alla conclusione che, in $d=4$, $(s+1)^{2}$ \`{e} il
numero di componenti indipendenti di un tensore ad $s$ indici
totalmente simmetrico e a traccia nulla. La condizione di
divergenza nulla della (\ref{SHs}) comprende altri $s^{2}$ vincoli
indipendenti, sicch\'{e} si rimane esattamente con $2s+1$
componenti indipendenti per $\Phi^{(s)}$ (per una dimostrazione
pi\`{u} precisa si veda, ad esempio, \cite{Riccioni}).

L'eq. (\ref{SHs}) \`{e} una diretta generalizzazione di quella di
Proca del fotone massivo,
\begin{equation}\label{proca}
\Box\Phi_{\mu}-\partial_{\mu}(\partial\cdot\Phi)-m^{2}\Phi_{\mu}=0
\ ,
\end{equation}
che si ottiene come condizione di stazionariet\`{a} della
lagrangiana
\begin{equation}
{\cal
L}=-\frac{1}{2}(\partial_{\mu}\Phi_{\nu})^{2}+\frac{1}{2}(\partial\cdot
\Phi)^{2}-\frac{m^{2}}{2}(\Phi_{\mu})^{2}\ .
\end{equation}
Poich\'{e} $m^{2}\neq 0$, prendere la divergenza della
(\ref{proca}) implica, per consistenza, il vincolo
$\partial\cdot\Phi=0$, e come conseguenza l'equazione di Proca si
riduce all'equazione di Klein-Gordon insieme al vincolo di
divergenza nulla per il campo del fotone, ovvero equivale alle
(\ref{SHs}) per $s=1$. Una generalizzazione ``ingenua''
dell'equazione di Proca al caso di spin $2$,
\begin{equation}\label{proca2}
\Box\Phi_{\mu\nu}-\partial_{\mu}(\partial\cdot\Phi_{\nu})-m^{2}\Phi_{\mu\nu}=0
\ ,
\end{equation}
sebbene conduca alle (\ref{SHs}), non corrisponde ad un'equazione
lagrangiana, dal momento che il secondo termine non ha la stessa
simmeria del campo (come dovrebbe, poich\'{e} in generale le
equazioni di Eulero-Lagrange si ottengono come derivata funzionale
dell'azione rispetto al campo stesso, $\frac{\delta
S}{\delta\Phi^{(s)}}$). La corretta equazione del moto per spin
$2$ \`{e}
\begin{equation}\label{SH2}
\Box\Phi_{\mu\nu}-\alpha\left[\frac{1}{2}\partial_{\mu}(\partial\cdot\Phi_{\nu})+\frac{1}{2}\partial_{\nu}(\partial\cdot\Phi_{\mu})-\frac{1}{4}\eta_{\mu\nu}(\partial\cdot\partial\cdot\Phi)\right]-m^{2}\Phi_{\mu\nu}=0
\ ,
\end{equation}
il cui secondo temine \`{e} simmetrico e a traccia nulla,
ottenibile dalla lagrangiana
\begin{equation}
{\cal
L}=-\frac{1}{2}(\partial_{\mu}\Phi_{\nu\rho})^{2}+\frac{\alpha}{2}(\partial\cdot
\Phi)_{\mu}^{2}-\frac{m^{2}}{2}(\Phi_{\mu\nu})^{2}\ .
\end{equation}
Prendendo la divergenza della (\ref{SH2}) si ottengono le
(\ref{SHs}) per $s=2$ soltanto se $\alpha=2$ e
$\partial\cdot\partial\cdot\Phi=0$, un vincolo che non pu\`{o}
essere ottenuto dall'eq. (\ref{SH2}) stessa.

L'osservazione cruciale \`{e} dunque che una formulazione
lagrangiana richiede in generale la presenza di ulteriori campi,
detti \emph{ausiliari}, le cui equazioni, combinate con quelle dei
campi fisici, forniscano i vincoli necessari per arrivare alle
(\ref{SHs}) ed annullino i campi ausiliari stessi. Illustriamo
questo procedimento nel semplice caso di spin $2$. Dal momento che
il vincolo che si vuole ottenere,
$\partial\cdot\partial\cdot\Phi=0$, \`{e} scalare, \`{e} naturale
introdurre in questo caso un campo scalare $\Phi$, il cui
accoppiamento col campo di spin $2$ \`{e} del tipo
$\Phi_{\mu\nu}\,\partial^{\mu}\partial^{\nu}\Phi$. In particolare,
si considera la lagrangiana
\begin{eqnarray}
{\cal L}^{\textrm{tot}} & = & {\cal L}+{\cal
L}^{\textrm{aux}}\,=\,-\frac{1}{2}(\partial_{\mu}\Phi_{\nu\rho})^{2}+(\partial\cdot
\Phi)_{\mu}^{2}-\frac{m^{2}}{2}(\Phi_{\mu\nu})^{2}\nonumber \\
&&+\frac{2}{3}\left[\frac{1}{2}(\partial_{\mu}\Phi)^{2}+m^{2}\Phi^{2}+\Phi_{\mu\nu}\partial^{\mu}\partial^{\nu}\Phi\right]
\ ,
\end{eqnarray}
da cui discendono le equazioni del moto per i campi
$\Phi_{\mu\nu}$ e $\Phi$, rispettivamente,
\begin{equation}\label{SHlagraux1}
(\Box-m^{2})\Phi_{\mu\nu}-2\{\partial_{\mu}(\cdot\Phi)_{\nu}\}_{S.T.}+\frac{2}{3}\{\partial_{\mu}\partial_{\nu}\Phi\}_{S.T.}=0
\ ,
\end{equation}
\begin{equation}\label{SHlagraux2}
(\Box-2m^{2})\Phi-\partial\cdot\partial\cdot\Phi=0 \ ,
\end{equation}
dove seguiamo la notazione di Singh e Hagen indicando con
$\{A_{\mu_{1}...\mu_{s}}\}_{S.T.}$ la parte simmetrica e a traccia
nulla di $A_{\mu_{1}...\mu_{s}}$. Contraendo la prima equazione
con $\partial^{\mu}\partial^{\nu}$ si ottiene l'equazione scalare
\begin{equation}
(-\frac{1}{2}\Box-m^{2})(\partial\cdot\partial\cdot\Phi)+\frac{1}{2}\Box^{2}\Phi=0
\ ,
\end{equation}
che, insieme alla (\ref{SHlagraux2}), costituisce un sistema di
due equazioni lineari nelle incognite $\Phi$ e
$\partial\cdot\partial\cdot\Phi$, a determinante algebrico e non
nullo,
\begin{equation}
\det \left( \begin{array}{cc} (\Box-2m^{2}) & -1 \\
\frac{1}{2}\Box^{2} & (-\frac{1}{2}\Box-m^{2})
\end{array}
    \right)=2m^{4}\neq 0\ ,
\end{equation}
in conseguenza della scelta dei coefficienti numerici dei termini
di ${\cal L}^{\textrm{aux}}$. L'unica soluzione del sistema \`{e}
quindi $\Phi=0$ e $\partial\cdot\partial\cdot\Phi=0$, come
richiesto, e le equazioni lagrangiane (\ref{SHlagraux1}) e
(\ref{SHlagraux2}) si riducono alla (\ref{SH2}).

Questa costruzione procede analogamente per spin arbitrario. Nel
caso bosonico \cite{Singh:qz}, le equazioni del moto (\ref{SHs})
sono ottenute tramite l'introduzione, accanto al campo
$\Phi^{(s)}$ di spin $s$ simmetrico e a traccia nulla, dei campi
ausiliari $\Phi^{(s-\lambda)}$, $\lambda=2,3,...,s$, rappresentati
da tensori di rango $s-\lambda$ simmetrici e a traccia nulla: le
loro equazioni, combinate con divergenze di ordine opportuno di
quella per $\Phi^{(s)}$, danno luogo ai vincoli
\begin{equation}\label{vincdiv}
\Phi^{(s-\lambda)}_{\mu_{\lambda+1}...\mu_{s}}\equiv
\partial^{\mu_{1}}...\partial^{\mu_{\lambda}}\Phi^{(s)}_{\mu_{1}...\mu_{s}}
\ ,
\end{equation}
oltre ad annullare i campi ausiliari stessi. La lagrangiana che
descrive questo sistema di campi \`{e}
\begin{eqnarray}\label{Ltots}
&&{\cal
L}^{\textrm{tot}}=-\frac{1}{2}(\partial_{\mu}\Phi^{(s)})^{2}+\frac{s}{2}(\partial\cdot
\Phi^{(s)})^{2}-\frac{m^{2}}{2}(\Phi^{(s)})^{2} \nonumber \\
&&+c(s)\biggl\{\Phi^{(s-2)}(\partial\cdot\partial\cdot\Phi^{(s)})+\frac{1}{2}(\partial_{\mu}\Phi^{(s-2)})^{2}+a_{2}(s)\frac{m^{2}}{2}(\Phi^{(s-2)})^{2}
\nonumber \\
&&+\frac{1}{2}b_{2}(s)(\partial\cdot\Phi^{(s-2)})^{2}-\sum_{q=3}^{s}\left(\prod_{k=2}^{q-1}c_{k}\right)\biggl[-\frac{1}{2}(\partial_{\mu}\Phi^{(s-q)})^{2}-\frac{1}{2}a_{q}(s)m^{2}(\Phi^{(s-q)})^{2}
\nonumber \\
&&-\frac{1}{2}b_{q}(s)(\partial\cdot\Phi^{(s-q)})^{2}-m\Phi^{(s-q)}(\partial\cdot\Phi^{(s-q+1)})\biggr]\biggr\}
\ ,
\end{eqnarray}
dove i coefficienti $c(s),a_{q}(s),b_{q}(s),c_{q}(s)$ dipendono
dallo spin, e vengono determinati in modo unico imponendo la
realizzazione dei vincoli (\ref{vincdiv}), come nei casi di spin
``basso'' (per la loro forma si veda \cite{Singh:qz}).

Analogamente procede l'analisi delle equazioni dei campi di gauge
di spin semi-intero arbitrario \cite{Singh:rc}. L'equazione che
propaga le $2s+1$ polarizzazioni indipendenti di un campo reale
massivo di spin $s=n+\frac{1}{2}$ \`{e}
\begin{equation}\label{SHsf}
(i\dsl-m)\Psi^{(n)}=0 \ , \qquad
\partial\cdot\Psi^{(n)}=0 \ ,
\end{equation}
dove $\dsl\equiv\gamma\cdot\partial$, e
$\Psi^{(n)}\equiv\Psi_{\mu_{1}...\mu_{n}}$ (l'indice spinoriale
\`{e} implicito) \`{e} uno spinor-tensore di rango $n$ totalmente
simmetrico negli indici vettoriali e che soddisfa la condizione di
$\gamma$-traccia nulla
\begin{equation}
\gamma\cdot\Psi^{(n)}_{\mu_{2}...\mu_{n}}=0 \ .
\end{equation}
La lagrangiana da cui derivano le equazioni (\ref{SHsf}) \`{e}
formulata in termini di $\Psi^{(n)}$, di un campo ausiliario
$\Psi^{(n-1)}$, anch'esso uno spinor-tensore totalmente simmetrico
negli $n-1$ indici vettoriali e a $\gamma$-traccia nulla, e due
insiemi di campi ausiliari $\Psi^{(n-\lambda)}$ e
$\chi^{(n-\lambda)}$, con $\lambda=2,3,...,n$, tutti
spinor-tensori totalmente simmetrici e a $\gamma$-traccia nulla.
Non riportiamo i dettagli della costruzione, simile ma pi\`{u}
complicata di quella del caso bosonico, rimandando a
\cite{Singh:rc} e \cite{Riccioni} per maggiori dettagli.

Fronsdal studi\`{o} il limite di massa nulla delle equazioni dei
campi di spin arbitrario. Facendo riferimento ai casi semplici di
spin $1$ e spin $2$, \`{e} evidente che per $m=0$ le equazioni
\begin{equation}\label{maxwell}
\Box\Phi_{\mu}-\partial_{\mu}(\partial\cdot\Phi)=0\ ,
\end{equation}
ovvero le equazioni di Maxwell, ammettono la simmetria di gauge
\begin{equation}\label{maxwell}
\delta\Phi_{\mu}=\partial_{\mu}\epsilon \ ,
\end{equation}
e analogamente, nel limite di massa nulla, le (\ref{SHlagraux1}) e
(\ref{SHlagraux2}),
\begin{eqnarray}
&&\Box\Phi_{\mu\nu}-\partial_{\mu}(\partial\cdot\Phi)_{\nu}-\partial_{\nu}(\partial\cdot\Phi)_{\mu}+\frac{1}{2}\eta_{\mu\nu}(\partial\cdot\partial\cdot\Phi)
\nonumber \\
&&+\frac{2}{3}(\partial_{\mu}\partial_{\nu}-\frac{1}{4}\eta_{\mu\nu}\Box)\Phi=0
 \ , \label{SH2m0}
\end{eqnarray}
\begin{equation}\label{SH2m0aux}
\Box\Phi-\partial\cdot\partial\cdot\Phi=0 \ ,
\end{equation}
acquistano l'invarianza sotto le trasformazioni di gauge
\begin{equation}
\delta\Phi_{\mu\nu}=\partial_{\mu}\epsilon_{\nu}+\partial_{\nu}\epsilon_{\mu}-\frac{1}{2}\eta_{\mu\nu}(\partial\cdot\epsilon)
\ ,
\end{equation}
\begin{equation}\label{eisteinlin}
\delta\Phi=\frac{3}{2}\,\partial\cdot\epsilon\ .
\end{equation}
Si noti che nel caso di spin $2$ \`{e} possibile ottenere una
semplificazione lavorando con un solo tensore simmetrico, definito
come la combinazione
\begin{equation}
h_{\mu\nu}=\Phi_{\mu\nu}+\alpha\,\eta_{\mu\nu}\Phi \ ,
\end{equation}
dove il coefficiente $\alpha$ viene determinato dalla richiesta
che la variazione di gauge di $h_{\mu\nu}$ sia
\begin{equation}
\delta\Phi_{\mu\nu}=\partial_{\mu}\epsilon_{\nu}+\partial_{\nu}\epsilon_{\mu}
\ ,
\end{equation}
sicch\'{e} $\alpha=\frac{1}{3}$ e
\begin{eqnarray}
\Phi_{\mu\nu} & = & h_{\mu\nu}-\frac{1}{4}\eta_{\mu\nu}h' \ ,\\
\Phi & = & \frac{3}{4}h' \ ,
\end{eqnarray}
dove $h'\equiv h^{\mu}_{\mu}$. Si noti inoltre che il campo
$h_{\mu\nu}$ \`{e} simmetrico, ma non ha traccia nulla. Scrivendo
le due equazioni (\ref{SH2m0}) e (\ref{SH2m0aux}) in termini di
$h_{\mu\nu}$ e combinandole si ottiene una nuova equazione, con lo
stesso contenuto fisico ma pi\`{u} semplice,
\begin{equation}
\Box h_{\mu\nu}-\partial_{\mu}(\partial\cdot
h)_{\nu}-\partial_{\nu}(\partial\cdot
h)_{\mu}+\partial_{\mu}\partial_{\nu}h'=0 \ ,
\end{equation}
che coincide con la linearizzazione delle equazioni di Einstein
nel vuoto $R_{\mu\nu}=0$.

Questo schema si generalizza a spin arbitrario. In particolare,
facendo riferimento alla lagrangiana bosonica di Singh e Hagen
(\ref{Ltots}), si pu\`{o} notare che nel limite di massa nulla i
campi $\Phi^{(s)}$ e $\Phi^{(s-2)}$ si disaccoppiano da tutti gli
altri campi ausiliari, e si ottiene la lagrangiana
\begin{eqnarray}\label{Ltotsm0}
&&{\cal
L}^{\textrm{tot}}=-\frac{1}{2}(\partial_{\mu}\Phi^{(s)})^{2}+\frac{s}{2}(\partial\cdot
\Phi^{(s)})^{2} \nonumber \\
&&+\frac{s(s-1)^{2}}{2s-1}\biggl\{\Phi^{(s-2)}(\partial\cdot\partial\cdot\Phi^{(s)})+\frac{1}{2}(\partial_{\mu}\Phi^{(s-2)})^{2}
\nonumber \\
&&\frac{1}{2}\frac{(s-2)^{2}}{2s-1}(\partial\cdot\Phi^{(s-2)})^{2}\biggr\}
\ ,
\end{eqnarray}
le cui equazioni del moto, opportunamente combinate, conducono
all'equazione di Fronsdal per un campo di gauge di spin $s$ intero
\begin{equation}\label{Fronsdals}
{\cal F}^{(s)}\equiv \Box
\varphi^{(s)}-\sum_{i=1}^{s}\partial_{\mu_{i}}(\partial\cdot
\varphi^{(s)})+\sum_{i<j}^{s}\partial_{\mu_{i}}\partial_{\mu_{j}}\varphi^{(s)\prime}=0
\ ,
\end{equation}
dove
\begin{equation}\label{defFronfield}
\varphi^{(s)}=\Phi^{(s)}+\frac{1}{2s-1}\sum_{i<j}^{s}\eta_{\mu_{i}\mu_{j}}\varphi^{(s)\prime}
\end{equation}
e
\begin{equation}
\varphi^{(s)\prime}=\varphi^{(s)\mu}_{\phantom{(s)\mu}\mu\mu_{3}...\mu_{s}}
\ .
\end{equation}
Il fattore $\frac{1}{2s-1}$ viene fissato in modo tale che la
trasformazione di gauge sotto cui (\ref{Fronsdals}) \`{e}
invariante sia
\begin{equation}\label{spinsgaugetr}
\delta\varphi^{(s)}=\sum_{i=1}^{s}\partial_{\mu_{i}}\epsilon^{(s-1)}
\ ,
\end{equation}
dove il parametro di gauge $\epsilon^{(s-1)}$ \`{e} un tensore
totalmente simmetrico di rango $s-1$ e a traccia nulla,
\begin{equation}\label{trparam}
\epsilon^{(s-1)\mu}_{\phantom{(s-1)\mu}\mu\mu_{3}...\mu_{s}}=0 \ .
\end{equation}
Questa propriet\`{a} del parametro di gauge \`{e} essenziale,
poich\'{e}, sotto la variazione (\ref{spinsgaugetr}),
\begin{equation}\label{dopfronsd}
\delta{\cal
F}^{(s)}_{\mu_{1}...\mu_{s}}=\frac{s(s-1)(s-2)}{2}\,\partial_{(\mu_{1}}\partial_{\mu_{2}}\partial_{\mu_{3}}\epsilon^{(s-1)\prime}_{\mu_{4}...\mu_{s})}
\ ,
\end{equation}
dove gli indici inclusi nelle parentesi tonde si intendono
totalmente simmetrizzati con normalizzazione all'unit\`{a},
sicch\'{e} per spin $s\geq 3$ l'equazione di Fronsdal ammette la
simmetria di gauge (\ref{spinsgaugetr}) soltanto se il parametro
ha traccia nulla. La (\ref{defFronfield}) implica che
\begin{equation}
\Phi^{(s)}=\varphi^{(s)}-\frac{1}{2s}\sum_{i<j}^{s}\eta_{\mu_{i}\mu_{j}}\varphi^{(s)\prime}
\ ,
\end{equation}
\begin{equation}
\Phi^{(s-2)}=\frac{2s-1}{2s}\varphi^{(s)\prime} \ .
\end{equation}
Inoltre, poich\'{e} $\Phi^{(s-2)}$ \`{e} a traccia nulla, il
tensore totalmente simmetrico $\varphi^{(s)}$ non \`{e} a traccia
nulla, ma a \emph{doppia traccia} nulla,
\begin{equation}\label{doubletr}
\varphi^{(s)\prime\prime}=\varphi^{(s)\mu\phantom{\mu\nu}\nu}_{\phantom{(s)\mu}\mu\phantom{\mu\nu}\nu\mu_{5}...\mu_{s}}=0
\ ,
\end{equation}
condizione preservata dalla trasformazione di gauge
(\ref{spinsgaugetr}), in conseguenza della (\ref{trparam}).

L'equazione di Fronsdal (\ref{Fronsdals}) propaga il corretto
numero di polarizzazioni indipendenti associate ad un campo di
spin $s$ e massa nulla. Introduciamo infatti la condizione di
gauge di de Donder generalizzata
\begin{equation}\label{gendD}
{\cal
D}_{\mu_{2}...\mu_{s}}=\partial\cdot\varphi^{(s)}_{\mu_{2}...\mu_{s}}-\frac{s-1}{2}\,\partial_{(\mu_{2}}\varphi^{(s)\prime}_{\mu_{3}...\mu_{s})}=0
\ ,
\end{equation}
che riduce la (\ref{Fronsdals}) all'equazione delle onde ordinaria
\begin{equation}
\Box\varphi^{(s)}=0 \ .
\end{equation}
La variazione della condizione (\ref{gendD}),
\begin{equation}\label{resgaufree}
\delta(\partial\cdot\varphi^{(s)}_{\mu_{2}...\mu_{s}}-\frac{s-1}{2}\,\partial_{(\mu_{2}}\varphi^{(s)\prime}_{\mu_{3}...\mu_{s})})=\Box\epsilon^{(s-1)}_{\mu_{2}...\mu_{s}}
\ ,
\end{equation}
\`{e} tale da consentire ulteriori trasformazioni locali che non
alterano il gauge di de Donder scelto, purch\'{e} il parametro
soddisfi l'equazione delle onde
\begin{equation}
\Box\epsilon^{(s-1)}_{\mu_{2}...\mu_{s}}=0 \ .
\end{equation}
Le componenti indipendenti di un tensore di rango $s$ totalmente
simmetrico e a doppia traccia nulla, in $d$ dimensioni, sono
${s+d-1 \choose d-1}-{s+d-5 \choose d-1}$. Si noti ora che la
condizione (\ref{gendD}) ha traccia nulla, in virt\`{u} della
(\ref{doubletr}),
\begin{equation}
{\cal
D}'_{\mu_{4}...\mu_{s}}=-\frac{s-3}{2}\,\partial_{(\mu_{4}}\varphi^{(s)\prime\prime}_{\mu_{5}...\mu_{s})}=0
\ ,
\end{equation}
e pertanto contiene un numero di vincoli indipendenti pari ai
gradi di libert\`{a} del parametro di gauge
$\epsilon^{(s-1)}_{\mu_{2}...\mu_{s}}$, ovvero ${s+d-2 \choose
d-1}-{s+d-4 \choose d-1}$. Le componenti del campo di gauge non
ancora fissate soddisfano l'equazione delle onde, e ci\`{o}
consente di utilizzare gli ${s+d-2 \choose d-1}-{s+d-4 \choose
d-1}$ parametri indipendenti che determinano la residua
libert\`{a} di gauge (\ref{resgaufree}) per annullare altrettante
componenti di $\varphi^{(s)}$ \cite{Riccioni}. In conclusione, si
trova che i gradi di libert\`{a} fisici di un campo di spin $s$ e
massa nulla in $d$ dimensioni sono pari a $2{s+d-5 \choose
d-4}-{s+d-5 \choose d-5}$. In quattro dimensioni, ad esempio, un
tensore totalmente simmetrico a traccia nulla ha $2(s^{2}+1)$
componenti indipendenti, $2s^2$ delle quali vengono fissate
utilizzando la condizione di gauge di de Donder e la residua
libert\`{a} di gauge, ottenendo infine $2s^{2}+2-2s^{2}=2$ gradi
di libert\`{a}, corrispondenti alle elicit\`{a} $\pm s$.

In termini del campo di Fronsdal $\varphi^{(s)}$ la lagrangiana
(\ref{Ltotsm0}) si scrive
\begin{eqnarray}\label{Ltotsm0Fr}
&&{\cal
L}^{\textrm{tot}}=-\frac{1}{2}(\partial_{\mu}\varphi^{(s)})^{2}+\frac{s}{2}(\partial\cdot
\varphi^{(s)})^{2}+\frac{s(s-1)}{4}(\partial_{\mu}\varphi^{(s)\prime})^{2} \nonumber \\
&&+\frac{s(s-1)(s-2)}{8}(\partial\cdot\varphi^{(s)\prime})^{2}+\frac{s(s-1)}{2}\varphi^{(s)\prime}(\partial\cdot\partial\cdot\varphi^{(s)})
\ ,
\end{eqnarray}
ed \`{e} interessante notare che la (\ref{Fronsdals}) non \`{e} la
condizione di stazionariet\`{a} di questa lagrangiana, ma una
opportuna combinazione della corrispondente equazione di
Eulero-Lagrange e della sua traccia. Ci\`{o} accade gi\`{a} per
spin $2$, poich\'{e} l'equazione che si ricava dalla variazione
della lagrangiana di Einstein-Hilbert linearizzata non \`{e}
$R^{(lin)}_{\mu\nu}=0$, ma
\begin{equation}\label{trueinstlin}
G^{(lin)}_{\mu\nu}=0 \ ,
\end{equation}
dove
\begin{equation}
G_{\mu\nu}=R^{(lin)}_{\mu\nu}-\frac{1}{2}\eta_{\mu\nu}R^{(lin)}
\end{equation}
\`{e} il tensore di Einstein linearizzato. Analogamente si pu\`{o}
introdurre un tensore di Einstein generalizzato per spin
arbitrario,
\begin{equation}
{\cal G}_{\mu_{1}...\mu_{s}}={\cal
F}_{\mu_{1}...\mu_{s}}-\frac{s(s-1)}{2}\eta_{(\mu_{1}\mu_{2}}{\cal
F}'_{\mu_{3}...\mu_{s})}
\end{equation}
(omettendo l'indice $(s)$), in termini del quale la variazione di
gauge della lagrangiana (\ref{Ltotsm0Fr}) si scrive
\begin{equation}
\delta{\cal
L}^{\textrm{tot}}=\delta\varphi^{\mu_{1}...\mu_{s}}{\cal
G}_{\mu_{1}...\mu_{s}} \ .
\end{equation}

\`{E} molto conveniente lavorare, d'ora in poi, con una notazione
introdotta in \cite{Francia:2002aa, Francia:2002pt}, nella quale
tutti gli indici impliciti sono da intendersi totalmente
simmetrizzati. Valgono le propriet\`{a}
\begin{eqnarray}
&& \left( \partial^{\; p} \ \varphi  \right)^{\; \prime} \ = \
\Box \ \partial^{\; p-2} \
\varphi \ + \ 2 \, \partial^{\; p-1} \  \partial\cdot\varphi \ + \ \partial^{\; p} \ \varphi^{\;'} \\
&& \partial^{\; p} \ \partial^{\; q} \ = \ \left( {p+q} \atop p
\right) \ \partial^{\; p+q} \label{usefulxnotaz}
\\
&& \partial\cdot \left( \partial^{\; p} \ \varphi \right) \ = \
\Box \ \partial^{\; p-1} \ \varphi \ + \
\partial^{\; p} \ \partial\cdot\varphi \\
&& \partial\cdot \eta^{\;k} \ = \ \partial \, \eta^{\;k-1}
\label{etak} \\
&& \left(\eta^{k}\,T_{(s)}\right)^{\;
\prime}=k\,[d+2(s+k-1)]\,\eta^{k-1}\,T_{(s)}+\eta^{k}\,T_{(s)}^{\;'}
\ ,
\end{eqnarray}
dove indichiamo, al solito, con $T_{(s)}$ un generico tensore
totalmente simmetrico di rango $s$. Per guadagnare familiarit\`{a}
con questa notazione riscriviamo alcuni dei risultati ottenuti.
L'equazione di Fronsdal per spin $s$ intero (\ref{Fronsdals})
diventa
\begin{equation}\label{Fronsdalscomp}
{\cal F}\equiv \Box \varphi-\partial\,\partial\cdot
\varphi+\partial^{2}\,\varphi'=0 \ ,
\end{equation}
mentre la trasformazione di gauge del campo $\varphi$
(\ref{spinsgaugetr}) si riscrive
\begin{equation}\label{sgaugetrcomp}
\delta\varphi=\partial\,\epsilon \ .
\end{equation}
Se $s=1,2$, l'operatore di Fronsdal ${\cal F}$ \`{e} invariante,
ma in generale
\begin{equation}\label{varopFr}
\delta\,{\cal F}=3\,\partial^{3}\,\epsilon' \ ,
\end{equation}
che coincide con la (\ref{dopfronsd}).

La propriet\`{a} cruciale che, per spin $2$, distingue il tensore
di Einstein da quello di Ricci \`{e} che il primo ha divergenza
nulla, e questo assicura l'invarianza della lagrangiana di
Einstein-Hilbert sotto trasformazioni di gauge, a meno di derivate
totali. Tuttavia, il suo analogo di spin $s$
\begin{equation}
{\cal G}={\cal F}-\frac{1}{2}\,\eta\,{\cal F}
\end{equation}
soddisfa invece un'identit\`{a} di Bianchi ``anomala'',
\begin{equation}
\partial\cdot{\cal G}=-\frac{3}{2}\,\partial^{3}\,\varphi''-\frac{1}{2}\,\eta\,\partial\cdot{\cal F} \ ,
\end{equation}
che fa s\`{i} che la variazione della lagrangiana
(\ref{Ltotsm0Fr}), a meno di derivate totali, sia data da
\begin{equation}
\delta{\cal L}^{\textrm{tot}}\sim
\epsilon\left[-\frac{3}{2}\,\partial^{3}\,\varphi''-\frac{1}{2}\,\eta\,\partial\cdot{\cal
F}\right] \ ,
\end{equation}
il cui ultimo termine genera la traccia del parametro di gauge. Si
scopre cos\`{i} che mentre l'invarianza di gauge dell'equazione
(non lagrangiana) di Fronsdal richiede che il parametro sia a
traccia nulla, la simmetria del corrispondente principio d'azione
richiede, in pi\`{u}, il vincolo di doppia traccia nulla sui campi
di gauge.

Simili considerazioni possono essere applicate anche al caso
fermionico. Nella corrispondente lagrangiana di Singh e Hagen, i
campi $\Psi^{(n)}$, $\Psi^{(n-1)}$ e $\Psi^{(n-2)}$ si
disaccoppiano da tutti gli altri campi ausiliari nel limite di
massa nulla, e le relative equazioni acquistano un'invarianza di
gauge sotto opportune trasformazioni con parametro locale a
$\gamma$-traccia nulla,
\begin{equation}\label{paramconstr}
\gamma\cdot\epsilon^{(n-1)}=0 \ .
\end{equation}
La combinazione lineare
\begin{equation}
\psi^{(n)}=\Psi^{(n)}+\alpha\sum_{i=1}^{n}\gamma_{\mu_{i}}\Psi^{(n-1)}+\beta\sum_{i<j}^{n}\eta_{\mu_{i}\mu_{j}}\Psi^{(n-2)}
\ ,
\end{equation}
con coefficienti scelti in modo tale che la sua variazione di
gauge sia
\begin{equation}\label{fermgautr}
\delta\psi^{(n)}=\sum_{i=1}^{n}\partial_{\mu_{i}}\epsilon^{(n-1)}
\ ,
\end{equation}
soddisfa il vincolo
\begin{equation}\label{analog2trlness}
(\gamma\cdot\psi)'=0 \ ,
\end{equation}
in virt\`{u} del vincolo di $\gamma$-traccia nulla imposto su
$\Psi^{(n-2)}$. L'equazione di Fang-Fronsdal per spin semi-intero
arbitrario, scritta nella notazione di \cite{Francia:2002aa,
Francia:2002pt}, \`{e}
\begin{equation}\label{eqFrsferm}
{\cal S}\equiv i\,[\dsl\psi-\partial\psisl]=0 \ ,
\end{equation}
dove $\psisl=\gamma\cdot\psi$, ed ammette l'invarianza di gauge
sotto la (\ref{fermgautr}), ovvero
\begin{equation}
\delta\psi=\partial\epsilon \ ,
\end{equation}
soltanto se il parametro di gauge soddisfa il vincolo
(\ref{paramconstr}), poich\'{e}
\begin{equation}
\delta{\cal S}=i\partial\,\epsilonsl \ .
\end{equation}

Si noti che la (\ref{eqFrsferm}) \`{e} una generalizzazione a spin
semi-intero arbitrario dell'equazione di Rarita-Schwinger per un
campo a massa nulla di spin $3/2$,
\begin{equation}\label{RS}
\gamma^{\mu\nu\rho}\,\partial_\nu \psi_\rho=0 \ ,
\end{equation}
che infatti, prendendone la $\gamma$-traccia ed usando la
relazione
\begin{equation}\label{3gamma}
\gamma^{\mu\nu\rho}=\gamma^\mu
\gamma^{\nu\rho}-\eta^{\mu\nu}\gamma^{\rho}+\eta^{\mu\rho}\gamma^\nu
\ ,
\end{equation}
conduce alla condizione
\begin{equation}
\gamma^{\mu\nu}\,\partial_\nu \psi_\rho=0 \ ,
\end{equation}
dove $\gamma^{\mu\nu}$ \`{e} antisimmetrica in $\mu$ e $\nu$, ed
equivale al prodotto $\gamma^\mu\gamma^\nu$ quando $\mu\neq\nu$.
Sostituendo infine quest'ultima nella (\ref{RS}), tenendo conto
della (\ref{3gamma}), si ottiene
\begin{equation}\label{RS2}
\dsl\psi_\mu -\partial_\mu\psisl=0 \ .
\end{equation}

La (\ref{paramconstr}) ed il vincolo sul campo
(\ref{analog2trlness}) giocano lo stesso ruolo dei vincoli
(\ref{trparam}) e (\ref{doubletr}) del caso bosonico: in
particolare, la (\ref{eqFrsferm}) non \`{e} l'equazione di
Eulero-Lagrange per la corrispondente lagrangiana di
Fang-Fronsdal, ma una combinazione opportuna di tale equazione con
la sua traccia e la sua $\gamma$-traccia, e si pu\`{o} definire un
analogo fermionico del tensore di Einstein generalizzato,
\begin{equation}
{\cal T}\equiv {\cal S}-\frac{1}{2}(\,\eta\,{\cal S}+\gamma\,\Ssl)
\ ,
\end{equation}
in termini del quale la variazione di gauge della lagrangiana
risulta
\begin{equation}
\delta{\cal L}=\delta\bar{\psi}\,{\cal T} \ .
\end{equation}
Tuttavia, la divergenza di quest'ultimo tensore \`{e} nulla solo
per spin semi-intero $s<7/2$, poich\'{e} vale
\begin{equation}
\partial\cdot{\cal S}-\frac{1}{2}\,\partial\,{\cal
S}-\frac{1}{2}\,\dsl\;\Ssl=i\,\partial^{2}\,\psisl' \ ,
\end{equation}
sicch\'{e} il vincolo (\ref{analog2trlness}) \`{e} necessario per
assicurare l'invarianza di gauge della lagrangiana di
Fang-Fronsdal fermionica.

Un'altra osservazione importante fatta in \cite{Francia:2002aa,
Francia:2002pt} \`{e} che la semplice relazione tra gli operatori
di Klein-Gordon e di Dirac si generalizza a spin arbitrario in
quella che lega l'operatore di Fronsdal bosonico ${\cal F}_{s}$ di
spin $s$ e quello fermionico ${\cal S}_{s+1/2}$ di spin $s+1/2$:
\begin{equation}\label{linkBF}
{\cal
S}_{s+1/2}\,-\,\frac{1}{2}\frac{\partial}{\Box}\,\dsl\,\Ssl_{s+1/2}\,=\,i\,\frac{\dsl}{\Box}\,{\cal
F}_{s}(\psi) \ .
\end{equation}

\section{Geometria delle equazioni libere di HS}
\label{deWFreed}

L'analogia tra le teorie di gauge di spin elevato e di spin
``basso'' pu\`{o} essere resa ancor pi\`{u} evidente attraverso
l'introduzione di alcuni tensori dalle semplici propriet\`{a} di
trasformazione sotto le (\ref{sgaugetrcomp}), che generalizzano i
simboli di Christoffel della gravit\`{a}. Seguendo \cite{deWit:pe}
possiamo anzitutto definire il simbolo di Christoffel di primo
ordine per spin $s$ come la quantit\`{a}
\begin{equation}\label{1passo}
\Gamma^{(1)}_{\alpha;\beta_{1}...\beta_{s}}=\frac{1}{2}\left(\partial_{\alpha}\varphi_{\beta_{1}...\beta_{s}}-s\,\partial_{(\beta_{1}|}\varphi_{\alpha|\beta_{2}...\beta_{s})}\right)
\ ,
\end{equation}
la cui variazione sotto una trasformazione di gauge del campo
$\varphi$ \`{e} data da
\begin{equation}
\delta\Gamma^{(1)}_{\alpha;\beta_{1}...\beta_{s}}=-\frac{s(s-1)}{2}\,\partial_{(\beta_{1}}\partial_{\beta_{2}|}\epsilon_{\alpha|\beta_{3}...\beta_{s})}
\ .
\end{equation}
Si noti che per spin $2$ tale definizione coincide con quella
degli ordinari simboli di Christoffel linearizzati, mentre \`{e}
possibile generare ricorsivamente un'intera gerarchia di simboli
di Christoffel generalizzati di rango pi\`{u} alto, come
\begin{equation}\label{Gammagamma}
\Gamma^{(m)}_{\alpha_{1}...\alpha_{m};\beta_{1}...\beta_{s}}=\frac{1}{m+1}\left(m\,\partial_{\alpha_{1}}\Gamma^{(m-1)}_{\alpha_{2}...\alpha_{m};\beta_{1}...\beta_{s}}-s\,\partial_{(\beta_{1}|}\Gamma^{(m-1)}_{\alpha_{2}...\alpha_{m};\alpha_{1}|\beta_{2}...\beta_{s})}\right)
\ ,
\end{equation}
simmetrico sia negli indici $\beta_{i}$ (manifestamente) che negli
$\alpha_{i}$, come si pu\`{o} vedere, ad esempio, sostituendo
iterativamente nella (\ref{Gammagamma}) sino ad ottenerne
l'espressione in termini del campo di gauge,
\begin{equation}\label{Gammaphi}
\Gamma^{(m)} \ = \ \frac{1}{m+1} \ \sum_{k=0}^{m}\
\frac{(-1)^{k}}{ \left( {{m} \atop {k}} \right) }\ \partial^{\;
m-k}\, \bigtriangledown^{\; k} \, \varphi\ ,
\end{equation}
che riportiamo, per brevit\`{a}, nella notazione con indici
simmetrici impliciti e dove, seguendo \cite{Francia:2002aa,
Francia:2002pt}, indichiamo con $\partial$ le derivate che portano
gli indici $\alpha_{i}$ (introdotti dalle derivate nel primo passo
(\ref{1passo})) e con $\bigtriangledown$ quelle che portano gli
indici $\beta_{i}$ (che provengono dal campo di gauge). La
trasformazione di gauge di $\Gamma^{(m)}$ \`{e}
\begin{equation}\label{genChristransf}
\delta\Gamma^{(m)}=(-)^{m}\bigtriangledown^{\; m+1}\,\epsilon \ ,
\end{equation}
dove il parametro porta tutti gli $m$ indici $\alpha_{i}$, insieme
ad ulteriori $s-(m+1)$ indici $\beta_{i}$. Il secondo membro
contiene tutte le simmetrizzazioni di questi ultimi indici con i
restanti $m+1$ indici $\beta_{i}$ delle derivate
$\bigtriangledown^{\; m+1}$, ed \`{e} pertanto chiaro che si
riduce ad un solo termine per $m=s-1$,
\begin{equation}
\delta\Gamma^{(s-1)}_{\alpha_{1}...\alpha_{s-1};\beta_{1}...\beta_{s}}=\partial_{\beta_{1}}...\partial_{\beta_{s}}\epsilon_{\alpha_{1}...\alpha_{s-1}}
\ ,
\end{equation}
caratterizzando in tal modo $\Gamma^{(s-1)}$ come l'analogo della
connessione di Christoffel per un campo di gauge di spin $s$
arbitrario. Non solo, ma la (\ref{genChristransf}) mostra anche
che tutte le $\Gamma^{(m)}$ della gerarchia con $m\geq s$ sono
invarianti di gauge, e che
$\Gamma^{(s)}_{\alpha_{1}...\alpha_{s};\beta_{1}...\beta_{s}}\equiv
{\cal R}_{\alpha_{1}...\alpha_{s};\beta_{1}...\beta_{s}}$,
\begin{equation}\label{Rphi}
\Gamma^{(s)} \ = \ \frac{1}{s+1} \ \sum_{k=0}^{s}\
\frac{(-1)^{k}}{ \left( {{s} \atop {k}} \right) }\ \partial^{\;
s-k}\, \bigtriangledown^{\; k} \, \varphi\ ,
\end{equation}
definisce una generalizzazione a spin arbitrario del tensore di
curvatura di Riemann. Tale curvatura di spin $s$ eredita la
propriet\`{a} dei $\Gamma^{(m)}$ di simmetria sotto lo scambio di
qualsiasi coppia di indici di tipo $\alpha$ o di tipo $\beta$,
separatamente, mentre sotto lo scambio dei due insiemi di indici
vale
\begin{equation}
{\cal R}_{\alpha_1 \cdots \alpha_s;\beta_1 \cdots \beta_s} \ = \
(-1)^s \ {\cal R}_{\beta_1 \cdots \beta_s;\alpha_1 \cdots
\alpha_s} \ .
\end{equation}
Si pu\`{o} inoltre verificare l'identit\`{a} ciclica
\begin{equation}
{\cal R}_{\alpha_1 \cdots \alpha_s;\beta_1 \cdots
\beta_s}\,+\,s{\cal R}_{(\beta_1|\alpha_2 \cdots
\alpha_s;\alpha_1|\beta_2 \cdots \beta_s)}=0 \ .
\end{equation}

Questa gerarchia di connessioni generalizzate suggerisce la
``geometria'' che governa le teorie di gauge di spin arbitrario,
un'estensione dei casi familiari di spin ``basso''. Per spin $1$,
la $\Gamma^{(1)}$, al primo ordine della gerarchia, \`{e} gi\`{a}
la curvatura gauge-invariante,
\begin{equation}
{\cal
R}_{\alpha;\beta}=\partial_{\alpha}\varphi_{\beta}-\partial_{\beta}\varphi_{\alpha}
\ ,
\end{equation}
che coincide con la usuale \emph{field-strength} di Maxwell.
Questo \`{e} il solo caso in cui si riesce a costruire un oggetto
gauge-invariante di primo ordine nelle derivate, e gi\`{a} per
spin $2$, come \`{e} ben noto, il tensore di Riemann contiene due
derivate del campo di gauge. In particolare, la curvatura ${\cal
R}_{\alpha_1 \alpha_2;\beta_1  \beta_2}$ \`{e} uguale ad una
combinazione simmetrica di due tensori di Riemann ordinari
(antisimmetrici sotto lo scambio degli indici entro ciascun
insieme, e simmetrici sotto lo scambio dei due insiemi di indici)
linearizzati,
\begin{equation}
{\cal R}_{\alpha_1 \alpha_2;\beta_1
\beta_2}=\frac{1}{2}(R^{(lin)}_{\alpha_1 \beta_1;\alpha_2
\beta_2}+R^{(lin)}_{\alpha_1 \beta_2;\alpha_2 \beta_1}) \ .
\end{equation}
Analogamente, per spin $s$ l'oggetto gauge-invariante ha $s$
derivate del campo, ed in tal senso il numero di ordini della
gerarchia che precedono la curvatura di spin $s$ misura la
``distanza'' che c'\`{e} tra quest'ultima ed il campo
fondamentale. Il fatto che gi\`{a} il primo caso non banale,
quello della gravit\`{a}, presenti (qualora si includano
interazioni) una struttura altamente non lineare, lascia pensare
che interazioni tra campi di HS diano luogo ad una teoria
estremamente ricca e complicata, come vedremo nei capitoli
seguenti.

Poich\'{e} le equazioni di campo devono essere gauge-invarianti,
esse conterranno, in generale, combinazioni della curvatura di
spin $s$. Tuttavia, richiedendo che il parametro di gauge abbia
traccia nulla (specializzando al caso bosonico) \`{e} possibile
ottenere invarianti di ordine pi\`{u} basso, come
\begin{equation}
W^{(m)}_{\alpha_1 \cdots \alpha_s;\beta_1 \cdots \beta_s}\equiv
\Gamma^{(m+2)\sigma}_{\phantom{(m+2)\sigma}\sigma\alpha_{1}...\alpha_{m};\beta_{1}...\beta_{s}}
\ ,
\end{equation}
che contiene, per $m=0$, ovvero al secondo livello della gerarchia
dei $\Gamma$, l'equazione di Fronsdal,
\begin{equation}
W=\Gamma^{(2)\prime}= \Box \varphi-\partial\,\partial\cdot
\varphi+\partial^{2}\varphi'=0 \ ,
\end{equation}
che, come abbiamo sottolineato nella sezione precedente, \`{e}
infatti gauge-invariante soltanto se il parametro \`{e} a traccia
nulla.

Nulla ci impedisce per\`{o}, in linea di principio, di mirare ad
ottenere equazioni gauge-invarianti senza porre alcun vincolo sui
parametri, al prezzo di introdurre derivate di ordine superiore al
secondo dei campi fondamentali, come suggerisce la gerarchia di
connessioni generalizzate. Vedremo nella sezione seguente come
ci\`{o} sia in effetti possibile e porti alle equazioni libere non
locali scritte in termini delle curvature generalizzate proposte
da Francia e Sagnotti \cite{Francia:2002aa, Francia:2002pt}.

\section{Equazioni ``geometriche'' nello spazio-tempo piatto}
\label{df/as}

Ci concentreremo nel seguito sul caso bosonico, per brevit\`{a}.
Tra le equazioni libere di spin $s=1,2$
\begin{eqnarray}
&& \Box A_{\mu}-\partial_{\mu}\partial\cdot A=0 \ ,\\
&& \Box h_{\mu\nu}-\partial_{\mu}\partial\cdot
h_{\nu}-\partial_{\nu}\partial\cdot
h_{\mu}+\partial_{\mu}\partial_{\nu}h'=0 \ ,
\end{eqnarray}
e la loro generalizzazione a spin arbitrario (\ref{Fronsdalscomp})
c'\`{e} un'importante differenza: le prime contengono infatti
tutti i costrutti di spin pi\`{u} basso ottenuti come divergenze e
tracce del campo di gauge, che sono invece assenti dall'equazione
di Fronsdal per spin $s>2$. Questo fatto \`{e} all'origine della
variazione (\ref{varopFr}) dell'operatore di Fronsdal, e dunque
del vincolo di traccia nulla sul parametro, che pu\`{o} quindi
essere eliminato modificando l'equazione di Fronsdal con
l'aggiunta di nuovi opportuni costrutti di spin pi\`{u} basso la
cui variazione di gauge cancelli quella (\ref{varopFr}) dei
termini ``canonici''. Evidentemente, per consistenza con questi
ultimi, tali nuovi costrutti dovranno contenere derivate del campo
di gauge di ordine pi\`{u} alto del secondo che ne portino gli
indici e sar\`{a} dunque necessario riscalarli con potenze inverse
dell'operatore Lorentz-invariante di d'Alembert per ristabilire le
giuste dimensioni, ottenendo cos\`{i} equazioni \emph{non locali}.

Ad esempio, nel caso semplice di spin $3$ si possono scrivere due
equazioni gauge-invarianti senza alcun vincolo sul parametro,
\begin{eqnarray}
&& {\cal F}_{\mu_1\mu_2\mu_3} \ - \ \frac{1}{3 \; \Box} \,
\left(\,
\partial_{\mu_1} \, \partial_{\mu_2} \, {\cal F}\;'_{\mu_3} +  \partial_{\mu_2} \,
\partial_{\mu_3} \, {\cal F}\;'_{\mu_1} + \partial_{\mu_3} \, \partial_{\mu_1} \,
{\cal F}\;'_{\mu_2} \, \right) \ = \ 0 \ ,  \label{spin3box} \\
&& {\cal F}_{\mu_1\mu_2\mu_3} \ - \ \frac{\partial_{\mu_1}
\partial_{\mu_2}
\partial_{\mu_3}}{\Box^2} \
\partial \cdot {\cal F}\;' \ = \ 0 \ , \label{spin3box2}
\end{eqnarray}
che sono, in realt\`{a}, due forme diverse di un'unica equazione,
l'una potendo essere ottenuta dall'altra combinandola con la sua
traccia. La variazione del secondo termine di entrambe le
equazioni (\ref{spin3box}) e (\ref{spin3box2}) produce esattamente
$3\,\partial^{3}\,\epsilon'$, rendendole pertanto
gauge-invarianti, come anticipato. Si noti inoltre che
\begin{equation}
\delta \left(\frac{\partial \cdot {\cal
F}\;'}{\Box^2}\right)\,=\,3\,\epsilon' \ ,
\end{equation}
che mostra come la non localit\`{a} di queste equazioni sia
\emph{pura gauge}, essendo sufficiente fissare opportunamente la
traccia del parametro di gauge per tornare alla forma locale delle
equazioni, conservando in tal modo la libert\`{a} di gauge
parametrizzata dalla sola parte a traccia nulla del parametro.

Le stesse conclusioni si possono raggiungere per spin arbitrario,
costruendo degli opportuni operatori di Fronsdal modificati,
analoghi di quelli che compaiono nelle (\ref{spin3box}) e
(\ref{spin3box2}), mediante la definizione ricorsiva
\begin{equation}
{\cal F}^{(n+1)} \ = \ {\cal F}^{(n)} \ + \ \frac{1}{(n+1) (2 n +
1)} \ \frac{\partial^{\;2}}{\Box} \, {{\cal F}^{(n)}}\;' \ - \
\frac{1}{n+1} \ \frac{\partial}{\Box} \ \partial\cdot  {\cal
F}^{(n)} \ ,
\end{equation}
dove ${\cal F}^{(1)}={\cal F}$, ed il cui termine generico ha la
variazione di gauge
\begin{equation}\label{tropFrmodif}
\delta {\cal F}^{(n)} \ = \ \left( 2 n + 1 \right) \
\frac{\partial^{\; 2 n + 1}} {\Box^{\; n-1}} \ \epsilon^{[n]} \ ,
\end{equation}
dove con $\epsilon^{[n]}$ indichiamo la traccia del parametro su
$n$ coppie di indici. Ma quest'ultima \`{e} non banale soltanto se
$s\geq 2n+1$, e dunque per ogni $s$ si ottiene un operatore
cinetico non locale gauge-invariante (sotto trasformazioni di
gauge di parametro completamente arbitrario, non vincolato) dopo
almeno $\left[ \frac{s-1}{2} \right]$ iterazioni.

Anche il vincolo di doppia traccia nulla sui campi scompare dopo
un numero sufficiente di iterazioni, poich\'{e} l'$n$-mo operatore
cinetico modificato soddisfa l'identit\`{a} di Bianchi
\begin{equation}
\partial\cdot {\cal F}^{(n)} \ - \ \frac{1}{2n} \ \partial {{\cal F}^{(n)}}{\;
'} \ = \ - \ \left( 1 + \frac{1}{2n}  \right) \ \frac{\partial^{\;
2n+1}}{\Box^{\; n-1}} \ \varphi^{[n+1]} \ , \label{bianchin}
\end{equation}
``anomala'' solo se $s\geq 2n+2$. Per $n$ sufficientemente grande,
la $k$-upla traccia della (\ref{bianchin}) fornisce
\begin{equation}
\partial\cdot {\cal F}^{(n)\, [k]} \ - \ \frac{1}{2(n-k)} \ \partial {{\cal
F}^{(n)\, [k+1]}} \ = \ 0 \ , \qquad ( k \leq n-1)
\label{bianchink}
\end{equation}
il cui secondo termine si annulla per $k=n-1$ e spin dispari
$s=2n-1$,
\begin{equation}
\partial\cdot {\cal F}^{(n)\, [k]}=0 \ .
\end{equation}
In base a questi risultati si pu\`{o} allora definire un tensore
di Einstein generalizzato completamente gauge-invariante,
combinando gli ${\cal F}^{(n)}$ con le loro tracce, come
\begin{equation}
{\cal G}^{(n)} \ = \ \sum_{p \leq n} \ \frac{(-1)^p}{2^p \ p! \
\left( {n \atop p} \right)} \ \eta^p \ {\cal F}^{(n)\, [p]} \ ,
\end{equation}
che ha divergenza nulla per $n$ sufficientemente grande,
caratteristica che assicura l'invarianza di gauge, senza alcun
vincolo sul campo, della lagrangiana di spin $s$ ottenibile
integrando
\begin{equation}
\delta{\cal L}\sim \delta\varphi\,{\cal G}^{(n)} \ .
\end{equation}

Le equazioni di spin $s$ completamente gauge-invarianti assumono
una forma compatta in termini delle variabili
\begin{equation}
\hat{\Phi}(x,\xi) \ = \ \frac{1}{s!}\ \xi^{\mu_1} \cdots
\xi^{\mu_s} \ \varphi_{\mu_1 \cdots \mu_s} \ ,
\end{equation}
ottenute contraendo gli indici del campo di gauge con i vettori
$\xi^{\mu_i}$, in termini delle quali tracce e divergenze di
$\varphi$ si realizzano con gli operatori
$\partial_{\xi}\cdot\partial_{\xi}$ e
$\partial_{\xi}\cdot\partial$, dove con $\partial_{\xi}$
indichiamo la derivata rispetto a $\xi$. In questo formalismo,
l'operatore di Fronsdal ha l'espressione
\begin{equation}
\hat{\cal F}(\hat{\Phi}) = \left[ \ \Box \ - \ \xi \cdot \partial
\ \ \partial\cdot \partial_\xi \ + \ (\xi \cdot \partial)^2 \
\partial_\xi \cdot \partial_\xi
  \ \right]
\hat{\Phi} \ , \label{fronsdalxi}
\end{equation}
e dopo l'$n$-ma iterazione si arriva ad equazioni della forma
\begin{equation}
\prod_{k=0}^{n-1} \left[\  1 \ +\  \frac{1}{(k+1)(2k+1)} \
\frac{(\xi \cdot \partial)^2}{\Box} \
 \partial_\xi \cdot \partial_\xi \ - \ \frac{1}{k+1} \ \frac{\xi \cdot \partial}{\Box} \ \partial_\xi \cdot \ \partial
 \  \right]  \ \hat{\cal F}(\hat{\Phi}) \ = \ 0 \ , \label{eqnonlocgen}
\end{equation}
che per $n=\left[ \frac{s+1}{2} \right]$ danno luogo all'
equazione completamente gauge-invariante per spin $s$.
Analogamente a quanto accade per spin $3$, questa equazione,
combinata con la sua traccia, ne genera una sempre riducibile alla
forma
\begin{equation}
{\cal F} \ =  \ \partial^{\; 3} \ {\cal H} \ , \label{fd3h}
\end{equation}
dove ${\cal H}$ contiene tutti i termini non locali che le
successive iterazioni aggiungono, e trasforma proporzionalmente
alla traccia del parametro,
\begin{equation}
\delta {\cal H} = 3 \ \epsilon{\; '} \ .
\end{equation}
Ci\`{o} dimostra che, in generale, i termini non locali delle
equazioni completamente invarianti sono pura gauge, e possono
quindi essere annullati fissando $\epsilon{\; '}=-\frac{1}{3}{\cal
H}$, il che equivale a scegliere una ``gauge di Fronsdal'' ${\cal
H}=0$ che riporta le equazioni in forma locale e lascia non
vincolata soltanto la parte a traccia nulla del parametro di
gauge.

Che esistesse, in linea di principio, la possibilit\`{a} di
giungere ad equazioni gauge-invarianti senza restrizioni sul campo
o sul parametro di gauge era stato sottolineato gi\`{a} alla fine
della precedente sezione, insieme al fatto che queste equazioni
sarebbero state naturalmente scritte in termini delle curvature
ivi introdotte, generalizzando cos\`{i} i casi di spin $1$ e spin
$2$. Il legame con la geometria emerge osservando che le
propriet\`{a} di trasformazione delle $\Gamma^{m}$ con $m=2n$ pari
implicano che la variazione della traccia $\Gamma^{(2n)\, [n]}$ su
tutte le coppie di indici $\alpha_{i}$ \`{e} data da
\begin{equation}
\delta \, \left( \, \frac{1}{\Box^{n-1}}\ \Gamma^{(2n)\, [n]} \,
\right) \ = \ \frac{\partial^{\; 2n+1}}{\Box^{\; n-1}} \
\epsilon^{[n]} \ ,
\end{equation}
che coincide, a meno di un fattore, con la legge di trasformazione
(\ref{tropFrmodif}) degli operatori di Fronsdal modificati con i
termini non locali ${\cal F}^{(n)}$! Questo significa che le
equazioni di spin arbitrario completamente gauge-invarianti
scritte in termini dei campi di gauge ammettono, come accade per
spin $1$ e $2$, una riformulazione in termini di curvature, che
non \`{e} invece manifesta a partire dalle equazioni di Fronsdal,
prive dei termini contenenti tutti i costrutti di spin pi\`{u}
basso. Infatti, per spin $s=2n$, $\Gamma^{(2n)\, [n]}$ \`{e} la
traccia su $n$ coppie di indici del tensore di Riemann
generalizzato di spin $s$, e l'equazione geometrica completamente
gauge-invariante per spin pari \`{e}
\begin{equation}
\frac{1}{\Box^{n-1}} \ {\cal R}^{[n]}{}_{;\mu_1 \cdots \mu_{2n}} \
= \ 0 \ , \label{geomeven}
\end{equation}
che contiene come caso particolare ($n=1$) le equazioni di
Einstein linearizzate, che costituiscono anche l'ultimo caso di
equazioni di spin pari locali.

Per spin intero dispari $s=2n+1$, tuttavia, gli indici della
curvatura $\Gamma^{(2n+1)}$ non vengono completamente saturati
dalla traccia su $n$ coppie di indici, ma per analogia con il
sottocaso delle equazioni di Maxwell si pu\`{o} saturare l'indice
rimanente mediante una divergenza, ottenendo quindi
\begin{equation}
\frac{1}{\Box^{n}} \ \partial\cdot{\cal R}^{[n]}{}_{;\mu_1 \cdots
\mu_{2n+1}} \ = \ 0 \ , \label{geomodd}
\end{equation}
di cui il caso elettromagnetico (n=0) \`{e} l'ultimo locale.

Una combinazione di ciascuna di queste due equazioni con la sua
traccia ha la forma (\ref{fd3h}), e si riduce quindi all'equazione
di Fronsdal tramite una scelta di gauge (\emph{gauge fixing}).
Questa procedura tuttavia non garantisce banalmente l'equivalenza
con la formulazione locale, poich\'{e} il \emph{gauge fixing}
$\epsilon{\; '}=-\frac{1}{3}{\cal H}$ porta all'equazione di
Fronsdal ma non implica alcun vincolo sui campi, mentre abbiamo
visto nella sezione \ref{Fr} che il vincolo di doppia traccia
nulla \`{e} essenziale affinch\'{e} tale equazione propaghi il
corretto numero di polarizzazioni associato ad un campo di massa
nulla. Inoltre, se $\varphi''\neq 0$ la condizione di de Donder
generalizzata
\begin{equation}\label{dDbis}
{\cal
D}\,\equiv\,\partial\cdot\varphi\,-\,\frac{1}{2}\,\partial\varphi'\,=\,0
\end{equation}
non \`{e} una buona condizione di gauge, nel senso che non \`{e}
possibile utilizzare la libert\`{a} di gauge residua, quella sotto
trasformazioni parametrizzate dalla sola parte a traccia nulla di
$\epsilon$, per far s\`{i} che essa possa sempre essere
soddisfatta. Infatti, in generale,
\begin{equation}
\delta{\cal D}\,=\,\Box\epsilon-\partial^{2}\,\epsilon' \ ,
\end{equation}
dove $\epsilon'$ \`{e} fissato, mentre la traccia della
(\ref{dDbis}), ${\cal D}'\,=\,-\frac{1}{2}\partial\varphi''$, non
lo \`{e} affatto, e viene quindi a mancare l'uguaglianza tra il
numero di condizioni indipendenti contenute nella gauge di de
Donder e quello delle componenti libere del parametro.

La via di uscita consiste nel modificare la condizione di de
Donder in modo che sia a traccia \emph{identicamente nulla}. Non
imponendo $\varphi''=0$, questo risultato pu\`{o} essere ottenuto
soltanto aggiungendo alla (\ref{dDbis}) dei costrutti di spin
pi\`{u} basso ottenuti mediante tracce multiple del campo, secondo
una procedura simile a quella utilizzata per giungere ad equazioni
del moto gauge-invarianti senza necessit\`{a} di alcun vincolo sul
parametro. Nel primo caso non banale, $s=4$, si pu\`{o} rendere
(\ref{dDbis}) identicamente a traccia nulla aggiungendo il termine
$\Delta$, ovvero imponendo la condizione (in $d$ dimensioni)
\begin{equation}
{\cal
D}+\Delta\,\equiv\,\partial\cdot\varphi\,-\,\frac{1}{2}\,\partial\varphi'\,+\,\frac{1}{2(d+2)}\,\eta\,\partial\varphi''=\,0
\ ,
\end{equation}
che riduce l'equazione di Fronsdal a
\begin{equation}
p^{2}\varphi_{\mu\nu\rho\sigma}\,+\,\frac{1}{(d+2)}\,[\eta_{\mu\nu}p_{\rho}p_{\sigma}+...]\,\varphi''=\,0
\ ,
\end{equation}
scritta nello spazio degli impulsi. Poich\'{e} il secondo termine
non contiene alcun parametro dimensionale, per invarianza di
Lorentz quest'ultima equazione evidentemente ammette come sole
soluzioni non banali quelle con autovalore $p^{2}=0$, e ci\`{o} a
sua volta implica $\varphi''=0$ \cite{Francia:2002pt}, ovvero
\`{e} l'equazione del moto stessa ad annullare la doppia traccia
del campo di gauge, facendo s\`{i} che le equazioni non locali
propaghino effettivamente il giusto numero di gradi di libert\`{a}
di un campo a massa nulla.

Questo procedimento si generalizza a spin arbitrario, definendo
$\Delta$ ricorsivamente richiedendo che soddisfi $({\cal
D}+\Delta)'=0$ \cite{Francia:2002pt}. Per ogni $s$, la condizione
di gauge modificata ${\cal D}+\Delta=0$, sostituita nelle
equazioni del moto, fornisce
\begin{equation}
\Box\varphi\,+\,\partial\Delta\,=\,0 \ ,
\end{equation}
e l'esistenza di soluzioni non banali richiede che $p^{2}=0$. Ma
allora, utilizzando ancora la condizione di de Donder modificata,
\begin{equation}
\partial{\cal D}=0 \ ,
\end{equation}
che implica \cite{Francia:2002pt}
\begin{equation}
{\cal D}=0 \ ,
\end{equation}
e quindi, per la traccia, $\varphi''=0$. A questo punto
l'equivalenza con la formulazione locale di Fronsdal, ottenuta
mediante un parziale \emph{gauge fixing}, \`{e} completa, ed il
conteggio dei gradi di libert\`{a} procede esattamente come nella
sezione \ref{Fr}.

Non ci occupiamo in dettaglio delle equazioni geometriche per spin
semi-intero, tuttavia sottolineamo il fatto che il legame
(\ref{linkBF}) tra gli operatori cinetici bosonici e fermionici si
estende anche agli operatori modificati ${\cal S}^{(n)}_{s+1/2}$ e
${\cal F}^{(n)}_{s}$, rendendo in tal modo evidente la
possibilit\`{a} di scrivere equazioni del moto completamente
gauge-invarianti anche per campi di gauge fermionici.

L'importanza delle equazioni geometriche sta, oltre che nel
possedere un'invarianza di gauge completa (ovvero sotto
trasformazioni di parametro completamente arbitrario, senza
necessit\`{a} dei vincoli algebrici della formulazione di
Fronsdal), anche nello stabilire un contatto con la Teoria delle
Stringhe nel suo limite di tensione nulla. Come accennato infatti
nell'Introduzione, le equazioni del campo di stringa danno luogo,
per $\alpha'\rightarrow\infty$ e nel settore dei soli tensori
simmetrici, ad un sistema di equazioni per un tripletto di campi
invarianti sotto trasformazioni di gauge senza alcun vincolo di
traccia e che, opportunamente combinate, producono le equazioni
non locali (\ref{fd3h}) \cite{Sagnotti:2003qa}. \`{E} notevole
anche il fatto che queste ultime possono sempre essere portate in
una forma locale \emph{senza dover introdurre vincoli sul
parametro} per mezzo di un \emph{compensatore}, ovvero un campo di
Stueckelberg che trasforma proporzionalmente alla traccia del
parametro,
\begin{equation}\label{trcompens}
\delta\,\alpha\,=\, \epsilon' \ .
\end{equation}
In questo modo, tutti i termini non locali, che, come gi\`{a}
detto, sono pura gauge, possono essere assorbiti in $\alpha$, e le
equazioni geometriche assumono la forma locale
\begin{equation}
{\cal F}(\varphi)\,=\,3\,\partial^{3}\,\alpha \ ,
\end{equation}
rimanendo invarianti sotto le trasformazioni di gauge
(\ref{trcompens}) e
\begin{equation}
\delta\,\varphi\,=\, \partial\,\epsilon \ .
\end{equation}

\section{Equazioni di Fronsdal in $(A)dS$}
\label{FrAdS}

Abbiamo gi\`{a} avuto modo di commentare l'importanza degli
spazi-tempo a curvatura costante per la costruzione di interazioni
consistenti fra campi di spin arbitrario. \`{E} utile quindi
esaminare la forma delle equazioni di Fronsdal in uno spazio-tempo
$(A)dS$, anche alla luce di quanto verr\`{a} esposto nei capitoli
seguenti.

L'interazione con il campo gravitazionale di background viene
introdotta, al solito, covariantizzando le derivate,
$\partial\rightarrow \nabla\equiv\partial+\Gamma$, dove $\Gamma$
rappresenta la connessione di Christoffel di $(A)dS$, introdotta
nel capitolo $2$. Inoltre,
\begin{equation}
\varphi'_{\mu_3...\mu_s}=g^{\mu_1\mu_2}\varphi_{\mu_1...\mu_s} \ ,
\end{equation}
dove $g$ \`{e} il tensore metrico di $(A)dS$, e assumiamo
$\varphi''=0$ ed $\epsilon'=0$.

Trattiamo in dettaglio il caso bosonico. Abbiamo mostrato che, in
uno spazio-tempo piatto, l'operatore cinetico di Fronsdal \`{e}
invariante sotto la trasformazione di gauge (\ref{spinsgaugetr}) o
(\ref{sgaugetrcomp}),
\begin{equation}
{\cal F}(\delta\varphi)\,=\,0 \ ,
\end{equation}
purch\'{e} $\epsilon'=0$. Questa propriet\`{a} non \`{e} pi\`{u}
verificata in $(A)dS$, a causa della non commutativit\`{a} delle
derivate covarianti: vale infatti la relazione
\begin{equation}
[ \nabla_\mu , \nabla_\nu ] \, \varphi_{\rho_1...\rho_s} \ = \
\frac{s}{L^2} \left( g_{\nu(\rho_1|} \, V_{\mu|\rho_2...\rho_s)} \
- \
 g_{\mu(\rho_1|} \, V_{\nu|\rho_2...\rho_s)} \right) \ , \label{noncommvect}
\end{equation}
dove $L$ rappresenta il raggio di curvatura di $AdS$, mentre
l'analoga relazione per lo spazio-tempo $dS$ pu\`{o} essere
ottenuta dalla (\ref{noncommvect}) cambiando il segno della
curvatura, ovvero continuando formalmente $L$ a valori immaginari.
La trasformazione di gauge del campo di spin $s$ diventa inoltre
\begin{equation}\label{trcov}
\delta\,\varphi_{\mu_1...\mu_s}\,=\,s\,\nabla_{(\mu_1}\,\epsilon_{\mu_2...\mu_s)}
\ ,
\end{equation}
e l'operatore di Fronsdal covariantizzato \`{e}
\begin{equation}\label{opcov}
{\cal
F}^{\textrm{cov}}_{\mu_1...\mu_s}(\varphi)=\Box\varphi_{\mu_1...\mu_s}-s\nabla_{(\mu_1}\nabla\cdot\varphi_{\mu_2...\mu_s}+\frac{s(s-1)}{4}\,\left\{\nabla_{(\mu_1},\nabla_{\mu_2}\right\}\,\varphi'_{\mu_3...\mu_s)}
\ ,
\end{equation}
dove $\Box=g^{\mu\nu}\nabla_\mu\nabla_\nu$. Ma questo operatore
non annulla la variazione di gauge del campo, e anzi la
sostituzione diretta di (\ref{trcov}) in ${\cal F}^{\textrm{cov}}$
produce i termini
\begin{equation}\label{stepinterm}
s\left[\Box,\nabla_{(\mu_1}\right]\,\epsilon_{\mu_2...\mu_s)}+\frac{1}{L^{2}}s(s-1)(d+s-3)\,\nabla_{(\mu_1}\,\epsilon_{\mu_2...\mu_s)}
\ .
\end{equation}
Per eliminarli \`{e} necessario quindi modificare l'operatore
cinetico di Fronsdal aggiungendo termini opportuni di ordine
$1/L^2$ che cancellino la variazione di (\ref{opcov}) e scompaiano
nel limite piatto $L\rightarrow\infty$. Calcolando il commutatore
che compare nella (\ref{stepinterm}) si pu\`{o} verificare che
l'operatore cinetico invariante in $AdS$ \`{e}
\begin{equation}\label{FropL}
{\cal F}_L \ = \ {\cal F}^{\textrm{cov}} \ - \ \frac{1}{L^2} \,
\left\{ \left[ (3-d-s)(2-s) - s \right]\, \varphi \ + \ 2 \, g \;
\varphi' \right\} \ ,
\end{equation}
scritto con indici simmetrici impliciti, e dove $d$ rappresenta la
dimensionalit\`{a} dello spazio-tempo. Si noti che, sebbene si
abbia a che fare con campi a massa nulla, la richiesta di
invarianza delle equazioni di Fronsdal su uno spazio-tempo curvo
ha come conseguenza l'introduzione di un ``termine di massa''
associato alla curvatura, una caratteristica che ritroveremo nel
capitolo successivo. Anche in questo caso, le equazioni
\begin{equation}
{\cal F}_L \ = \ 0
\end{equation}
non sono equazioni lagrangiane, e si pu\`{o} definire un
corrispondente tensore di Einstein generalizzato
\begin{equation}
{\cal G}_L \ = \ {\cal F}_L \ - \ \frac{1}{2}\,g\,{\cal F}'_L \ ,
\end{equation}
in termini del quale si scrivono le equazioni ${\cal G}_L=0$, che
seguono da un principio variazionale.

Analoghe considerazioni valgono anche per il caso fermionico.
L'operatore di Fang-Fronsdal covariantizzato in modo ``ingenuo'',
\begin{equation}\label{naiveFF}
{\cal S}^{\textrm{cov}}\,=\,i\,(\nablasl\,\psi-\nabla\,\psisl)
\end{equation}
non \`{e} invariante sotto la trasformazione di un fermione di
spin $s=n+\frac{1}{2}$,
\begin{equation}
\delta \psi \ = \ \nabla \, \epsilon \ + \ \frac{1}{2 L} \, \gamma
\; \epsilon \ ,
\end{equation}
e deve quindi venire anch'esso modificato, con l'aggiunta di
termini che assorbano la variazione di gauge di ordine $1/L^2$ di
${\cal S}^{\textrm{cov}}$, dovuta ai commutatori di derivate
covarianti, che agiscono su uno spinore come
\begin{equation}
[ \nabla_\mu , \nabla_\nu ] \, \eta \ = \ - \ \frac{1}{2 L^2} \
\gamma_{\mu\nu} \ \eta \ . \label{noncommspin}
\end{equation}
Il risultato \`{e}
\begin{equation}
{\cal S}_L  \ =  \ {\cal S}^{\textrm{cov}} \ + \ \frac{i}{2 L} \,
\left[ d\ + \ 2( n \, - \, 2) \right] \psi \ + \ \frac{i}{2L} \,
\gamma \, \psisl \ .
\end{equation}

Anche le equazioni completamente gauge-invarianti di Francia e
Sagnotti, nella formulazione che utilizza i compensatori, sono
state scritte in $(A)dS$ \cite{Sagnotti:2003qa}, mentre la
corrispondente versione non locale delle equazioni non \`{e}
attualmente nota.

\chapter{Higher spins: teoria ``unfolded'' lineare}

\section{$\star$-prodotto e algebre di Higher Spin}
\label{4.1}

\`{E} conveniente riproporre quanto esposto nel capitolo 3 circa
la gravit\`{a} su $AdS_{4}$ nella formulazione di MacDowell e
Mansouri nel formalismo a due componenti, sostituendo gli indici
di Lorentz $a,b=0,1,2,3$ con gli indici spinoriali
$\alpha,\beta=1,2$ e $\dot{\alpha},\dot{\beta}=1,2$ della
rappresentazione fondamentale del ricoprimento universale del
gruppo di Lorentz, $SL(2,C)$, e della coniugata. Questo riflette
l'isomorfismo esistente a livello delle corrispondenti algebre
(vedi Appendice \ref{appendice A}). La connessione tra gli indici
di $so(3,1)$ e di $sl(2,C)$ \`{e} resa manifesta con l'uso dei
simboli di van der Waerden $(\sigma^{a})_{\alpha\dot{\alpha}}$
definiti nell'Appendice A, che comprendono le tre matrici di Pauli
e la matrice identit\`{a}, poich\'{e}
\begin{equation}
V_{a}\longleftrightarrow
V_{a}(\sigma^{a})_{\alpha\dot{\alpha}}\equiv
V_{\alpha\dot{\alpha}}\ .
\end{equation}
Ad una coppia di indici di Lorentz antisimmetrici corrispondono
inoltre due coppie di indici, una non puntata e l'altra puntata,
ciascuna simmetrica (dato che le rappresentazioni irriducibili di
$sl(2,C)$ corrispondono a multispinori separatamente simmetrici
nei loro indici di ciascun tipo),
\begin{equation}
J_{[ab]}\longleftrightarrow
\varepsilon^{\alpha\beta}\bar{J}^{(\dot{\alpha}\dot{\beta})}+\varepsilon^{\dot{\alpha}\dot{\beta}}J^{(\alpha\beta)}
\ ,
\end{equation}
dove $\varepsilon^{\alpha\beta}$ \`{e} il tensore invariante di
$sl(2,C)$, totalmente antisimmetrico e proporzionale a
$\sigma_{2}$, mentre $J^{\alpha\beta}\equiv
-\frac{1}{2}J_{[ab]}(\sigma^{ab})^{\alpha\beta}$ e
$\bar{J}^{(\dot{\alpha}\dot{\beta})}\equiv
-\frac{1}{2}J_{[ab]}(\bar{\sigma}^{ab})^{\dot{\alpha}\dot{\beta}}$
( $(\sigma^{ab})^{\alpha\beta}$ e
$(\bar{\sigma}^{ab})^{\dot{\alpha}\dot{\beta}}$ sono definite in
Appendice \ref{appendice A}).

Riscriviamo dunque alcuni dei risultati ottenuti nella precedente
sezione nel formalismo a due componenti.

Ai generatori di Lorentz $M_{ab}$ corrisponde la coppia
$M_{\alpha\beta},\bar{M}_{\dot{\alpha}\dot{\beta}}$, mentre ai
generatori di traslazione $AdS$ corrispondono i
$P_{\alpha\dot{\alpha}}$. In questa notazione l'algebra di
$so(3,2)$ diventa
\begin{eqnarray}
\left[M_{\alpha\beta},M_{\gamma\delta}\right] & = &
-i(\varepsilon_{\alpha\gamma}M_{\beta\delta}+\varepsilon_{\beta\delta}M_{\alpha\gamma}+\varepsilon_{\alpha\delta}M_{\beta\gamma}+\varepsilon_{\beta\gamma}M_{\alpha\delta})
\ , \nonumber \\
\left[ P_{\alpha\dot{\alpha}},M_{\beta\gamma} \right]  & = &
-i(\varepsilon_{\alpha\beta}P_{\gamma\dot{\alpha}}+\varepsilon_{\alpha\gamma}P_{\beta\dot{\alpha}})
\ , \\
\left[ P_{\alpha\dot{\alpha}},P_{\beta\dot{\beta}} \right] & = &
-iL^{-2}(\varepsilon_{\alpha\beta}\bar{M}_{\dot{\alpha}\dot{\beta}}+\varepsilon_{\dot{\alpha}\dot{\beta}}M_{\alpha\beta})
\ , \nonumber
\end{eqnarray}
insieme con i coniugati di questi commutatori,
\begin{eqnarray}
\left[\bar{M}_{\dot{\alpha}\dot{\beta}},\bar{M}_{\dot{\gamma}\dot{\delta}}\right]
& = &
-i(\varepsilon_{\dot{\alpha}\dot{\gamma}}\bar{M}_{\dot{\beta}\dot{\delta}}+\varepsilon_{\dot{\beta}\dot{\delta}}\bar{M}_{\dot{\alpha}\dot{\gamma}}+\varepsilon_{\dot{\alpha}\dot{\delta}}\bar{M}_{\dot{\beta}\dot{\gamma}}+\varepsilon_{\dot{\beta}\dot{\gamma}}\bar{M}_{\dot{\alpha}\dot{\delta}})
\ , \nonumber \\
\left[ P_{\alpha\dot{\alpha}},\bar{M}_{\dot{\beta}\dot{\gamma}}
\right] & = &
-i(\varepsilon_{\dot{\alpha}\dot{\beta}}P_{\gamma\dot{\alpha}}+\varepsilon_{\dot{\alpha}\dot{\gamma}}P_{\beta\dot{\alpha}})
\ , \\
\left[ P_{\alpha\dot{\alpha}},P_{\beta\dot{\beta}} \right] & = &
-iL^{-2}(\varepsilon_{\alpha\beta}\bar{M}_{\dot{\alpha}\dot{\beta}}+\varepsilon_{\dot{\alpha}\dot{\beta}}M_{\alpha\beta})
\ , \nonumber
\end{eqnarray}
mentre tutte gli altri commutatori sono nulli.

Ai generatori di Lorentz e delle traslazioni di $AdS_{4}$
associamo le 1-forme di connessione
$\omega_{\mu}^{\alpha\beta},\bar{\omega}_{\mu}^{\dot{\alpha}\dot{\beta}}$
ed $e_{\mu}^{\alpha\dot{\alpha}}$, rispettivamente, e con esse
costruiamo la 2-forma di curvatura
\begin{eqnarray}\label{genRiem}
\Re & = & d(e+\omega)+(e+\omega)\wedge (e+\omega) {} \nonumber\\
& & {} = \frac{1}{2i}\ dx^{\mu}\wedge
dx^{\nu}\,[T^{\alpha\dot{\alpha}}_{\mu\nu}P_{\alpha\dot{\alpha}}+\frac{1}{2}(R^{\alpha\beta}_{\mu\nu}+2L^{-2}e^{\alpha\dot{\alpha}}_{\mu}e_{\nu}\,^{\beta}_{\dot{\alpha}})M_{\alpha\beta}{} \nonumber\\
& & {}
+\frac{1}{2}(\bar{R}^{\dot{\alpha}\dot{\beta}}_{\mu\nu}+2L^{-2}e^{\alpha\dot{\alpha}}_{\mu}e_{\nu}\,_{\alpha}^{\dot{\beta}})\bar{M}_{\dot{\alpha}\dot{\beta}}]
\ ,
\end{eqnarray}
dove
\begin{eqnarray}
T^{\alpha\dot{\alpha}}_{\mu\nu} & = & \partial_{\mu}
e^{\alpha\dot{\alpha}}_{\nu}+\omega_{\mu}^{\phantom{\mu}\alpha\beta}\,
e_{\nu\beta}^{\phantom{\nu\beta}\dot{\alpha}}+\bar{\omega}_{\mu}^{\phantom{\mu}\dot{\alpha}\dot{\beta}}\,
e^{\alpha}_{\nu\dot{\beta}} - (\mu\leftrightarrow\nu) \ ,
\label{deftorsion2cp} \\
R^{\alpha\beta}_{\mu\nu} & = &
\partial_{\mu}\omega_{\nu}^{\alpha\beta}+\omega_{\mu}^{\phantom{\mu}\alpha\gamma}\,\omega_{\nu\gamma}^{\phantom{\nu\gamma}\beta}-
(\mu\,\leftrightarrow\,\nu) \ , \label{defriem1}\\
\bar{R}^{\dot{\alpha}\dot{\beta}}_{\mu\nu} & = &
\partial_{\mu}\bar{\omega}_{\nu}^{\dot{\alpha}\dot{\beta}}+\bar{\omega}_{\mu}^{\phantom{\mu}\dot{\alpha}\dot{\gamma}}\,\bar{\omega}_{\nu\dot{\gamma}}\,^{\phantom{\nu\gamma}\dot{\beta}}-
(\mu\,\leftrightarrow\,\nu) \label{defriem2} \ .
\end{eqnarray}
La curvatura $\Re\equiv d(e+\omega)+(e+\omega)\wedge (e+\omega)$
soddisfa l'identit\`{a} di Bianchi
\begin{equation}\label{idB}
d\Re=[\Re,e+\omega] \ ,
\end{equation}
ovvero $D\Re=0$, che segue dalla condizione $d^{2}=0$.

Il vincolo di torsione \`{e}
\begin{equation}\label{torsion2cp}
T^{\alpha\dot{\alpha}}_{\mu\nu}=0 \ ,
\end{equation}
che, con le equazioni
\begin{eqnarray}
\Re^{\alpha\beta}_{\mu\nu} & = &
R^{\alpha\beta}_{\mu\nu}+2L^{-2}e^{\alpha\dot{\alpha}}_{\mu}\,e_{\nu}\,^{\beta}_{\dot{\alpha}}=0  \label{defads1} \ , \\
\bar{\Re}^{\dot{\alpha}\dot{\beta}}_{\mu\nu} & = &
\bar{R}^{\dot{\alpha}\dot{\beta}}_{\mu\nu}+2L^{-2}e^{\alpha\dot{\alpha}}_{\mu}\,e_{\nu}\,_{\alpha}^{\dot{\beta}}=0
\ , \label{defads2}
\end{eqnarray}
descrive uno spaziotempo $AdS$.

I generatori di $so(3,2)$ ammettono una realizzazione come
bilineari in oscillatori (simile in forma alla cosiddetta
realizzazione di Schwinger delle relazioni di commutazione di
$su(2)$), per la quale
\begin{equation}
M_{\alpha\beta}=-\frac{1}{4}\{\hat{y}_{\alpha},\hat{y}_{\beta}\}\
,\qquad
\bar{M}_{\dot{\alpha}\dot{\beta}}=-\frac{1}{4}\{\hat{\bar{y}}_{\dot{\alpha}},\hat{\bar{y}}_{\dot{\beta}}\}\
,\qquad
P_{\alpha\dot{\alpha}}=\frac{1}{2}\hat{y}_{\alpha}\hat{\bar{y}}_{\dot{\alpha}}\
,
\end{equation}
dove gli $\hat{y}_{\alpha},
\hat{\bar{y}}_{\dot{\alpha}}=(\hat{y}_{\alpha})^{\dag}$ sono
spinori di Weyl commutanti, a valori nelle rappresentazioni
$(1/2,0)$ e $(0,1/2)$ del gruppo di Lorentz, che soddisfano
l'algebra di Heisenberg definita dalle regole di commutazione
\begin{equation}\label{Heisalg}
[\hat{y}_{\alpha},\hat{y}_{\beta}]=2i\varepsilon_{\alpha\beta}\
,\qquad
[\hat{\bar{y}}_{\dot{\alpha}},\hat{\bar{y}}_{\dot{\beta}}]=2i\varepsilon_{\dot{\alpha}\dot{\beta}}\
,\qquad [\hat{y}_{\alpha},\hat{\bar{y}}_{\dot{\beta}}]=0 \ .
\end{equation}
In questo modo i campi gravitazionali sono 1-forme bilineari negli
oscillatori:
\begin{equation}
e_{\mu}+\omega_{\mu}=\frac{1}{2}e_{\mu}^{\alpha\dot{\beta}}\,\hat{y}_{\alpha}\hat{\bar{y}}_{\dot{\beta}}+\omega_{\mu}^{\alpha\beta}\,\frac{1}{4}\{\hat{y}_{\alpha},\hat{y}_{\beta}\}+\bar{\omega}_{\mu}^{\dot{\alpha}\dot{\beta}}\,\frac{1}{4}\{\hat{\bar{y}}_{\dot{\alpha}},\hat{\bar{y}}_{\dot{\beta}}\}
\ .
\end{equation}

Il formalismo fin qui sviluppato consente una naturale estensione
ai campi di HS ed alle relative simmetrie: per includerli nella
teoria, infatti, si comincia con il considerare potenze
arbitrariamente elevate degli oscillatori
$\hat{y}_{\alpha},\hat{\bar{y}}_{\dot{\alpha}}$ e quindi l'algebra
associativa infinito-dimensionale generata da polinomi di grado
arbitrario negli oscillatori stessi. In particolare
\begin{equation}\label{ply1}
\hat{P}(\hat{y},
\hat{\bar{y}})=\sum_{n,m=0}^{\infty}\frac{i}{2n!m!}P^{\alpha_{1}...\alpha_{n}\dot{\alpha_{1}}...\dot{\alpha_{m}}}\hat{y}_{\alpha_{1}}...\hat{y}_{\alpha_{n}}\hat{\bar{y}}_{\dot{\alpha_{1}}}...\hat{\bar{y}}_{\dot{\alpha_{m}}}
\ ,
\end{equation}
dove i coefficienti
$P^{\alpha_{1}...\alpha_{n}\dot{\alpha_{1}}...\dot{\alpha_{m}}}$
sono totalmente simmetrici separatamente negli indici $\alpha_{i}$
e $\dot{\alpha}_{i}$.

Questo implica che il prodotto degli oscillatori
$\hat{y}_{\alpha_{i}}$ \`{e} sempre Weyl-ordinato e lo stesso vale
per gli oscillatori $\hat{\bar{y}}_{\dot{\alpha_{j}}}$. Il
prodotto Weyl-ordinato di n oscillatori $\hat{y}_{\alpha_{i}}$
\`{e}, per definizione, la parte totalmente simmetrica del
prodotto ordinario normalizzata all'unit\`{a}, ovvero
\begin{equation}
(\hat{y}_{\alpha_{1}}...\hat{y}_{\alpha_{n}})_{W}\equiv
\hat{y}_{(\alpha_{1}}...\hat{y}_{\alpha_{n})}=\frac{1}{n!}(\hat{y}_{\alpha_{1}}...\hat{y}_{\alpha_{n}}+\textrm{permutazioni})
\end{equation}
e analogamente per gli $\hat{\bar{y}}_{\dot{\alpha_{j}}}$.

Nella (\ref{ply1}) il pedice W \`{e} omesso per brevit\`{a}.
Inoltre spesso nel seguito useremo, per brevit\`{a}, le notazioni
$\hat{y}_{\alpha_{1}}...\hat{y}_{\alpha_{n}}\equiv
\hat{y}_{\alpha_{1}...\alpha_{n}}\equiv \hat{y}_{\alpha (n)}$.

Si noti che il prodotto di due funzioni Weyl-ordinate degli
oscillatori non \`{e} Weyl-ordinato. Se ad esempio consideriamo il
caso pi\`{u} semplice, il prodotto di due singoli oscillatori,
abbiamo
\begin{equation}
\hat{y}_{\alpha}\hat{y}_{\beta}=\frac{1}{2}\{\hat{y}_{\alpha},\hat{y}_{\beta}\}+\frac{1}{2}[\hat{y}_{\alpha},\hat{y}_{\beta}]=(\hat{y}_{\alpha}\hat{y}_{\beta})_{W}+i\epsilon_{\alpha\beta}
\ .
\end{equation}
In generale, il prodotto di due funzioni Weyl-ordinate degli
oscillatori produce un termine Weyl-ordinato, associato
alla potenza pi\`{u} alta negli y, seguito da ulteriori termini Weyl-ordinati, 
non nulli per il fatto che l'algebra degli oscillatori
(\ref{Heisalg}) \`{e} non commutativa, contenenti potenze via via
pi\`{u} basse di due unit\`{a} per volta.

Possiamo a questo punto dare una prima definizione delle algebre
di HS $shs(4)$ (dove l'argomento si riferisce al loro ruolo in
teorie in quattro dimensioni), in termini dell'algebra di Lie
associativa ${\cal A}$ la cui operazione di composizione \`{e} il
commutatore di funzioni Weyl-ordinate degli oscillatori
$\hat{y}_{\alpha}, \hat{\bar{y}}_{\dot{\alpha}}$, soddisfacenti le
relazioni di commutazione (\ref{Heisalg}), che possono essere
espresse come formali serie di potenze (\ref{ply1}). \footnote
{Sono state costruite estensioni anche supersimmetriche delle
algebre di HS, estendendo superalgebre di Lie, la cui operazione
di composizione \`{e} il supercommutatore
$[\hat{P}_{1},\hat{P}_{2}\}=\hat{P}_{1}\hat{P}_{2}-(-1)^{\pi_{1}\pi_{2}}\hat{P}_{2}\hat{P}_{1}$.}.
Preciseremo in seguito tale definizione.

\`{E} molto conveniente ai fini pratici del calcolo, ed \`{e}
ormai una procedura standard nella letteratura, sostituire agli
operatori i loro simboli. Questo significa che al posto di ciascun
$\hat{P}(\hat{y}, \hat{\bar{y}}) \in {\cal A}$, che si espande in
potenze degli operatori $\hat{y}_{\alpha},
\hat{\bar{y}}_{\dot{\alpha}}$, si utilizzano i simboli
$P(y,\bar{y})$ che sono, per definizione, funzioni delle variabili
commutanti $y_{\alpha}, \bar{y}_{\dot{\alpha}}$ della stessa forma
(e ammettono quindi un'espansione in serie di potenze degli
$y_{\alpha}, \bar{y}_{\dot{\alpha}}$ con identici coefficienti):
\begin{equation}\label{ply2}
P(y,\bar{y})=\sum_{n,m=0}^{\infty}\frac{i}{2n!m!}P^{\alpha_{1}...\alpha_{n}\dot{\alpha_{1}}...\dot{\alpha_{m}}}y_{\alpha_{1}}...y_{\alpha_{n}}\bar{y}_{\dot{\alpha_{1}}}...\bar{y}_{\dot{\alpha_{m}}}
\ .
\end{equation}
Tuttavia, per riprodurre l'algebra ${\cal A}$ in termini di
simboli degli operatori \`{e} necessario riprodurre anzitutto le
(\ref{Heisalg}), introducendo una nuova definizione di prodotto,
dal momento che $y_{\alpha}$ e $\bar{y}_{\dot{\alpha}}$ sono
variabili commutanti nell'ordinario prodotto tra funzioni.

A tale scopo si definisce lo $\star$-prodotto in modo tale che,
date due qualsiasi funzioni $P_{1},P_{2}$, simboli degli operatori
$\hat{P}_{1},\hat{P}_{2}$, $P_{1}\star P_{2}$ sia il simbolo
dell'operatore $\hat{P}_{1}\hat{P}_{2}$. Si pu\`{o} mostrare che
questo conduce a
\begin{equation}\label{defdiff}
P(y,\bar{y})\star
Q(y,\bar{y})=P(y,\bar{y})e^{-i(\overleftarrow{\partial}^{\alpha}\overrightarrow{\partial}_{\alpha}+\overleftarrow{\partial}^{\dot{\alpha}}\overrightarrow{\partial}_{\dot{\alpha}})}Q(y,\bar{y})
\ ,
\end{equation}
dove $\partial_{\alpha}\equiv \frac{\partial}{\partial
y^{\alpha}}$, e $\partial^{\alpha}\equiv
\varepsilon^{\alpha\beta}\partial_{\beta}=-\frac{\partial}{\partial
y_{\alpha}}$, semplicemente perch\'{e} i termini successivi
dell'esponenziale producono le successive contrazioni.

Si noti che lo $\star$-prodotto \`{e} evidentemente non locale
negli oscillatori, ovvero nelle coordinate interne, poich\'{e}
include derivate di ordine arbitrariamente elevato rispetto a
questi ultimi. Questa legge di prodotto \`{e} inoltre associativa
\begin{equation}
(P\star (Q\star R))=((P\star Q)\star R) \ ,
\end{equation}
ed \`{e} normalizzata in modo tale che $1$ sia l'elemento neutro
dell'algebra, ovvero $P\star 1=1\star P=P$. La definizione
(\ref{defint}) implica inoltre che, dati due polinomi
$P(y,\bar{y})$ e $Q(y,\bar{y})$, $(P\star Q)(y,\bar{y})$ \`{e}
anch'esso un polinomio, ovvero che la suddetta definizione d\`{a}
luogo ad una legge di prodotto regolare.

Nel caso pi\`{u} semplice, $P(y,\bar{y})=y_{\alpha}$,
$Q(y,\bar{y})=y_{\beta}$, si ottiene:
\begin{equation}\label{contrule1}
y_{\alpha}\star
y_{\beta}=y_{\alpha}y_{\beta}+i\varepsilon_{\alpha\beta} \ ,
\end{equation}
che implica
\begin{equation}
[y_{\alpha},y_{\beta}]_{\star}=y_{\alpha}\star
y_{\beta}-y_{\beta}\star y_{\alpha}=2i\varepsilon_{\alpha\beta}\ ,
\end{equation}
e analogamente si possono provare le relazioni
\begin{equation}\label{contrule2}
\bar{y}_{\dot{\alpha}}\star\bar{y}_{\dot{\beta}}=\bar{y}_{\dot{\alpha}}\bar{y}_{\dot{\beta}}+i\varepsilon_{\dot{\alpha}\dot{\beta}}
\ ,\qquad
y_{\alpha}\star\bar{y}_{\dot{\beta}}=y_{\alpha}\bar{y}_{\dot{\beta}}
\ ,
\end{equation}
che conducono alle
\begin{equation}
[\bar{y}_{\dot{\alpha}},\bar{y}_{\dot{\beta}}]_{\star}=2i\varepsilon_{\dot{\alpha}\dot{\beta}}
\ ,\qquad [y_{\alpha},\bar{y}_{\dot{\beta}}]_{\star}=0 \ ,
\end{equation}
tutti casi particolari delle regole di contrazione generali (cfr.
anche (\ref{compactcontrule}))
\begin{eqnarray}\label{contrule}
y_{\alpha_{1}...\alpha_{n}}\star
y_{\beta_{1}...\beta_{m}} = y_{\alpha_{1}...\alpha_{n}\beta_{1}...\beta_{m}}+i\,n\,m\,y_{\alpha_{1}...\alpha_{n-1}\beta_{1}...\beta_{m-1}}\,\varepsilon_{\alpha_{n}\beta_{m}}{}\nonumber\\
 {}+i^{2}\frac{n(n-1)m(m-1)}{2!}\,y_{\alpha_{1}...\alpha_{n-2}\beta_{1}...\beta_{m-2}}\,\varepsilon_{\alpha_{n-1}\beta_{m-1}}\varepsilon_{\alpha_{n}\beta_{m}}+...{}\nonumber\\
 {}+i^{k}k!\,{n \choose k}\,{m \choose
k}\,y_{\alpha_{1}...\alpha_{n-k}\beta_{1}...\beta_{m-k}}\,\varepsilon_{\alpha_{n-k+1}\beta_{m-k+1}}...\varepsilon_{\alpha_{n}\beta_{m}}+...
\ ,
\end{eqnarray}
\begin{equation}
(y_{\alpha_{1}...\alpha_{n}}\bar{y}_{\dot{\alpha}_{1}...\dot{\alpha}_{m}})\star(y_{\beta_{1}...\beta_{p}}\bar{y}_{\dot{\beta}_{1}...\dot{\beta}_{q}})=(y_{\alpha_{1}...\alpha_{n}}\star
y_{\beta_{1}...\beta_{p}})(\bar{y}_{\dot{\alpha}_{1}...\dot{\alpha}_{m}}\star
\bar{y}_{\dot{\beta}_{1}...\dot{\beta}_{q}})\ .
\end{equation}
Lo $\star$-prodotto descrive dunque il prodotto di polinomi
Weyl-ordinati negli oscillatori in termini dei simboli degli
operatori, come testimoniato dalla struttura del secondo membro di
(\ref{contrule}): questo presenta infatti un termine Weyl-ordinato
(il prodotto ordinario degli $y,\bar{y}$ \`{e} totalmente
simmetrico per costruzione) che contiene la potenza pi\`{u} alta,
seguito da altri termini Weyl-ordinati di grado via via pi\`{u}
basso derivanti da tutte le possibili contrazioni degli n
oscillatori $y_{\alpha_{i}}$ con gli m $y_{\beta_{j}}$ a coppie,
secondo le (\ref{contrule1}) e (\ref{contrule2}).

In altri termini, la $\star$-algebra ${\cal A}^{\star}$ di
funzioni $P(y,\bar{y})$ delle variabili commutanti $y,\bar{y}$
\`{e} isomorfa all'algebra ${\cal A}$ di funzioni Weyl-ordinate
$\hat{P}(\hat{y},\hat{\bar{y}})$ degli operatori
$\hat{y},\hat{\bar{y}}$, a loro volta soddisfacenti l'algebra di
Heisenberg (\ref{Heisalg}), ottenute dalle $P(y,\bar{y})$
sostituendo $y\rightarrow\hat{y}$.

Possiamo ora mostrare come l'algebra ${\cal A}$ e le algebre di HS
siano infinito-dimensionali. Notiamo anzitutto che lo spin $s$ del
generatore che costituisce il termine generico dell'espansione
(\ref{ply2}), $y_{\alpha (n)}\bar{y}_{\dot{\beta}(m)}$, \`{e} dato
da $s=\frac{n+m}{2}$. Dati allora due generatori $P_{s}$ e
$P_{s'}$, di spin $s,s'\geq 1$ rispettivamente \footnote{Il caso
in cui almeno uno dei due generatori ha spin $0$ \`{e} invece
banale, poich\'{e} non d\`{a} luogo a contrazioni ed il
commutatore \`{e} nullo.}, il loro $\star$-commutatore d\`{a}
\begin{equation}\label{infdim}
[P_{s},P_{s'}]_{\star}=P_{s+s'-1}+P_{s+s'-3}+P_{s+s'-5}+...+P_{|s-s'|+1}
\ ,
\end{equation}
come \`{e} evidente da (\ref{contrule}) considerando che nel
commutatore si elidono i termini con un numero pari di
contrazioni. Questo implica che:
\begin{enumerate}
\item il commutatore di due generatori almeno uno dei quali abbia
spin $s>1$ produce generatori di spin superiore e inferiore a
quelli di partenza, ovvero l'algebra non si chiude se non
includendo tutti gli spin. Questo fatto, ricavato qui per via
puramente algebrica, \`{e} importantissimo perch\'{e} riflette i
risultati di \cite{Coleman:ad, Bengtsson:1983pd,
Bengtsson:1983pg}: per costruire una teoria di HS consistente
\`{e} necessario includere campi di spin arbitrariamente elevato.
Inoltre, questo risultato mostra una stimolante connessione con la
Teoria delle Stringhe, che descrive naturalmente infinite
eccitazioni (massive) di spin via via crescente.

\item se $s,s'\leq 1$ il commutatore non produce spin pi\`{u}
alti: le simmetrie di gauge di HS, infinito-dimensionali,
contengono quindi le simmetrie di gauge ordinarie come
sottoalgebra massimale finita. I generatori di tale sottoalgebra
hanno spin $0$, $1/2$ ed $1$, rispettivamente corrispondenti a
simmetrie di gauge interne, sia abeliane che non abeliane
\footnote {\`{E} possibile infatti dotare i generatori di spin $0$
di ulteriori indici interni legati ad opportune algebre di
simmetria non abeliane compatibili con le simmetrie di HS, e
questo d\`{a} luogo alle teorie di YM come generalizzazioni della
teoria di Maxwell. Tuttavia, come per le stringhe aperte
\cite{Marcus:1982fr, Marcus:1986cm} (per una rassegna si veda
\cite{Angelantonj:2002ct}), questa estensione delle algebre di HS
si realizza mediante matrici di Chan-Paton e d\`{a} luogo solo
alle algebre classiche, ovvero ai gruppi di gauge $U(n)$, $O(n)$ e
$USp(2n)$.}, alla supersimmetria locale e all'invarianza sotto
$so(3,2)$.
\end{enumerate}
I generatori di spin $1/2$, le cariche di supersimmetria, sono
proporzionali ad un singolo oscillatore, $Q_{\alpha}\sim
y_{\alpha}$ \footnote {Questo vale per $N=1$. Si noti tuttavia il
ruolo importante giocato da altri due operatori, chiamati da
Vasiliev \emph{operatori di Klein} $k$ e $\bar{k}$, con le
propriet\`{a}
\begin{displaymath}
k^{2}=\bar{k}^{2}=1 \ ,\quad
\{k,\hat{y}_{\alpha}\}=\{\bar{k},\hat{\bar{y}}_{\dot{\alpha}}\}=0
\ ,
\end{displaymath}
nella costruzione di algebre di supersimmetria estese. Ad esempio,
le cariche di supersimmetria $N=2$ sono realizzate come
\begin{displaymath}
Q_{\alpha}^{1}=\hat{y}_{\alpha}\ ,\quad
Q_{\alpha}^{2}=ik\bar{k}\hat{y}_{\alpha}\ ,\quad
\bar{Q}_{\dot{\alpha}}^{1}=\hat{\bar{y}}_{\bar{\alpha}}\ ,\quad
\bar{Q}_{\dot{\alpha}}^{2}=ik\bar{k}\hat{\bar{y}}_{\bar{\alpha}} \
.
\end{displaymath}
}, mentre quelli di $so(3,2)$, in termini di simboli degli
operatori, diventano
\begin{equation}\label{genso3,2}
M_{\alpha\beta}=-\frac{1}{2}y_{\alpha}y_{\beta}\ ,\qquad
M_{\dot{\alpha}\dot{\beta}}=-\frac{1}{2}\bar{y}_{\dot{\alpha}}\bar{y}_{\dot{\beta}}\
,\qquad
P_{\alpha\dot{\alpha}}=\frac{1}{2}y_{\alpha}\bar{y}_{\dot{\alpha}}\
.
\end{equation}

Lo $\star$-prodotto ha anche una realizzazione integrale, come
\begin{equation}\label{defint}
(P\star Q)(Y)=\frac{1}{(2\pi)^{4}}\int d^{4}U d^{4}V
P(Y+U)Q(Y+V)\exp
i(U^{\underline{\alpha}}C_{\underline{\alpha\beta}}V^{\underline{\beta}})
\ ,
\end{equation}
dove $Y,U,V$ sono spinori di Majorana di $so(3,2)$ tali che
$Y_{\underline{\alpha}}=(y_{\alpha},\bar{y}^{\dot{\alpha}})$, e
analogamente per gli altri, mentre $C$ \`{e} la matrice di
coniugazione di carica,
\begin{eqnarray}
C_{\underline{\alpha\beta}}= \left(
\begin{array}{ccc} \varepsilon_{\alpha\beta} & 0  \\
0 & \varepsilon^{\dot{\alpha}\dot{\beta}} \end{array} \right) \ .
\end{eqnarray}
Le definizioni (\ref{defdiff}) e (\ref{defint}) sono completamente
equivalenti. Dimostriamo, ad esempio, come la seconda segua dalla
prima, esprimendo anzitutto $P(Y)$ e $Q(Y)$ in forma integrale,
\begin{eqnarray}
P(Y) & = &
\int\frac{d^{4}\xi}{(2\pi)^{4}}e^{i\xi^{\underline{\alpha}}C_{\underline{\alpha\beta}}Y^{\underline{\beta}}}\tilde{P}(\xi)
\ , \nonumber \\
Q(Y) & = &
\int\frac{d^{4}\xi}{(2\pi)^{4}}e^{i\xi^{\underline{\alpha}}C_{\underline{\alpha\beta}}Y^{\underline{\beta}}}\tilde{Q}(\xi)
\ , \label{antiFT}
\end{eqnarray}
dove $\tilde{P}(\xi)$ e $\tilde{Q}(\xi)$ sono le loro trasformate
di Fourier nella variabile $\xi$, uno spinore di Majorana
anch'esso,
\begin{eqnarray}
\tilde{P}(\xi) & = & \int d^{4}Y
e^{-i\xi^{\underline{\alpha}}C_{\underline{\alpha\beta}}Y^{\underline{\beta}}}P(Y)
\ , \nonumber \\
\tilde{Q}(\xi) & = & \int d^{4}Y
e^{-i\xi^{\underline{\alpha}}C_{\underline{\alpha\beta}}Y^{\underline{\beta}}}Q(Y)
\ . \label{FT}
\end{eqnarray}
Sostituendo le (\ref{antiFT}) nella (\ref{defdiff}) si ottiene
\begin{equation}
P(Y)\star Q(Y)=\int\frac{d^{4}\xi
d^{4}\eta}{(2\pi)^{8}}e^{i\xi^{\underline{\alpha}}C_{\underline{\alpha\beta}}Y^{\underline{\beta}}}e^{-i\xi^{\underline{\gamma}}C_{\underline{\gamma\delta}}Y^{\underline{\delta}}}e^{i\eta^{\underline{\rho}}C_{\underline{\rho\sigma}}Y^{\underline{\sigma}}}
\tilde{P}(\xi)\tilde{Q}(\eta) \ .
\end{equation}
Inserendo in quest'ultima espressione le (\ref{FT}), con variabili
di integrazione $U$ e $V$, e facendo uso di
\begin{equation}
\int\frac{d^{4}\xi}{(2\pi)^{4}}e^{i\xi^{\underline{\alpha}}C_{\underline{\alpha\beta}}(Y^{\underline{\beta}}-Z^{\underline{\beta}})}=\delta(Y-Z)
\ ,
\end{equation}
\begin{equation}
\int d^{4}Z\delta(Y-Z)F(Z)=F(Y) \ ,
\end{equation}
si arriva infine a
\begin{equation}
P(Y)\star Q(Y)=\int\frac{d^{4}U
d^{4}V}{(2\pi)^{4}}e^{i(Y-U)^{\underline{\alpha}}C_{\underline{\alpha\beta}}(Y-V)^{\underline{\beta}}}
P(U)Q(V) \ ,
\end{equation}
riconducibile alla (\ref{defint}) attraverso il cambio di
variabili $U\rightarrow U+Y$, $V\rightarrow V+Y$.

Possiamo introdurre la 1-forma di connessione dell'algebra di HS
$shs(4)$ (\emph{master 1-form})
\begin{eqnarray}\label{master 1-form}
w(x|y,\bar{y})& = & dx^{\mu}w_{\mu}(x|y,\bar{y}) \nonumber \\
& = &
\sum_{n,m=0}^{\infty}\frac{i}{2n!m!}dx^{\mu}w_{\mu}\,^{\alpha_{1}...\alpha_{n}\dot{\alpha}_{1}...\dot{\alpha}_{m}}(x)y_{\alpha_{1}}...y_{\alpha_{n}}\bar{y}_{\dot{\alpha}_{1}}...\bar{y}_{\dot{\alpha}_{m}}
\ ,
\end{eqnarray}
dove la componente
$w_{\mu}\,^{\alpha_{1}...\alpha_{n}\dot{\alpha_{1}}...\dot{\alpha_{m}}}(x)$
corrisponde, come vedremo in dettaglio in seguito, ad un campo di
gauge di spin $1+\frac{n+m}{2}$, intero se $n+m$ \`{e} pari e
semi-intero se \`{e} dispari. I campi con spin semi-intero,
fermionici, sono anticommutanti, secondo l' usuale relazione tra
spin e statistica.

Costruiamo anche la 2-forma di curvatura
\begin{equation}
R(x|y,\bar{y})=dw(x|y,\bar{y})+w(x|y,\bar{y})\wedge\star
w(x|y,\bar{y})\ ,
\end{equation}
dove l'ultimo termine a secondo membro \`{e} generato dalla non
commutativit\`{a} dello $\star$-prodotto e si traduce
automaticamente in un anticommutatore, anzich\'{e} in un
commutatore, quando coinvolge due componenti fermioniche dello
sviluppo in oscillatori.

Possiamo ora dare una definizione pi\`{u} precisa dell'algebra di
HS $shs(4)$: essa \`{e} la $\star$-(super)algebra di Lie
costituita dal sottospazio di ${\cal A}^{\star}$ formato dagli
elementi $P_{i}$ antihermitiani,
\begin{equation}
P^{\dagger}_{i}=-P_{i} \ ,
\end{equation}
(dove la coniugazione hermitiana agisce come un'anti-involuzione
della $\star$-algebra, $(F\star G)^{\dag}=G^{\dag}\star F^{\dag}$)
e la cui parentesi di Lie \`{e} data dallo $\star$-commutatore
\begin{equation}
[P_{i},P_{j}]_{\star}=P_{i}\star P_{j}-P_{j}\star P_{i} \ .
\end{equation}

Si \`{e} visto come l'algebra di HS sia un'estensione
infinito-dimensionale dell'algebra di $AdS_{4}$ che ha per
generatori polinomi di grado arbitrario negli oscillatori. La
(\ref{infdim}) indica in particolare che, per descrivere un campo
di spin $s$, si deve introdurre l'intera collezione di 1-forme
$w^{\alpha_{1}...\alpha_{n}\dot{\alpha_{1}}...\dot{\alpha_{m}}}(x)$
con spin arbitrario, legato ad $n$ ed $m$ da: $n+m=2(s-1)$. Non
tutti questi campi portano gradi di libert\`{a} fisici: tramite
alcuni vincoli, \`{e} infatti possibile determinare algebricamente
alcuni campi, detti \emph{ausiliari}, in termini di altri, detti
\emph{fisici} o \emph{dinamici}, analogamente a quanto accade in
gravit\`{a}, dove il vincolo di torsione (\ref{torsion}) consente
di esprimere la connessione di spin in termini del vielbein.

Questo complica notevolmente la dinamica, e tuttora la
formulazione di una teoria di gauge di HS coerente \`{e} nota
soltanto a livello delle equazioni del moto, formulate in termini
degli infiniti campi inclusi in $w(x|y,\bar{y})$ come un sistema
di infinite equazioni differenziali di primo ordine tra loro
consistenti, secondo lo schema delle cosiddette \emph{free
differential algebras} (o \emph{sistemi integrabili di Cartan}).

Esaminiamo tale formulazione partendo dalla teoria libera e
seguendo l'analogia con la gravit\`{a} in $AdS_{4}$ da cui siamo
partiti.

\section{Gravit\`a con costante cosmologica}
\label{4.2}

Le equazioni di Einstein che descrivono il campo gravitazionale
con costante cosmologica implicano che il tensore di Ricci si
annulla a meno di un termine costante proporzionale alla costante
cosmologica stessa, ovvero che le sole componenti non nulle del
tensore di Riemann appartengono al tensore di Weyl. Nel formalismo
a due componenti, il tensore di Riemann generalizzato \emph{\'{a}
la} MacDowell-Mansouri come nella (\ref{genRiem}) si spezza nelle
$2$-forme di curvatura
\begin{eqnarray}
\Re_{\alpha_{1}\alpha_{2}} & = &
d\omega_{\alpha_{1}\alpha_{2}}+\omega_{\alpha_{1}}\,^{\gamma}\wedge\omega_{\alpha_{2}\gamma}+L^{-2}\,
e_{\alpha_{1}}\,^{\dot{\gamma}}\wedge e_{\alpha_{2}\dot{\gamma}} \ ,\\
\bar{\Re}_{\dot{\alpha}_{1}\dot{\alpha}_{2}} & = &
d\bar{\omega}_{\dot{\alpha}_{1}\dot{\alpha}_{2}}+\bar{\omega}_{\dot{\alpha}_{1}}\,^{\dot{\gamma}}\wedge\bar{\omega}_{\dot{\alpha}_{2}\dot{\gamma}}+L^{-2}\,
e^{\gamma}\,_{\dot{\alpha}_{1}}\wedge e_{\gamma\dot{\alpha}_{2}}\
,
\end{eqnarray}
mentre il tensore di Weyl \`{e} descritto dai due multispinori
($0$-forme) $C_{\alpha_{1}\alpha_{2}\alpha_{3}\alpha_{4}}$ e
$\bar{C}_{\dot{\alpha}_{1}\dot{\alpha}_{2}\dot{\alpha}_{3}\dot{\alpha}_{4}}$,
tra loro complessi coniugati e corrispondenti rispettivamente alle
rappresentazioni $(2,0)$ e $(0,2)$ del gruppo di Lorentz.

Le equazioni di Einstein con costante cosmologica assumono quindi
la forma:
\begin{eqnarray}
T_{\alpha\dot{\beta}} & = & 0 \ , \label{torsion2}\\
\Re_{\alpha_{1}\alpha_{2}} & = & e^{\beta_{1}\dot{\gamma}}\wedge
e^{\beta_{2}}\,_{\dot{\gamma}}\,C_{\alpha_{1}\alpha_{2}\beta_{1}\beta_{2}} \ , \label{einstein2a}\\
\bar{\Re}_{\dot{\alpha}_{1}\dot{\alpha}_{2}} & = &
e^{\gamma\dot{\beta}_{1}}\wedge
e_{\gamma}\,^{\dot{\beta}_{2}}\,\bar{C}_{\dot{\alpha}_{1}\dot{\alpha}_{2}\dot{\beta}_{1}\dot{\beta}_{2}}
\ . \label{einstein2b}
\end{eqnarray}
Affinch\'{e} queste siano tra loro consistenti, \`{e} necessario
che le $0$-forme $C_{\alpha(4)}$ e $\bar{C}_{\dot{\alpha}(4)}$
soddisfino alcune relazioni differenziali, come conseguenza delle
identit\`{a} di Bianchi (\ref{idB}) per le curvature $\Re$ e
$\bar{\Re}$ che compaiono a primo membro,
\begin{eqnarray}
\nabla C_{\alpha(4)} & = &
e^{\beta\dot{\gamma}}\,C_{\alpha(4)\beta\dot{\gamma}} \ , \label{1aofchain}\\
\nabla\bar{C}_{\dot{\alpha}(4)} & = &
e^{\beta\dot{\gamma}}\,\bar{C}_{\beta\dot{\alpha}(4)\dot{\gamma}}
\ , \label{1bofchain}
\end{eqnarray}
dove $C_{\alpha(4)\beta\dot{\gamma}}\equiv
C_{\alpha(5)\dot{\gamma}}$ e
$\bar{C}_{\beta\dot{\alpha}(4)\dot{\gamma}}\equiv
\bar{C}_{\beta\dot{\alpha}(5)}$ sono nuovi campi multispinoriali
totalmente simmetrici separatamente negli indici puntati e non
puntati , ovvero appartenenti alle rappresentazioni $(5/2, 1/2)$ e
$(1/2,5/2)$, rispettivamente. In queste equazioni
$\nabla=e^{\alpha\dot{\alpha}}\nabla_{\alpha\dot{\alpha}}$ \`{e}
la derivata esterna Lorentz-covariante, che agisce su un generico
tensore di Lorentz $A_{\alpha...\dot{\beta}...}$ come
\begin{equation}
\nabla
A_{\alpha...\dot{\beta}...}=dA_{\alpha...\dot{\beta}...}+\omega_{\alpha}\,^{\gamma}\wedge
A_{\gamma...\dot{\beta}...}+...+\bar{\omega}_{\dot{\beta}}\,^{\dot{\gamma}}\wedge
A_{\alpha...\dot{\gamma}...}+...\ .
\end{equation}
Anche le nuove $0$-forme $C_{\alpha(5)\dot{\gamma}}$ e
$\bar{C}_{\beta\dot{\alpha}(5)}$ devono essere soggette ad un
vincolo, per mantenere la consistenza del primo membro delle
(\ref{1aofchain}) e (\ref{1bofchain}) con l'identit\`{a} di
Bianchi. Tenendo presente che
\begin{equation}
\nabla\wedge\nabla
A_{\alpha...\dot{\beta}...}=\frac{1}{2}R_{\alpha}\,^{\gamma}A_{\gamma...\dot{\beta}...}+...+\frac{1}{2}\bar{R}_{\dot{\beta}}\,^{\dot{\gamma}}A_{\alpha...\dot{\gamma}...}+...
\ ,
\end{equation}
dove $R$ ed $\bar{R}$ sono stati definiti nelle (\ref{defriem1}) e
(\ref{defriem2}), esprimendo questi ultimi in termini di $\Re$ e
$\bar{\Re}$ come
\begin{eqnarray}
R_{\alpha\beta} & = &
\Re_{\alpha\beta}-2L^{-2}e_{\alpha}\,^{\dot{\gamma}}\wedge
e_{\beta\dot{\gamma}} \ ,\\
\bar{R}_{\dot{\alpha}\dot{\beta}} & = &
\bar{\Re}_{\dot{\alpha}\dot{\beta}}-2L^{-2}e^{\gamma}\,_{\dot{\alpha}}\wedge
e_{\gamma\dot{\beta}} \ ,
\end{eqnarray}
dove $\Re_{\alpha\beta}$ ed $\bar{\Re}_{\dot{\alpha}\dot{\beta}}$
non contribuiscono che a livello non lineare, ed usando infine le
equazioni di Einstein, tale vincolo pu\`{o} essere scritto
\begin{eqnarray}
\nabla C_{\alpha(5)\dot{\beta}} & = &
e^{\beta\dot{\gamma}}\,C_{\alpha(5)\beta\dot{\beta}\dot{\gamma}}-5L^{-2}e_{\{\alpha\dot{\beta}}\,C_{\alpha(4)\}_{\alpha\dot{\beta}}}\,+O(C^{2}) \ ,\\
\nabla \bar{C}_{\alpha\dot{\beta}(5)} & = &
e^{\beta\dot{\gamma}}\,\bar{C}_{\alpha\beta\dot{\beta}(5)\dot{\gamma}}-5L^{-2}e_{\{\alpha\dot{\beta}}\,\bar{C}_{\dot{\beta}(4)\}_{\alpha\dot{\beta}}}\,+O(\bar{C}^{2})
\ ,
\end{eqnarray}
dove sono state introdotte le variabili
$C_{\alpha(5)\beta\dot{\beta}\dot{\gamma}}\equiv
C_{\alpha(6)\dot{\beta}(2)}$ e
$\bar{C}_{\alpha\beta\dot{\beta}(5)\dot{\gamma}}\equiv
\bar{C}_{\alpha(2)\dot{\beta}(6)}$ e sono stati trascurati termini
non lineari, dal momento che siamo per ora interessati a costruire
la teoria linearizzata. La notazione $\{...\}_{\alpha\dot{\beta}}$
indica che gli indici $\alpha$ e $\dot{\beta}$ sono simmetrizzati
con gli altri indici dello stesso tipo inclusi all'interno della
parentesi. La normalizzazione all'unit\`{a} che la
simmetrizzazione porta con s\'{e} d\`{a} ragione del fattore che
compare davanti al secondo termine a secondo membro.

Iterando tale procedimento si giunge alla catena di infinite
equazioni differenziali
\begin{eqnarray}
\nabla C_{\alpha(n+4)\dot{\beta}(n)} & = &
e^{\beta\dot{\gamma}}\,C_{\alpha(n+4)\beta\dot{\beta}(n)\dot{\gamma}} \nonumber \\
 & & {} -n(n+4)L^{-2}e_{\{\alpha\dot{\beta}}\,C_{\alpha(n+3)\dot{\beta}(n-1) \}_{\alpha\dot{\beta}}}+O(C^{2}) \ , \label{naofchain}\\
\nabla \bar{C}_{\alpha(n)\dot{\beta}(n+4)} & = &
e^{\beta\dot{\gamma}}\,\bar{C}_{\alpha(n)\beta\dot{\beta}(n+4)\dot{\gamma}}
\nonumber \\
 & & {} -n(n+4)L^{-2}e_{\{\alpha\dot{\beta}}\,\bar{C}_{\alpha(n-1)\dot{\beta}(n+3)\}_{\alpha\dot{\beta}}}+O(\bar{C}^{2})\ .
\label{nbofchain}
\end{eqnarray}
Ogni equazione rappresenta la condizione di consistenza della
precedente, in virt\`{u} dell'identit\`{a} di Bianchi, e non
aggiunge quindi informazioni sulla dinamica a quelle contenute
nelle equazioni di Einstein. Il sistema di equazioni
(\ref{torsion2}), (\ref{einstein2a}), (\ref{einstein2b}),
(\ref{naofchain}) e (\ref{nbofchain}) \`{e} quindi dinamicamente
equivalente alle equazioni di Einstein con costante cosmologica.

Le (\ref{naofchain}) e (\ref{nbofchain}) legano tra loro le
$0$-forme $C_{\alpha(n)\dot{\beta}(m)}$ e
$\bar{C}_{\alpha(n)\dot{\beta}(m)}$ con $|n-m|=4$ in modo ben
preciso, e rendono possibile esprimere i multispinori di rango
pi\`{u} alto in termini di derivate di $C_{\alpha(4)}$ e
$\bar{C}_{\dot{\beta}(4)}$, da cui parte la catena. Tutte le
$0$-forme di diverso rango non sono pertanto indipendenti da
queste ultime, ma anzi formano una rappresentazione di dimensione
infinita dell'algebra $so(3,2)$, detta \emph{twisted adjoint}.

\section{Equazioni di HS linearizzate in $AdS_{4}$}
\label{4.3}

Vasiliev ha mostrato che la dinamica libera dei campi di HS
pu\`{o} essere codificata da una generalizzazione del sistema di
equazioni che descrive la gravit\`{a} in $AdS_{4}$ appena
esaminato, ottenuta sostituendo all'algebra di $so(3,2)$ l'algebra
$shs(4)$.

Riscriviamo anzitutto le equazioni di Einstein (\ref{torsion2}),
(\ref{einstein2a}) e (\ref{einstein2b}) in modo compatto come
\begin{equation}\label{eincompact}
\Re_{\alpha(n)\dot{\beta}(m)}=\delta_{n0}e^{\gamma\dot{\delta}}\wedge
e^{\gamma}\,_{\dot{\delta}}C_{\alpha(n)\gamma(2)}+\delta_{m0}e^{\eta\dot{\delta}}\wedge
e_{\eta}\,^{\dot{\delta}}\bar{C}_{\dot{\beta}(m)\dot{\delta}(2)} \
,
\end{equation}
con $|n+m|=2(s-1)=2$. Come abbiamo visto, il tensore di Weyl ha
$2s=4$ indici spinoriali, e prendendo $k$ derivate
Lorentz-covarianti del secondo membro vengono introdotte, nella
catena di equazioni (\ref{naofchain}) e (\ref{nbofchain}), nuove
$0$-forme con ulteriori coppie di indici $\alpha,\dot{\alpha}$,
del tipo $C_{\alpha(2s+k)\dot{\beta}(k)}$ e
$\bar{C}_{\alpha(k)\dot{\beta}(2s+k)}$. Questo schema si
generalizza a tutti gli spin $s\geq1$.

A tale scopo estendiamo la connessione per $so(3,2)$, definita
come
\begin{equation}\label{omega0}
\omega_{0}(x|y,\bar{y})=\frac{i}{8}[\omega_{0\alpha\beta}y^{\alpha}y^{\beta}+\bar{\omega}_{0\dot{\alpha}\dot{\beta}}\bar{y}^{\dot{\alpha}}\bar{y}^{\dot{\beta}}+2L^{-1}e_{0\alpha\dot{\alpha}}y^{\alpha}\bar{y}^{\dot{\alpha}}]
\end{equation}
ad una generica connessione $w$ a valori nella $\star$-algebra
$shs(4)$, come nella definizione (\ref{master 1-form}).

Inoltre, la \emph{master $0$-form}
\begin{equation}\label{master $0$-form}
C(x|y,\bar{y})=\sum_{n,m=0}^{\infty}\frac{1}{n!m!}C^{\alpha_{1}...\alpha_{n}\dot{\alpha}_{1}...\dot{\alpha}_{m}}(x)\,y_{\alpha_{1}}...y_{\alpha_{n}}\bar{y}_{\dot{\alpha}_{1}}...\bar{y}_{\dot{\alpha}_{n}}
\end{equation}
contiene il tensore di Weyl di spin $2$, i campi fisici di spin
$0$ e $1/2$, i tensori di Weyl generalizzati $C_{\alpha(2s)}$,
$\bar{C}_{\dot{\alpha}(2s)}$ e le loro derivate. Un'espansione
perturbativa in potenze di $C$, ovvero attorno alla soluzione di
vuoto $AdS$, pu\`{o} essere ottenuta ponendo
\begin{equation}
w  =  w_{0}+w_{1} \ ,
\end{equation}
dove $w_{0}\equiv\omega_{0}$ e $w_{1}\equiv\omega$ contiene le
fluttuazioni di tutti i campi di spin arbitrario rispetto al
background.

Si noti che $\omega_{0}$ \`{e} una soluzione particolare
dell'equazione di curvatura nulla
\begin{equation}\label{0curv}
d\Omega+\Omega\star\Omega=0 \ ,
\end{equation}
(dove il prodotto esterno $\wedge$ \`{e} sottinteso), che
generalizza le (\ref{torsion2cp}), (\ref{defads1}) e
(\ref{defads2}). La (\ref{0curv}) \`{e} invariante sotto le
trasformazioni di gauge di HS
\begin{equation}
\delta \Omega=D\epsilon=d\epsilon+[\Omega,\epsilon]_{\star} \ ,
\end{equation}
con $\epsilon(x|y,\bar{y})$ arbitrario elemento dell'algebra
locale di HS, sicch\'{e} la soluzione di vuoto $AdS$ \`{e}
invariante sotto trasformazioni di gauge di parametro
$\epsilon_{0}$ tali che
\begin{equation}\label{genKillingeq}
\delta\omega_{0}=D_{0}\epsilon_{0}=d\epsilon_{0}+[\omega_{0},\epsilon_{0}]_{\star}=0
\ ,
\end{equation}
un'equazione di Killing generalizzata. Si pu\`{o} mostrare
\cite{Sezgin:1998gg} che le soluzioni di questa equazione generano
un'algebra isomorfa a $shs(4)$, ovvero che la soluzione di vuoto
$AdS$ ammette in effetti l'intera simmetria $shs(4)$ e non
soltanto $so(3,2)$.

Le equazioni linearizzate di HS che generalizzano le equazioni di
Einstein (\ref{eincompact}) sono riassunte in
\begin{equation}\label{cplinconstr1}
R_{1\,\alpha(n)\dot{\beta}(m)}=\delta_{m0}e^{\gamma\dot{\delta}}\wedge
e^{\gamma}\,_{\dot{\delta}}C_{\alpha(n)\gamma(2)}+\delta_{n0}e^{\eta\dot{\delta}}\wedge
e_{\eta}\,^{\dot{\delta}}\bar{C}_{\dot{\beta}(m)\dot{\delta}(2)}\
,
\end{equation}
con $n+m=2(s-1)$. Le curvature $R_{1}^{\alpha(n)\dot{\beta}(m)}$
sono le componenti della curvatura di HS linearizzata
\begin{eqnarray}\label{firstlincurv}
R_{1}(x|y,\bar{y}) & \equiv &
d\omega(x|y,\bar{y})+\omega_{0}(x|y,\bar{y})\star\omega(x|y,\bar{y})+\omega(x|y,\bar{y})\star\omega_{0}(x|y,\bar{y}){}\nonumber\\
 {} & = &
 \sum_{n,m=0}^{\infty}\frac{i}{2n!m!}R_{1}\,^{\alpha_{1}...\alpha_{n}\dot{\beta}_{1}...\dot{\beta}_{m}}(x)\,y_{\alpha_{1}}...y_{\alpha_{n}}\bar{y}_{\dot{\beta}_{1}}...\bar{y}_{\dot{\beta}_{n}}\ ,
\end{eqnarray}
che soddisfa l'identit\`{a} di Bianchi
\begin{equation}\label{Bidsxlincurv}
dR_{1}=[R_{1},\omega_{0}]_{\star} \ .
\end{equation}
Sviluppando gli $\star$-prodotti (vedi Appendice \ref{appendice
B}), si ottiene
\begin{equation}\label{lincurv}
R_{1}(x|y,\bar{y})=\nabla_{0}\omega(x|y,\bar{y})-L^{-1}e_{0}\,^{\alpha\dot{\beta}}(y_{\alpha}\bar{\partial}_{\dot{\beta}}+\bar{y}_{\dot{\beta}}\partial_{\alpha})\omega(x|y,\bar{y})
\ ,
\end{equation}
dove $\nabla_{0}$ contiene la sola connessione di Lorentz. In
analogia con il caso di spin $2$, la condizione di
compatibilit\`{a} delle (\ref{cplinconstr1}) \`{e} costituita
dalla catena infinita di equazioni
\begin{equation}\label{cplinconstr2}
\nabla_{0} C_{\alpha(n)\dot{\beta}(m)}=
\frac{1}{2}e_{0}^{\beta\dot{\gamma}}\,C_{\alpha(n)\beta\dot{\beta}(m)\dot{\gamma}}-\frac{mn}{2}L^{-2}e_{0\{\alpha\dot{\beta}}\,C_{\alpha(n-1)\dot{\beta}(m-1)
\}_{\alpha\dot{\beta}}}\ ,
\end{equation}
con $|n-m|=2s$ (il valore assoluto include automaticamente la
catena ottenuta per complessa coniugazione). Come nel caso della
gravit\`{a}, tutte le $0$-forme sono esprimibili come derivate di
quelle che compaiono a secondo membro delle (\ref{cplinconstr1})
$C_{\alpha(2s)}$ e $\bar{C}_{\dot{\beta}(2s)}$, interpretabili
come tensori di Weyl generalizzati. La \emph{master 0-form} $C$,
che li contiene unitamente alle loro derivate e ai campi di
materia, realizza una rappresentazione dell'algebra $shs(4)$ detta
\emph{twisted adjoint}, che generalizza quella vista in precedenza
per il caso di spin $2$.

In termini della \emph{master $0$-form} $C$ le
(\ref{cplinconstr2}) si scrivono, in modo compatto,
\begin{equation}\label{linconstr2}
{\cal D}_{0}C\equiv dC+\omega_{0}\star
C-C\star\bar{\pi}(\omega_{0})=0 \ ,
\end{equation}
dove ${\cal D}_{0}$ \`{e} la derivata covariante per la
rappresentazione \emph{twisted adjoint}, caratterizzata dal
\emph{twisting} dato dalla mappa $\bar{\pi}$.

Le mappe $\pi$ e $\bar{\pi}$ possono essere definite in termini
della loro azione su un generico elemento dell'algebra $shs(4)$
\begin{equation}
\pi(P(y,\bar{y}))=P(-y,\bar{y})\ ,\quad
\bar{\pi}(Q(y,\bar{y}))=Q(y,-\bar{y})\ .
\end{equation}
Sono entrambe automorfismi involutivi della $\star$-algebra,
ovvero mappe iniettive e suriettive di ${\cal A}^{\star}$ in se
stessa, che rispettano la struttura di $\star$-algebra,
\begin{equation}
\pi(P\star Q)=\pi(P)\star\pi(Q)
\end{equation}
(analogamente per $\bar{\pi}$), e sono di ordine $2$,
\begin{equation}
\pi^{2}=\bar{\pi}^{2}=1 \ .
\end{equation}
I \emph{twists} $\pi$ e $\bar{\pi}$ invertono il segno dei
generatori delle traslazioni $AdS$, dispari in $y$ e $\bar{y}$, e
trasformano cos\`{i} il commutatore con il vielbein in un
anticommutatore, peculiarit\`{a} della \emph{twisted adjoint}
rispetto all'aggiunta:
\begin{equation}\label{d0}
{\cal D}_{0}C=\nabla_{0} C+\frac{i}{4}L^{-1} \{
e_{0\alpha\dot{\alpha}}y^{\alpha}\bar{y}^{\dot{\alpha}},C
\}_{\star}\ ,
\end{equation}
e hanno un simile effetto su tutti i generatori di $shs(4)$
associati a potenze dispari delle $y$ o $\bar{y}$. Calcolando gli
$\star$-prodotti il vincolo diventa
\begin{equation}\label{D0}
{\cal D}_{0}C=\nabla_{0}
C+\frac{i}{2}L^{-1}e_{0}\,^{\alpha\dot{\beta}}\left(y_{\alpha}\bar{y}_{\dot{\beta}}-\frac{\partial}{\partial
y^{\alpha}}\frac{\partial}{\partial \bar{y}^{\dot{\beta}}} \right
) C=0 \ ,
\end{equation}
poich\'{e} l'anticommutatore con due oscillatori produce soltanto
due contrazioni, e sostituendo infine lo sviluppo in componenti di
$C$ si ottiene
\begin{equation}\label{altercplinconstr2}
\nabla_{0} C_{\alpha(n)\dot{\beta}(m)}=\frac{i}{2}L^{-1}
e_{0}^{\beta\dot{\gamma}}\,C_{\alpha(n)\beta\dot{\beta}(m)\dot{\gamma}}-i\frac{mn}{2}L^{-1}e_{0\{\alpha\dot{\beta}}\,C_{\alpha(n-1)\dot{\beta}(m-1)
\}_{\alpha\dot{\beta}}}\ ,
\end{equation}
che differisce dalla (\ref{cplinconstr2}) nei soli fattori di
proporzionalit\`{a} $i$ ed $L$, e quindi \`{e} nuovamente la
condizione di compatibilit\`{a} delle (\ref{cplinconstr1})
\footnote {Come preciseremo ulteriormente in seguito, quest'ultima
propriet\`{a} dipende infatti dalla struttura indiciale comune
alle (\ref{cplinconstr2}) e (\ref{altercplinconstr2}), e non dai
fattori che moltiplicano i termini che le costituiscono.}.
Bench\'{e} in questo senso equivalente alla (\ref{cplinconstr2}),
riportiamo la (\ref{altercplinconstr2}) poich\'{e} la utilizzeremo
per ricavare le usuali equazioni del moto in $AdS_{4}$ dei diversi
spin. La prima pu\`{o} comunque essere ottenuta dalla
(\ref{linconstr2}) sostituendovi uno sviluppo in componenti della
\emph{master $0$-form} che includa opportunamente fattori di $L$,
come \cite{Bolotin:1999fa}
\begin{equation}\label{altermaster $0$-form}
C(x|y,\bar{y})=\sum_{n,m=0}^{\infty}\frac{L^{-\left(2-\frac{n+m}{2}\right)}}{n!m!}C^{\alpha_{1}...\alpha_{n}\dot{\alpha}_{1}...\dot{\alpha}_{m}}(x)\,y_{\alpha_{1}}...y_{\alpha_{n}}\bar{y}_{\dot{\alpha}_{1}}...\bar{y}_{\dot{\alpha}_{n}}
\ ,
\end{equation}
accompagnato da
\begin{equation}
\omega(x|y,\bar{y})=\sum_{n,m=0}^{\infty}\frac{iL^{-\left(1-\left|\frac{n-m}{2}\right|\right)}}{2n!m!}\omega^{\alpha_{1}...\alpha_{n}\dot{\alpha}_{1}...\dot{\alpha}_{m}}(x)\,y_{\alpha_{1}}...y_{\alpha_{n}}\bar{y}_{\dot{\alpha}_{1}}...\bar{y}_{\dot{\alpha}_{n}}
\ .
\end{equation}

La (\ref{linconstr2}) \`{e} invariante sotto la simmetria di
Killing
\begin{equation}\label{twKillingsym}
\delta C=-\epsilon_{0}\star C+C\star\bar{\pi}(\epsilon_{0})\ ,
\end{equation}
che costituisce un'altra definizione della \emph{twisted adjoint},
se il parametro $\epsilon_{0}$ soddisfa la (\ref{genKillingeq}).

Notiamo ancora che, oltre alle simmetrie di Killing
(\ref{genKillingeq}) e (\ref{twKillingsym}), le equazioni
(\ref{cplinconstr1}) e (\ref{linconstr2}) ammettono anche le
simmetrie di gauge dei vari campi fluttuanti rispetto al fondo
$AdS$,
\begin{equation}\label{gaugetransf}
\delta\omega=d\epsilon+[\omega_{0},\epsilon]_{\star}\ ,\qquad
\delta C=0 \ ,
\end{equation}
dove $\epsilon$ \`{e} un parametro locale arbitrario a valori
nell'algebra $shs(4)$.

Come osservato nell'esempio della gravit\`{a}, il sistema di
equazioni linearizzate (\ref{linconstr2}) si decompone in
sottosistemi indipendenti con $|n-m|$ fissato; ciascuno di questi
\`{e} costituito da infinite equazioni che, come verificheremo,
descrivono la dinamica di un campo di spin $s$ e massa nulla. Per
$s>1$, la (\ref{linconstr2}) \`{e} la condizione di consistenza
del sistema (\ref{cplinconstr1}). Per $s=1$ la (\ref{linconstr2})
costituisce invece un vincolo indipendente e contiene le equazioni
di Maxwell. Per i campi di materia di spin $s=0,1/2$ ovviamente
non esiste un analogo delle equazioni di Einstein, poich\'{e} non
esiste un campo di gauge ad essi associato; tuttavia le loro
equazioni sono contenute nei settori con $|n-m|=0,1$\,,
rispettivamente, della (\ref{linconstr2}).

\section{Esempi e classificazione dei campi HS}
\label{4.4}

Il modo in cui le usuali equazioni di moto libere di un campo di
spin $s$ sono incorporate in un sistema di infinite equazioni
coinvolgenti infiniti campi $C_{\alpha(n)\dot{\beta}(m)}$ con
$|n-m|=2s$, generalizzando quanto visto per spin $2$, \`{e}
singolare e merita di essere esaminato in dettaglio
\cite{Vasiliev:1999ba, Sezgin:1998eh, Sezgin:1998gg}.

Consideriamo anzitutto il caso di spin $s=0$, ovvero il settore
$|n-m|=0$ della (\ref{linconstr2}). Per $n=m=0$ la
(\ref{altercplinconstr2}) diventa (omettendo d'ora in poi l'indice
$0$)
\begin{equation}\label{1ofspin0}
C_{\alpha\dot{\beta}}=iLe^{\mu}_{\alpha\dot{\beta}}\nabla_{\mu}C \
,
\end{equation}
che semplicemente esprime $C_{\alpha\dot{\beta}}$ come derivata
covariante della $0$-forma $C$ di rango pi\`{u} basso del settore
in questione. Qui $e^{\mu}_{\alpha\dot{\beta}}$ \`{e} il vielbein
inverso, che ha le usuali propriet\`{a}
\begin{equation}\label{Vbprops}
e^{\mu}_{\alpha\dot{\beta}}\,e_{\mu,\gamma\dot{\delta}}=-2\varepsilon_{\alpha\gamma}\varepsilon_{\dot{\beta}\dot{\delta}}
\ ,\qquad
g^{\mu\nu}=-\frac{1}{2}e^{\mu}_{\alpha\dot{\beta}}\,e^{\nu,\alpha\dot{\beta}}
\ .
\end{equation}
La seconda equazione della catena, con $n=m=1$, \`{e}
\begin{equation}\label{2ofspin0}
\nabla_{\mu}
C_{\alpha\dot{\beta}}=\frac{i}{2}L^{-1}e_{\mu}^{\gamma\dot{\delta}}C_{\alpha\gamma\dot{\beta}\dot{\delta}}-\frac{i}{2}L^{-1}e_{\mu,\alpha\dot{\beta}}C
\ ,
\end{equation}
la quale, contraendo ovunque gli indici con
$e^{\mu,\alpha\dot{\beta}}$ e ricordando la totale simmetria delle
$0$-forme negli indici di ciascun tipo, conduce a
\begin{equation}
e^{\mu,\alpha\dot{\beta}}\nabla_{\mu}C_{\alpha\dot{\beta}}-4iL^{-1}C=0
\ .
\end{equation}
Usando infine la (\ref{1ofspin0}) otteniamo l'equazione di
Klein-Gordon in $AdS_{4}$
\begin{equation}
\Box C+2L^{-2}C=0 \ ,
\end{equation}
dove il termine di massa \`{e} quello associato ad un campo
scalare di ``massa nulla'' in $AdS_{4}$. La (\ref{2ofspin0}) non
contiene ulteriori informazioni sulla dinamica, ma lega soltanto
il campo $C_{\alpha\alpha\dot{\beta}\dot{\beta}}$ alla derivata
seconda di $C$:
\begin{equation}
C_{\alpha\alpha\dot{\beta}\dot{\beta}}=\left(iL\right)^{2}e^{\nu}_{\alpha\dot{\beta}}\nabla_{\nu}e^{\mu}_{\alpha\dot{\beta}}\nabla_{\mu}C
\ .
\end{equation}
Allo stesso modo, tutte le altre equazioni del settore scalare con
$n=m>1$ esprimono le $0$-forme di rango pi\`{u} alto della catena
in termini di derivate di ordine $n$ del campo $C$, senza imporre
su quest'ultimo nuove condizioni:
\begin{equation}
C_{\alpha(n)\dot{\beta}(n)}=\left(iL\right)^{n}e^{\mu_{1}}_{\alpha\dot{\beta}}\nabla_{\mu_{1}}...e^{\mu_{n}}_{\alpha\dot{\beta}}\nabla_{\mu_{n}}C
\ .
\end{equation}
Questo mostra come il sottosistema $n=m$ della (\ref{linconstr2})
descriva effettivamente un campo scalare ed abbia esattamente lo
stesso contenuto fisico dell'equazione di Klein-Gordon.
Riassumendo, un campo di spin $0$ \`{e} descritto da una
collezione infinita di $0$-forme totalmente simmetriche
separatamente nei due tipi di indici, soggette al vincolo di
costanza covariante (\ref{linconstr2}). Tale vincolo ha inoltre la
sua condizione di consistenza nell' equazione di curvatura nulla
(\ref{0curv}), la quale da un lato definisce, attraverso la sua
soluzione di vuoto $\omega_{0}$, la connessione che compare in
${\cal D}_{0}$, e dall'altro implica $({\cal D}_{0})^{2}=0$,
sicch\'{e} le successive equazioni della catena non forniscono
ulteriori condizioni. Questo equivale a dire che la collezione di
0-forme che tale catena coinvolge genera una rappresentazione
dell'intera algebra $shs(4)$ (ci\`{o} risulta a maggior ragione
evidente dal solo fatto che la (\ref{0curv}) implica la
consistenza formale di (\ref{linconstr2}) indipendentemente dalla
sua soluzione particolare $AdS$).

Il fatto che le $0$-forme che descrivono il campo siano soggette
al vincolo, puramente algebrico e non differenziale, di totale
simmetria negli indici spinoriali interni di ciascun tipo svolge
un ruolo cruciale nel codificare la dinamica del campo stesso.
\`{E} infatti soltanto in virt\`{u} di questa condizione che la
(\ref{2ofspin0}) contiene l'equazione di Klein-Gordon, come sopra
sottolineato. Questo fatto \`{e} importante e completamente
generale: senza questo vincolo algebrico, l'intero sistema
(\ref{linconstr2}) sarebbe ancora consistente ma dinamicamente
vuoto! I suoi diversi settori non conterrebbero altro che
equazioni che esprimono multispinori di rango sempre pi\`{u} alto
come derivate di alcune variabili dinamiche indipendenti. In
questo senso si pu\`{o} affermare che in questa formulazione i
vincoli algebrici sulle $0$-forme contengono l'informazione
dinamica fondamentale.

\vspace{0.5cm}

In modo analogo, un fermione di spin $1/2$ viene descritto dalla
catena di campi $C_{\alpha(n)\dot{\beta}(m)}$ con $|n-m|=1$. La
prima equazione \`{e} data dalla (\ref{altercplinconstr2}) per
$n=1,m=0$:
\begin{equation}\label{1ofspin1/2}
\nabla_{\mu}C_{\alpha}=\frac{i}{2}L^{-1}e_{\mu}^{\gamma\dot{\delta}}C_{\alpha\gamma\dot{\delta}}
\ .
\end{equation}
Grazie alla totale simmetria di $C_{\alpha\gamma\dot{\delta}}$
negli indici spinoriali $\alpha$ e $\gamma$, essa contiene
l'equazione di Dirac (o meglio, l'equazione di Weyl di uno spinore
sinistro $C_{\alpha}$, mentre quella per lo spinore destro
$C_{\dot{\alpha}}$ segue dalla complessa coniugata della
(\ref{1ofspin1/2})) in $AdS_{4}$, come diviene evidente contraendo
con $e^{\mu}_{\alpha\dot{\beta}}$ (in virt\`{u} della
(\ref{Vbprops}))
\begin{equation}
e^{\mu}_{\alpha\dot{\beta}}\nabla_{\mu}C^{\alpha}=0 \ ,
\end{equation}
mentre le restanti equazioni della catena non aggiungono ulteriori
condizioni sul campo fisico $C_{\alpha}$, limitandosi ad esprimere
attraverso derivate di ordine via via crescente di quest'ultimo i
tensori di rango pi\`{u} alto della catena.

\vspace{0.5cm}

Il caso di spin $s=1$ \`{e} il primo in cui le equazioni del moto
seguono dall'analisi combinata di (\ref{altercplinconstr2}) e
(\ref{cplinconstr1}), quest'ultima non essendo altro che la
definizione delle curvature linearizzate di spin $1$. La prima,
nel settore $|n-m|=2$ ed in particolare all'inizio della catena
($n=2$, $m=0$), fornisce
\begin{equation}\label{1ofspin1}
\nabla_{\alpha\dot{\alpha}}C_{\beta_{1}\beta_{2}}=-iL^{-1}C_{\alpha\beta_{1}\beta_{2}\dot{\alpha}}
\ ,
\end{equation}
insieme alla complessa coniugata ($n=0$, $m=2$), mentre la seconda
d\`{a}
\begin{equation}
R^{1}_{\mu\nu}=e^{\alpha\dot{\delta}}_{[\mu}
e^{\beta}_{\nu]}\,_{\dot{\delta}}C_{\alpha\beta}+e^{\eta\dot{\alpha}}_{[\mu}
e_{\nu]\eta}\,^{\dot{\beta}}\bar{C}_{\dot{\alpha}\dot{\beta}}\ .
\end{equation}
Prendendo la derivata Lorentz-covariante di ambo i membri,
ricordando il vincolo di torsione nulla e usando la
(\ref{1ofspin1}) si ottiene
\begin{equation}\label{eqinmezzo}
\nabla_{\rho}R^{1}_{\mu\nu}=-iL^{-1}
\{e^{\alpha\dot{\delta}}_{[\mu}
e^{\beta}_{\nu]}\,_{\dot{\delta}}e^{\gamma\dot{\gamma}}_{\rho}C_{\alpha\beta\gamma\dot{\gamma}}+e^{\eta\dot{\alpha}}_{[\mu}
e_{\nu]\eta}\,^{\dot{\beta}}e^{\gamma\dot{\gamma}}_{\rho}\bar{C}_{\gamma\dot{\alpha}\dot{\beta}\dot{\gamma}}\}
\ .
\end{equation}
Contraendo ambo i membri con $g^{\mu\rho}$ ed usando le
identit\`{a}
\begin{equation}\label{membraneids1}
g^{\mu\rho}e^{(\alpha |\dot{\delta}}_{[\mu}
e^{|\beta}_{\nu]}\,_{\dot{\delta}}e^{\gamma)\dot{\gamma}}_{\rho}=0
\ , \qquad g^{\mu\rho}e^{\eta (\dot{\alpha}}_{[\mu}
e_{\nu]\eta}\,^{\dot{\beta}|}e^{\gamma |\dot{\gamma})}_{\rho}=0 \
,
\end{equation}
si giunge infine all'equazione di moto per la curvatura
$R_{1}^{\mu\nu}$ di spin $1$,
\begin{equation}\label{eomspin1}
\nabla^{\mu}R^{1}_{\mu\nu}=0 \ .
\end{equation}
Notiamo che anche in questo caso la totale simmetria delle
componenti della \emph{master $0$-form} $C$ negli indici
spinoriali di ciascun tipo gioca un ruolo cruciale, rendendo
possibile l'utilizzo delle identit\`{a} sopra citate che annullano
il secondo membro delle (\ref{eqinmezzo}). Dalla (\ref{lincurv})
per $s=1$, tenendo conto che il generatore corrispondente al campo
di gauge di spin $1$ ha spin $0$, si ottiene
\begin{equation}
R^{1}_{\mu\nu}=\nabla_{[\mu}\omega_{\nu]} \ ,
\end{equation}
che, sostituita in (\ref{eomspin1}), determina le equazioni del
moto per la connessione di spin $1$ in $AdS_{4}$,
\begin{equation}
\Box\omega_{\nu}-\nabla_{\nu}(\nabla\cdot\omega)+\frac{3}{L^{2}}\omega_{\nu}=0
\ ,
\end{equation}
ove $\Box=\nabla^{\mu}\nabla_{\mu}$ e l'ultimo termine, contenente
la scala caratteristica di $AdS$, proviene da
$-[\nabla^{\mu},\nabla_{\nu}]\omega_{\mu}$, in cui il commutatore
produce il tensore di Ricci di $AdS_{4}$,
$R^{\mu}_{\nu}=-\frac{3}{L^{2}}\delta^{\mu}_{\nu}$. Il termine di
massa \`{e}, nuovamente, quello corretto per un campo di spin 1 e
``massa nulla'' in $AdS_{4}$. Contraendo ambo i membri della
(\ref{eqinmezzo}) con il tensore totalmente antisimmetrico
$\varepsilon^{\sigma\rho\mu\nu}$ ed usando le identit\`{a}
\begin{equation}\label{membraneids2}
\varepsilon^{\sigma\mu\nu\rho}e^{(\alpha |\dot{\delta}}_{[\mu}
e^{|\beta}_{\nu]}\,_{\dot{\delta}}e^{\gamma)\dot{\gamma}}_{\rho}=0
\ , \qquad \varepsilon^{\sigma\mu\nu\rho}e^{\eta
(\dot{\alpha}}_{[\mu} e_{\nu]\eta}\,^{\dot{\beta}|}e^{\gamma
|\dot{\gamma})}_{\rho}=0 \ ,
\end{equation}
si ottengono anche le identit\`{a} di Bianchi
\begin{equation}\label{Maxwomogenee}
\varepsilon^{\sigma\rho\mu\nu}\nabla_{\rho}R^{1}_{\mu\nu}=0 \ ,
\end{equation}
che completano le equazioni di Maxwell in $AdS_{4}$. Al solito, le
successive equazioni della catena non forniscono ulteriori
condizioni indipendenti sul campo fisico.

\vspace{0.5cm}

Da spin $s\geq 3/2$ in poi, le (\ref{linconstr2}), come osservato
nel caso della gravit\`{a}, non contengono vincoli indipendenti
dalle (\ref{cplinconstr1}), ma esprimono soltanto i multispinori
$C_{\alpha(n)\dot{\beta}(m)}$ come derivate dei tensori di Weyl di
HS, definiti dalle (\ref{cplinconstr1})
\begin{equation}\label{aux1}
C_{\alpha(n)\dot{\beta}(m)}=\left( iL
\right)^{\frac{1}{2}(n+m-2s)}
e^{\mu_{1}}_{\alpha\dot{\beta}}\nabla_{\mu_{1}}\ldots
e^{\mu_{\frac{1}{2}(n+m-2s)}}_{\alpha\dot{\beta}}\nabla_{\mu_{\frac{1}{2}(n+m-2s)}}C_{\alpha(2s)}
\ ,\quad n\geq m
\end{equation}
o
\begin{equation}\label{aux2}
C_{\alpha(n)\dot{\beta}(m)}=\left( iL
\right)^{\frac{1}{2}(n+m-2s)}
e^{\mu_{1}}_{\alpha\dot{\beta}}\nabla_{\mu_{1}}\ldots
e^{\mu_{\frac{1}{2}(n+m-2s)}}_{\alpha\dot{\beta}}\nabla_{\mu_{\frac{1}{2}(n+m-2s)}}C_{\dot{\beta}(2s)}\
,\quad m\geq n \ .
\end{equation}
Per spin $s\geq 3/2$ si procede nello stesso modo, anche se
l'analisi dei vincoli \`{e} in generale pi\`{u} complicata, e
l'eliminazione dei campi ausiliari richiede l'uso combinato dei
vincoli di torsione generalizzati (contenuti nelle
(\ref{cplinconstr1}), come visto nel caso della gravit\`{a}),
delle simmetrie di gauge e dell'identit\`{a} di Bianchi.

Esaminiamo questo procedimento nel caso di spin $s=3$. \`{E}
conveniente anzitutto decomporre i campi di gauge, le loro
derivate, le curvature e le trasformazioni di gauge in tensori
Lorentz-irriducibili $(n,m)$. I campi di gauge di spin $3$
contengono:
\begin{eqnarray}
\omega_{\mu\,\alpha(2)\dot{\beta}(2)} & = &
\omega(3,3)+\omega(3,1)+\omega(1,3)+\omega(1,1) \ ,\label{IRfields1} \\
\omega_{\mu\,\alpha(3)\dot{\beta}(1)} & = &
\omega(4,2)+\omega(4,0)+\omega(2,2)+\omega(2,0) \ ,\label{IRfields2} \\
\omega_{\mu\,\alpha(4)\dot{\beta}(0)} & = &
\omega(5,1)+\omega'(3,1) \label{IRfields3} \ ,\\
\omega_{\mu\,\alpha(1)\dot{\beta}(3)} & = &
\overline{\omega_{\mu\,\alpha(3)\dot{\beta}(1)}} \ ,\label{IRfields4} \\
\omega_{\mu\,\alpha(0)\dot{\beta}(4)} & = &
\overline{\omega_{\mu\,\alpha(4)\dot{\beta}(0)}} \
,\label{IRfields5}
\end{eqnarray}
dove $\overline{\omega_{\mu\,\alpha(n)\dot{\beta}(m)}}$ indica
naturalmente il complesso coniugato di
$\omega_{\mu\,\alpha(n)\dot{\beta}(m)}$ e la complessa
coniugazione scambia le etichette $(n,m)$ nelle rappresentazioni
irriducibili in cui decomponiamo i campi. Procediamo analogamente
con le curvature linearizzate di spin $3$, ricordando che
\begin{equation}\label{da appendix}
R^{1}_{\mu\nu\,\alpha(n)\dot{\beta}(m)}=-\frac{1}{4}(\sigma_{\mu\nu})^{\alpha\beta}R^{1}_{\alpha\beta,\alpha(n)\dot{\beta}(m)}-\frac{1}{4}(\bar{\sigma}_{\mu\nu})^{\dot{\alpha}\dot{\beta}}R^{1}_{\dot{\alpha}\dot{\beta},\alpha(n)\dot{\beta}(m)}\
,
\end{equation}
come segue dalla (\ref{utilexspin3b}). Si ottiene quindi
\begin{eqnarray}
R^{1}_{\alpha\beta,\alpha(2)\dot{\beta}(2)} & = &
R(4,2)+R(2,2)+R(0,2) \ ,\label{IRcurvs1} \\
R^{1}_{\dot{\alpha}\dot{\beta},\alpha(2)\dot{\beta}(2)} & = &
\overline{R^{1}_{\alpha\beta,\alpha(2)\dot{\beta}(2)}}\ ,
\label{IRcurvs2} \\
R^{1}_{\alpha\beta,\alpha(3)\dot{\beta}(1)} & = &
R(5,1)+R(3,1)+R(1,1) \ ,\label{IRcurvs3} \\
R^{1}_{\dot{\alpha}\dot{\beta},\alpha(3)\dot{\beta}(1)} & = &
R(3,3)+R'(3,1) \ ,\label{IRcurvs4} \\
R^{1}_{\alpha\beta,\alpha(4)\dot{\beta}(0)} & = &
R(6,0)+R(4,0)+R'(2,0) \ ,\label{IRcurvs5} \\
R^{1}_{\dot{\alpha}\dot{\beta},\alpha(4)\dot{\beta}(0)} & = &
R'(4,2) \ .\label{IRcurvs6}
\end{eqnarray}
Iniziamo la nostra analisi dalla componente $n=m=s-1=2$ della
(\ref{cplinconstr1}), ovvero dal vincolo
\begin{equation}
R^{1}_{\mu\nu\,\alpha(2)\dot{\beta}(2)}=2\nabla_{[\mu}\omega_{\nu]\alpha(2)\dot{\beta}(2)}+2e_{[\mu|\alpha}\,^{\dot{\alpha}}\omega_{|\nu]\alpha(1)\dot{\beta}(2)\dot{\alpha}}+2e_{[\mu|\dot{\alpha}}\,^{\beta}\omega_{|\nu]\alpha(2)\beta\dot{\beta}(1)}=0
\end{equation}
(ove stiamo sottintendendo, per semplicit\`{a} di notazione, la
totale simmetria tra indici spinoriali dello stesso tipo). Tenendo
presenti le (\ref{IRcurvs1}) e (\ref{IRcurvs2}) e chiamando
$\lambda(n,m)$ le rappresentazioni irriducibili contenute in
$\nabla_{[\mu}\omega_{\nu]\alpha(2)\dot{\beta}(2)}$, $\omega(n,m)$
ed $\bar{\omega}(n,m)$ quelle contenute in
$\omega_{\alpha\dot{\alpha},\beta(3)\dot{\beta}(1)}$ e
$\omega_{\alpha\dot{\alpha},\beta(1)\dot{\beta}(3)}$,
rispettivamente, possiamo riscrivere in modo schematico tale
vincolo come
\begin{eqnarray}
R(4,2) & = & \lambda(4,2)+\omega(4,2)=0 \ ,\\
R(2,2) & = & \lambda(2,2)+\omega(2,2)+\bar{\omega}(2,2)=0 \ ,\\
R(0,2) & = & \lambda(0,2)+\bar{\omega}(0,2)=0 \ ,
\end{eqnarray}
insieme alle complesse coniugate. Questo significa che il vincolo
$R^{1}_{\mu\nu\,\alpha(2)\dot{\beta}(2)}=0$ determina le
componenti Lorentz-irriducibili $\omega(4,2)$, $\omega(2,2)$,
$\omega(2,0)$ di $\omega_{\mu\,\alpha(3)\dot{\beta}(1)}$ (e le
corrispondenti complesse coniugate di
$\omega_{\mu\,\alpha(1)\dot{\beta}(3)}$) in termini di quelle del
rotore Lorentz-covariante di
$\omega_{\mu\,\alpha(2)\dot{\beta}(2)}$. Resta tuttavia
indeterminata la componente $\omega(4,0)$ ($\omega(0,4)$) di
$\omega_{\mu\,\alpha(3)\dot{\beta}(1)}$
($\omega_{\mu\,\alpha(1)\dot{\beta}(3)}$), che non compare nel
vincolo. Ma le equazioni (\ref{cplinconstr1}) e
(\ref{altercplinconstr2}) ammettono la simmetria di gauge
(\ref{gaugetransf}) e la trasformazione delle $1$-forme si
riscrive esplicitamente
\begin{equation}\label{cpgaugetransf}
\delta\omega_{\mu\,\alpha(n)\dot{\beta}(m)}=\nabla_{\mu}\epsilon_{\alpha(n)\dot{\beta}(m)}+n
e_{\mu\,\alpha}\,^{\dot{\alpha}}\,\epsilon_{\alpha(n-1)\dot{\beta}(m)\dot{\alpha}}+m
e_{\mu\,\dot{\alpha}}\,^{\beta}\,\epsilon_{\alpha(n)\beta\dot{\beta}(m-1)}
\end{equation}
dove la simmetrizzazione degli indici dello stesso tipo \`{e}
nuovamente implicita. Risulta quindi chiaro che la componente
indeterminata $\omega(4,0)$ ha una trasformazione di gauge
proporzionale al parametro $\epsilon(4,0)$, il quale, d'altra
parte, lascia inalterato il campo
$\omega_{\mu\,\alpha(2)\dot{\beta}(2)}$ (poich\'{e} quest'ultimo
non contiene alcuna rappresentazione irriducibile del tipo
$(4,0)$). In altri termini, $\omega(4,0)$ \`{e} un campo di
Stueckelberg, pura gauge, e pu\`{o} essere eliminato utilizzando
$\epsilon(4,0)$, il che equivale a porsi in una gauge ``fisica''.
Identiche considerazioni valgono per il caso complesso coniugato,
e tralasceremo d'ora in poi riferimenti espliciti ad esso.

A questo punto possiamo dire di aver interamente determinato il
campo $\omega_{\mu\,\alpha(3)\dot{\beta}(1)}$ in termini del
rotore del campo $\omega_{\mu\,\alpha(2)\dot{\beta}(2)}$, ovvero
di aver eliminato il campo ausiliario
$\omega_{\mu\,\alpha(3)\dot{\beta}(1)}$.

Ulteriori vincoli vengono forniti dall'identit\`{a} di Bianchi,
\begin{eqnarray}\label{explicitBid}
D_{[\mu}R^{1}_{\nu\rho]\,\alpha(n)\dot{\beta}(m)} & = &
\nabla_{[\mu}R^{1}_{\nu\rho]\,\alpha(n)\dot{\beta}(m)}+n
e_{[\mu|\,\alpha}\,^{\dot{\alpha}}R^{1}_{|\nu\rho]\,\alpha(n-1)\dot{\beta}(m)\dot{\alpha}}\nonumber \\
 &  & {}+m
e_{[\mu|\,\dot{\alpha}}\,^{\beta}R^{1}_{|\nu\rho]\,\alpha(n)\beta\dot{\beta}(m-1)}=0
\ ,
\end{eqnarray}
usando il precedente vincolo di curvatura
$R^{1}_{\mu\nu\,\alpha(2)\dot{\beta}(2)}=0$. Decomponiamo anche
l'identit\`{a} di Bianchi in termini di tensori irriducibili.
Moltiplicando ambo i membri per
$\varepsilon^{\lambda\mu\nu\rho}(\sigma_{\lambda})_{\gamma\dot{\gamma}}$
ed utilizzando le identit\`{a} (\ref{da appendix}),
(\ref{utilexspin3}) e (\ref{duality}), la (\ref{explicitBid}) si
pu\`{o} anzitutto riscrivere come
\begin{eqnarray}\label{spinorialBid}
& & \nabla_{\dot{\alpha}}\,^{\gamma}
R^{1}_{\gamma\alpha,\beta(n)\dot{\beta}(m)}-\nabla_{\alpha}\,^{\dot{\gamma}}R^{1}_{\dot{\gamma}\dot{\alpha},\beta(n)\dot{\beta}(m)}
\nonumber \\
 & = & {}
 n\left[R^{1}_{\alpha\beta(1),\beta(n-1)\dot{\alpha}\dot{\beta}(m)}-\varepsilon_{\alpha\beta(1)}R^{1}_{\dot{\alpha}}\,^{\dot{\gamma}}\,_{,\beta(n-1)\dot{\gamma}\dot{\beta}(m)}\right]\nonumber \\
 &  & {}
 -m\left[R^{1}_{\dot{\alpha}\dot{\beta}(1),\alpha\beta(n)\dot{\beta}(m-1)}-\varepsilon_{\dot{\alpha}\dot{\beta}(1)}R^{1}_{\alpha}\,^{\gamma}\,_{,\gamma\beta(n)\dot{\beta}(m-1)}\right] \
 .
\end{eqnarray}
Inoltre
\begin{eqnarray}
(DR^{1})_{\gamma\dot{\gamma},\alpha(2)\dot{\beta}(2)} & = &
DR(3,3)+DR(3,1)+\overline{DR}(1,3) \nonumber \\
 & & {} +DR(1,1)\ , \\
(DR^{1})_{\gamma\dot{\gamma},\alpha(3)\dot{\beta}(1)} & = &
DR(4,2)+DR(4,0)+DR(2,2) \nonumber \\
 & & {} +DR(2,0) \ ,\\
(DR^{1})_{\gamma\dot{\gamma},\alpha(4)\dot{\beta}(0)} & = &
DR(5,1)+DR(3,1)\ ,
\end{eqnarray}
e, come possiamo leggere dalla (\ref{spinorialBid}) sostituendovi
il vincolo di curvatura suddetto, le condizioni fornite
dall'identit\`{a} di Bianchi sono, schematicamente,
\begin{equation}
DR(3,3)=0\Rightarrow R'(3,3)-\bar{R}'(3,3)=0 \ ,\\
\end{equation}
\begin{equation}\label{clinBid}
DR(3,1)=0\Rightarrow R'(3,1)+R(3,1)=0 \ ,\\
\end{equation}
\begin{equation}
DR(1,1)=0\Rightarrow R(1,1)-\bar{R}(1,1)=0 \ ,\\
\end{equation}
da cui deduciamo, tra l'altro, che $R'(3,3)$ ed $R(1,1)$ sono
entrambi reali.

La decomposizione in tensori irriducibili dell'ulteriore vincolo
di curvatura \\
$R^{1}_{\mu\nu\,\alpha(3)\dot{\beta}(1)}=0$ si traduce nelle
condizioni
\begin{eqnarray}
R'(3,3) & = & 0 \ ,\label{eomdyn1} \\
R(1,1) & = & 0 \ ,\label{eomdyn2}\\
R'(3,1)+R(3,1) & = & 0 \ ,\label{eomdyn3}\\
R(5,1) & = & 0 \ .
\end{eqnarray}
Si pu\`{o} mostrare inoltre che la (\ref{eomdyn3}) annulla una
combinazione lineare di $R'(3,1)$ ed $R(3,1)$ indipendente da
quella che appare nell'identit\`{a} di Bianchi (in particolare
nella (\ref{clinBid})), rendendo in tal modo possibile annullare
separatamente i due tensori irriducibili. Esplicitando allora il
contenuto in rappresentazioni irriducibili dei campi di gauge dei
vincoli appena ottenuti,
\begin{eqnarray}
R(5,1) & = & \lambda(5,1)+\omega(5,1)=0 \ ,\\
R(1,1) & = & \lambda(1,1)+\omega(1,1)=0 \ ,\\
R'(3,1) & = & \lambda'(3,1)+\omega'(3,1)=0 \ ,\\
R(3,1) & = & \lambda(3,1)+\omega(3,1)=0 \ ,
\end{eqnarray}
\`{e} evidente che possiamo esprimere entrambe le componenti
irriducibili del campo $\omega_{\mu\,\alpha(4)\dot{\beta}(0)}$ in
termini di quelle del rotore di
$\omega_{\mu\,\alpha(3)\dot{\beta}(1)}$, e che queste ultime sono
a loro volta esprimibili in termini di quelle di
$\omega_{\mu\,\alpha(2)\dot{\beta}(2)}$. Scopriamo cos\`{i} che
anche $\omega_{\mu\,\alpha(4)\dot{\beta}(0)}$ \`{e} un campo
ausiliario, e $\omega_{\mu\,\alpha(2)\dot{\beta}(2)}$ \`{e} il
solo campo fisico, sul quale i rimanenti vincoli di curvatura non
pongono ulteriori condizioni. Tuttavia, non tutte le
rappresentazioni irriducibili in esso contenute corrispondono a
gradi di libert\`{a} fisici: basta infatti osservare la
trasformazione di gauge di
$\omega_{\mu\,\alpha(2)\dot{\beta}(2)}$, secondo la
(\ref{cpgaugetransf}), per notare che la componente $\omega(3,1)$
trasforma proporzionalmente al parametro $\epsilon(3,1)$ e pu\`{o}
dunque essere annullata.

Troviamo dunque che le componenti fisiche corrispondono ai tensori
irriducibili $\omega(3,3)$ ed $\omega(1,1)$, entrambi reali. Le
equazioni del moto per tali componenti dinamiche sono le
(\ref{eomdyn1}) e (\ref{eomdyn2}), che possono esser riscritte
come
\begin{equation}\label{eom}
e_{\{\dot{\beta}_{1}|}^{(0)\,\beta}\wedge
R^{1}_{\beta\alpha_{1}\alpha_{2}|\dot{\beta}_{2}\}_{\dot{\beta}}}=0
\ ,
\end{equation}
dove abbiamo esplicitato l'indice $0$ sinora soppresso.

Si pu\`{o} mostrare che la (\ref{eom}) corrisponde effettivamente
all'equazione di Fronsdal su $AdS_{4}$ per il campo di spin $3$
$\omega_{\mu\nu\rho}$, reale e totalmente simmetrico, dato da
\begin{equation}\label{totsymfield}
\omega_{\mu\nu\rho}=e_{(\nu}^{(0)\,\alpha_{1}\dot{\beta}_{1}}e_{\rho}^{(0)\,\alpha_{2}\dot{\beta}_{2}}\omega_{\mu)\,\alpha_{1}\alpha_{2}\dot{\beta}_{1}\dot{\beta}_{2}}\
,
\end{equation}
vale a dire
\begin{equation}
\Box\omega_{\mu\nu\rho}-3\nabla_{(\mu}\nabla\cdot\omega_{\nu\rho)}+\frac{3}{2}\{\nabla_{(\mu},\nabla_{\nu}\}\omega^{\sigma}_{\phantom{\sigma}\sigma\rho)}-\frac{1}{L^{2}}\omega_{\mu\nu\rho}+\frac{2}{L^{2}}g_{(\mu\nu}\omega^{\sigma}_{\phantom{\sigma}\sigma\rho)}=0
\ .
\end{equation}
Si noti che il campo $\omega_{\mu\nu\rho}$ costruito come in
(\ref{totsymfield}) non \`{e} a traccia nulla. Inoltre, la sua
variazione sotto la trasformazione di gauge data dall'unico
parametro di spin $s-1=2$ non ancora fissato, $\epsilon(2,2)$,
corrisponde alla simmetria locale dell'equazione di Fronsdal
covariantizzata (rispetto al background $AdS$) per spin $3$,
\begin{equation}
\delta\omega_{\mu\nu\rho}=3\nabla_{(\mu}\epsilon_{\nu\rho)}\ ,
\end{equation}
dove $\epsilon_{\nu\rho}=e_{\nu}^{(0)\,\alpha_{1}\dot{\beta}_{1}}
e_{\rho}^{(0)\,\alpha_{2}\dot{\beta}_{2}}
\epsilon_{\alpha_{1}\alpha_{2}\dot{\beta}_{1}\dot{\beta}_{2}}$
\`{e} simmetrico e a traccia nulla, in conseguenza, al solito,
della totale simmetria tra gli $\alpha_{1}\alpha_{2}$ e
$\dot{\beta}_{1}\dot{\beta}_{2}$, separatamente.

L'analisi procede analogamente per spin arbitrario. Le componenti
dei vincoli (\ref{cplinconstr1}) che non vengono usate per
determinare i campi ausiliari in termini di quelli fisici
contengono le equazioni del moto per questi ultimi. Per ogni
$s\geq 3/2$ le equazioni del moto linearizzate hanno quindi la
forma
\begin{equation}\label{eomgenint}
e_{\{\dot{\beta}_{1}|}^{(0)\,\beta}\wedge
R^{1}_{\beta\alpha_{1}...\alpha_{s-1}|\dot{\beta}_{2}...\dot{\beta}_{s-1}\}_{\dot{\beta}}}=0
\ , \qquad s= 2,3,\ldots\ ,
\end{equation}
\begin{equation}\label{eomgenhalfint}
e_{\{\dot{\beta}_{1}|}^{(0)\,\beta}\wedge
R^{1}_{\beta\alpha_{1}...\alpha_{s-1/2}|\dot{\beta}_{2}...\dot{\beta}_{s-3/2}\}_{\dot{\beta}}}=0
\ , \qquad s= \frac{3}{2},\frac{5}{2},\ldots\ .
\end{equation}
Per spin intero, le (\ref{eomgenint}) sono equazioni di secondo
ordine nei campi fisici
$\omega_{\mu\,\alpha(s-1)\dot{\beta}(s-1)}$, come si vede notando
che la curvatura linearizzata $R^{1}_{\alpha(s)\dot{\beta}(s-2)}$
contiene il rotore del campo ausiliario
$\omega_{\mu\,\alpha(s)\dot{\beta}(s-2)}$, che a sua volta \`{e}
esprimibile attraverso il rotore del campo fisico
$\omega_{\mu\,\alpha(s-1)\dot{\beta}(s-1)}$ per mezzo del vincolo
di curvatura $R^{1}_{\alpha(s-1)\dot{\beta}(s-1)}=0$. Le equazioni
(\ref{eomgenhalfint}) sono invece di primo ordine per i campi
fisici $\omega_{\mu\,\alpha(s-1/2)\dot{\beta}(s-3/2)}$, di spin
semi-intero.

Da spin $s=4$ in poi (nel settore bosonico) la condizione di
doppia traccia nulla, caratteristica dei campi di spin arbitrario
nella formulazione di Fronsdal, \`{e} non banale. I campi fisici
$\omega_{\mu\,\alpha(s-1)\dot{\beta}(s-1)}$ danno luogo,
analogamente a quanto visto nell'esempio di spin $3$, ai campi
totalmente simmetrici
\begin{equation}
\omega_{\mu_{1}\ldots\mu_{s}}=e_{(\mu_{2}}^{(0)\,\alpha_{1}\dot{\beta}_{1}}\ldots
e_{\mu_{s}}^{(0)\,\alpha_{s-1}\dot{\beta}_{s-1}}\omega_{\mu_{1})\,\alpha_{1}\ldots\alpha_{s-1}\dot{\beta}_{1}\ldots\dot{\beta}_{s-1}}
\ ,
\end{equation}
che sono effettivamente a doppia traccia nulla:
\begin{equation}
g^{\mu\nu}g^{\rho\sigma}\omega_{\mu\nu\rho\sigma\lambda(s-4)}=0\ .
\end{equation}
Le loro equazioni del moto (\ref{eomgenint}) contengono le
equazioni di Fronsdal e sono invarianti sotto le trasformazioni di
gauge
\begin{equation}
\delta\omega_{\mu_{1}\ldots\mu_{s}}=s\nabla_{(\mu_{1}}\epsilon_{\mu_{2}\ldots\mu_{s})}
\end{equation}
dove i parametri $\epsilon_{\mu_{1}\ldots\mu_{s-1}}$ hanno tracce
nulle.

 \vspace{0.5cm}

Le equazioni del moto libere per tutti i campi di massa nulla e
spin arbitrario in $AdS_{4}$ possono dunque essere raccolte nelle
tre condizioni
\begin{equation}\label{COMSTh0}
d\omega_{0}+\omega_{0}\star\omega_{0}=0 \ ,
\end{equation}
\begin{equation}\label{COMSTh1}
R_{1}(x|y,\bar{y})=\frac{i}{4}\left[e^{\alpha\dot{\gamma}}\wedge 
e^{\beta}\,_{\dot{\gamma}}\frac{\partial}{\partial
y^{\alpha}}\frac{\partial}{\partial
y^{\beta}}C(x|y,0)+e^{\gamma\dot{\alpha}}\wedge
e_{\gamma}\,^{\dot{\beta}}\frac{\partial}{\partial
\bar{y}^{\dot{\alpha}}}\frac{\partial}{\partial
\bar{y}^{\dot{\beta}}}C(x|0,\bar{y})\right] \ ,
\end{equation}
\begin{equation}\label{COMSTh2}
{\cal D}_{0}C(x|y,\bar{y})=0 \ ,
\end{equation}
dove la seconda equazione \`{e} una riscrittura della
(\ref{cplinconstr1}) in termini della \emph{master 2-form} $R_{1}$
e della \emph{master 0-form} $C$. Questo \`{e} il contenuto del
cosiddetto \emph{Central On-Mass-Shell Theorem} di Vasiliev
\cite{Vasiliev:1999ba}.

Come abbiamo osservato, ciascuna di queste tre equazioni \`{e}
consistente. In particolare, (\ref{COMSTh0}) costituisce la
condizione di compatibilit\`{a} di se stessa e di (\ref{COMSTh2}),
mentre quest'ultima garantisce la consistenza del sistema
(\ref{COMSTh1}). L'identit\`{a} di Bianchi (\ref{Bidsxlincurv})
implica infatti che il secondo membro sia covariantemente
costante,
\begin{eqnarray}\label{step1}
\!0 &=& D_{0}\left(e^{\alpha\dot{\gamma}}\wedge
e^{\beta}\,_{\dot{\gamma}}\frac{\partial}{\partial
y^{\alpha}}\frac{\partial}{\partial
y^{\beta}}C(x|y,0)+\textrm{h.c.}\right) \nonumber\\
&=& 2(de^{\alpha\dot{\gamma}}\wedge
e^{\beta}\,_{\dot{\gamma}})\partial_{\alpha}\partial_{\beta}C(x|y,0)+e^{\alpha\dot{\gamma}}\wedge
e^{\beta}\,_{\dot{\gamma}}D_{0}\partial_{\alpha}\partial_{\beta}
C(x|y,0)+\textrm{h.c.} \ ,
\end{eqnarray}
dove h.c. indica l'hermitiano coniugato. Utilizzando il vincolo di
torsione nulla nel primo termine a secondo membro e le formule di
contrazione (\ref{sympldiff}) nel secondo, si dimostra che
\begin{equation}\label{step2}
\nabla_{0}\left(e^{\alpha\dot{\gamma}}\wedge
e^{\beta}\,_{\dot{\gamma}}\frac{\partial}{\partial
y^{\alpha}}\frac{\partial}{\partial
y^{\beta}}C(x|y,0)+\textrm{h.c.}\right)=e^{\alpha\dot{\gamma}}\wedge
e^{\beta}\,_{\dot{\gamma}}\frac{\partial}{\partial
y^{\alpha}}\frac{\partial}{\partial
y^{\beta}}\nabla_{0}C(x|y,0)+\textrm{h.c.}\ ,
\end{equation}
mentre il termine contenente lo $\star$-commutatore con il
vielbein, presente in
$D_{0}\partial_{\alpha}\partial_{\beta}C(x|y,0)$, non
contribuisce, in virt\`{u} dell'identit\`{a} (\ref{membraneids2}).
In conclusione, l'identit\`{a} di Bianchi (\ref{Bidsxlincurv})
implica l'annullamento dell'ultimo termine della (\ref{step2}), un
vincolo differenziale sulle $0$-forme puramente olomorfe o
anti-olomorfe che costituisce il punto di partenza della catena di
equazioni (\ref{altercplinconstr2}), che si ottiene prendendo
successive derivate Lorentz-covarianti del vincolo suddetto,
analogamente a quanto fatto per la gravit\`{a} (ovvero per il
settore $|n-m|=4$) nella sezione \ref{4.2}.

Le (\ref{COMSTh1}) e (\ref{COMSTh2}) esprimono inoltre tutti i
campi ausiliari attraverso derivate di quelli fisici, secondo le
(\ref{aux1}) e (\ref{aux2}) per quanto riguarda le $0$-forme e
analogamente per le $1$-forme.

Le equazioni (\ref{COMSTh1}) del \emph{Central On-Mass-Shell
Theorem} possono anche esser scritte come
\begin{equation}
R^{1}_{\alpha\beta,\gamma_{1}\gamma_{2}...\gamma_{2s-2}} =
C_{\alpha\beta\gamma_{1}\gamma_{2}...\gamma_{2s-2}}\ , \quad
s=1,\frac{3}{2},2... \ ,
\end{equation}
\begin{equation}\label{gentorsion}
R^{1}_{\alpha\beta,\gamma_{1}...\gamma_{k}\dot{\gamma}_{k+1}...\dot{\gamma}_{2s-2}}
= 0\ , \quad s=\frac{3}{2},2,\frac{5}{2}..., \quad k=0,1,...2s-3 \
,
\end{equation}
\cite{Sezgin:1998eh, Sezgin:1998gg} e analogamente per le
complesse coniugate. I vincoli di torsione generalizzati
(\ref{gentorsion}) annullano dunque tutte le componenti delle
curvature linearizzate eccetto i tensori di Weyl generalizzati. Il
risultato \`{e} che soltanto i campi di gauge rappresentati dalle
$1$-forme $\omega_{\alpha(n)\dot{\beta}(m)}$ con $|n-m|\leq 1$
sono dinamici, mentre quelli con $|n-m|\geq 2$ sono ausiliari, e
tutti i tensori che compaiono nel loro sviluppo in
rappresentazioni irriducibili di Lorentz possono essere espressi
in termini di derivate dei campi fisici o sono pura gauge. Si noti
che non esistono campi di gauge ausiliari per spin
$s\leq\frac{3}{2}$, mentre i campi ausiliari
$\omega_{\alpha(s-2)\dot{\beta}(s)},\quad s=2,3,...$ possono
essere pensati come \emph{connessioni di Lorentz generalizzate}.

I campi di gauge fisici si suddividono in:
\begin{itemize}
\item Campo di gauge di spin $s=1$
\begin{equation}
\omega_{\mu} \ ,
\end{equation}
reale.

\item \emph{Vielbein generalizzati}
\begin{equation}
\omega_{\mu\,\alpha(n)\dot{\beta}(n)}\ , \quad n=s-1\ ,\quad
s=2,3,... \ ,
\end{equation}
reali anch'essi.

\item \emph{Gravitini generalizzati}
\begin{equation}
\omega_{\mu\,\alpha(n)\dot{\beta}(m)}\ , \quad n=s-\frac{3}{2}\
,m=s-\frac{1}{2}\ ,\quad s=\frac{3}{2},\frac{5}{2},...\ ,
\end{equation}
insieme con i loro complessi coniugati.
\end{itemize}
I campi di materia sono infine descritti dalle seguenti $0$-forme:
\begin{itemize}

\item Campo scalare
\begin{equation}
C \ ,\qquad s=0 \ .
\end{equation}
\item Campo di spin $1/2$
\begin{equation}
C_{\alpha}\oplus C_{\dot{\alpha}} \ , \qquad s=1/2 \ .
\end{equation}
\end{itemize}
\`{E} analogamente possibile distinguere tra simmetrie di gauge
dinamiche ed ausiliarie. Definiamo \emph{simmetrie di gauge
dinamiche} quelle che agiscono in modo non banale su un campo di
gauge dinamico, e \emph{simmetrie di gauge ausiliarie} quelle che
non agiscono su alcun campo di gauge dinamico. Le prime
costituiscono dunque le simmetrie locali delle equazioni del moto
dei campi fisici, mentre le seconde giocano il ruolo di simmetrie
di Stueckelberg per le componenti tensoriali irriducibili dei
campi ausiliari che non vengono determinate in termini dei campi
fisici dalle (\ref{gentorsion}). Tali componenti indeterminate
risultano cos\`{i} pura gauge e possono essere eliminate con
opportune scelte dei parametri di gauge ausiliari.

Le simmetrie di gauge ausiliarie corrispondono ai parametri
\begin{equation}
\epsilon_{\alpha(n)\beta(m)}\ , \qquad |n-m|\geq 4
\end{equation}
ed ai relativi complessi coniugati. Le simmetrie di gauge
dinamiche possono essere invece classificate come segue:
\begin{itemize}
\item \emph{Trasformazioni di gauge di tipo YM}
\begin{equation}
\epsilon \ ,
\end{equation}
reali.

\item \emph{Trasformazioni di Lorentz generalizzate}
\begin{equation}
\epsilon_{\alpha(s-1)\beta(s)}\ , \qquad s=2,3,...\ ,
\end{equation}
con i loro complessi coniugati.

\item \emph{Riparametrizzazioni generalizzate }
\begin{equation}
\epsilon_{\alpha(s-1)\beta(s-1)}\ ,  \qquad s=1,2,...\ ,
\end{equation}
reali.

\item \emph{Trasformazioni fermioniche locali}
\begin{equation}
\epsilon_{\alpha(s-\frac{5}{2})\beta(s+\frac{1}{2})}\ , \qquad
s=\frac{5}{2},\frac{7}{2}...\ ,
\end{equation}
con i loro complessi coniugati. Sono l'analogo fermionico delle
trasformazioni di Lorentz generalizzate.

\item \emph{Supersimmetrie locali generalizzate}
\begin{equation}
\epsilon_{\alpha(s-\frac{3}{2})\beta(s-\frac{1}{2})}\ , \qquad
s=\frac{3}{2},\frac{5}{2}...\ ,
\end{equation}
con i loro complessi coniugati.
\end{itemize}

\section{Alcune osservazioni}
\label{4.5}

Il \emph{Central On-Mass-Shell Theorem} contiene un'altra
importante informazione: le (\ref{COMSTh1}) e (\ref{COMSTh2})
legano infatti le derivate rispetto alle coordinate
spazio-temporali $x^{\mu}$ e alle coordinate spinoriali interne
$y_{\alpha}$ e $\bar{y}_{\dot{\alpha}}$. Ad esempio, tenendo conto
dell'espressione (\ref{D0}) per ${\cal D}_{0}$, \`{e} evidente che
le derivate nelle variabili spinoriali interne agiscono sulle
$0$-forme $C$ come una sorta di radice quadrata delle derivate
spazio-temporali,
\begin{equation}\label{linkderivs}
\frac{\partial}{\partial x^{\mu}}C(x|y,\bar{y})\sim L^{-1}
e_{\mu}^{\alpha\dot{\beta}}\frac{\partial}{\partial
y^{\alpha}}\frac{\partial}{\partial
\bar{y}^{\dot{\beta}}}C(x|y,\bar{y})\ ,
\end{equation}
e ricordando inoltre l'espressione delle curvature linearizzate
data in (\ref{lincurv}), per le $1$-forme si ottiene
\begin{equation}
\frac{\partial}{\partial x^{\mu}}\omega(x|y,\bar{y})\sim L^{-1}
e_{\mu}^{\alpha\dot{\beta}}\left(
\bar{y}_{\dot{\beta}}\frac{\partial}{\partial
y^{\alpha}}\omega(x|y,\bar{y})+y_{\alpha}\frac{\partial}{\partial
\bar{y}^{\dot{\beta}}}\omega(x|y,\bar{y})\right)\ .
\end{equation}

Questo risultato \`{e} importante, poich\'{e} collega la
dipendenza dalle variabili interne a quella dalle variabili di
spazio-tempo. Abbiamo gi\`{a} notato, commentando la
(\ref{defdiff}), come lo $\star$-prodotto sia non locale negli
oscillatori. Alla luce di quanto appena detto, questo significa
che la presenza di $\star$-prodotti nelle equazioni della teoria,
ad esempio nei termini di accoppiamento delle derivate covarianti,
introduce automaticamente dei termini con un numero arbitrario di
derivate spazio-temporali!

Notiamo tuttavia che, nel caso linearizzato finora trattato, le
connessioni che compaiono nelle equazioni sono sempre quelle del
background $AdS$, quadratiche negli oscillatori: questo significa
che le espressioni come $C\star\omega_{0}$ non contengono pi\`{u}
di due derivate spinoriali e che dunque la teoria, a livello
linearizzato, \`{e} locale. Naturalmente questa \`{e} una
conseguenza dell'espansione perturbativa sopra utilizzata, mentre
la teoria completa potrebbe facilmente ammettere non localit\`{a}.

Un'altra notevole caratteristica delle teorie di gauge di HS
emerge dal \emph{Central On-Mass-Shell Theorem}. Le derivate nelle
variabili interne sono infatti collegate dalle (\ref{COMSTh1}) e
(\ref{COMSTh2}) a quelle spazio-temporali pesate da un fattore
$L\sim\frac{1}{\sqrt{\Lambda}}$. Ci\`{o} comporta la presenza di
potenze inverse della costante cosmologica nei termini
alto-derivativi che compaiono nello sviluppo degli
$\star$-prodotti delle interazioni, termini che dunque non hanno
senso nel limite piatto $\Lambda\rightarrow0$.

La teoria di Vasiliev recupera cos\`{i} tutte le caratteristiche
di una teoria consistente di HS menzionate nell'introduzione, ivi
suggerite dalla richiesta di descrivere interazioni consistenti
dei campi di spin arbitrario con il campo gravitazionale. Possiamo
aggiungere due ulteriori osservazioni:
\begin{enumerate}
\item La formulazione della teoria di HS a partire dal
\emph{gauging} di una $\star$-algebra fa emergere naturalmente e
lega tra loro le seguenti caratteristiche:
\begin{itemize}
\item rilevanza del background $AdS$, che compare come soluzione
di vuoto delle equazioni di curvatura nulla della teoria;

\item potenziale non localit\`{a} delle interazioni di HS, legata
alla presenza di derivate di ordine superiore al secondo dei campi
dinamici, a livello non lineare. I termini alto-derivativi sono
dimensionalmente permessi grazie alla presenza del raggio di $AdS$
o, equivalentemente, della costante cosmologica. Allo stesso
tempo, le teorie di gauge di HS restano locali a livello
linearizzato.
\end{itemize}
\item Abbiamo pi\`{u} volte sottolineato che il rappresentare i
campi mediante tensori irriducibili di Lorentz, totalmente
simmetrici negli indici spinoriali di ciascun tipo, incorpora un
vincolo puramente algebrico nella teoria, che \`{e} cruciale per
recuperare le equazioni del moto dei campi fisici dalla catena di
equazioni del \emph{Central On-Mass-Shell Theorem}. Il formalismo
di Vasiliev qui presentato conduce dunque ad una formulazione
intrinsecamente on-shell!  
Tuttavia, a livello linearizzato la formulazione qui esposta
risulta equivalente a quella di Fronsdal e dunque ammette
un'azione quadratica nei campi. 
Questo per\`{o} non \`{e} pi\`{u} vero a livello non lineare, ove
tuttora non si conosce un principio d'azione da cui ricavare la
teoria interagente di HS. \`{E} chiaro comunque che un primo passo
verso questa formulazione a livello \emph{off-shell} richiede
opportune modifiche dell'algebra HS in modo da includere in
qualche maniera tracce nei generatori.
\end{enumerate}

\section{Formulazione ``unfolded''}
\label{4.6}

La formulazione che abbiamo riassunto, in cui il contenuto fisico
dell'equazione del moto di un campo di spin $s$ \`{e}, per
cos\`{i} dire, dipanato in un sistema formalmente consistente di
infinite equazioni, viene denominata nella letteratura
formulazione ``\emph{unfolded}''
\cite{Vasiliev:1995dn,Vasiliev:1999ba}. La propriet\`{a}
fondamentale \`{e} la consistenza (o integrabilit\`{a}) del
sistema: essa consente infatti di esprimere infiniti campi
ausiliari, cui non sono associati gradi di libert\`{a} fisici, in
funzione dei campi dinamici, evitando una proliferazione di
variabili indipendenti. Tuttavia, come osservato, sono i vincoli
algebrici sulle variabili del sistema a far s\`{i} si riproducano
effettivamente le equazioni del moto del campo di spin $s$.

Naturalmente, vien da chiedersi quale possa essere il vantaggio di
questo modo di procedere rispetto alla formulazione usuale. La
risposta \`{e} che finch\'{e} si ha a che fare con la sola teoria
libera la  formulazione \emph{unfolded} non rappresenta che un
approccio equivalente in grado di contenere, in modo sintetico,
tutte le equazioni dei campi di spin arbitrario entro alcune
\emph{master equations} di primo ordine, come le (\ref{COMSTh1}) e
(\ref{COMSTh2}), al prezzo per\`{o} di introdurre infiniti campi
ausiliari, eliminabili attraverso un'accurata analisi dei vincoli.
Questo formalismo consente anche una naturale realizzazione
dell'algebra infinito-dimensionale di HS: le simmetrie di HS
legano tra loro derivate dei campi fisici di ordine
arbitrariamente alto, ed \`{e} pertanto notevole la presenza di
multipletti infinito-dimensionali, in grado di contenere i campi
fisici e tutte le loro derivate compatibili con le equazioni del
moto. D'altro canto, si \`{e} visto sin dall'equazione
(\ref{infdim}) che, per introdurre interazioni consistenti con
campi di spin $s\geq2$, \`{e} necessario lavorare con infiniti
campi di massa nulla e spin arbitrario. Ma questo non \`{e} tutto.

Nella formulazione \emph{unfolded} le equazioni del moto di tutti
gli spin sono formulate in termini di condizioni di curvatura
nulla (equazioni di Maurer-Cartan) come (\ref{0curv}) e
(\ref{linconstr2}) e questo offre la possibilit\`{a} di esprimere
le corrispondenti soluzioni come pura gauge:
\begin{equation}\label{puregauge1}
\Omega(x|y,\bar{y})=g^{-1}(x|y,\bar{y})\star dg(x|y,\bar{y})\ ,
\end{equation}
\begin{equation}\label{puregauge2}
C(x|y,\bar{y})=g^{-1}(x|y,\bar{y})\star
C_{0}(y,\bar{y})\star\bar{\pi}(g)(x|y,\bar{y}) \ ,
\end{equation}
dove $g(x|y,\bar{y})$ \`{e} un arbitrario elemento invertibile
della $\star$-algebra ${\cal A}$ dipendente dalle coordinate
spaziotemporali $x$ e $C_{0}(y,\bar{y})$ \`{e} un'arbitraria
funzione delle sole variabili interne. Naturalmente, se la
soluzione $\Omega$ della (\ref{0curv}) deve rappresentare un
background $AdS$, $g$ dovr\`{a} esser fissata in modo tale da
riprodurre la (\ref{omega0}). A questo punto, la
(\ref{puregauge2}) fornisce la soluzione generale delle equazioni
per i campi di massa nulla e spin arbitrario (\ref{linconstr2}).
Supponendo che $g(x_{0}|y,\bar{y})=I$ per qualche $x=x_{0}$ (il
che pu\`{o} sempre essere ottenuto riscalando i $g(x)\rightarrow
g'(x)=g(x)\star g^{-1}(x_{0})$ in modo da non modificare la forma
della (\ref{0curv})), si trova che
$C_{0}(y,\bar{y})=C(x_{0}|y,\bar{y})$, sicch\'{e}, nella
(\ref{puregauge2}), $C_{0}$ fa da condizione iniziale su cui
$g(x)$ agisce come una sorta di operatore di evoluzione.

Alla luce di queste osservazioni, la presenza di infiniti campi
ausiliari nella teoria acquista un significato pi\`{u} profondo.
L'analisi delle equazioni per $C$ ha infatti mostrato come le sue
componenti $C_{\alpha(n)\dot{\beta}(m)}$ parametrizzino tutte e
sole le combinazioni delle derivate Lorentz-covarianti dei campi
fisici non banali on-shell. Questo significa che
$C_{0}(y,\bar{y})$ ha il ruolo di funzione generatrice per tutte
le derivate dei campi fisici che non vengono annullate dalle
equazioni del moto nel punto $x_{0}$. Ma allora diventa possibile
ricostruire i campi fisici stessi in un intorno di $x_{0}$
attraverso un'espansione in serie di Taylor (covariantizzata), e
questo \`{e} proprio ci\`{o} che la (\ref{puregauge2}) realizza.
Evidentemente, questa procedura richiede di fatto che le $0$-forme
generino una rappresentazione infinito-dimensionale, perch\'{e}
esse devono contenere tutte le derivate (fino ad ordine
arbitrariamente elevato) dei campi fisici compatibili con le
equazioni del moto.

Chiariamo queste considerazioni con un esempio semplice, l'analisi
delle equazioni \emph{unfolded} che descrivono il campo di
Klein-Gordon in uno spazio piatto di dimensione arbitraria $d$.
L'algebra di simmetria della soluzione di vuoto banale della
(\ref{0curv}) corrispondente allo spazio-tempo di Minkowski \`{e}
l'algebra di Poincar\'{e} $iso(d-1,1)$. Introduciamo la
connessione $\Omega_{\mu}$, che, espandendo sui generatori
$(P_{a},M_{ab})$, d\`{a} luogo alle componenti
$(e_{\mu}^{a},\omega_{\mu}^{ab}),\,a,b=0,...,d-1$. Le condizioni
di curvatura nulla per il vielbein $e_{\mu}^{a}$ e per la
connessione di Lorentz $\omega_{\mu}^{ab}$,
\begin{equation}
R_{\mu\nu}^{a}=0\ , \qquad R_{\mu\nu}^{ab}=0\ ,
\end{equation}
descrivono uno spazio-tempo piatto, e dalla prima in generale si
pu\`{o} ricavare $\omega$. Sullo spazio piatto possiamo
identificare gli indici di Lorentz $a,b$ con gli indici di
$GL(d,R)$, ovvero possiamo fissare una gauge opportuna in cui
\begin{equation}
e^{a}_{\mu}=\delta^{a}_{\mu}\ , \qquad \omega_{\mu}^{ab}=0 \ .
\end{equation}
In altri termini, possiamo scegliere un riferimento cartesiano
globale, la cui esistenza \`{e} garantita dalla piattezza dello
spazio-tempo, in conformit\`{a} con il principio di equivalenza di
Einstein.

Descriviamo la dinamica del campo di spin $0$ $\phi(x)$
introducendo una collezione infinita di $0$-forme
$\phi_{a_{1}...a_{n}}(x)$, legate alle $0$-forme
$C_{\alpha(n)\dot{\beta}(n)}$ incluse nella \emph{master $0$-form}
$C$ da
\begin{equation}
\phi_{a_{1}...a_{n}}=e_{a_{1}}^{\,\alpha_{1}\dot{\beta}_{1}}\ldots
e_{a_{n}}^{\,\alpha_{n}\dot{\beta}_{n}}C_{\alpha_{1}\ldots\alpha_{n}\dot{\beta}_{1}\ldots\dot{\beta}_{n}}
\ .
\end{equation}
Ricordando le (\ref{Vbprops}), \`{e} evidente che la condizione di
totale simmetria negli indici di ciascun tipo delle $0$-forme
$C_{\alpha(n)\dot{\beta}(n)}$ corrisponde alla condizione di
totale simmetria e traccia nulla delle $\phi_{a_{1}...a_{n}}$,
\begin{equation}\label{tracelessness}
\eta^{bc}\phi_{bca_{3}...a_{n}}=0 \ .
\end{equation}
Come vedremo, questo \`{e} il vincolo algebrico sulle $0$-forme
che codifica la dinamica nel caso del background piatto. Nel
limite piatto, $L^{-1}\rightarrow 0$ e
$\nabla_{0\,\mu}=\partial_{\mu}$, il sistema (\ref{cplinconstr2})
diventa
\begin{equation}\label{flatKGunf}
\partial_{\mu}\phi_{a_{1}...a_{n}}(x)=e_{\mu}^{\,b}\phi_{a_{1}...a_{n}b}(x)
\ .
\end{equation}
Tale sistema \`{e} consistente, poich\'{e} prendendone successive
derivate $\partial_{\nu}$ e antisimmetrizzando rispetto allo
scambio $\nu\leftrightarrow \mu$ non si ottengono ulteriori
condizioni indipendenti, ma soltanto equazioni vuote tipo $0=0$,
in virt\`{u} delle ipotesi fatte. Questo equivale a dire che le
$\phi_{a_{1}...a_{n}}(x)$ generano una rappresentazione
infinito-dimensionale dell'algebra di Poincar\'{e}. Le prime due
equazioni della catena (\ref{flatKGunf}) sono
\begin{equation}
\partial_{\mu}\phi=\phi_{\mu} \ , \qquad
\partial_{\mu}\phi_{\nu}=\phi_{\mu\nu} \ .
\end{equation}
Sostituendo la prima nella seconda, e prendendo la traccia di
quest'ultima, si ottiene
\begin{equation}\label{flatKG}
\Box\phi=0 \ ,
\end{equation}
l'equazione di Klein-Gordon nello spazio-tempo piatto, grazie alla
condizione di traccia nulla (\ref{tracelessness}). Tutte le altre
equazioni della catena non impongono ulteriori condizioni su
$\phi$, ma si riducono alle identificazioni
\begin{equation}\label{higherderivs}
\phi_{\mu_{1}...\mu_{n}}=\partial_{\mu_{1}}...\partial_{\mu_{n}}\phi
\ ,
\end{equation}
automaticamente consistenti con il vincolo di traccia nulla
(\ref{tracelessness}) in virt\`{u} della (\ref{flatKG}).

Costruiamo ora la funzione generatrice delle $0$-forme
$\phi_{a_{1}...a_{n}}(x)$,
\begin{equation}
\Phi(x,u)=\sum_{n=0}^{\infty}\frac{1}{n!}\phi_{a_{1}...a_{n}}(x)u^{a_{1}}...u^{a_{n}}
\ ,
\end{equation}
espandendo in una formale serie di potenze nelle coordinate
ausiliarie $u^{a}$ e prendendo convenzionalmente
$\Phi(x,0)=\phi(x)$. La condizione di traccia nulla
(\ref{tracelessness}) sui coefficienti dello sviluppo si traduce
in termini della funzione generatrice in
\begin{equation}
\Box_{u}\Phi(x,u)\equiv \frac{\partial}{\partial
u^{a}}\frac{\partial}{\partial u_{a}}\Phi=0 \ .
\end{equation}
Si noti che le coordinate spinoriali interne $y$ ed $\bar{y}$
delle sezioni precedenti possono essere viste come una sorta di
radice quadrata delle coordinate ausiliarie $u$ introdotte per
questo esempio sullo spazio piatto (qualitativamente $u\sim
y\bar{y}$). Ricordando le (\ref{D0}) e (\ref{linkderivs}), non
\`{e} pertanto sorprendente il fatto che le equazioni
(\ref{flatKGunf}) si possano riscrivere come
\begin{equation}
\frac{\partial}{\partial
x^{\mu}}\Phi(x,u)=\frac{\partial}{\partial u^{\mu}}\Phi(x,u)\ .
\end{equation}
Queste equazioni mostrano che, per il multipletto
infinito-dimensionale dell'algebra di Poincar\'{e} costituito
dalle $0$-forme $\phi_{a_{1}...a_{n}}(x)$, le traslazioni
spazio-temporali sono realizzate come traslazioni nelle variabili
ausiliarie $u^{a}$. In altri termini, le loro soluzioni sono tali
che
\begin{equation}
\Phi(x,u)=\Phi(x+u,0)=\Phi(0,x+u)\ ,
\end{equation}
e dunque, per $u=0$,
\begin{equation}\label{Taylorexp}
\phi(x)=\Phi(0,x)=\sum_{n=0}^{\infty}\frac{1}{n!}\phi_{\mu_{1}...\mu_{n}}(0)x^{\mu_{1}}...x^{\mu_{n}}
\ .
\end{equation}
Tenendo presenti le (\ref{higherderivs}), questa \`{e} una vera
espansione in serie di Taylor, che ricostruisce il campo di
Klein-Gordon $\phi(x)$ a partire dal valore di tutte le $0$-forme
$\phi_{a_{1}...a_{n}}$ in $x=0$, come preannunciato.

Quanto appena detto mostra che, nella formulazione
\emph{unfolded}, il problema dinamico \`{e} ben definito e
risolubile, almeno localmente, una volta assegnati i valori di
tutte le $0$-forme in un punto dello spazio-tempo. Bench\'{e}
equivalenti nel caso libero, la formulazione \emph{unfolded} e la
formulazione canonica di una teoria di campo relativistica sono
dunque notevolmente diverse nell'impostazione del problema.
Nell'approccio ordinario, infatti, ci si riduce ad un problema di
Cauchy, in cui compaiono soltanto i campi fisici ed i loro
impulsi, e la dinamica \`{e} completamente definita una volta noti
i loro valori ad un istante iniziale. La formulazione
\emph{unfolded} introduce invece, partendo dal \emph{gauging} di
un'algebra di simmetria spazio-temporale, i campi fisici e
un'infinit\`{a} di campi ausiliari, questi ultimi inclusi in una
collezione di $0$-forme che alcune equazioni vincolano ad esser
uguali a derivate di ordine arbitrariamente elevato dei primi:
\`{e} poi necessario specificare i valori dei campi fisici e di
tutte le loro derivate in un singolo punto $x_{0}$ dello
spazio-tempo, e non su di una ipersuperficie spaziale ad un certo
$t=t_{0}$.

Si noti inoltre il ruolo cruciale svolto dal \emph{twist}
$\bar{\pi}$ nel far s\`{i} che la (\ref{linconstr2}) codifichi una
dinamica non banale. Se non ci fosse, quest'ultima equazione si
ridurrebbe infatti alla
\begin{equation}\label{auxd0}
{\cal D}_{0}C=\nabla_{0} C+\frac{i}{2}L^{-1} [
e_{0\alpha\dot{\alpha}}y^{\alpha}\bar{y}^{\dot{\alpha}},C
]_{\star}=0 \ ,
\end{equation}
dove l'anticommutatore col vielbein viene rimpiazzato dal
commutatore. Ci\`{o} ha conseguenze importanti: si pu\`{o} notare
infatti fin dalla (\ref{infdim}) che lo $\star$-commutatore di un
generatore di spin $s$ arbitrario con un generatore di spin $1$
non altera lo spin del primo, ovvero si chiude su generatori di
spin $s$, ed in particolare il commutatore col vielbein
sostituisce soltanto un oscillatore $y$ con un $\bar{y}$, come
\`{e} evidente dall'esempio (\ref{lincurv}) (o dalle formule di
contrazione (\ref{sympldiff})). Questo significa che la
(\ref{auxd0}) lega tra loro soltanto componenti di $C$ con lo
stesso spin, e pi\`{u} in particolare, per ogni $s=\frac{n+m}{2}$,
soltanto i coefficienti delle potenze $y^{n}\bar{y}^{m}$,
$y^{n+1}\bar{y}^{m-1}$ e $y^{n-1}\bar{y}^{m+1}$, che ne
costituiscono settori indipendenti. Ma ciascuno di questi contiene
dunque un numero finito di elementi, ovvero un numero finito di
derivate del campo di spin $s$, e non \`{e} quindi possibile
ricostruire, secondo quanto appena esposto, i campi di diverso
spin contenuti in $C$ attraverso la loro serie di Taylor
nell'intorno di un punto arbitrario dello spazio-tempo. Al
contrario, l'anticommutatore presente nella (\ref{linconstr2}) non
si chiude su ciascuno spin, come visto, e ci\`{o} implica che, per
ogni spin, i settori indipendenti contengono infiniti tensori,
tutti i coefficienti delle potenze $y^{n}\bar{y}^{m}$ con
$|n-m|=2s$. In altre parole, la (\ref{auxd0}) ammette soltanto
soluzioni ``rigide'', con un numero finito di gradi di libert\`{a}
\footnote {Per gradi di libert\`{a} si intende qui il numero di
condizioni iniziali da fissare per impostare il problema dinamico,
vale a dire il numero di componenti della ``condizione iniziale''
$C_{0}$.}, mentre il \emph{twist} $\bar{\pi}$ della
(\ref{linconstr2}) rende possibile una dinamica non banale,
descrivendo sistemi con un numero infinito di gradi di
libert\`{a}.

In linea di principio, \`{e} sempre possibile riformulare la 
dinamica di un sistema descritto in modo canonico attraverso un
\emph{unfolding}, aggiungendo un opportuno insieme di infinite
nuove variabili, analoghe ai multispinori
$C_{\alpha(n)\dot{\beta}(m)}$, che possa costituire la funzione
generatrice per tutte le derivate dei campi fisici che le
equazioni del moto non vincolano ad esser nulle
\cite{Vasiliev:1999ba}. Quest'ultima procedura \`{e} tuttavia
pi\`{u} generale, ed in questo modo sar\`{a} possibile introdurre
le interazioni come deformazioni non lineari di questi sistemi che
non violino la condizione di consistenza. Di fatto, questo \`{e}
al momento il solo approccio noto alla teoria interagente dei
campi di HS. Nella sucessiva sezione inquadreremo la formulazione
qui esposta nel contesto pi\`{u} generale delle \emph{free
differential algebras} o \emph{sistemi integrabili di Cartan}.

\section{Free Differential Algebras}
\label{4.7}

Le equazioni di curvatura nulla (\ref{COMSTh0}) e (\ref{COMSTh2})
realizzano un particolare esempio di \emph{free differential
algebra} (FDA) \cite{D'Auria:nx, Castellani:kd, Castellani:ke,
D'Auria:1982my, Fre:1984pc, D'Auria:1982pm, Vasiliev:1999ba},
importanti sistemi introdotti da D'Auria e Fre in
supergravit\`{a}, di cui vogliamo presentare una definizione
completamente generale.

Dato un insieme di $p$-forme differenziali $W^{A}(x)$, con $p\geq
0$, definiamo le curvature generalizzate $R^{A}(x)$ come
\begin{equation}\label{defgencurv}
R^{A}=dW^{A}+F^{A}(W) \ ,
\end{equation}
dove $d=dx^{\mu}\frac{\partial}{\partial x^{\mu}}$ \`{e}, al
solito, la derivata esterna. Indichiamo inoltre con $F^{A}(W)$ una
qualsiasi funzione delle $p$-forme $W^{B}(x)$, costruita
utilizzando il solo prodotto esterno, e tale che
\begin{equation}\label{genJacobi}
F^{B}\wedge\frac{\delta F^{A}}{\delta W^{B}}\equiv 0 \ ,
\end{equation}
dove la derivata funzionale rispetto a $W$ agisce a sinistra, che
corrisponde ad una identit\`{a} di Jacobi generalizzata. Definiamo
allora \emph{free differential algebra} qualunque insieme di
$p$-forme che rispettino l'identit\`{a} (\ref{genJacobi}), che
\`{e} la condizione di consistenza del sistema di equazioni di
curvatura nulla
\begin{equation}\label{0gencurv}
R^{A}=0 \ .
\end{equation}
Prendendo la derivata esterna della (\ref{defgencurv}) ed
imponendo tale identit\`{a}, \`{e} infatti immediato verificare
che essa implica l'identit\`{a} di Bianchi generalizzata
\begin{equation}\label{genBianchi}
dR^{A}=R^{B}\wedge\frac{\delta F^{A}}{\delta W^{B}}\ ,
\end{equation}
che a sua volta implica la consistenza del sistema
(\ref{0gencurv}).

L'identit\`{a} di Jacobi generalizzata garantisce anche
l'invarianza delle equazioni (\ref{0gencurv}) sotto le
trasformazioni di gauge
\begin{equation}\label{gengauge1}
\delta W^{A}=d\epsilon^{A}-\epsilon^{B}\wedge\frac{\delta
F^{A}}{\delta W^{B}}\ ,\qquad p\geq 1 \ ,
\end{equation}
dove $\epsilon^{A}(x)$ \`{e} una $(p-1)$-forma se $W^{A}$ \`{e}
una $p$-forma. Le $0$-forme non introducono invece alcun parametro
di gauge e non possono che avere una legge di trasformazione
omogenea:
\begin{equation}\label{gengauge0}
\delta W^{A}=-\epsilon^{B}\wedge\frac{\delta F^{A}}{\delta W^{B}}
\ ,\qquad p=0\ .
\end{equation}
L'invarianza sotto le (\ref{gengauge1}) e (\ref{gengauge0})
pu\`{o} essere facilmente verificata notando che le curvature
trasformano come
\begin{equation}
\delta R^{A}=-R^{C}\wedge \frac{\delta}{\delta
W^{C}}\left(\epsilon^{B}\wedge\frac{\delta F^{A}}{\delta
W^{B}}\right)\ .
\end{equation}
In virt\`{u} della (\ref{genJacobi}) il sistema di equazioni di
curvatura nulla (\ref{0gencurv}) che costituisce la FDA ammette
dunque la simmetria di gauge (\ref{gengauge1}) e
(\ref{gengauge0}).

Essendo costruite in termini di forme differenziali, inoltre, le
(\ref{0gencurv}) sono anche esplicitamente invarianti sotto
diffeomorfismi. Ritroviamo qui, in ambito pi\`{u} generale, una
caratteristica incontrata nel capitolo $3$ a proposito della
formulazione di Stelle-West della gravit\`{a}: una delle
conseguenze delle equazioni di curvatura nulla, ed una delle
caratteristiche delle FDA, \`{e} l'automatica inclusione dei
diffeomorfismi tra le simmetrie di gauge! L'effetto di un
diffeomorfismo infinitesimo $\delta x^{\mu}=\xi^{\mu}(x)$, sulla
$p$-forma $W^{A}$ \`{e} dato dalla derivata di Lie agente su
$W^{A}$, come
\begin{equation}\label{Liederiv}
\delta_{\xi} W^{A}= {\cal L}_{\xi}W^{A}=\{d,i_{\xi}\}W^{A} \ ,
\end{equation}
nella quale $i_{\xi}$ rappresenta l'operatore prodotto interno,
che opera su una $p$-forma sostituendo un differenziale per volta
col vettore $-\xi^{\mu}$. Ad esempio, per una $2$-forma
$R_{\mu\nu}$ si ha
\begin{equation}
i_{\xi}\,R_{\mu\nu}\,dx^{\mu}\wedge
dx^{\nu}=-R_{\mu\nu}\,\xi^{\mu}\,dx^{\nu}+R_{\mu\nu}\,dx^{\mu}\xi^{\nu}
\ .
\end{equation}
Una trasformazione di gauge (\ref{gengauge1}) di parametro
$\epsilon^{A}(x)=i_{\xi}W^{A}(x)$ dipendente dai campi \`{e}
equivalente alla variazione sotto diffeomorfismi (\ref{Liederiv})
a meno di termini che sono nulli sulle equazioni (\ref{0gencurv}):
\begin{equation}
\delta W^{A}=\delta_{\xi} W^{A}-i_{\xi}R^{A}\ .
\end{equation}
Resta dunque dimostrato che l'invarianza sotto diffeomorfismi
delle (\ref{0gencurv}) \`{e} realizzata, in una FDA, come una
particolare simmetria di gauge dipendente dai campi.

Se l'insieme delle $W^{A}$ di una particolare FDA contiene
soltanto $1$-forme $w^{i}$, la funzione $F^{A}(W)$, che deve
allora essere una $2$-forma, sar\`{a} data dal bilineare
\begin{equation}\label{Ffor1form}
F^{i}=f^{i}_{jk}\,w^{j}\wedge w^{k}\ ,
\end{equation}
caratteristico delle teorie di Yang-Mills, ove le $f^{i}_{jk}$
sono le costanti di struttura dell'algebra di simmetria coinvolta.
In questo caso, la (\ref{genJacobi}) si riduce all'usuale
identit\`{a} di Jacobi per un'algebra di Lie scritta in termini
delle costanti di struttura,
\begin{equation}
f^{i}_{[jk}\,f^{l}_{i\,m]}=0\ ,
\end{equation}
e le (\ref{0gencurv}) sono le equazioni di curvatura nulla per le
$1$-forme di connessione a valori nell'algebra di simmetria $g$ in
questione.

Supponenendo invece che tra le $W^{A}$ vi siano anche $p$-forme
$C^{\alpha}$ ($p\neq 1$), e che le corrispondenti funzioni
$F^{\alpha}$ siano lineari in $C^{\alpha}$,
\begin{equation}\label{Ffor0form}
F^{\alpha}=(t_{i})^{\alpha}_{\,\beta}\,w^{i}\wedge C^{\beta}\ ,
\end{equation}
la (\ref{genJacobi}) implica allora che le matrici
$(t_{i})^{\alpha}_{\,\beta}$ corrispondono ad una rappresentazione
$t$ dell'algebra di simmetria $g$. Per stabilire un contatto con
le equazioni linearizzate del \emph{Central On-Mass-Shell Theorem}
prendiamo ad esempio il caso in cui le $C^{\alpha}$ siano
$0$-forme: la (\ref{genJacobi}),
\begin{equation}
F^{\gamma}\wedge\frac{\delta F^{\alpha}}{\delta
C^{\gamma}}+F^{k}\wedge\frac{\delta F^{\alpha}}{\delta
w^{k}}\equiv 0 \ ,
\end{equation}
implica infatti le condizioni
\begin{equation}
[(t_{i})^{\gamma}_{\,\beta},(t_{j})^{\alpha}_{\,\gamma}]=f^{k}_{ij}(t_{k})^{\alpha}_{\,\beta}
\ .
\end{equation}
Come gi\`{a} messo in evidenza precedentemente, le (\ref{COMSTh0})
e (\ref{COMSTh2}) sono esattamente della forma (\ref{0gencurv}):
le prime sono equazioni di curvatura nulla contenenti termini tipo
(\ref{Ffor1form}), e le seconde equazioni di costanza covariante
per i campi $C$ contenenti termini tipo (\ref{Ffor0form}). Quanto
sopra detto mostra, ancora una volta, che la condizione di
consistenza (\ref{genJacobi}) assicura che i campi $C$ formino una
rappresentazione dell'algebra in questione. Si pu\`{o} anzi dire
di pi\`{u}, come accennato nella sezione \ref{4.3}. Poich\'{e} le
(\ref{COMSTh0}) e (\ref{COMSTh2}) sono formalmente consistenti
indipendentemente dalla particolare soluzione di vuoto $AdS$ della
prima, le $0$-forme $C$ generano in effetti
una rappresentazione dell'intera algebra $shs(4)$, 
la rappresentazione \emph{twisted adjoint}.

\chapter{Higher spins: teoria ``unfolded'' non lineare}

\section{Algebre di HS estese e deformazioni della FDA lineare}
\label{deformalg}

Nel capitolo precedente abbiamo definito un'algebra di HS,
costruito una teoria di gauge che la ammette come simmetria locale
e ottenuto le equazioni del moto della teoria linearizzata
attraverso un'espansione attorno ad un vuoto gravitazionale $AdS$
(che, come si \`{e} visto, appare naturalmente nelle teorie di
gauge di HS come soluzione delle equazioni di curvatura nulla
(\ref{COMSTh0})), tenendo solo l'ordine lineare delle fluttuazioni
attorno ad esso (ovvero, consentendo interazioni dei campi di spin
arbitrario soltanto con i campi di background). Resta ancora da
sfruttare la natura non abeliana dell'algebra $shs(4)$ per
generare interazioni tra i campi di spin arbitrario inclusi nei
\emph{master fields} $w_{\mu}(x|y,\bar{y})$ e $C(x|y,\bar{y})$,
definiti nelle (\ref{master 1-form}) e (\ref{master $0$-form}),
rispettivamente. Facendo riferimento ad una teoria non lineare ben
nota, quella della gravit\`{a}, possiamo paragonare la situazione
in cui ci troviamo a quella in cui volessimo ricostruire la
dinamica completa del campo gravitazionale a partire dalle
equazioni libere, lineari nel campo $h_{\mu\nu}$ che descrive le
deviazioni dalla geometria piatta,
\begin{equation}
g_{\mu\nu}=\eta_{\mu\nu}+h_{\mu\nu}\ .
\end{equation}
Sebbene molto laboriosa, questa procedura \`{e} stata descritta in
\cite{Deser:uk}, ma molti anni dopo che la soluzione completa era
stata ottenuta da Einstein con metodi geometrici. In
Relativit\`{a} Generale si giunge direttamente alla forma completa
delle equazioni di campo facendo leva sul principio di gauge che
sottende alla teoria, l'invarianza per diffeomorfismi, e
stabilendo il contatto con la gravitazione newtoniana nel limite
non relativistico e di campo debole, ottenendo cos\`{i}, in un sol
colpo, tutti i termini non lineari. Qui non abbiamo invece altro
punto di partenza che la teoria libera in $AdS$, cui ci si deve
ridurre all'ordine lineare in $w_{1}$. Qual \`{e} il giusto
criterio guida verso la costruzione di interazioni consistenti?
Anzitutto, queste ultime non dovranno alterare il numero di
simmetrie di gauge della teoria libera: come accade nelle teorie
di Yang-Mills e in Relativit\`{a} Generale, le interazioni possono
deformare le simmetrie di gauge abeliane, ma non romperle. Il
formalismo delle FDA ci viene in aiuto, fornendo vincoli
sufficienti per determinare la forma dei termini non lineari, a
meno di alcune ambiguit\`{a} che costituiranno l'oggetto della
sezione \ref{5.4}. Inoltre, come vedremo, sar\`{a} cruciale
richiedere che le equazioni complete mantengano l'invarianza sotto
trasformazioni di Lorentz locali, ovvero che la simmetria di
Lorentz locale rimanga una sottoalgebra dell'algebra di HS
infinito-dimensionale.

Cerchiamo dunque una deformazione delle equazioni (\ref{COMSTh1})
e (\ref{COMSTh2}) del tipo
\begin{equation}\label{deformFDA}
R(w)=F_{2}(w,C)\ ,\qquad {\cal D}C=F_{1}(w,C)\ ,
\end{equation}
dove $F_{2}$ ed $F_{1}$ sono, rispettivamente, una $2$-forma e una
$1$-forma, $R$ \`{e} la curvatura di HS completa e ${\cal D}$ la
derivata covariante della rappresentazione \emph{twisted adjoint}
contenente l'intera connessione di HS $w$. Tale deformazione deve
risultare in una FDA, vale a dire che il sistema di equazioni
(\ref{deformFDA}) deve essere consistente nel senso di
(\ref{genJacobi}), e deve riprodurre le equazioni linearizzate
(\ref{COMSTh1}) e (\ref{COMSTh2}) al primo ordine nell'espansione
attorno al vuoto $AdS$ (che appare come soluzione particolare
della prima delle (\ref{deformFDA}) all'ordine zero), ovvero
nell'espansione in potenze di $C$ e $w_{1}$. $C$ appare come il
naturale parametro d'espansione della teoria, poich\'{e} descrive,
come esaminato nel capitolo precedente, le deviazioni dei campi di
materia e dei tensori di Weyl generalizzati di spin arbitrario dai
loro valori di vuoto $C=0$. \`{E} chiaro che, trattandosi di una
FDA, la deformazione consistente (\ref{deformFDA}) si tradurr\`{a}
automaticamente nella deformazione delle simmetrie di gauge della
teoria libera, contenenti la sola covariantizzazione rispetto ai
campi gravitazionali di background. La struttura delle equazioni
complete, cos\`{i} come delle trasformazioni di gauge, sar\`{a}
completamente determinata una volta che lo siano le funzioni
$F_{2}(w,C)$ ed $F_{1}(w,C)$. Per costruzione inoltre il sistema
non lineare (\ref{deformFDA}) sar\`{a} formalmente consistente,
invariante per trasformazioni generali di coordinate, e si
ridurr\`{a} alle equazioni libere del \emph{Central On-Mass-Shell
Theorem} ``spegnendo'' le interazioni.

\`{E} la condizione di integrabilit\`{a} stessa ad implicare che
il sistema (\ref{deformFDA}) contenga termini non lineari: facendo
agire la derivata covariante esterna ${\cal D}$ sulla seconda
equazione si ottiene infatti
\begin{equation}
{\cal D}^{2}C\sim RC\sim C^{2}+\textrm{ordini superiori} \ ,
\end{equation}
come risulta dal fatto che l'equazione (\ref{COMSTh1}) vincola la
curvatura linearizzata $R_{1}$ ad essere di ordine $C$. Questo
significa che $F_{1}$ parte almeno con termini quadratici in $C$,
il che \`{e} coerente con la (\ref{COMSTh2}), la quale implica
che, al primo ordine, $F_{1}(w,C)=0$ \cite{Vasiliev:1989yr}.

La consistenza dei vincoli (\ref{deformFDA}) assicura inoltre che,
note $F_{2}(w,C)$ ed $F_{1}(w,C)$, almeno in linea di principio il
nostro sistema sar\`{a} completamente risolvibile una volta noti i
valori delle $0$-forme in un punto arbitrario dello spazio-tempo,
secondo quanto esposto nella sezione \ref{4.6}. La prima equazione
fornisce $w$ in funzione di $C$, a meno di trasformazioni di
gauge. Dalla seconda otteniamo poi $C$ in termini di una
``condizione iniziale'' $C(x_{0})$, dove $x_{0}$ \`{e} un punto
nello spazio-tempo.

Nonostante la potenza del formalismo delle FDA, grazie al quale
\`{e} sufficiente richiedere l'integrabilit\`{a} del sistema per
ottenere automaticamente l'invarianza sotto trasformazioni di
gauge di spin arbitrario, ricavare la forma delle $F_{2}(w,C)$ e
$F_{1}(w,C)$, controllando ordine per ordine che preservino la
consistenza delle equazioni, \`{e} tutt'altro che semplice.
Tuttavia Vasiliev ha sviluppato un formalismo
\cite{Vasiliev:en,Vasiliev:1990bu,Vasiliev:vu,Vasiliev:1992av} che
consente di deformare la FDA linerizzata e ottenere le
(\ref{deformFDA}) partendo dalle equazioni di una FDA non
deformata (analoghe in forma alle (\ref{0gencurv})) ma formulata
in uno spazio-tempo esteso $\widehat{{\cal M}}$, che ne
costituisce la variet\`{a} di base, dato dal prodotto dello
spazio-tempo ordinario ${\cal M}$ per uno spazio complesso di
coordinate spinoriali
$Z_{\underline{\alpha}}=(z_{\alpha},-z^{\dot{\alpha}})$ (dove
$Z_{\underline{\alpha}}$ \`{e} uno spinore di Majorana puramente
immaginario) non commutanti. Definiamo dunque un'immersione
$i:\,{\cal M}\hookrightarrow\widehat{{\cal M}}$: il sottospazio
$Z=0$ di $\widehat{{\cal M}}$ corrisponde allo spazio-tempo
ordinario, $\widehat{{\cal M}}|_{Z=0}={\cal M}$. Tale procedimento
comporta anche il raddoppiamento delle variabili spinoriali
ausiliarie $y,\bar{y}$ nella \emph{master $1$-form} di connessione
e nella \emph{master $0$-form} di HS
\begin{equation}
w(x|Y)\rightarrow \hat{w}(x|Y,Z)\ , \qquad C(x|Y)\rightarrow
\hat{C}(x|Y,Z)\ .
\end{equation}
Seguendo le notazioni di Vasiliev, nel resto di questo capitolo
denoteremo con $W(x|Y,Z)$ la \emph{master $1$-form} estesa
$\hat{w}(x|Y,Z)$ e con $B(x|Y,Z)$ la \emph{master $0$-form} estesa
$\hat{C}(x|Y,Z)$. Ciascun coefficiente dello sviluppo di $W$ e $B$
in potenze degli oscillatori $Y$ ammette a sua volta un'espansione
in serie di potenze delle variabili $Z$, e viceversa ciascun
coefficiente di quest'ultima \`{e} funzione delle $Y$. Tale
raddoppiamento di variabili ausiliarie consente di immergere
equazioni di HS interagenti in semplici condizioni di curvatura
nulla per una $\star$-(super)algebra estesa, rendendone evidente
la consistenza \cite{Vasiliev:1990bu}. Affinch\'{e} tali
condizioni in $\widehat{{\cal M}}$ effettivamente diano luogo ad
una FDA deformata nel sottospazio $\widehat{{\cal M}}|_{Z=0}$
\`{e} cruciale richiedere che le coordinate spinoriali
$Z_{\underline{\alpha}}$ siano non commutative. Si tratta quindi
di estendere l'algebra associativa ${\cal A}^{\star}$ in
un'algebra $\hat{{\cal A}}^{\star}$, definendo anzitutto
l'operazione di composizione di quest'ultima,
\begin{eqnarray}\label{defdiffext}
(P\star Q)(Y,Z) & = &
P\exp\biggl[i\biggl(\frac{\overleftarrow{\partial}}{\partial
z_{\alpha}}+\frac{\overleftarrow{\partial}}{\partial y_{\alpha}}
\biggr)\biggl(\frac{\overrightarrow{\partial}}{\partial
z^{\alpha}}-\frac{\overrightarrow{\partial}}{\partial y^{\alpha}}
\biggr) \nonumber\\
&&+i \biggl(\frac{\overleftarrow{\partial}}{\partial
\bar{z}_{\dot{\alpha}}}-\frac{\overleftarrow{\partial}}{\partial
\bar{y}_{\dot{\alpha}}} \biggr) \biggl(
\frac{\overrightarrow{\partial}}{\partial
\bar{z}^{\dot{\alpha}}}+\frac{\overrightarrow{\partial}}{\partial
\bar{y}^{\dot{\alpha}}} \biggr) \biggr] Q \ ,
\end{eqnarray}
manifestamente non locale nelle variabili spinoriali $Y$ e $Z$.
Tale generalizzazione mantiene tutte le propriet\`{a} dello
$\star$-prodotto precedentemente definito, in particolare la
regolarit\`{a} \footnote {Il fatto che $\star$ sia una legge di
prodotto regolare \`{e} fondamentale, poich\'{e} assicura che
tutte le equazioni della teoria che lo contengono hanno senso per
i coefficienti dello sviluppo dei \emph{master fields} in potenze
degli oscillatori, e dunque per i campi relativistici di spin
arbitrario ad essi corrispondenti.}, e dalla (\ref{defdiffext})
discendono, come casi particolari, le di contrazioni
\begin{equation}
z_{\alpha}\star z_{\beta}  =
z_{\alpha}z_{\beta}-i\varepsilon_{\alpha\beta}\ , \qquad
\bar{z}_{\dot{\alpha}}\star\bar{z}_{\dot{\beta}}  =
\bar{z}_{\dot{\alpha}}\bar{z}_{\dot{\beta}}-i\varepsilon_{\dot{\alpha}\dot{\beta}}
\ ,\nonumber
\end{equation}
\begin{equation}
y_{\alpha}\star z_{\beta}  =
y_{\alpha}z_{\beta}-i\varepsilon_{\alpha\beta}\ , \qquad
\bar{y}_{\dot{\alpha}}\star\bar{z}_{\dot{\beta}}  =
\bar{y}_{\dot{\alpha}}\bar{z}_{\dot{\beta}}+i\varepsilon_{\dot{\alpha}\dot{\beta}}
\ , \nonumber
\end{equation}
\begin{equation}
z_{\alpha}\star y_{\beta}  =
z_{\alpha}y_{\beta}+i\varepsilon_{\alpha\beta}\ , \qquad
\bar{z}_{\dot{\alpha}}\star\bar{y}_{\dot{\beta}}  =
\bar{z}_{\dot{\alpha}}\bar{y}_{\dot{\beta}}-i\varepsilon_{\dot{\alpha}\dot{\beta}}
\ , \label{contrulez}
\end{equation}
che danno origine a regole generali analoghe alle
(\ref{contrule}), a parte un ulteriore fattore $(-1)$ per ogni
contrazione del tipo $z\star z$, $\bar{z}\star\bar{z}$, $y\star z$
e $\bar{y}\star\bar{z}$. Tutte le altre contrazioni sono nulle.
Dati due polinomi $P(Y,Z),Q(Y,Z)\in \hat{{\cal A}}^{\star}$, la
realizzazione integrale dello $\star$-prodotto generalizzato \`{e}
data da
\begin{eqnarray}
&&\!(P\star Q)(Y,Z) =  \nonumber\\
&&\!\frac{1}{(2\pi)^{4}}\int d^{4}U d^{4}V
P(Y+U,Z+U)Q(Y+V,Z-V)\exp
i(u_{\alpha}v^{\alpha}+\bar{u}_{\dot{\alpha}}\bar{v^{\dot{\alpha}}})
\ . \label{defintext}
\end{eqnarray}

Per semplicit\`{a},  nel seguito ci ridurremo alla sottoalgebra
$hs(4)\subset shs(4)$ costituita dai soli campi bosonici di spin
pari $s=0,2,4,6,...$ (per un'analisi della teora non lineare
completamente generale cfr. \cite{Vasiliev:1992av}) . \`{E}
infatti evidente da (\ref{infdim}) che l'insieme dei generatori di
spin dispari costituisce effettivamente una sottoalgebra,
poich\'{e} lo $\star$-commutatore si chiude su di essi. Il modello
corrispondente viene detto \emph{modello bosonico
minimale}\footnote {Naturalmente i soli generatori di spin dispari
rendono conto della presenza, nel modello, di campi di gauge ad
essi associati di spin $s$ pari e maggiore o uguale a $2$. Si
pu\`{o} mostrare per\`{o} \cite{Sezgin:1998eh, Sezgin:1998gg} che
la costruzione di una teoria unitaria basata su $hs(4)$ richiede
l'inclusione del campo scalare, di $s=0$. L'argomento si fonda
sulla considerazione che una teoria di gauge basata su $hs(4)$
\`{e} fisicamente consistente se lo spettro degli stati fisici che
essa coinvolge corrisponde ad una rappresentazione unitaria
dell'algebra stessa. Lo spettro del modello bosonico minimale,
cos\`{i} come emerge da un'analisi simile a quella della sezione
\ref{4.4}, contiene esattamente gli stati provenienti dalla
decomposizione in rappresentazioni irriducibili di $so(3,2)$ della
parte totalmente simmetrica del prodotto tensoriale di due
rappresentazioni unitarie di $so(3,2)$ dette singletoni
\cite{Flato:1978qz, Sezgin:1998gg}. Si pu\`{o} dimostrare che
queste ultime sono rappresentazioni unitarie anche dell'intera
algebra $hs(4)$ e che tale \`{e} anche il loro prodotto totalmente
simmetrizzato. La decomposizione in rappresentazioni irriducibili
di quest'ultimo ne include necessariamente una di spin $s=0$
insieme ad infinite altre di spin $2,4,6...$, sicch\'{e} il
settore scalare deve essere incluso per unitariet\`{a}.} e pu\`{o}
essere definito formalmente facendo riferimento all'insieme dei
polinomi negli oscillatori $P$ (vedi (\ref{ply2})) che soddisfano
le condizioni
\begin{equation}\label{minbosmod}
\tau(P)=-P\ , \qquad P^{\dag}=-P\ ,
\end{equation}
dove l'azione della mappa $\tau$ \`{e} definita da
\begin{equation}\label{tau}
\tau(y_{\alpha})=iy_{\alpha}\ ,\qquad
\tau(\bar{y}_{\dot{\alpha}})=i\bar{y}_{\dot{\alpha}}\ .
\end{equation}
Dalla definizione segue che $\tau$ agisce come un'anti-involuzione
della $\star$-algebra,
\begin{equation}
\tau(F\star G)=\tau(G)\star\tau(F)\ ,\qquad F,G\in{\cal
A}^{\star}\ .
\end{equation}
Inoltre
\begin{equation}
\tau([F,G]_{\star})=-[\tau(F),\tau(G)]_{\star}\ ,
\end{equation}
sicch\'{e} la condizione $\tau(P)=-P$ proietta effettivamente su
un sottospazio invariante di $shs(4)$, denominato nella
letteratura $hs(4)$, costituito da tutti i $P$ della forma
\begin{equation}\label{hsply}
P(y,\bar{y})=\sum_{n+m=2\,\textrm{mod}\,
4}\frac{i}{2n!m!}P^{\alpha_{1}...\alpha_{n}\dot{\alpha_{1}}...\dot{\alpha_{m}}}y_{\alpha_{1}}...y_{\alpha_{n}}\bar{y}_{\dot{\alpha_{1}}}...\bar{y}_{\dot{\alpha_{m}}}
\ .
\end{equation}
Cos\`{i}, ad esempio, lo sviluppo (\ref{master 1-form}) sui
generatori della connessione $w(x|Y)$ si adatta al modello
bosonico minimale con
\begin{eqnarray}\label{hsmaster 1-form}
&& w(x|y,\bar{y}) = dx^{\mu}w_{\mu}(x|y,\bar{y}) \nonumber \\
& = & \sum_{n+m=2\,\textrm{mod}\,
4}\frac{i}{2n!m!}dx^{\mu}w_{\mu}\,^{\alpha_{1}...\alpha_{n}\dot{\alpha_{1}}...\dot{\alpha_{m}}}(x)y_{\alpha_{1}}...y_{\alpha_{n}}\bar{y}_{\dot{\alpha_{1}}}...\bar{y}_{\dot{\alpha_{m}}}
\ ,
\end{eqnarray}
e analogamente per la curvatura di HS. Il campo scalare viene
incluso, al solito, nella \emph{master $0$-form} $C$, che ha
valori nella rappresentazione \emph{twisted adjoint}. Nell'ambito
del modello bosonico minimale essa pu\`{o} esser definita tramite
le condizioni \cite{Sezgin:1998gg, Sezgin:1998eh, Sezgin:2002ru}
\begin{equation}\label{hstwadj}
\tau(C)=\bar{\pi}(C)\ , \qquad C^{\dag}=\pi(C)\ ,
\end{equation}
la cui soluzione generale ha la forma
\begin{equation}\label{hsmaster0}
C=c+\pi(c^{\dag})\ ,
\end{equation}
dove $c$ ammette l'espansione
\begin{equation}\label{hsmaster$0$-form}
c(x|y,\bar{y})=\sum_{n-m=0\,\textrm{mod}\,4}\frac{1}{n!m!}c^{\alpha_{1}...\alpha_{n}\dot{\alpha}_{1}...\dot{\alpha}_{m}}(x)\,y_{\alpha_{1}}...y_{\alpha_{n}}\bar{y}_{\dot{\alpha}_{1}}...\bar{y}_{\dot{\alpha}_{n}}
\ ,
\end{equation}
e contiene tutte le soluzioni di $\tau(c)=\bar{\pi}(c)$ che hanno
$n\geq m$. Per $c^{\dag}$ vale la disuguaglianza opposta, ma si
noti come il $\pi$-\emph{twist} che compare in (\ref{hsmaster0})
impedisca di scrivere $C$ in modo compatto cambiando in
$|n-m|=0\,\textrm{mod}\,4$ il pedice della somma di
(\ref{hsmaster$0$-form}). Si noti anche che la condizione
$\tau(C)=\bar{\pi}(C)$ implica $\bar{\pi}(C)=\pi(C)$ e che,
pi\`{u} in generale, nel modello bosonico minimale vale
$\pi(P)=\bar{\pi}(P)$, come conseguenza del fatto che la prima
condizione in (\ref{minbosmod}) seleziona i soli generatori con
$n+m$ pari. Alternativamente, \`{e} sufficiente osservare l'azione
delle mappe $\tau$ e $\pi,\bar{\pi}$ per rendersi conto che
$\tau^{2}(P)=\pi\bar{\pi}(P)=P,\,\forall P\in hs(4)$.

Si pu\`{o} a questo punto estendere tale algebra di HS bosonica,
definendo l'immersione $i:\,hs(4)\hookrightarrow\widehat{hs}(4)$,
prendendo il sottoinsieme di $\hat{{\cal {A}}}^{\star}$
corrispondente ai generatori $\hat{P}$ anti-hermitiani,
$\hat{P}^{\dag}=-\hat{P}$ e tali che
$\tau(\hat{P})^{\dag}=-\hat{P}$. Naturalmente, tale
$\star$-algebra estesa si riduce ad $hs(4)$ per $Z=0$,
\begin{equation}
hs(4)=\widehat{hs}(4)|_{Z=0}\ .
\end{equation}
Estendiamo anzitutto la mappa $\tau$, definendone l'azione sulle
variabili spinoriali $Z$:
\begin{equation}\label{tauz}
\tau(z_{\alpha})=-iz_{\alpha}\ ,\qquad
\tau(\bar{z}_{\dot{\alpha}})=-i\bar{z}_{\dot{\alpha}}\ .
\end{equation}
Promuoviamo $\widehat{hs}(4)$ a simmetria locale introducendo in
$\widehat{{\cal M}}$ la $1$-forma di connessione a valori in
$\widehat{hs}(4)$ (\emph{master $1$-form} estesa totale)
\begin{equation}\label{calW}
{\cal
W}=W+V=dx^{\mu}W_{\mu}+dz^{\alpha}V_{\alpha}-d\bar{z}^{\dot{\alpha}}\bar{V}_{\dot{\alpha}}
\ ,
\end{equation}
dove $V$ gioca il ruolo di $1$-forma di connessione nello spazio
$Z$, la componente puramente spinoriale della connessione totale
${\cal W}$. Naturalmente
\begin{equation}
\tau({\cal W})=-{\cal W}\ ,\qquad {\cal W}^{\dag}=-{\cal W}\ ,
\end{equation}
che implica identiche condizioni su $W$ e $V$. Si noti tuttavia
che, in virt\`{u} delle (\ref{tauz}), per la componente
$V_{\alpha}$ si ha $\tau(V_{\alpha})=-iV_{\alpha}$, e analogamente
per $\bar{V}_{\dot{\alpha}}$. Introduciamo inoltre la derivata
esterna totale sullo spazio prodotto (x,Z) $\hat{d}$, definita
come
\begin{equation}
\hat{d}\equiv
dx^{\mu}\partial_{\mu}+dz^{\alpha}\partial_{\alpha}+d\bar{z}^{\dot{\alpha}}\bar{\partial}_{\dot{\alpha}}\equiv
d+\partial+\bar{\partial}\equiv d+d_{Z}\ ,
\end{equation}
dove $\partial_{\alpha}=\frac{\partial}{\partial z^{\alpha}}$, e
analogamente per $\bar{\partial}_{\dot{\alpha}}$. \`{E} allora
possibile definire una $2$-forma di curvatura totale come
\begin{equation}\label{totalmastercurv}
{\cal R}=\hat{d}{\cal W}+{\cal W}\star{\cal W}\ ,
\end{equation}
che soddisfa l'identit\`{a} di Bianchi
\begin{equation}\label{totalBid}
\hat{D}{\cal R}\equiv \hat{d}{\cal R}+[{\cal W},{\cal
R}]_{\star}=0 \ .
\end{equation}
Sotto le trasformazioni di gauge di $\widehat{hs}(4)$,
\begin{equation}\label{extgaugetransfW}
\delta_{\hat{\epsilon}}{\cal
W}=\hat{D}\hat{\epsilon}=\hat{d}\hat{\epsilon}+[{\cal
W},\hat{\epsilon}]_{\star} \ ,
\end{equation}
${\cal R}$ trasforma covariantemente,
\begin{equation}
\delta_{\hat{\epsilon}}{\cal R}=[{\cal R},\hat{\epsilon}]_{\star}
\ ,
\end{equation}
dove $\hat{\epsilon}(x|Y,Z)$ \`{e} un parametro locale a valori in
$\widehat{hs}(4)$. Conformemente alle assunzioni fatte, la
\emph{master $1$-form} di connessione $w(x|Y)$ e la \emph{master
$0$-form} $C(x|Y)$ di $hs(4)$ si ottengono come
\begin{eqnarray}
w(x|Y) & = & i^{\ast}{\cal W}(x|Y,Z)\equiv W(x|Y,Z)|_{Z=0}\ , \\
C(x|Y) & = & i^{\ast}B(x|Y,Z)\equiv B(x|Y,Z)|_{Z=0}\ ,
\end{eqnarray}
dove con $i^{\ast}$ indichiamo il \emph{pull-back} della mappa $i$
\footnote {Data l'immersione $i:\,{\cal
M}\hookrightarrow\widehat{{\cal M}}$, definiamo $i^{\ast}$ come la
mappa lineare che manda la funzione $f$ definita su
$\widehat{{\cal M}}$ nella funzione $i^{\ast}f$ definita su ${\cal
M}$ e corrispondente alla funzione il cui valore sul punto $p\in
{\cal M}$ \`{e} il valore di $f$ su $i(p)\in \widehat{{\cal M}}$,
ovvero $i^{\ast}f(p)=f(i(p))$.}. Si noti tuttavia che
\begin{equation}\label{observ1}
i^{\ast}{\cal R}=\frac{1}{2}dx^{\mu}\wedge dx^{\nu}{\cal
R}_{\mu\nu}|_{Z=0}\neq R(x|Y) \ ,
\end{equation}
dove ${\cal R}_{\mu\nu}$ denota le componenti puramente
spazio-temporali della $2$-forma di curvatura totale dell'algebra
$\widehat{hs}(4)$ definita in (\ref{totalmastercurv}), ovvero
\begin{equation}
\frac{1}{2}dx^{\mu}\wedge dx^{\nu}{\cal
R}_{\mu\nu}(x|Y,Z)=dW(x|Y,Z)+W(x|Y,Z)\star W(x|Y,Z)\ ,
\end{equation}
che trasformano covariantemente sotto le componenti puramente
spazio-temporali delle (\ref{extgaugetransfW}),
\begin{equation}\label{extgaugetransfw}
\delta_{\hat{\epsilon}}
W=D\hat{\epsilon}=d\hat{\epsilon}+[W,\hat{\epsilon}]_{\star} \ .
\end{equation}
Si pu\`{o} inoltre definire
\begin{equation}\label{twtotalcovdev}
\hat{{\cal D}}B=\hat{d}B+{\cal W}\star B-B\star\bar{\pi}({\cal W})
\ ,
\end{equation}
dove l'azione delle mappe $\pi,\bar{\pi}$ viene estesa ad
$\widehat{hs}(4)$ con
\begin{equation}
\pi(z_{\alpha}) = -z_{\alpha}\ , \qquad \bar{\pi}(z_{\alpha}) =
z_{\alpha} \ ,\nonumber
\end{equation}
\begin{equation}
\pi(\bar{z}_{\dot{\alpha}}) = \bar{z}_{\dot{\alpha}}\ , \qquad
\bar{\pi}(\bar{z}_{\dot{\alpha}}) = -\bar{z}_{\dot{\alpha}}\ ,
\end{equation}
involuzioni dell'algebra estesa. La (\ref{twtotalcovdev})
trasforma covariantemente sotto (\ref{extgaugetransfW}),
\begin{equation}
\delta_{\hat{\epsilon}}\hat{{\cal
D}}B=-\hat{\epsilon}\star\hat{{\cal D}}B+\hat{{\cal
D}}B\star\bar{\pi}(\hat{\epsilon})\ ,
\end{equation}
se $B$, a sua volta, trasforma come
\begin{equation}\label{twextgaugetransf}
\delta_{\hat{\epsilon}} B=-\hat{\epsilon}\star
B+B\star\bar{\pi}(\hat{\epsilon})\ ,
\end{equation}
cio\`{e} se appartiene alla \emph{twisted adjoint} di
$\widehat{hs}(4)$, definita dalle condizioni
\begin{equation}\label{exthstwadj}
\tau(B)=\bar{\pi}(B)\ , \qquad B^{\dag}=\pi(B)\ .
\end{equation}
Si trova che
\begin{equation}\label{observ2}
i^{\ast}\hat{{\cal D}}B=dx^{\mu}(\hat{{\cal D}}_{\mu}B)|_{Z=0}\neq
{\cal D}C(x|Y)\ ,
\end{equation}
dove $\hat{{\cal D}}_{\mu}B$ corrisponde alla componente puramente
spazio-temporale della derivata covariante totale \emph{twisted}
definita in (\ref{twtotalcovdev}), ovvero
\begin{eqnarray}
dx^{\mu}\hat{{\cal D}}_{\mu}B(x|Y,Z) & = & d
B(x|Y,Z)+W(x|Y,Z)\star B(x|Y,Z) \nonumber \\
&&-B(x|Y,Z)\star\bar{\pi}(W(x|Y,Z))\ ,
\end{eqnarray}
mentre
\begin{equation}
{\cal D}C(x|Y)=dC(x|Y)+w(x|Y)\star
C(x|Y)-C(x|Y)\star\bar{\pi}(w(x|Y))\ .
\end{equation}
La (\ref{linconstr2}) costituisce la linearizzazione di
quest'ultima equazione rispetto al background $AdS$.

Le equazioni (\ref{observ1}) e (\ref{observ2}) mostrano per quale
motivo \`{e} necessario che le coordinate ausiliarie $Z$ siano non
commutative, al fine di ottenere equazioni di HS non banali nello
spazio-tempo ordinario. Se le $Z$ fossero state variabili
commutanti, partire da condizioni di curvatura nulla per una FDA
estesa  ed imporne la restrizione al sottospazio $Z=0$ avrebbe
automaticamente implicato le condizioni
\begin{eqnarray}
i^{\ast}{\cal R} & = & R(x|Y)=dw(x|Y)+w(x|Y)\star w(x|Y)=0 \ ,\nonumber \\
i^{\ast}\hat{{\cal D}}B & = & {\cal D}C(x|Y)=0 \ ,
\end{eqnarray}
senza alcuna deformazione non banale del tipo (\ref{deformFDA})
cercato. Viceversa, le relazioni (\ref{contrulez}) fanno s\`{i}
che $i^{\ast}{\cal R}$ e $i^{\ast}\hat{{\cal D}}B$ contengano
contributi, provenienti dai termini quadratici, che hanno origine
dalle contrazioni dei termini di ordine arbitrariamente elevato
dell'espansione in serie di Taylor nelle variabili $z$ e $\bar{z}$
di $W(x|Y,Z)$ e $B(x|Y,Z)$ (in particolare, poich\'{e} le $Y$ e le
$Z$ soddisfano algebre isomorfe, esistono termini di contrazione
non banali anche tra $z$ ed $y$ e tra $\bar{z}$ ed $\bar{y}$, come
abbiamo visto), e ci\`{o} si traduce in una dipendenza da tutti i
coefficienti di espansione, e non solo da quelli di ordine zero
identificabili con $w(x|Y)$ e $C(x|Y)$. Ma come nascono le non
linearit\`{a} in $C$ del sistema (\ref{deformFDA})?

Le equazioni della FDA estesa avranno la forma generale
\begin{equation}\label{extgenericFDA}
{\cal R}=\hat{F}_{2}({\cal W},B)\ , \qquad \hat{{\cal
D}}B=\hat{F}_{1}({\cal W},B)\ ,
\end{equation}
e, per definizione, saranno tra loro consistenti sia rispetto alle
variabili $x$ che alle $Z$ (ovvero saranno compatibili con la
condizione $\hat{d}^{2}=0$). Questo implica che potremo risolvere
la prima equazione per ${\cal W}(x|Y,Z)$ in termini di $B(x|Y,Z)$,
ottenendo poi dalla seconda $B(x|Y,Z)$ in termini della condizione
iniziale $B(x_{0}|Y,0)=C(x_{0}|Y)$, a meno di una trasformazione
di gauge di $\widehat{hs}(4)$. Si noti che, secondo la discussione
della sezione \ref{4.6}, \`{e} questa propriet\`{a} che consente
di affermare che la FDA estesa (\ref{extgenericFDA}) \`{e}
equivalente alla (\ref{deformFDA}), poich\'{e} entrambe hanno gli
stessi dati iniziali.

Si \`{e} detto che il raddoppiamento delle variabili spinoriali
offre la possibilit\`{a} di immergere un complicato sistema non
lineare come (\ref{deformFDA}) 
in vincoli di curvatura non pi\`{u} complicati, in forma, di
quelli linearizzati del \emph{Central On-Mass-Shell Theorem}, ma
definiti sullo spazio $(x,Z)$. Questo perch\'{e}, come vedremo,
tale artificio consente di esprimere complicate espressioni come
$F_{2}(w,C)$ e $F_{1}(w,C)$ come soluzioni di alcuni semplici
vincoli differenziali rispetto alle variabili ausiliarie $Z$,
inclusi nelle componenti spinoriali (quelle che hanno almeno un
indice di spazio $Z$) delle (\ref{extgenericFDA}). La consistenza
di queste ultime implica infatti che si possa anzitutto ottenere
la dipendenza dalle $Z$ di $W$, $V$ e $B$ perturbativamente in
termini delle condizioni iniziali $W(x|Y,0)=w(x|Y)$ e
$B(x|Y,0)=C(x|Y)$, legando in questo modo i coefficienti delle
varie potenze di $z,\bar{z}$ ad espressioni contenenti potenze
sempre pi\`{u} alte di $C$, come vedremo in dettaglio nella
sezione \ref{esppert}. Vedremo inoltre che sar\`{a} possibile
utilizzare parte della libert\`{a} di gauge
(\ref{extgaugetransfW}) per imporre le condizioni
$V_{\alpha}|_{Z=0}=\bar{V}_{\dot{\alpha}}|_{Z=0}=0$, preservando
l'intera simmetria di gauge $hs(4)$.

Sostituendo le soluzioni ottenute $W=W(w,C)$ e $B=B(C)$ nelle
componenti puramente spazio-temporali delle (\ref{extgenericFDA})
si ottiene infine una FDA deformata del tipo (\ref{deformFDA}),
contenente potenze arbitrariamente elevate di $C$. Ci\`{o} in
virt\`{u} del fatto che, come evidenziato in precedenza, anche per
$Z=0$ sopravvivono termini provenienti da contrazioni di ordine
arbitrariamente elevato, corrispondenti a derivate di ordine
arbitrariamente elevato rispetto a $z$ e $\bar{z}$ (cfr. Appendice
\ref{appendice B}) di $B$ e $W$ in $Z=0$: ma abbiamo detto che le
componenti spinoriali dei vincoli di curvatura riesprimono ogni
derivata di $W$ o $B$ rispetto alle variabili ausiliarie in
termini delle condizioni iniziali, e questo implica le infinite
non linearit\`{a} in $C$. Da questa discussione segue che
deformazioni non lineari consistenti delle equazioni del
\emph{Central On-Mass-Shell Theorem} esistono, ed in particolare
che
\begin{eqnarray}\label{dots1}
\frac{1}{2}dx^{\mu}\wedge dx^{\nu}{\cal
R}_{\mu\nu}|_{Z=0} & = & R(x|Y)+... \ ,\nonumber \\
dx^{\mu}(\hat{{\cal D}}_{\mu}B)|_{Z=0} & = & {\cal D}C(x|Y)+...\
,\label{dots1}
\end{eqnarray}
e
\begin{eqnarray}\label{dots2}
F_{2}(w,C) & = & \hat{F}_{2}(w,C)+...\ ,\nonumber \\
F_{1}(w,C) & = & \hat{F}_{1}(w,C)+...\ ,\label{dots2}
\end{eqnarray}
dove con ... sottintendiamo gli infiniti termini di ordine
superiore in $C$ provenienti dalle contrazioni di potenze
arbitrariamente alte in $z$, $\bar{z}$ delle espansioni di Taylor
di ${\cal W}$ e $B$.

L'estensione tramite coordinate non commutative \`{e} quindi un
trucco matematico che consente di ottenere complicate equazioni
non lineari come restrizione al sottospazio $Z=0$ di equazioni
semplici di una FDA estesa, un po' come un modello sigma non
lineare ``nasconde'' una dinamica altamente non lineare in una
lagrangiana identica in forma a quella di campi liberi attraverso
un vincolo che restringe questi ultimi su una certa variet\`{a}.
Non \`{e} chiaro attualmente quale interpretazione ammettano le
coordinate $Z$ nella corrispondenza con la Teoria delle Stringhe.

\section{Equazioni di HS non lineari nello spazio esteso (x,Z)}
\label{eqnonlin}

Tenendo conto delle (\ref{dots1}) e (\ref{dots2}), oltre che delle
equazioni linearizzate (\ref{COMSTh1}) e (\ref{COMSTh2}) alle
quali vogliamo ridurci, possiamo proporre un ansatz sulla forma
delle (\ref{extgenericFDA}):
\begin{eqnarray}
{\cal R} & = & i\,dz^{\alpha}\wedge dz_{\alpha}{\cal V}(B) +
i\,d\bar{z}^{\dot{\alpha}}\wedge d\bar{z}_{\dot{\alpha}}({\cal
V}(B))^{\dag}\ ,\nonumber \\
\hat{{\cal D}}B & = & 0\ , \label{extgenericFDA2}
\end{eqnarray}
dove ${\cal V}(B)$ \`{e} una funzione di $B$ compatibile con la
condizione di consistenza del sistema. Daremo tra breve condizioni
ulteriori che ne specificheranno la forma, a meno di alcune
ambiguit\`{a} che discuteremo in seguito.

Abbiamo gi\`{a} introdotto i differenziali $dz^{\alpha}$ e
$d\bar{z}^{\dot{\alpha}}$ dello spazio $Z$, e possiamo aggiungere
le condizioni
\begin{equation}
(z^{\alpha})^{\dag}=\bar{z}^{\dot{\alpha}}\ , \qquad
(dz^{\alpha})^{\dag}=d\bar{z}^{\dot{\alpha}}\ .
\end{equation}
Tenendo conto delle (\ref{contrulez}) (o delle (\ref{sympldiff}))
e del fatto che lo $\star$-prodotto di un qualsiasi elemento
$P\in\hat{{\cal A}}^{\star}$ con $dZ^{\underline{\alpha}}$ non
d\`{a} luogo a contrazioni, dovrebbe essere chiaro che la
definizione
\begin{equation}\label{S0}
S_{0}\equiv
dz^{\alpha}z_{\alpha}+d\bar{z}^{\dot{\alpha}}\bar{z}_{\dot{\alpha}}=(S_{0})^{\dag}
\end{equation}
d\`{a} luogo ad una $1$-forma rispetto ai differenziali di base
$dz^{\alpha}$ e $d\bar{z}^{\dot{\alpha}}$ che realizza l'azione
della derivata esterna $d_{Z}$ nello spazio $Z$ come
\begin{equation}\label{propS0}
S_{0}\star F_{p}-(-1)^{p}F_{p}\star S_{0}=-2id_{Z}F_{p}\ , \qquad
d_{Z}\equiv
dz^{\alpha}\frac{\partial}{z^{\alpha}}+d\bar{z}^{\dot{\alpha}}\frac{\partial}{\partial\bar{z}^{\dot{\alpha}}}
\ ,
\end{equation}
dove $F_{p}$ \`{e} una forma di grado totale $p$ nello spazio
(x,Z) a valori in $\widehat{hs}(4)$ e $dx^{\mu}\wedge
dz^{\alpha}=-dz^{\alpha}\wedge dx^{\mu}$, $dx^{\mu}\wedge
d\bar{z}^{\dot{\alpha}}=-d\bar{z}^{\dot{\alpha}}\wedge dx^{\mu}$.
La propriet\`{a} associativa dello $\star$-prodotto implica la
validit\`{a} della regola di Leibniz
\begin{equation}
d_{Z}(A_{p}\star B_{q})=d_{Z}A_{p}\star B_{q}+(-1)^{q}A_{p}\star
d_{Z}B_{q}\ .
\end{equation}
Definiamo inoltre l'azione di $\pi$ sui differenziali
$dz^{\alpha},d\bar{z}^{\dot{\alpha}}$ attraverso
\begin{equation}
d_{Z}(\pi(F(Y,Z))=\pi(d_{Z}(F(Y,Z))\ ,
\end{equation}
che implica $\pi(dz^{\alpha})=-dz^{\alpha}$ ed analogamente per
$d\bar{z}^{\dot{\alpha}}$ e per $\bar{\pi}$.

Torniamo alla forma delle (\ref{extgenericFDA2}). Per confronto
con la (\ref{COMSTh1}) deduciamo che la funzione ${\cal V}(B)$
deve possedere una caratteristica singolare: sebbene il suo
argomento $B$ sia una funzione di $Y^{\underline{\alpha}}$ e
$Z^{\underline{\beta}}$, la sua restrizione al sottospazio $Z=0$
deve in effetti dipendere soltanto dalle $y^{\alpha}$ e non dalle
$\bar{y}^{\dot{\alpha}}$, e viceversa per l'hermitiano coniugato
$({\cal V}(B))^{\dag}$. Tali condizioni vengono soddisfatte
costruendo ${\cal V}(B)$ con la speciale funzione
$\kappa(y,z)\in\hat{{\cal A}}^{\star}$, definita come
\begin{equation}
\kappa(y,z)\equiv \exp(iz_{\alpha}y^{\alpha})=\tau(\kappa(y,z))\ ,
\end{equation}
e $({\cal V}(B))^{\dag}$ con il suo hermitiano coniugato
\begin{equation}
\bar{\kappa}(\bar{y},\bar{z})\equiv
\exp(-i\bar{z}_{\dot{\alpha}}\bar{y}^{\dot{\alpha}})=(\kappa(y,z))^{\dag}=\tau(\bar{\kappa}(\bar{y},\bar{z}))
\ .
\end{equation}
Usando la (\ref{defintext}) \`{e} possibile verificare le seguenti
propriet\`{a}:
\begin{equation}
\kappa\star F(z,\bar{z};y,\bar{y})=\kappa F(y,\bar{z};z,\bar{y})\
, \qquad \bar{\kappa}\star
F(z,\bar{z};y,\bar{y})=\bar{\kappa}F(z,-\bar{y};y,-\bar{z})\
,\nonumber
\end{equation}
\begin{equation}\label{propk}
F(z,\bar{z};y,\bar{y})\star\kappa=\kappa F(-y,\bar{z};-z,\bar{y})\
, \qquad
F(z,\bar{z};y,\bar{y})\star\bar{\kappa}=\bar{\kappa}F(z,\bar{y};y,\bar{z})\
.
\end{equation}
\`{E} immediato controllare che esse implicano
\begin{equation}
\pi(F)=\kappa\star F\star\kappa\ , \qquad
\bar{\pi}(F)=\bar{\kappa}\star F\star\bar{\kappa}
\end{equation}
(con $\kappa\star\kappa=1$ e $\bar{\kappa}\star\bar{\kappa}=1$),
ovvero gli automorfismi involutivi esterni $\pi,\bar{\pi}$
diventano in effetti interni in $\hat{{\cal A}}^{\star}$, essendo
generati dal coniugio con $\kappa,\bar{\kappa}$, rispettivamente
\footnote {In generale, un automorfismo $\omega$ di un'algebra di
Lie $g$ \`{e} detto \emph{interno} se pu\`{o} essere ottenuto come
prodotto di automorfismi del tipo $\textrm{Ad}_{x}\equiv
\exp(\textrm{ad}_{x})$, dove con $\textrm{ad}_{x}$ si intende la
\emph{mappa aggiunta} associata ad un qualsiasi elemento $x\in g$,
definita come $y\mapsto\textrm{ad}_{x}(y)\equiv [x,y]$, $y\in g$,
mentre, data una generica mappa $\varphi$,  $\exp(\varphi)\equiv
\sum_{n=0}^{\infty}\frac{1}{n!}\varphi^{n}$. Ogni automorfismo che
non soddisfi questa condizione si dice \emph{esterno}. Nel nostro
caso l'operazione di composizione non \`{e} il commutatore ma lo
$\star$-commutatore, e $\textrm{Ad}_{x}$ corrisponde al coniugio
di un generico elemento della $\star$-algebra con
$\kappa(y,z)\equiv \exp(iz_{\alpha}y^{\alpha})$, elemento del
corrispondente gruppo.}. Le (\ref{propk}) fanno inoltre s\`{i}
che, almeno a livello linearizzato, $\kappa$ e $\bar{\kappa}$
agiscano su $B$ come proiettori sulla sua parte anti-chirale
(indipendente dalle $y$) e chirale (indipendente dalle $\bar{y}$),
rispettivamente:
\begin{eqnarray}
B(z,\bar{z};y,\bar{y})\star\kappa\,|_{Z=0} &=& C(0,\bar{y})+...\ ,\nonumber \\
B(z,\bar{z};y,\bar{y})\star\bar{\kappa}\,|_{Z=0}&=& C(y,0)+... \ ,
\label{kprojects}
\end{eqnarray}
dove $...$ indica i termini di ordine superiore al primo in $C$
corrispondenti ai coefficienti delle varie potenze di $z,\bar{z}$.
Per ottenere il vincolo linearizzato (\ref{COMSTh1}) nello
spazio-tempo ordinario \`{e} dunque necessario che ${\cal V}(B)$
dipenda da $B$ attraverso il costrutto $B\star\kappa$, e
analogamente $({\cal V}(B))^{\dag}$ attraverso
$B\star\bar{\kappa}$.

Abbiamo precedentemente introdotto la $1$-forma anti-hermitiana
$V_{\underline{\alpha}}=(V_{\alpha},\bar{V}^{\dot{\alpha}})$ di
connessione nelle direzioni ausiliarie $Z$, $V^{\dag}=-V$. Si
pu\`{o} definire una derivata covariante $S$ nello spazio $Z$ come
\begin{equation}\label{defS}
S=S_{0}-2iV \ , \qquad \tau(S)=-S\ , \qquad S^{\dag}=S \ , 
\end{equation}
con $S_{0}$ definito in (\ref{S0}). Assumiamo che $S$ trasformi
nell'aggiunta di $\widehat{shs}(4)$,
\begin{equation}
\delta_{\hat{\epsilon}} S=S\star\hat{\epsilon}-\hat{\epsilon}\star
S \ .
\end{equation}
Segue, utilizzando (\ref{propS0}), che $V$ trasforma
effettivamente come una $1$-forma di connessione di spazio $Z$,
\begin{equation}\label{Vgaugetransf}
\delta_{\hat{\epsilon}}V=d_{Z}\hat{\epsilon}+[V,\hat{\epsilon}]_{\star}\
,
\end{equation}
la cui curvatura \`{e} data da
\begin{equation}
d_{Z}V+V\star V=\frac{1}{4}S\star S \ .
\end{equation}

Siamo a questo punto in grado di scrivere una forma del sistema di
equazioni totalmente consistenti di Vasiliev (che mostreremo
essere equivalente a quella dell'ansatz (\ref{extgenericFDA2}))
nello spazio esteso $(x,Z)$:
\begin{equation}\label{Vas1}
dW+W\star W = 0 \ ,
\end{equation}
\begin{equation}\label{Vas2}
d B+W\star B-B\star\bar{\pi}(W) = 0 \ ,
\end{equation}
\begin{equation}\label{Vas3}
d S+W\star S+ S\star W = 0 \ ,
\end{equation}
\begin{equation}\label{Vas4}
S\star B-B\star\bar{\pi}(S) = 0 \ ,
\end{equation}
\begin{equation}\label{Vas5}
S\star S = i\,dz^{\alpha}\wedge dz_{\alpha}(1+B\star\kappa) +
i\,d\bar{z}^{\dot{\alpha}}\wedge
d\bar{z}_{\dot{\alpha}}(1+B\star\bar{\kappa}) \ .
\end{equation}
Queste equazioni, che nel seguito chiameremo equazioni di
Vasiliev, sono compatibili con le trasformazioni di $W$, $B$, $S$,
$\kappa$ sotto le mappe $\tau$ e $\pi,\bar{\pi}$, e con le
condizioni di hermiticit\`{a} dei medesimi. Esse sono inoltre tra
loro consistenti, sicch\'{e} l'intero sistema pu\`{o} in effetti
essere derivato dalla sola (\ref{Vas5}) insieme alla (\ref{Vas2})
oppure (\ref{Vas3}). Eq. (\ref{Vas1}) rappresenta infatti la
condizione di consistenza delle (\ref{Vas2}) e (\ref{Vas3}), come
si verifica facilmente operando, ad esempio, con $d$ su queste
ultime ed imponendo $d^{2}=0$, oppure, il che \`{e} lo stesso,
prendendone la derivata covariante ed imponendo la condizione
${\cal D}^{2}=0$ e $D^{2}=0$, rispettivamente. La (\ref{Vas4})
segue invece dalla (\ref{Vas5}), facendo uso della propriet\`{a}
associativa $(S\star S)\star S=S\star(S\star S)$\footnote {Tenendo
conto della (\ref{propS0}) e della definizione di $S$, \`{e}
chiaro che questa propriet\`{a} \`{e} equivalente all'identit\`{a}
di Bianchi per la derivata covariante $S$ nelle direzioni
ausiliarie $Z$, sicch\'{e} (\ref{Vas4}) gioca effettivamente il
ruolo di condizione di consistenza di (\ref{Vas5}). Si noti
inoltre che, sebbene $\kappa$ anticommuti con le potenze dispari
di $dz^{\alpha}$, $S$ commuta con i termini proporzionali a
$B\star\kappa$ di (\ref{Vas5}) in virt\`{u} del fatto che
$dz^{\alpha}\wedge dz^{\beta}\wedge dz^{\gamma}=0$, poich\'{e} gli
indici spinoriali assumono solo due valori (analoghe
considerazioni valgono per gli hermitiani coniugati).}, e delle
identit\`{a}
\begin{eqnarray}
\bar{\pi}(S)\star\kappa & = & \kappa\star S \ , \nonumber \\
\bar{\pi}(S)\star\bar{\kappa} & = & \bar{\kappa}\star S \ .
\end{eqnarray}
Infine, (\ref{Vas3}) segue dalla (\ref{Vas2}) (o viceversa)
prendendo la derivata covariante spazio-temporale $D$ della
(\ref{Vas5}) ed imponendo (\ref{Vas2}).

In termini di ${\cal W}$, \emph{master $1$-form} dello spazio
$(x,Z)$ definita nella (\ref{calW}), e della derivata esterna
totale $\hat{d}=d+d_{Z}$, il sistema (\ref{Vas1}-\ref{Vas5}) si
riscrive in modo compatto come
\begin{equation}\label{Vastot1}
{\cal R}  =  \frac{i}{4}(dz^{\alpha}\wedge dz_{\alpha}B\star\kappa
+ d\bar{z}^{\dot{\alpha}}\wedge
d\bar{z}_{\dot{\alpha}}B\star\bar{\kappa})\ ,
\end{equation}
\begin{equation}\label{Vastot2}
\hat{d}B+{\cal W}\star B-B\star\bar{\pi}({\cal W}) = 0 \ ,
\end{equation}
dove ${\cal R}$ \`{e} la $2$-forma di curvatura totale definita in
(\ref{totalmastercurv}). Questa forma delle equazioni di Vasiliev
\`{e} la pi\`{u} diretta realizzazione dell'ansatz
(\ref{extgenericFDA2}). La consistenza del sistema nello spazio
$(x,Z)$ si dimostra facilmente notando che la (\ref{Vastot2})
\`{e} il vincolo differenziale sulla $0$-forma $B$ necessario
affinch\'{e} la (\ref{Vastot1}) sia compatibile con l'identit\`{a}
di Bianchi (\ref{totalBid})
\begin{equation}\label{condintegr1}
\hat{D}{\cal R}=0=\frac{i}{4}(dz^{\alpha}\wedge
dz_{\alpha}\hat{D}(B\star\kappa))-\textrm{h.c.}\ .
\end{equation}
Si trova infatti che
\begin{equation}
\hat{D}(B\star\kappa)=\hat{{\cal
D}}B\star\kappa+B\star\partial\kappa \ ,
\end{equation}
con $\partial=dz^{\alpha}\frac{\partial}{\partial z^{\alpha}}$, ma
il secondo termine non contribuisce in (\ref{condintegr1})
poich\'{e} $dz^{\alpha}\wedge dz^{\beta}\wedge dz^{\gamma}=0$. La
(\ref{Vastot1}), a sua volta, \`{e} la condizione di
compatibilit\`{a} della (\ref{Vastot2}) con l'identit\`{a}
$\hat{{\cal D}}^{2}B=0$:
\begin{equation}
0=\hat{{\cal D}}^{2}B={\cal R}\star B-B\star\bar{\pi}({\cal R})\ ,
\end{equation}
e la consistenza si verifica sostituendo la (\ref{Vastot1}) in
quest'ultima equazione e usando $\pi(B)=\bar{\pi}(B)$ e $\pi({\cal
R})=\bar{\pi}({\cal R})$. Il vincolo di curvatura (\ref{Vastot1})
pu\`{o} essere riscritto in componenti come
\begin{eqnarray}
{\cal R}_{\mu\nu} & = & 0 \ ,\\
{\cal R}_{\alpha\mu} & = & 0 \ ,\\
{\cal R}_{\dot{\alpha}\mu} & = & 0 \ ,\\
{\cal R}_{\alpha\dot{\alpha}} & = & 0 \ ,\\
{\cal R}_{\alpha\beta} & = &
-\frac{i}{2}\epsilon_{\alpha\beta}B\star\kappa \ ,\\
{\cal R}_{\dot{\alpha}\dot{\beta}} & = &
-\frac{i}{2}\epsilon_{\dot{\alpha}\dot{\beta}}B\star\bar{\kappa} \
,
\end{eqnarray}
evidenziandone pi\`{u} direttamente il contenuto: le componenti
$xx$ e $xZ$ della curvatura totale sono annullate dalle equazioni
del moto, mentre non lo sono le componenti $ZZ$, proporzionali
alla \emph{master $0$-form} proiettata con $\kappa,\bar{\kappa}$.
Ricordando le definizioni di $W$ ed $S$, \`{e} chiaro che questa
\`{e} esattamente l'informazione contenuta nei vincoli di
curvatura (\ref{Vas1}), (\ref{Vas3}) e (\ref{Vas5}), mentre
(\ref{Vas2}) e (\ref{Vas4}) indicano che la $0$-forma $B$ \`{e}
covariantemente costante sia nelle direzioni spazio-temporali $x$
che in quelle ausiliarie non commutative $Z$, analogamente a
quanto fa (\ref{Vastot2}). L'equivalenza dei due sistemi di
equazioni si pu\`{o} dimostrare in modo pi\`{u} rigoroso
attraverso l'analisi delle (\ref{Vastot1}-\ref{Vastot2}) e della
(\ref{totalBid}) nelle loro componenti $(p,q,r)$, dove con $p$,
$q$ ed $r$ si intende il grado nelle direzioni $x$, $z$ e
$\bar{z}$, rispettivamente, delle forme differenziali che in esse
compaiono. Facendo uso di (\ref{propS0}), si verifica infatti che
le componenti $(1,1,0)$ e $(1,0,1)$ di (\ref{Vastot1}) danno
\begin{equation}
dV+d_{Z}W+W\star V+V\star W=0 \ ,
\end{equation}
equivalente alla (\ref{Vas3}). Le componenti $(0,2,0)$, $(0,1,1)$
e $(0,0,2)$ forniscono invece la condizione
\begin{equation}
d_{Z}V+V\star V=\frac{i}{4}(dz^{\alpha}\wedge
dz_{\alpha}B\star\kappa + d\bar{z}^{\dot{\alpha}}\wedge
d\bar{z}_{\dot{\alpha}}B\star\bar{\kappa})\ ,
\end{equation}
che corrisponde esattamente alla (\ref{Vas5}), come si vede
ricordando la definizione di $S$ (\ref{defS}). Per definizione di
${\cal R}_{\mu\nu}$, le componenti $(2,0,0)$ corrispondono alla
(\ref{Vas1}). Le componenti $(1,2,0)$ dell'identit\`{a} di Bianchi
(\ref{totalBid}) danno inoltre luogo alla (\ref{Vastot2}), mentre
dalle componenti $(0,2,1)$ e $(0,1,2)$ scaturisce il vincolo
\begin{equation}
d_{Z}B+V\star B-B\star\bar{\pi}(V)=0 \ ,
\end{equation}
equivalente alla (\ref{Vas4}).

La FDA estesa realizzata nei vincoli (\ref{Vastot1}-\ref{Vastot2})
ammette naturalmente le simmetrie di gauge (\ref{extgaugetransfW})
e (\ref{twextgaugetransf}).  Le prime si traducono, in termini
delle variabili $W$ ed $S$, nelle
\begin{eqnarray}
\delta_{\hat{\epsilon}} W & = &
d\hat{\epsilon}+[W,\hat{\epsilon}]_{\star} \ ,\\
\delta_{\hat{\epsilon}} S & = & [S,\hat{\epsilon}]_{\star} \ ,
\label{Sgaugetr}
\end{eqnarray}
che insieme alla (\ref{twextgaugetransf}) costituiscono le
simmetrie di gauge del sistema (\ref{Vas1}-\ref{Vas5}). Come
gi\`{a} sottolineato nel caso lineare, le equazioni sono anche
manifestamente invarianti sotto diffeomorfismi spazio-temporali,
essendo formulate in termini di forme differenziali. Le
trasformazioni generali di coordinate $dx^{\mu}=\xi^{\mu}$ sono
infatti incorporate nel gruppo di gauge, tramite i parametri
dipendenti dai campi $\hat{\epsilon}(\xi)=i_{\xi}W$.

Se nello spazio esteso $(x,Z)$ le (\ref{Vas1}-\ref{Vas5})
realizzano una FDA (tenendo presente che nelle ultime due
equazioni del sistema i termini derivativi di spazio $Z$ sono
inclusi in $S$), \`{e} altres\`{i} evidente che nello spazio-tempo
ordinario, considerando le $Z$ come un raddoppiamento delle
variabili spinoriali interne $Y$, le equazioni di Vasiliev hanno
la forma \emph{unfolded}: la (\ref{Vas1}) \`{e} un'equazione di
curvatura nulla, le (\ref{Vas2}) e (\ref{Vas3}) stabiliscono che
due $0$-forme (dal punto di vista spazio-temporale $S$ \`{e}
infatti una $0$-forma) sono covariantemente costanti, e le
(\ref{Vas4}) e (\ref{Vas5}) hanno la funzione di vincoli puramente
algebrici. Questa osservazione suggerisce la possibilit\`{a} di
esprimere la dipendenza dei campi $W$, $B$ ed $S$ dalle coordinate
$x$ come pura libert\`{a} di gauge, risolvendo le
(\ref{Vas1}-\ref{Vas3}) in termini delle condizioni iniziali
$B(x_{0}|Y,Z)\equiv b(Y,Z)$ ed $S(x_{0}|Y,Z)\equiv s(Y,Z)$,
secondo la discussione generale della sezione \ref{4.6}:
\begin{equation}\label{puregaugesoln1}
W(x|Y,Z)=\hat{g}(x|Y,Z)\star d\hat{g}(x|Y,Z)\ ,
\end{equation}
\begin{equation}
B(x|Y,Z)=\hat{g}^{-1}(x|Y,Z)\star
B(x_{0}|Y,Z)\star\bar{\pi}(\hat{g})(x|Y,Z)\ ,
\end{equation}
\begin{equation}\label{puregaugesoln3}
S(x|Y,Z)=\hat{g}^{-1}(x|Y,Z)\star S(x_{0}|Y,Z)\star\hat{g}(x|Y,Z)\
,
\end{equation}
dove $\hat{g}(x|Y,Z)$ \`{e} un elemento invertibile di
$\widehat{hs}(4)$. Questo implica che le soluzioni
gauge-inequivalenti dell'intero sistema (\ref{Vas1}-\ref{Vas5})
sono tutte e sole le soluzioni delle equazioni differenziali di
primo ordine nelle coordinate ausiliarie $Z$ che si ottengono
sostituendo le (\ref{puregaugesoln1}-\ref{puregaugesoln3}) nelle
rimanenti equazioni (\ref{Vas4}) e (\ref{Vas5})
\cite{Sezgin:1998gg}:
\begin{eqnarray}
s\star b & = &
b\star\bar{\pi}(s) \ ,\nonumber \\
s\star s & = & i\,dz^{\alpha}\wedge dz_{\alpha}(1+B\star\kappa) +
i\,d\bar{z}^{\dot{\alpha}}\wedge
d\bar{z}_{\dot{\alpha}}(1+B\star\bar{\kappa}) \ ,
\end{eqnarray}
indipendenti dalle coordinate $x$ e tuttavia contenenti tutta
l'informazione sulla dinamica dei campi di spin arbitrario.
Incontriamo quindi nuovamente, nell'ambito della teoria non
lineare, la peculiarit\`{a} che alcuni vincoli puramente algebrici
codificano interamente la  
dinamica, a riprova del fatto che il sistema
(\ref{Vas1}-\ref{Vas5}) \`{e} un esempio di formulazione
\emph{unfolded}. Le condizioni
(\ref{puregaugesoln1}-\ref{puregaugesoln3}) sono invarianti sotto
trasformazioni di gauge di $\widehat{hs}(4)$ indipendenti dalle
coordinate spazio-temporali $x$, ed il loro studio potrebbe dar
luogo a soluzioni di vuoto della teoria differenti dal background
$AdS$, come sottolineato in \cite{Sezgin:1998gg}.

\section{Espansione perturbativa da $AdS$}
\label{esppert}

Si tratta a questo punto di stabilire un contatto tra le
(\ref{Vas1}-\ref{Vas5}) e le equazioni linearizzate trattate nel
capitolo precedente, che abbiamo visto corrispondere esattamente
alle equazioni del moto libere dei campi di massa nulla e spin
arbitrario in $AdS_{4}$. A tale scopo, come spiegato nella sezione
\ref{deformalg}, sar\`{a} necessario seguire una strategia opposta
a quella che ci ha condotto alle
(\ref{puregaugesoln1}-\ref{puregaugesoln3}), ottenendo anzitutto,
dalle (\ref{Vas3}-\ref{Vas5}), la dipendenza dalle coordinate
ausiliarie $Z$ di $W$, $B$ ed $S$ in termini delle ``condizioni
iniziali'' $w(x|Y)$ e $C(x|Y)$ ($0$-forme di spazio $Z$), ed
inserendo quindi le espressioni ottenute nelle rimanenti
(\ref{Vas1}-\ref{Vas2}). Queste ultime, all'ordine lineare
nell'espansione attorno al background $AdS$, coincideranno con le
equazioni (\ref{COMSTh1}) e (\ref{COMSTh2}) del \emph{Central
On-Mass-Shell Theorem}.

Si pu\`{o} verificare che la scelta
\begin{eqnarray}
B_{0} & = & 0 \ ,\nonumber \\
S_{0} & = &
dz^{\alpha}z_{\alpha}+d\bar{z}^{\dot{\alpha}}\bar{z}_{\dot{\alpha}}\
, \nonumber \\
W_{0} & = &
\frac{i}{8}[\omega_{0\alpha\beta}y^{\alpha}y^{\beta}+\bar{\omega}_{0\dot{\alpha}\dot{\beta}}\bar{y}^{\dot{\alpha}}\bar{y}^{\dot{\beta}}+2L^{-1}e_{0\alpha\dot{\alpha}}y^{\alpha}\bar{y}^{\dot{\alpha}}]
= \omega_{0} \ , \label{bg}
\end{eqnarray}
\`{e} una soluzione di vuoto delle (\ref{Vas1}-\ref{Vas5}), e
identifica il fondo $AdS_{4}$. Essa rompe la simmetria
$\widehat{hs}(4)$ di queste ultime alla sottoalgebra che la
stabilizza, corrispondente all'insieme dei parametri
$\hat{\epsilon}_{0}$ che soddisfano le equazioni di Killing
generalizzate
\begin{equation}\label{extgenKilling1}
\delta_{\hat{\epsilon}_{0}}W_{0}=d\hat{\epsilon}_{0}+[W_{0},\hat{\epsilon}_{0}]_{\star}=0
\ ,
\end{equation}
\begin{equation}\label{extgenKilling2}
\delta_{\hat{\epsilon}_{0}}B_{0}=-\hat{\epsilon}\star
B_{0}+B_{0}\star\bar{\pi}(\hat{\epsilon})=0 \ ,
\end{equation}
\begin{equation}\label{extgenKilling3}
\delta_{\hat{\epsilon}_{0}}S_{0}=[S_{0},\hat{\epsilon}_{0}]_{\star}=d_{Z}\hat{\epsilon}_{0}=0
\ .
\end{equation}
L'equazione (\ref{extgenKilling2}) \`{e} soddisfatta banalmente,
la (\ref{extgenKilling3}) implica che $\hat{\epsilon}_{0}$ sia
indipendente dalle $Z$, mentre la (\ref{extgenKilling1}), come
accennato nel capitolo precedente, ammette soluzioni
$\epsilon_{0}\in hs(4)$. In particolare, quest'ultima corrisponde
ad un'equazione di costanza covariante per il parametro
$\epsilon_{0}$, indipendente da $Z$, la cui consistenza \`{e}
garantita dal fatto che $W_{0}$ risolve un'equazione di curvatura
nulla come la (\ref{Vas1}). Ci\`{o} comporta
\cite{Vasiliev:1999ba} che la (\ref{extgenKilling1}) ammette
localmente una soluzione unica del tipo
\begin{equation}
\epsilon_{0}(x|Y)=g^{-1}(x|Y)\star \epsilon_{0}(x_{0}|Y)\star
g(x|Y) \ ,
\end{equation}
con $\epsilon_{0}(x_{0}|Y)=\epsilon_{0}(Y)$ elemento arbitrario e
costante dell'algebra $hs(4)$, che corrisponde dunque alla
simmetria globale della teoria (ovvero al sottoinsieme delle
trasformazioni di gauge di $\widehat{hs}(4)$ che lascia invarianti
i campi di background (\ref{bg})).

Si procede quindi ad un'analisi perturbativa
\cite{Vasiliev:1992av} delle equazioni (\ref{Vas1}-\ref{Vas5})
espandendo in potenze di $B$, ovvero intorno alla soluzione di
vuoto (\ref{bg}):
\begin{eqnarray}
B & = & B_{1}+B_{2}+...  \ ,\nonumber \\
S & = & S_{0}+S_{1}+S_{2}+...\ ,\nonumber \\
W & = & W_{0}+W_{1}+W_{2}+...\ .
\end{eqnarray}
Inserendo tali espansioni nelle (\ref{Vas4}-\ref{Vas5}) e
ricordando la (\ref{propS0}), si ottiene
\begin{eqnarray}
d_{Z}B_{n} & = &
\frac{i}{2}\sum_{j=1}^{n-1}\left(B_{j}\star\bar{\pi}(S_{n-j})-S_{j}\star
B_{n-j}\right)\ , \label{4ordn}\\
d_{Z}S_{n} & = & -\frac{1}{2}dz^{\alpha}\wedge
dz_{\alpha}B_{n}\star\kappa-\frac{1}{2}d\bar{z}^{\dot{\alpha}}\wedge
d\bar{z}_{\dot{\alpha}}B_{n}\star\bar{\kappa}
\nonumber \\
&&-\frac{i}{2}\sum_{j=1}^{n-1}S_{j}\star S_{n-j}\ , \qquad
n=1,2,... \ ,\label{5ordn}
\end{eqnarray}
un sistema di equazioni differenziali lineari di primo ordine
nelle variabili spinoriali $Z$ consistente ordine per ordine in
teoria delle perturbazioni, come si pu\`{o} verificare applicando
$d_{Z}$ ad ambo i membri di (\ref{4ordn}-\ref{5ordn}) ed imponendo
che $d_{Z}^{2}=0$ e che le equazioni di ordine $i<n$ siano
soddisfatte. Avendo dunque integrato tali equazioni fino
all'ordine $n-1$, la soluzione di ordine $n$ si ottiene ricavando
anzitutto $B_{n}$ dalla (\ref{4ordn}),
\begin{eqnarray}
B_{n}(x|Y,Z) & = & C_{n}(x|Y) \nonumber \\
&&+\frac{i}{2}\sum_{j=1}^{n-1}\int_{0}^{1}dt\biggl\{z^{\alpha}(B_{j}\star\bar{\kappa}\star
S_{\alpha}^{n-j}\star\bar{\kappa}-S_{\alpha}^{j}\star
B_{n-j})(x|Y,t
Z)  \nonumber \\
&&-\bar{z}^{\dot{\alpha}}(B_{j}\star\bar{\kappa}\star
\bar{S}_{\dot{\alpha}}^{n-j}\star\bar{\kappa}+\bar{S}_{\dot{\alpha}}^{j}\star
B_{n-j})(x|Y,t Z)\biggr\}\ , \label{int4ordn}
\end{eqnarray}
sostituendolo successivamente in (\ref{5ordn}), e risolvendo per
$S_{n}$,
\begin{eqnarray}
S^{n}_{\alpha}(x|Y,Z) & = & \frac{\partial}{\partial
z^{\alpha}}\xi_{n}(x|Y,Z) \nonumber \\
&&+\int_{0}^{1}dt\,t\biggl\{z_{\alpha}(B_{n}\star\kappa+\frac{i}{2}\sum_{j=1}^{[\frac{n-1}{2}]}\left[S_{j}^{\beta},S_{\beta}^{n-j}\right]_{\star})(x|Y,t
Z) \nonumber \\
&&+\bar{z}^{\dot{\alpha}}\frac{i}{2}\sum_{j=1}^{n-1}\left[S^{j}_{\alpha},\bar{S}_{\dot{\alpha}}^{n-j}\right]_{\star})(x|Y,t
Z)\biggr\} \ , \label{int5ordna}
\end{eqnarray}
\begin{eqnarray}
\bar{S}^{n}_{\dot{\alpha}}(x|Y,Z) & = & \frac{\partial}{\partial
\bar{z}^{\dot{\alpha}}}\xi_{n}(x|Y,Z)  \nonumber \\
&&+\int_{0}^{1}dt\,t\biggl\{\bar{z}_{\dot{\alpha}}(B_{n}\star\bar{\kappa}+\frac{i}{2}\sum_{j=1}^{[\frac{n-1}{2}]}\left[\bar{S}_{j}^{\dot{\beta}},\bar{S}_{\dot{\beta}}^{n-j}\right]_{\star})(x|Y,t
Z) \nonumber \\
&&+z^{\alpha}\frac{i}{2}\sum_{j=1}^{n-1}\left[S^{j}_{\alpha},\bar{S}_{\dot{\alpha}}^{n-j}\right]_{\star})(x|Y,t
Z)\biggr\} \ , \label{int5ordnb}
\end{eqnarray}
dove $[\frac{n-1}{2}]$ indica la parte intera di $\frac{n-1}{2}$.
Per l'integrazione di queste equazioni sono state applicate le
formule generali (\ref{symplint1}) e (\ref{symplint2}) date in
appendice \ref{appendice B}. Tutti gli $\star$-prodotti si
intendono calcolati prima che le variabili $Z$ siano riscalate in
$tZ$.

Si noti che, a seguito dell'integrazione, a secondo membro della
(\ref{int4ordn}) compare la condizione iniziale
\begin{equation}
B_{n}(x|Y,Z)|_{Z=0}\equiv C_{n}(x|Y)\ , \qquad n=1,2,... \ ,
\end{equation}
che eredita le propriet\`{a} (\ref{exthstwadj}) di $B_{n}$ nel
sottospazio $Z=0$ dello spazio esteso, e trasforma quindi nella
rappresentazione \emph{twisted adjoint} di $hs(4)$. Anche le
(\ref{int5ordna}) e (\ref{int5ordnb}) sono definite a meno di una
$1$-forma di spazio $Z$ esatta, $d_{Z}\xi_{n}(x|Y,Z)$, dove
$\xi_{n}(x|Y,Z)$ \`{e} una $0$-forma arbitraria a valori in
$\widehat{hs}(4)$. Tuttavia quest'ultima \`{e} pura gauge, come
risulta chiaro osservando l'ordine $n$ della trasformazione di
gauge di $S$ (\ref{Sgaugetr}), di parametro
$\hat{\epsilon}=\hat{\epsilon}_{1}+\hat{\epsilon}_{2}+...$,
\begin{equation}
\delta
S_{n}=-2id_{Z}\hat{\epsilon}_{n}+\sum_{j=1}^{n-1}\left[S_{j},\hat{\epsilon}_{n-j}\right]_{\star}\
, \qquad n=1,2,... \ ,
\end{equation}
e pu\`{o} sempre essere posta uguale a zero, scegliendo
opportunamente $\hat{\epsilon}_{n}$. In altri termini, ordine a
ordine in $B$ si fissa
\begin{equation}\label{Vgaugechoice}
\xi_{n}=0 \ ,\qquad n=1,2,... \ ,
\end{equation}
e le simmetrie di gauge che preservano tale scelta sono tutte e
sole quelle di parametro
$\hat{\epsilon}_{i}(x|Y,Z)=\epsilon_{i}(Y,Z),\;i=1,...,n$,
indipendenti da $Z$ e a valori in $hs(4)$. Come conseguenza,
ordine per ordine il campo $S$ pu\`{o} essere interamente espresso
in termini di $B$, sicch\'{e} resta dimostrato, come preannunciato
nella sezione \ref{deformalg}, che esso non descrive gradi di
libert\`{a} fisici, ma ha il ruolo di un compensatore.

Avendo risolto le (\ref{Vas4}-\ref{Vas5}) per $B$ ed $S$, si passa
a ricavare la dipendenza dalle $Z$ di $W$ integrando la
(\ref{Vas3}), che all'ordine $n$ fornisce
\begin{equation}\label{3ordn}
d_{Z}W_{n}=-\frac{i}{2}dS_{n}-\frac{i}{2}\sum_{j=0}^{n-1}(W_{j}\star
S_{n-j}+S_{n-j}\star W_{j}) \ , \quad n=1,2,... \ .
\end{equation}
Introducendo la condizione iniziale
\begin{equation}
W_{n}(x|Y,Z)|_{Z=0}\equiv w_{n}(x|Y)\ , \qquad n=1,2,... \ ,
\end{equation}
ed applicando (\ref{symplint1}), si ottiene
\begin{eqnarray}
W_{n}(x|Y,Z) & = & w_{n}(x|Y) \nonumber \\
&&+\frac{i}{2}\int_{0}^{1}dt\biggl\{z^{\alpha}\left(\sum_{j=0}^{n-1}\left[W_{j}^{\beta},S_{\alpha}^{n-j}\right]_{\star}\right)(x|Y,t
Z) \nonumber \\
&&+\bar{z}^{\dot{\alpha}}\left(\sum_{j=0}^{n-1}\left[W_{j}^{\beta},\bar{S}_{\dot{\alpha}}^{n-j}\right]_{\star}\right)(x|Y,t
Z)\biggr\}\ , \label{int3ordn}
\end{eqnarray}
dove il termine proporzionale a $z^{\alpha}dS^{n}_{\alpha}$ ed il
complesso coniugato, a priori inclusi nell'integrando, sono nulli
grazie alle (\ref{int5ordna}) e (\ref{int5ordnb}) con $\xi_{n}=0$,
tenendo conto del fatto che
$z^{\alpha}z_{\alpha}=\bar{z}^{\dot{\alpha}}\bar{z}_{\dot{\alpha}}=0$. 

A questo punto abbiamo ricavato $W$ e $B$ completamente in termini
delle condizioni iniziali indipendenti
\begin{eqnarray}
C(x|Y) & = & B(x|Y,Z)|_{Z=0}\,=\,C_{1}(x|Y)+C_{2}(x|Y)+... \ ,
\nonumber \\
w(x|Y) & = &
W(x|Y,Z)|_{Z=0}\,=\,w_{0}(x|Y)+w_{1}(x|Y)+w_{2}(x|Y)+... \ ,
\end{eqnarray}
e non resta che sostituire le espressioni ottenute nelle
(\ref{Vas1}-\ref{Vas2}). Possiamo limitarci a studiare queste
ultime nel sottospazio $Z=0$, come conseguenza della consistenza
delle equazioni (\ref{Vas3}-\ref{Vas4}), che garantisce la
validit\`{a} delle condizioni
\begin{equation}
[S,dW+W\star W]_{\star}=0\ ,
\end{equation}
\begin{equation}
S\star{\cal D}B+{\cal D}B\star\bar{\pi}(S)=0 \ ,
\end{equation}
se si tengono presenti le (\ref{Vas3}-\ref{Vas4}) stesse. Ma tali
condizioni implicano che le (\ref{Vas1}-\ref{Vas2}) sono
covariantemente costanti, ovvero che la loro dipendenza da $Z$
pu\`{o} essere assorbita in una trasformazione di gauge di
$\widehat{hs}(4)$ (in modo analogo a quanto accadeva, in un
contesto diverso, per la dipendenza dalle $x$ nelle
(\ref{puregaugesoln1}-\ref{puregaugesoln3})), sicch\'{e} le
(\ref{Vas1}-\ref{Vas2}) sono soddisfatte per ogni $Z$ se lo sono
per $Z=0$.

Possiamo dunque verificare che all'ordine lineare in $B$ si
ottengano effettivamente le equazioni del \emph{Central
On-Mass-Shell Theorem}. Anzitutto, trascurando i termini di ordine
superiore al primo, le (\ref{int4ordn}),(\ref{int5ordna}),
(\ref{int5ordnb}) e (\ref{int3ordn}) forniscono
\begin{eqnarray}
B_{1}(x|Y,Z) & = & C(x|Y) \ , \\
S_{1}(x|Y,Z) & = & \int_{0}^{1}dt\,t
\bigl(dz^{\alpha}z_{\alpha}C(x|-t z,\bar{y})\kappa(t
z,y) \nonumber \\
&&+d\bar{z}^{\dot{\alpha}}\bar{z}_{\dot{\alpha}}C(x|y,t
\bar{z})\bar{\kappa}(t \bar{z},\bar{y})\bigr) \ , \\
W_{1}(x|Y,Z) & = & \omega(x|Y)+\Omega_{1}(x|Y,Z) \ ,
\end{eqnarray}
dove, con un leggero abuso di notazione, abbiamo posto
$C_{1}\rightarrow C$ e $w_{1}\rightarrow \omega$, e abbiamo
definito
\begin{eqnarray}
\Omega_{1}(x|Y,Z) & = &
-\frac{1}{2}\int_{0}^{1}dt'\int_{0}^{1}dt\,t\biggl\{\left(itt'\,\omega_{0}^{\phantom{0}\alpha\beta}z_{\alpha}z_{\beta}+e_{0}^{\phantom{0}\alpha\dot{\beta}}z_{\alpha}\bar{\partial}_{\dot{\beta}}\right)C(x|-tt'
z,\bar{y})\kappa(tt'z,y) \nonumber \\
&&+\left(itt'\,\omega_{0}^{\phantom{0}\dot{\alpha}\dot{\beta}}\bar{z}_{\dot{\alpha}}\bar{z}_{\dot{\beta}}-e_{0}^{\phantom{0}\alpha\dot{\beta}}\bar{z}_{\dot{\beta}}\partial_{\alpha}\right)C(x|y,
tt'z)\bar{\kappa}(tt'\bar{z},\bar{y})\biggr\} \ ,
\end{eqnarray}
dove $\partial_{\alpha}=\frac{\partial}{\partial y^{\alpha}}$.
Inseriamo ora $B_{1}$ e $W_{1}$ nelle (\ref{Vas1}-\ref{Vas2})
linearizzate e calcolate in $Z=0$. La prima diventa
\begin{equation}\label{Vas1lin}
R_{1}=-\{\omega_{0},\Omega_{1}\}_{\star}|_{Z=0} \ ,
\end{equation}
dove $R_{1}$ \`{e} la curvatura linearizzata rispetto al
background $AdS$ definita nelle (\ref{firstlincurv}) e
(\ref{lincurv}). Si noti che, sebbene $\Omega_{1}|_{Z=0}=0$, il
secondo membro della (\ref{Vas1lin}) \`{e} diverso da zero, a
causa della presenza di contrazioni non banali tra le $Y$ e le
$Z$. La (\ref{Vas2}) linearizzata fornisce invece
\begin{equation}\label{Vas2lin}
{\cal D}_{0}C=dC+\omega_{0}\star C-C\star\bar{\pi}(\omega_{0})=0 \
,
\end{equation}
che coincide evidentemente con la (\ref{COMSTh2}). Sostituendo le
definizioni di $\omega_{0}$ ed $\Omega_{1}$ nel secondo membro
della (\ref{Vas1lin}), calcolando gli $\star$-prodotti e prendendo
infine $Z=0$ si ottiene
\begin{equation}
R_{1}=\frac{i}{4}\left[e_{0}^{\phantom{0}\alpha\dot{\beta}}\wedge
e_{0\phantom{\gamma}\dot{\beta}}^{\phantom{0}\gamma}\partial_{\alpha}\partial_{\gamma}C(x|y,0)+e_{0}^{\phantom{0}\beta\dot{\alpha}}\wedge
e_{0\beta}^{\phantom{0\beta}\dot{\gamma}}\partial_{\alpha}\partial_{\gamma}C(x|y,0)\right]
\ ,
\end{equation}
ovvero la (\ref{COMSTh1}), come preannunciato.

Ci\`{o} dimostra che il sistema (\ref{Vas1}-\ref{Vas5}),
totalmente consistente e, di conseguenza (secondo lo schema delle
FDA), invariante sotto le trasformazioni di gauge di HS oltre che
esplicitamente invariante sotto diffeomorfismi, contiene tutta
l'informazione fisica del \emph{Central On-Mass-Shell Theorem } e
generalizza la dinamica dei campi di massa nulla e spin arbitrario
a tutti gli ordini nelle loro mutue interazioni. Ciononostante,
estrarre delle equazioni del moto interagenti che coinvolgano i
soli campi fisici (la generalizzazione delle (\ref{eom})),
eliminando tutti i campi ausiliari attraverso gli appropriati
vincoli di curvatura contenuti nelle (\ref{Vas1}-\ref{Vas5})
stesse (o, equivalentemente, nel sistema
(\ref{Vastot1}-\ref{Vastot2})) secondo la procedura esaminata nel
caso lineare, \`{e} una questione altamente non banale e tuttora
oggetto di indagine (si veda, ad esempio, \cite{Sezgin:2002ru} per
uno studio della struttura dei termini quadratici).

\section{Unicit\`a dei termini non lineari}
\label{5.4}

La richiesta di consistenza e la necessit\`{a} di riprodurre la
corretta dinamica a livello linearizzato non fissano univocamente
la forma delle equazioni non lineari. \`{E} dunque interessante
esaminare fino a che punto ulteriori vincoli ``naturali'' possano
ridurre l'arbitrariet\`{a} dei termini del sistema
(\ref{Vas1}-\ref{Vas5}) e in quali casi esista una effettiva
ambiguit\`{a} nei vertici di interazione dei campi di gauge
inclusi nei \emph{master fields} $W_{\mu}$, $B$ ed
$S_{\underline{\alpha}}$. Concentreremo la nostra analisi sul
vincolo di curvatura (\ref{Vas5}), che d\`{a} luogo ad interazioni
non banali decretando che la curvatura $ZZ$ sia proporzionale ad
un termine di sorgente dipendente da $B$. Analoga funzione svolge
la sua generalizzazione \cite{Vasiliev:1999ba}
\begin{eqnarray}
S\star S & = &  i\,dz^{\alpha}\wedge dz_{\alpha}\left(F(B)+{\cal
V}(B\star\kappa)\right) + i\,d\bar{z}^{\dot{\alpha}}\wedge
d\bar{z}_{\dot{\alpha}}\left(\bar{F}(B)+\overline{{\cal
V}}(B\star\bar{\kappa})\right) \nonumber \\
&& {} -i\,dz^{\alpha} \wedge
dz^{\dot{\beta}}H_{\alpha\dot{\beta}}(B)\ , \label{Vas5gen}
\end{eqnarray}
dove $F(X)$ e ${\cal V}(X)$ sono funzioni complesse arbitrarie, la
seconda delle quali costruita attraverso lo $\star$-prodotto come
${\cal V}(X)=b_{0}+b_{1}X+b_{2}X\star X+...$, mentre
$H_{\alpha\dot{\beta}}(X)$ \`{e} una funzione hermitiana,
$(H_{\alpha\dot{\beta}})^{\dag}=H_{\dot{\beta}\alpha}$, anch'essa
espandibile in serie di $\star$-potenze del suo argomento. La
(\ref{Vas5gen}) \`{e} compatibile con le identit\`{a} di Bianchi,
sia di spazio $x$ che di spazio $Z$, del primo membro, se si
assume la validit\`{a} delle rimanenti equazioni
(\ref{Vas1}-\ref{Vas4}), sicch\'{e} tale generalizzazione non
rovina la consistenza del sistema.

Notiamo anzitutto che, purch\'{e} $F(0)\neq 0$, $F(B)$ pu\`{o}
sempre essere riassorbita in una ridefinizione dei campi ausiliari
inclusi in $S$
\begin{equation}
S\rightarrow
S'=[F(B)]^{-\frac{1}{2}}\,dz^{\alpha}S_{\alpha}+[\bar{F}(B)]^{-\frac{1}{2}}\,d\bar{z}^{\dot{\alpha}}\bar{S}_{\dot{\alpha}}
\ ,
\end{equation}
del tutto innocua per le (\ref{Vas1}-\ref{Vas4}). Ci\`{o} implica
che \`{e} sempre possibile porre $F(B)=1$ e che, di conseguenza,
non c'\`{e} alcuna ambiguit\`{a} in questo termine.

Si pu\`{o} inoltre dimostrare \cite{Sezgin:1998eh} che la
possibilit\`{a} di ridefinire $B$ attraverso una funzione reale
$B\rightarrow B'=f(B)$ fa s\`{i} che ${\cal V}(X)$ sia definito a
meno della relazione di equivalenza
\begin{equation}
{\cal V}\sim{\cal V}'\quad \textrm{se}\quad{\cal V}(X)={\cal
V}'(f(X))\ , \quad f(-X)=-f(X)\ ,
\end{equation}
e che, come conseguenza, assumendo che ${\cal V}(X)$ parta con
termini lineari in $X$, si pu\`{o} in generale prendere
\begin{equation}
|b_{1}|=1\ ,\qquad b_{1}=e^{i\theta_{1}}\ ,
\end{equation}
ridefinendo $B\rightarrow |b_{1}|^{-1}B$. Vedremo nel seguito che
la richiesta di simmetria sotto trasformazioni di Lorentz locali
vincoler\`{a} ${\cal V}(X)$ ad essere una funzione dispari del suo
argomento,
\begin{equation}
{\cal V}(X)=b_{1}X+b_{3}X\star X\star X+...
\end{equation}
sicch\'{e} l'ipotesi $b_{0}=0$ \`{e} verificata. Abbiamo visto in
questo capitolo che anche la scelta pi\`{u} semplice, ${\cal
V}(X)=b_{1}X$, d\`{a} origine ad infiniti termini non lineari e ad
una dinamica altamente non banale, mentre gli ordini superiori di
${\cal V}(B\star\kappa)$ introdurrebbero correzioni alle
interazioni di ordine pi\`{u} alto, senza alterare il primo ordine
in $B$, che riproduce le corrette equazioni del moto dei campi di
gauge di spin arbitrario in $AdS_{4}$. I parametri liberi inclusi
in ${\cal V}(X)$ costituiscono dunque delle vere ambiguit\`{a},
non eliminabili attraverso alcuna ridefinizione dei campi. Resta
inoltre del tutto indeterminata la funzione
$H_{\alpha\dot{\beta}}(X)$.

Tuttavia, come anticipato, esiste una richiesta piuttosto naturale
ma molto restrittiva che specifica ulteriormente la forma della
(\ref{Vas5gen}): si tratta di imporre che le trasformazioni di
Lorentz locali siano una simmetria delle equazioni interagenti
(\ref{Vas1}-\ref{Vas5}), ovvero che le correzioni non lineari
legate all'evoluzione lungo le direzioni non commutative $Z$ non
deformino la simmetria di Lorentz locale. Tale richiesta trova
giustificazione nel fatto che \`{e} proprio questa simmetria
locale che garantisce l'equivalenza della formulazione di Einstein
(metrica) e di quella del formalismo del vielbein della
Relativit\`{a} Generale, assicurando che i parametri in pi\`{u}
del tensore fondamentale non simmetrico $e^{a}_{\mu}(x)$ rispetto
al tensore simmetrico $g_{\mu\nu}(x)$ ($6$, in $d=4$) possano
sempre essere riassorbiti in un'opportuna trasformazione di gauge
di $so(3,1)$, di parametro $\Lambda^{ab}(x)$ antisimmetrico
(contenente dunque esattamente $6$ parametri liberi, in $d=4$).
Ci\`{o} consente anche di introdurre gli spinori, come
rappresentazioni a due valori del gruppo di Lorentz locale, nel
contesto di una teoria invariante sotto diffeomorfismi, e appare
quindi ragionevole richiedere che tale simmetria non venga rotta
nel passaggio alla teoria non lineare di HS.

Una trasformazione di Lorentz locale infinitesima di parametro
$\Lambda^{\alpha\beta}(x)$ agisce su una generica funzione
$F(y,\bar{y})\in {\cal A}^{\star}$ come
\begin{equation}
\delta
F(y,\bar{y})=[F(y,\bar{y}),(\frac{1}{4i}\Lambda^{\alpha\beta}(x)M_{\alpha\beta})-\textrm{h.c.}]_{\star}
\ ,
\end{equation}
dove il generatore $M_{\alpha\beta}=-y_{\alpha}y_{\beta}$ e
l'hermitiano coniugato ruotano gli indici spinoriali interni di
$F$ attraverso lo $\star$-commutatore come $\delta
y_{\alpha}=\Lambda_{\alpha}^{\phantom{\alpha}\beta}y_{\beta}$ e
analogamente per $\bar{y}_{\dot{\alpha}}$. Avendo introdotto nuove
direzioni $z,\bar{z}$, per avere invarianza di Lorentz locale
delle equazioni (\ref{Vas1}-\ref{Vas5}) dobbiamo anzitutto
estendere i generatori di Lorentz definiti nelle prime due
equazioni della (\ref{genso3,2}), poich\'{e} \`{e} ora necessario
ruotare anche queste nuove variabili spinoriali. Tenendo conto del
fatto che le $Z$ soddisfano un'algebra identica a quella delle $Y$
ma con il segno della contrazione opposto, \`{e} chiaro che il
generatore
\begin{equation}
\widehat{M}_{\alpha\beta}=z_{\alpha}z_{\beta}-y_{\alpha}y_{\beta}
\end{equation}
ruota correttamente le funzioni $\hat{F}(Y,Z)\in\widehat{{\cal
A}}^{\star}$ mediante lo $\star$-commutatore con il parametro
\begin{equation}\label{genlorext}
\hat{\epsilon}_{0}(\Lambda)=\frac{1}{4i}\Lambda^{\alpha\beta}\widehat{M}_{\alpha\beta}-\textrm{h.c.}
\ .
\end{equation}
In particolare
\begin{equation}
\delta
y_{\alpha}=\Lambda_{\alpha}^{\phantom{\alpha}\beta}y_{\beta}\ ,
\qquad \delta
z_{\alpha}=\Lambda_{\alpha}^{\phantom{\alpha}\beta}z_{\beta}\ ,
\end{equation}
e analogamente per gli hermitiani coniugati.

Le trasformazioni di Lorentz locali generate dal parametro
(\ref{genlorext}) sono effettivamente una simmetria delle
equazioni (\ref{Vas1}-\ref{Vas5}), ma non della loro soluzione di
vuoto! La forma della (\ref{Vas5}) mostra infatti che
$S_{\underline{\alpha}}$ deve avere un valore di vuoto non banale,
$S_{0\,\underline{\alpha}}\sim Z_{\underline{\alpha}}$,
evidentemente non invariante sotto lo $\star$-commutatore con
(\ref{genlorext}). In altri termini, la simmetria delle equazioni
non lineari sotto trasformazioni di Lorentz locali \`{e}
spontaneamente rotta dalla loro soluzione di vuoto, poich\'{e} il
parametro (\ref{genlorext}) non soddisfa la
(\ref{extgenKilling3}). Si tratta quindi di trovare, se esistono,
trasformazioni di Lorentz locali che siano una simmetria delle
equazioni non lineari ma non alterino il vuoto. Mantenendo la
notazione $S_{0}$ per la soluzione di vuoto (\ref{bg}), tenendo
presente la definizione di $S$ (\ref{defS}) e ricordando che nella
sezione precedente abbiamo trovato che, a meno di una
trasformazione di gauge, $V$ \`{e} completamente esprimibile in
termini di $B$, e dunque \`{e} almeno di primo ordine
nell'espansione perturbativa attorno al background $AdS$, possiamo
riformulare il problema chiedendoci se esista una generalizzazione
dei generatori (\ref{genlorext}) della simmetria di Lorentz locale
che non alteri la scelta di gauge (\ref{Vgaugechoice}), ovvero
$V_{\alpha}|_{Z=0}=\bar{V}_{\dot{\alpha}}|_{Z=0}=0$, ovvero
$V_{0}=0$. Facendo uso di (\ref{Vgaugetransf}) si trova infatti
che, sotto (\ref{genlorext}),
\begin{equation}\label{trnonomog1}
\delta
V_{\alpha}=-\frac{1}{2i}\Lambda_{\alpha}^{\phantom{\alpha}\beta}z_{\beta}+[V_{\alpha},\hat{\epsilon}_{0}(\Lambda)]_{\star}
\ ,
\end{equation}
il cui termine non omogeneo altera evidentemente il valore di
vuoto. \`{E} necessario quindi che le trasformazioni di Lorentz
locali generalizzate ruotino in modo corretto anche l'indice
spinoriale libero di $V$, dando luogo ad una legge di
trasformazione omogenea. Riscriviamo anzitutto la
(\ref{trnonomog1}) in termini di $S$,
\begin{equation}\label{trnonomog2}
\delta
V_{\alpha}=-\frac{1}{2i}\Lambda_{\alpha}^{\phantom{\alpha}\beta}S_{\beta}-\Lambda_{\alpha}^{\phantom{\alpha}\beta}V_{\beta}+[V_{\alpha},\hat{\epsilon}_{0}(\Lambda)]_{\star}
\ ,
\end{equation}
e notiamo che a secondo membro abbiamo gi\`{a} la trasformazione
di $V$ che cerchiamo, a parte un termine proporzionale ad $S$, che
deve scomparire. L'osservazione cruciale \`{e} che ci\`{o} \`{e}
possibile se $H_{\alpha\dot{\beta}}(X)=0$! In questo caso infatti
il vincolo di curvatura (\ref{Vas5}), che, esplicitando gli indici
e guardando la sola parte olomorfa, si scrive
\begin{equation}\label{cpVas5}
S_{\alpha}\star S_{\beta}-S_{\beta}\star
S_{\alpha}=-2i\varepsilon_{\alpha\beta}(1+B\star\kappa)\ ,
\end{equation}
ha la forma di un'algebra di Heisenberg deformata,
\begin{equation}\label{defoscalg}
[\hat{y}_{\alpha},\hat{y}_{\beta}]=2i\varepsilon_{\alpha\beta}(1+\nu
k)\ , \quad \{\hat{y}_{\alpha},k\}=0\ , \quad k^{2}=1 \ ,
\end{equation}
dove $\nu$ \`{e} un numero reale. La propriet\`{a} importante di
quest'algebra \`{e} che, per ogni $\nu$, \`{e} possibile costruire
gli operatori
\begin{equation}
T_{\alpha\beta}=-\frac{1}{2}\{\hat{y}_{\alpha},\hat{y}_{\beta}\} \
,
\end{equation}
che soddisfano l'algebra di Lorentz
\begin{equation}
[T_{\alpha\beta},T_{\gamma\delta}]=-2i(\varepsilon_{\alpha\gamma}T_{\beta\delta}+\varepsilon_{\beta\delta}T_{\alpha\gamma}+\varepsilon_{\alpha\delta}T_{\beta\gamma}+\varepsilon_{\beta\gamma}T_{\alpha\delta})
\ ,
\end{equation}
\begin{equation}
[T_{\alpha\beta},\hat{y}_{\gamma}]=-2i(\varepsilon_{\alpha\gamma}\hat{y}_{\beta}+\varepsilon_{\beta\gamma}\hat{y}_{\alpha})
\ .
\end{equation}
Con l'aiuto di (\ref{Vas4}), che implica
\begin{equation}
S_{\alpha}\star B\star\kappa+B\star\kappa\star S_{\alpha}=0 \ ,
\end{equation}
otteniamo dunque immediatamente, a condizione che
$H_{\alpha\dot{\beta}}(X)=0$, che il generatore che ruota nel modo
corretto l'indice spinoriale di $S$ \`{e} dato da
\begin{equation}\label{defgenextra}
\hat{\epsilon}_{\textrm{extra}}(\Lambda)=\frac{1}{8i}\Lambda^{\alpha\beta}\left\{S_{\alpha},S_{\beta}\right\}_{\star}-\textrm{h.c.}
\ .
\end{equation}
Si noti inoltre che questa conclusione resta valida anche nel caso
in cui si generalizzi il termine di sorgente in ${\cal
V}(B\star\kappa)$, purch\'{e} ${\cal V}(X)$ sia una funzione
dispari del suo argomento, come anticipato, in modo tale che la
(\ref{Vas4}) continui ad implicare
\begin{equation}
S_{\alpha}\star {\cal V}(B\star\kappa)+{\cal V}(B\star\kappa)\star
S_{\alpha}=0 \ ,
\end{equation}
il che non sarebbe verificato se ${\cal V}(X)$ fosse pari, per le
propriet\`{a} delle mappe $\pi,\bar{\pi}$. Il segno opposto del
secondo membro della (\ref{cpVas5}) rispetto alla
(\ref{defoscalg}) fa s\`{i} che il generatore
$-\frac{1}{2}\left\{S_{\alpha},S_{\beta}\right\}_{\star}$ ruoti
l'indice spinoriale libero di $S$ con un fattore $(-1)$ in
pi\`{u},
\begin{equation}
\left[-\frac{1}{2}\left\{S_{\alpha},S_{\beta}\right\}_{\star},S_{\gamma}\right]=+2i(\varepsilon_{\alpha\gamma}S_{\beta}+\varepsilon_{\beta\gamma}S_{\alpha})
\ ,
\end{equation}
da cui segue che
\begin{equation}
\delta_{\textrm{extra}}V_{\alpha}=[V_{\alpha},\hat{\epsilon}_{\textrm{extra}}(\Lambda)]_{\star}=\frac{1}{2i}\Lambda_{\alpha}^{\phantom{\alpha}\beta}S_{\beta}
\ ,
\end{equation}
che cancella il termine proporzionale ad $S$ della
(\ref{trnonomog2}). Concludiamo dunque che il generatore della
vera simmetria di Lorentz locale delle equazioni dei campi di HS
non lineari, a tutti gli ordini nelle interazioni, \`{e}
\begin{equation}
L_{\alpha\beta}=\widehat{M}_{\alpha\beta}-\frac{1}{2}\left\{S_{\alpha},S_{\beta}\right\}_{\star}
\ ,
\end{equation}
insieme all'hermitiano coniugato, con il quale d\`{a} luogo alla
trasformazione
\begin{equation}\label{trueLL}
\hat{\epsilon}_{L}(\Lambda)=\hat{\epsilon}_{0}(\Lambda)-\hat{\epsilon}_{\textrm{extra}}(\Lambda)=\frac{1}{4i}\Lambda^{\alpha\beta}L_{\alpha\beta}-\textrm{h.c.}
\ .
\end{equation}
Esaminiamone l'azione sui \emph{master fields} della teoria nello
spazio esteso. La variazione di $B$ sotto (\ref{trueLL}) \`{e}
data da
\begin{equation}
\delta_{L}B=B\star\bar{\pi}(\hat{\epsilon}_{L})-\hat{\epsilon}_{L}\star
B \ ,
\end{equation}
che, tenendo conto della (\ref{Vas4}) e del fatto che
$\hat{\epsilon}_{L}$ \`{e} bilineare nelle $Y$ nelle $Z$, fornisce
\begin{equation}
\delta_{L}B=[B,\hat{\epsilon}_{0}(\Lambda)]_{\star} \ .
\end{equation}
Su $V_{\alpha}$ abbiamo ottenuto
\begin{equation}\label{varV}
\delta_{L}V_{\alpha}=-\Lambda_{\alpha}^{\phantom{\alpha}\beta}V_{\beta}+[V_{\alpha},\hat{\epsilon}_{0}(\Lambda)]_{\star}
\ ,
\end{equation}
sicch\'{e} la scelta di gauge $V_{0}=0$ resta preservata e con
essa, come sopra spiegato, la soluzione di vuoto delle equazioni
(\ref{Vas1}-\ref{Vas5}). Equivalentemente si pu\`{o} mostrare
\cite{Vasiliev:1999ba} che $L_{\alpha\beta}$ ruota soltanto i
termini di $S$ contenenti $B$,
\begin{equation}
\delta_{L}S_{\underline{\alpha}}=\frac{\delta
S_{\underline{\alpha}}}{\delta B}\delta B \ ,
\end{equation}
lasciando dunque intatto il valore di vuoto $S_{0}$. Infine,
l'azione su $W_{\mu}$ \`{e}
\begin{eqnarray}
\delta_{L}W_{\mu} & = & D_{\mu}\hat{\epsilon}_{L}(\Lambda)
\nonumber \\
 & = &
[W_{\mu},\hat{\epsilon}_{0}(\Lambda)]_{\star}+\frac{1}{4i}(\partial_{\mu}\Lambda^{\alpha\beta})L_{\alpha\beta}-\textrm{h.c.}
\ ,
\end{eqnarray}
dove si \`{e} tenuto conto della (\ref{Vas3}). I vincoli di
curvatura che realizzano la FDA non lineare nello spazio-tempo
ordinario, vale a dire le (\ref{Vas1}-\ref{Vas2}) in cui siano
stati sostituiti $W(w,C)$ e $B(C)$ ottenuti risolvendo la
dipendenza dalle $Z$ di $W$, $B$ ed $S$ dalle rimanenti equazioni
e si sia posto $Z=0$, sono dunque invarianti sotto le
trasformazioni di Lorentz locali
\begin{eqnarray}
\delta_{L}C & = & [C,\epsilon_{0}(\Lambda)]_{\star} \ , \\
\delta_{L}w_{\mu} & = &
[w_{\mu},\epsilon_{0}(\Lambda)]_{\star}+\frac{1}{4i}(\partial\mu\Lambda^{\alpha\beta})\left[y_{\alpha}y_{\beta}-4\left(V_{\alpha}\star
V_{\beta}\right)_{Z=0}\right]-\textrm{h.c.} \ ,
\end{eqnarray}
dove
\begin{equation}\label{lingen}
\epsilon_{0}(\Lambda)=\frac{1}{4i}\Lambda^{\alpha\beta}y_{\alpha}y_{\beta}-\textrm{h.c.}
\ .
\end{equation}
Si vede dunque che tutti i campi inclusi in $C$ sono tensori di
Lorentz, ovvero hanno trasformazioni di Lorentz locali omogenee.
Le trasformazioni di $w_{\mu}$ non sono invece omogenee, e vi
compaiono termini complicati, dipendenti dai campi. Questi termini
vengono tuttavia naturalmente assorbiti in una ridefinizione della
connessione di Lorentz $\omega_{L\,\mu}^{\phantom{L}\alpha\beta}$,
che, per definizione, trasforma in modo non omogeneo,
\begin{equation}\label{Lorconn}
\delta_{L}\omega_{L\,\mu}^{\phantom{L}\alpha\beta}=\partial\mu\Lambda^{\alpha\beta}+\Lambda^{\alpha}_{\phantom{\alpha}\gamma}\omega_{L\,\mu}^{\phantom{L}\gamma\beta}+\Lambda^{\beta}_{\phantom{\beta}\gamma}\omega_{L\,\mu}^{\phantom{L}\gamma\alpha}
\ .
\end{equation}
Seguendo \cite{Sezgin:2002ru}, mostriamo infatti che, tenendo
conto delle (\ref{varV}) e (\ref{Lorconn}), la quantit\`{a}
$\omega_{L\,\mu}+K_{\mu}$, con
\begin{eqnarray}
\omega_{L\,\mu} & = &
\frac{1}{4i}\omega_{L\,\mu}^{\phantom{L}\alpha\beta}y_{\alpha}y_{\beta}-\textrm{h.c.}
\ , \\
K_{\mu} & = &
i\omega_{L\,\mu}^{\phantom{L}\alpha\beta}\left(V_{\alpha}\star
V_{\beta}\right)_{Z=0}-\textrm{h.c.} \ ,
\end{eqnarray}
trasforma come
\begin{equation}
\delta_{L}(\omega_{L\,\mu}+K_{\mu})=[\omega_{L\,\mu}+K_{\mu},\epsilon_{0}(\Lambda)]_{\star}+\frac{1}{4i}\partial\mu\Lambda^{\alpha\beta}\left[y_{\alpha}y_{\beta}-4\left(V_{\alpha}\star
V_{\beta}\right)_{Z=0}\right]-\textrm{h.c.} \ ,
\end{equation}
sicch\'{e}
\begin{equation}
\delta_{L}(w_{\mu}-\omega_{L\,\mu}-K_{\mu})=[w_{\mu}-\omega_{L\,\mu}-K_{\mu},\epsilon_{0}(\Lambda)]_{\star}\
,
\end{equation}
ovvero tutti i campi di spin arbitrario contenuti in $w_{\mu}$
diversi da $\omega_{L\,\mu}^{\phantom{L}\alpha\beta}$ trasformano
come tensori di Lorentz, mentre, come diretta conseguenza della
generalizzazione (\ref{trueLL}), la connessione di Lorentz viene
ridefinita con un termine non lineare legato alla curvatura di
spazio $Z$. Tuttavia, la conclusione notevole che possiamo trarre
da questi risultati \`{e} che \emph{la simmetria di Lorentz locale
resta indeformata}, nel passaggio alla teoria non lineare: come
abbiamo visto, infatti, tutti i campi hanno trasformazioni
omogenee sotto un generatore che, utilizzando le equazioni del
moto (\ref{Vas3}-\ref{Vas5}), si riduce a quello della teoria
linearizzata (\ref{lingen}), mentre l'effetto delle interazioni
viene completamente riassorbito nella connessione di Lorentz.
Questo \`{e} un fatto assolutamente singolare, dato che, anche in
un modello non minimale, la sola altra simmetria locale che resta
indeformata \`{e} quella di Yang-Mills \cite{Vasiliev:vu}, mentre
tutte le altre vengono ridefinite mediante termini dipendenti
dalle curvature. Tale risultato comporta inoltre che la simmetria
di Lorentz locale resta una sottoalgebra dell'algebra
infinito-dimensionale di HS, il che non sarebbe potuto accadere se
il generatore di Lorentz avesse subito una ridefinizione in
termini di curvature di HS, con componenti sui generatori di spin
$s>1$.

\appendix

\chapter{Il teorema di Coleman-Mandula}
\label{CM}

Dopo la scoperta della simmetria di sapore $SU(3)$, che spiegava
in modo sorprendente le relazioni tra adroni di identico spin, ci
si chiese se esistesse un modo non banale di combinare le
simmetrie interne e quelle spazio-temporali di una teoria di campo
all'interno di un gruppo di simmetria pi\`{u} ampio. In altri
termini, detta $g$ un'algebra di simmetria interna e ${\cal P}$
l'algebra di Poincar\'{e}, le teorie di campo ``ordinarie'' sono
invarianti sotto la somma diretta ${\cal P}\oplus g$, il che
significa che i generatori delle simmetrie interne commutano con i
generatori delle traslazioni e delle trasformazioni di Lorentz
$P_\mu$ e $J_{\mu\nu}$ dell'algebra di Poincar\'{e}, e non
connettono quindi stati corrispondenti a diversi autovalori di
impulso e spin. L'idea era cercare un'algebra di simmetria $G$
pi\`{u} ampia che includesse ${\cal P}\oplus g$ come sottoalgebra
propria, in modo tale che, in una base opportuna, i suoi elementi
fossero rappresentati mediante matrici diagonali a blocchi,
\begin{eqnarray}
G\supset{\cal P}\oplus g\,\ni\, m=\left(\begin{array}{c|c} \ast & 0 \\
\hline 0 & \ast \end{array}\right) \ ,
\end{eqnarray}
dove $\ast$ sottintende gli elementi di generiche sottomatrici
appartenenti a ${\cal P}$ o $g$.

Il teorema di Coleman e Mandula \cite{Coleman:ad} mostr\`{o}
invece che, sotto ipotesi piuttosto naturali per qualsiasi teoria
di campo ``ordinaria'' (ma, come sottolineato nell'Introduzione e
nel capitolo $5$, sistematicamente evitate nell'ambito delle
teorie di gauge di HS), questo non \`{e} possibile, e la pi\`{u}
generale algebra di Lie di operatori di simmetria deve consistere
della somma diretta dell'algebra di Poincar\'{e} e di un'algebra
di simmetria interna, i cui elementi sono diagonali sugli
autostati di $P_\mu$ e $J_{\mu\nu}$ e indipendenti dalle variabili
di impulso e spin, vale a dire
\begin{equation}
G={\cal P}\oplus g \ .
\end{equation}

Per effettuare un'analisi il pi\`{u} possibile generale, Coleman e
Mandula lavorarono nell'ambito della teoria della matrice $S$ di
scattering, i cui elementi corrispondono alle ampiezze di
transizione $_{\textrm{out}}\langle p_1 p_2 ...|k_1
k_2...\rangle_{\textrm{in}}$ tra stati fisici, in generale a molte
particelle, di impulso definito e asintotici, ovvero definiti in
regioni di ``ingresso'' e di ``uscita'' lontane dalla regione di
interazione. La matrice $S$ contiene quindi tutta l'informazione
sull'evoluzione degli stati fisici nel tempo. Prima di dare la
precisa formulazione del teorema \`{e} conveniente rivedere alcune
definizioni fondamentali.

Anzitutto, lo spazio di Hilbert ${\cal H}$ degli stati rilevanti
per la teoria dello scattering si decompone nella somma diretta
infinita
\begin{equation}
{\cal H}=\bigoplus_{n=1}^{\infty}{\cal H}^{(n)} \ ,
\end{equation}
dove ${\cal H}^{(n)}$ \`{e} il sottospazio degli stati ad $n$
particelle. Quest'ultimo \`{e} a sua volta un sottospazio,
determinato dalle simmetrizzazioni o antisimmetrizzazioni legate
alla relazione spin-statistica, del prodotto diretto di $n$ spazi
di Hilbert a singola particella, ciascuno isomorfo ad ${\cal
H}^{(1)}$. La matrice $S$ \`{e} un operatore unitario su ${\cal
H}$, che collega stati fisici etichettati da impulsi sul
\emph{mass-shell}. Si dice che l'operatore hermitiano $A_\alpha$
su ${\cal H}$ \`{e} un  \emph{generatore di simmetria} se:
\begin{enumerate}
\item $A_\alpha$ commuta con $S$, per ogni $\alpha$;

\item $A_\alpha$ trasforma stati ad una particella in stati ad una
particella;

\item l'azione di $A_\alpha$ sugli stati a molte particelle
corrisponde alla somma diretta dell'azione sugli stati a singola
particella;

\item i commutatori di due operatori $A_\alpha$ sono a loro volta
generatori di simmetria.
\end{enumerate}
Il gruppo di trasformazioni generato dagli $A_\alpha$ definisce
quindi il gruppo di simmetria della teoria di campo in questione.
La matrice $S$ si dice \emph{invariante di Lorentz} se possiede
un'algebra di simmetria una cui sottoalgebra \`{e} isomorfa a
${\cal P}$. Si pu\`{o} quindi introdurre una base per ${\cal
H}^{(1)}$ in termini degli stati $|p \ m \rangle$, dove $p$ \`{e}
l'impulso dello stato ed $m$ un indice discreto collettivo, che
comprende sia la componente $z$ dello spin che il tipo di
particelle di massa definita $\sqrt{-p_\mu p^\mu}$ (e dunque
contiene, ad esempio, gli indici di sapore, nell'ipotesi di una
simmetria di sapore esatta, ovvero di degenerazione in massa delle
diverse generazioni di quarks). Una simmetria si dice inoltre
\emph{interna} se commuta con i generatori dell'algebra di
Poincar\'{e}, e dunque se i corrispondenti generatori agiscono in
modo non banale soltanto sugli indici che etichettano il tipo di
particella, mentre non hanno elementi di matrice tra stati con
diverso quadrimpulso o diverso spin. Infine, si separa la parte
banale della matrice $S$, che tiene conto dell'eventualit\`{a} che
nessuna interazione abbia luogo, come
\begin{equation}
S \ = \ 1 \ + \ (2\pi)^4 \ \delta^{4}(p-p')\,T \ ,
\end{equation}
dove $\delta^{(4)}(p-p')$ impone la conservazione del quadrimpulso
tra stato iniziale e finale, e gli elementi di matrice di $T$
corrispondono a genuine ampiezze di scattering.

Siamo a questo punto in grado di enunciare il
\begin{CM}
\emph{Se $G$ \`{e} un'algebra di simmetria della matrice S, e se
valgono le condizioni:}
\begin{enumerate}
\item $G$ contiene una sottoalgebra isomorfa a ${\cal P}$
(invarianza di Lorentz).

\item Per ogni $M$ finito, esiste soltanto un numero finito di
tipi di particelle con massa minore di M. \label{finiteness}

\item La matrice $S$ \`{e} non banale, nel senso che qualsiasi
stato a due particelle $|p,p'\rangle$ subisce uno scattering,
\begin{equation}
T|p,p'\rangle\neq 0 \ ,
\end{equation}
per quasi tutti i valori della variabile di Mandelstam
$\textrm{s}$, ovvero per tutti i valori dell'energia nel sistema
del centro di massa escluso, eventualmente, un insieme di misura
nulla. \label{nontrivialscatt}

\item Le ampiezze di scattering elastico \footnote {Si intende per
scattering elastico il processo di diffusione nel quale sia la
massa delle particelle che l'energia cinetica totale vengono
conservate.} a due corpi sono funzioni analitiche di $\textrm{s}$
e del momento trasferito invariante $\textrm{t}$, ovvero
dell'energia e dell'angolo di scattering $\theta$, per quasi tutti
i valori dell'energia e dell'angolo di scattering stessi.
\label{analyticity}

\item Supponendo che l'azione pi\`{u} generale dei generatori di
$G$ sugli stati a singola particella $|p\,n\rangle$ sia data da
\begin{equation}\label{actioncompl}
A_\alpha |p\,n\rangle \ = \ \sum_{n'}\int d^4 p'\left({\cal
A}_\alpha(p,p')\right)_{n,n'}|p'\,n'\rangle \ ,
\end{equation}
dove i nuclei integrali (\emph{kernels}) ${\cal A}(p,p')$ sono non
nulli soltanto se $p$ e $p'$ sono sul \emph{mass shell}, si assume
che tali nuclei siano \emph{distribuzioni}, ovvero che contengano
al pi\`{u} un numero finito $D_\alpha$ di derivate di
$\delta^{4}(p-p')$ \footnote {Il motivo per cui questa ipotesi
\`{e} necessaria diverr\`{a} chiaro nel corso della dimostrazione.
Sottolineamo inoltre che la richiesta che i \emph{kernels} di
integrazione siano non nulli solo se gli impulsi da cui dipendono
sono sul \emph{mass shell} segue dalla definizione della matrice
$S$ come operatore che collega stati iniziali e finali asintotici
entrambi sul \emph{mass shell}, ovvero stati fisici rivelabili in
un esperimento.}. \label{ugly}
\end{enumerate}

\emph{Segue allora la tesi,}
\begin{equation}
G \ \sim \ {\cal P}\oplus g \ ,
\end{equation}
\emph{ovvero segue che la pi\`{u} generale algebra di simmetria
della matrice $S$ \`{e} isomorfa alla somma diretta dell'algebra
di Poincar\'{e} e di un'algebra di simmetria interna.}
\end{CM}

\vspace{0.5cm}

Riportiamo nel seguito la dimostrazione del teorema, seguendo in
gran parte la versione data da Weinberg in \cite{Weinberg}, in
alcuni passaggi pi\`{u} esplicita di quella originale di Coleman e
Mandula. Poich\'{e} la dimostrazione \`{e} piuttosto elaborata,
\`{e} opportuno presentarne anzitutto lo schema, per poi
sviluppare i singoli passi in dettaglio.
\begin{enumerate}
\renewcommand{\labelenumi}{\Alph{enumi}}
\renewcommand{\labelenumii}{\roman{enumii}}
\item Si dimostra il teorema per la sottoalgebra di $G$ costituita
dai generatori di simmetria $B_\alpha$ che commutano con i
generatori delle traslazioni $P_\mu$. Questo viene fatto nei
seguenti passi:
\begin{enumerate}
\item Si dimostra che l'azione dei $B_\alpha$ sugli stati ad $n$
particelle, realizzata attraverso l'azione di matrici hermitiane
$b_{\alpha}(p)$ di dimensione finita agenti sugli stati a singola
particella (per definizione di operatore di simmetria e per
l'ipotesi $2$), definisce un isomorfismo $I: \ B_\alpha\rightarrow
b_\alpha$.

\item Si pu\`{o} quindi applicare alle matrici $b_{\alpha}(p)$ (o
 meglio alle corrispondenti matrici a traccia nulla)
 il teorema generale secondo cui qualsiasi algebra di Lie di
 matrici hermitiane di dimensione finita \`{e} una somma diretta
 ${\cal C}\oplus u(1)$ , dove ${\cal C}$ \`{e} un'algebra di Lie
 semisemplice e compatta (o, pi\`{u} in generale, di ${\cal C}$ e
 di pi\`{u} algebre $u(1)$), ed estenderlo ai generatori di simmetria
 $B_\alpha$ in virt\`{u} dell'isomorfismo $I$. Ci si riduce in tal
 modo ad analizzare un'algebra di simmetria ${\cal C}\oplus u(1)$.

\item Si dimostra che i generatori di $u(1)$ commutano con i
generatori $J_{\mu\nu}$ del gruppo di Lorentz, sicch\'{e}
commutano con tutti i generatori dell'algebra di Poincar\'{e}
${\cal P}$.

\item Si dimostra che i generatori di ${\cal C}$ commutano con i
generatori di Lorentz $J_{\mu\nu}$, e quindi con tutti i
generatori di ${\cal P}$. Resta a questo punto dimostrato che i
$B_\alpha\in G$ tali che $[B_\alpha,P_\mu]=0$ generano una
simmetria interna o sono combinazioni lineari dei $P_\mu$, ovvero
il teorema \`{e} dimostrato per la sottoalgebra generata dai
$B_\alpha$.
\end{enumerate}

\item Si considerano tutti i generatori di simmetria $A_\alpha$, e
si dimostra che quelli che non commutano con $P_\mu$ (ovvero
quelli con $D_\alpha\geq 1$, contenenti almeno una derivata di
$\delta^{(4)}(p-p')$) o sono nulli ($D_\alpha\geq 2$) o si
riducono ad una combinazione lineare dei generatori di Lorentz
$J_{\mu\nu}$ e dei generatori $B_\alpha$ trattati nei passi
precedenti ($D_\alpha=1$), completando in tal modo la
dimostrazione.
\end{enumerate}

Per compattezza, segnaleremo con commenti in nota i passaggi in
cui svolgono un ruolo cruciale le ipotesi che sono evitate nelle
teorie di gauge di HS: in particolare, in quel contesto l'ipotesi
1 \`{e} invalidata, poich\'{e} l'algebra di simmetria
spazio-temporale della teoria non \`{e} pi\`{u} quella di
Poincar\'{e} ma quella di $AdS$, $so(d-1,2)$, cos\`{i} come
l'ipotesi $2$, dal momento che si introducono simultaneamente
infiniti campi di massa nulla, e corrispondenti generatori, di
spin arbitrariamente elevato.

Iniziamo quindi dal passo A, mostrando l'azione dei generatori di
simmetria $B_\alpha$ sugli stati a molte particelle, data per
definizione da
\begin{eqnarray}
B_\alpha |p\,m,\;q\,n,\;...\rangle & = &
\sum_{m'}\left(b_\alpha(p)\right)_{m'm} |p\,m',\;q\,n,\;...\rangle
\nonumber \\
&&+\sum_{m'}\left(b_\alpha(q)\right)_{n'n}
|p\,m,\;q\,n',\;...\rangle+... \ ,\label{action}
\end{eqnarray}
dove le matrici $b_\alpha$ sono matrici hermitiane di dimensione
finita che definiscono l'azione dei generatori di simmetria sugli
stati a singola particella. La (\ref{action}) definisce una mappa
$I: \ B_\alpha\rightarrow b_\alpha(p)$, per ogni $p$ fissato, che
preserva la struttura dell'algebra generata dai $B_\alpha$,
poich\'{e} implica che le relazioni di commutazione
\begin{equation}\label{alg}
[B_\alpha,B_\beta]=i\sum_{\gamma}C^{\gamma}_{\alpha\beta}B_\gamma
\end{equation}
si traducano in identiche relazioni soddisfatte dalle matrici
$b_\alpha(p)$ in ciascun sottospazio ${\cal H}^{(1)}$,
\begin{equation}\label{algrappr}
[b_\alpha(p),b_\beta(p)]=i\sum_{\gamma}C^{\gamma}_{\alpha\beta}\,b_\gamma(p)
\ ,
\end{equation}
ovvero implica che la mappa $I$ sia un \emph{omomorfismo}.
Tuttavia, per poter inferire da questa osservazione che l'algebra
generata dai $B_\alpha$ sia del tipo ${\cal C}\oplus u(1)$ si deve
dimostrare che $I$ \`{e} un \emph{isomorfismo}, vale a dire un
omomorfismo iniettivo e suriettivo, ovvero si deve dimostrare che,
se esistono dei coefficienti $c^\alpha$ e un impulso $p$ tali che
$\sum_{\alpha}c^\alpha b_\alpha(p)=0$, deve necessariamente essere
$\sum_{\alpha}c^\alpha B_\alpha=0$, che significa
$\sum_{\alpha}c^\alpha b_\alpha(k)=0, \ \forall k$ in virt\`{u}
della definizione (\ref{action}).

Coleman e Mandula considerarono l'omomorfismo tra i $B_\alpha$ e
le matrici finito-dimensionali $b_\alpha(p,q)$ che ne definiscono
l'azione sugli stati a due particelle. In termini delle
$b_\alpha(p)$ si ha
\begin{equation}\label{def2pt1pt}
b_\alpha(p,q)=\left[b_\alpha(p)\otimes
1_q\right]\oplus\left[1_p\otimes b_\alpha(q)\right] \ ,
\end{equation}
dove $1_k$ denota la matrice identit\`{a} agente nel sottospazio
${\cal H}^{(1)}(k)$. L'invarianza della matrice $S$ sotto le
trasformazioni rappresentate dalle $b_\alpha(p,q)$,
\begin{eqnarray}
&&\langle p'\,m', q'\,n'|p\,m,q\,n\rangle=\langle p'\,m',
q'\,n'|b_\alpha(p',q')|p\,m,q\,n\rangle \nonumber \\
&&=\langle p'\,m', q'\,n'|b_\alpha(p,q)|p\,m,q\,n\rangle \ ,
\label{startingpt}
\end{eqnarray}
implica che
\begin{equation}\label{coniugio}
b_\alpha(p',q')\;T(p'\,q',p\,q)=T(p'\,q',p\,q)\;b_\alpha(p,q) \ ,
\end{equation}
come si pu\`{o} verificare inserendo un set completo di stati
asintotici ad ambo i membri della (\ref{startingpt}).
$T(p'\,q',p\,q)$ \`{e} una matrice della stessa dimensionalit\`{a}
delle $b_\alpha(p,q)$  e legata all'elemento di matrice
$T(pm,qn\rightarrow p'm',q'n')$, corrispondente all'ampiezza di
scattering elastico o quasi elastico di due particelle con
quadrimpulso $p$ e $q$ in due particelle con quadrimpulso $p'$ e
$q'$ sullo stesso \emph{mass-shell}, ovvero con masse
$\sqrt{-p_\mu p^\mu}=\sqrt{-p'_\mu p^{\prime\mu}}$ e $\sqrt{-q_\mu
q^\mu}=\sqrt{-q'_\mu q^{\prime\mu}}$, da
\begin{equation}
T(pm,qn\rightarrow
p'm',q'n')=\delta^4\,(p'+q'-p-q)\left(T(p'\,q',p\,q)\right)_{m'n',mn}
\ .
\end{equation}
Il teorema ottico, che segue dall'unitariet\`{a} della matrice
$S$, implica che la parte immaginaria dell'ampiezza di scattering
in avanti (con angolo di scattering nullo, e stato iniziale
$|i\rangle$ uguale allo stato finale $|f\rangle$) \`{e}
proporzionale alla sezione d'urto totale,
\begin{equation}
\sum_{n}|\langle n|T|i\rangle|^2=2\;\textrm{Im}\;\langle
i|T|i\rangle \ ,
\end{equation}
dove la somma corre su tutti i possibili stati finali. L'ipotesi
$3$ assicura che il primo membro \`{e} non nullo, e di conseguenza
\`{e} non nulla l'ampiezza di probabilit\`{a} di scattering in
avanti. Ma allora l'ipotesi $4$ di analiticit\`{a} della matrice
$S$ in $\theta$ garantisce, per continuit\`{a}, che la matrice
$T(p'\,q',p\,q)$ \`{e} non singolare per quasi tutti i valori di
$p'$ e $q'$ che soddisfano la condizione di conservazione del
quadrimpulso $p'+q'=p+q$ e sono sugli stessi iperboloidi di massa
degli stati iniziali. In questo caso la (\ref{coniugio}) \`{e} un
coniugio tramite $T$, e questo comporta che, se
$\sum_{\alpha}c^\alpha b_\alpha(p,q)=0$, per quasi tutti i $p$ e
$q$ fissati, allora $\sum_{\alpha}c^\alpha b_\alpha(p',q')=0$, per
quasi tutti i $p'$ e $q'$ che rispettano le condizioni appena
indicate. Non \`{e} tuttavia lecito da qui concludere che
$\sum_{\alpha}c^\alpha b_\alpha(p)=0$ e $\sum_{\alpha}c^\alpha
b_\alpha(q)=0$ separatamente, ma al pi\`{u}, come suggerisce la
(\ref{def2pt1pt}), che sono proporzionali alla matrice
identit\`{a}, e con coefficienti opposti.

\`{E} quindi conveniente estrarre preventivamente le tracce dai
generatori di simmetria e dai loro rappresentativi $b_\alpha(p)$
sugli stati ad una particella e $b_\alpha(p,q)$ sugli stati a due
particelle. Poich\'{e} la traccia \`{e} invariante per coniugio,
dalla (\ref{coniugio}) segue che
\begin{equation}
\textrm{Tr}\,b_\alpha(p',q')=\textrm{Tr}\,b_\alpha(p,q) \ ,
\end{equation}
che, utilizzando la (\ref{def2pt1pt}) e tenendo presente l'ipotesi
di elasticit\`{a} dell'urto, fornisce
\begin{eqnarray}
&&N(\sqrt{-q_\mu q^\mu})\,\textrm{tr}\,b_\alpha(p')+N(\sqrt{-p_\mu
p^\mu})\,\textrm{tr}\,b_\alpha(q') \nonumber \\
&&=N(\sqrt{-q_\mu q^\mu})\,\textrm{tr}\,b_\alpha(p)+N(\sqrt{-p_\mu
p^\mu})\,\textrm{tr}\,b_\alpha(q) \ , \label{traces}
\end{eqnarray}
dove $N(m)$ \`{e} la molteplicit\`{a} dei tipi di particelle con
massa $m$, e distinguiamo la traccia sugli stati a due particelle
``$\textrm{Tr}$'' dalla traccia sugli stati a singola particella
``$\textrm{tr}$''. Ma la (\ref{traces}) pu\`{o} essere soddisfatta
per tutti i $p,q,p',q'$ che soddisfino $p'+q'=p+q$ soltanto se
$\textrm{tr}\,b_\alpha(p)$ \`{e} una funzione lineare di $p$,
\begin{equation}\label{traces2}
\textrm{tr}\,b_\alpha(p)=N(\sqrt{-p_\mu p^\mu})\,a_\alpha^\mu
p_\mu \ ,
\end{equation}
dove gli $a_\alpha^\mu$ sono coefficienti costanti, indipendenti
da $p$ \footnote {Non si include un termine costante in
(\ref{traces2}) perch\'{e} pu\`{o} sempre essere riassorbito in un
cambiamento dell'azione delle simmetrie interne sugli stati
fisici.}.

Si possono ora definire nuovi generatori di simmetria, a traccia
nulla, sottraendo ai $B_\alpha$ termini lineari nell'impulso, come
\begin{equation}\label{tracelessgen}
B^\ast_\alpha\equiv B_\alpha-a_\alpha^\mu P_\mu \ ,
\end{equation}
che agiscono sugli stati a singola particella come le matrici
\begin{equation}\label{tracelessmatr}
\left(b^\ast_\alpha(p)\right)_{n'n}=\left(b_\alpha(p)\right)_{n'n}-\frac{\textrm{tr}\,b_\alpha(p)}{N(\sqrt{-p_\mu
p^\mu})}\,\delta_{n'n} \ ,
\end{equation}
a traccia nulla anch'esse. I generatori $B^\ast_\alpha$ ed i loro
rappresentativi $b^\ast_\alpha(p)$ soddisfano evidentemente le
stesse relazioni di commutazione dei $B_\alpha$ e $b_\alpha(p)$,
(\ref{alg}) e (\ref{algrappr}), dal momento che la matrice
identit\`{a} commuta banalmente con $B_\alpha$ e
$[B_\alpha,P_\mu]=0$ per ipotesi. Quindi, ad esempio,
\begin{equation}
[B^\ast_\alpha,B^\ast_\beta]=i\sum_{\gamma}C^{\gamma}_{\alpha\beta}\left[B^\ast_\gamma+a_\gamma^\mu
P_\mu\right] \ ,
\end{equation}
e
\begin{equation}\label{falsecommrel}
[b^\ast_\alpha(p),b^\ast_\beta(p)]=i\sum_{\gamma}C^{\gamma}_{\alpha\beta}\left[b^\ast_\gamma(p)+a_\gamma^\mu
p_\mu\right] \ .
\end{equation}
Ma per l'ipotesi $2$ le matrici $b^\ast_\alpha(p)$ hanno
dimensione finita, e il commutatore di due matrici di dimensione
finita ha traccia nulla \footnote {Si noti che le teorie di gauge
di HS contraddicono l'ipotesi $2$ (si veda il cap. $5$), non
consentendo quindi il passaggio a generatori a traccia nulla.
Inoltre, il teorema di Coleman e Mandula prende in considerazione
soltanto algebre di Lie, chiuse sotto il commutatore, e trascura
la possibilit\`{a} di algebre di simmetria pi\`{u} generali, come
le superalgebre, o algebre di Lie graduate, chiuse sotto il
supercommutatore
\begin{equation}
t_a\,t_b-(-1)^{\pi_a \pi_b}t_b\,t_a=i\sum_c C^{c}_{ab}\,t_c \ ,
\end{equation}
dove $\pi_a=0,1$ per i generatori bosonici e fermionici,
rispettivamente. \`{E} chiaro che un anticommutatore non ha
traccia nulla, e quindi il teorema di Coleman-Mandula non esclude,
come \`{e} ben noto, la possibilit\`{a} di combinare simmetrie
interne e spazio-temporali in modo non banale mediante
superalgebre di simmetria, su cui si basano le teorie
supersimmetriche. La generalizzazione del teorema al caso
supersimmetrico, dovuta ad Haag, Lopuszanski e Sohnius
\cite{Haag:1974qh}, \`{e} di notevole importanza, perch\'{e} le
corrispondenti cariche centrali sono alla base della definizione
dei multipletti BPS saturati. Le teorie supersimmetriche sono
inoltre inglobate in modo naturale nelle teorie di gauge di HS,
come esposto nel capitolo $5$, che si fondano su una
generalizzazione ulteriore delle algebre di simmetria prese in
considerazione da Coleman e Mandula: queste ultime coinvolgono
infatti soltanto generatori di spin $0$ e $1$, mentre le teorie di
gauge di HS richiedono l'introduzione simultanea di tutti gli
spin, violando l'ipotesi $2$ del teorema.}, sicch\'{e} anche il
secondo membro della (\ref{falsecommrel}) deve essere a traccia
nulla, ovvero $C^{\gamma}_{\alpha\beta}a_\gamma^\mu=0$, e
\begin{equation}\label{truecommrel}
[B^\ast_\alpha,B^\ast_\beta]=i\sum_{\gamma}C^{\gamma}_{\alpha\beta}B^\ast_\gamma
\ ,
\end{equation}
Ci\`{o} mostra che \`{e} lecito prendere i $B^\ast_\alpha$ come
nuovi generatori dell'algebra $G$ di simmetria. Ripetendo passaggi
identici a quelli fatti in precedenza si arriva alla
(\ref{coniugio})
\begin{equation}\label{coniugio2}
b^\ast_\alpha(p',q')\;T(p'\,q',p\,q)=T(p'\,q',p\,q)\;b^\ast_\alpha(p,q)
\ ,
\end{equation}
con le matrici a traccia nulla $b^\ast_\alpha(p)$ al posto delle
$b_\alpha(p)$, e l'invertibilit\`{a} di $T$ implica nuovamente che
$\sum_{\alpha}c^\alpha b^\ast_\alpha(p,q)=0 \Rightarrow
\sum_{\alpha}c^\alpha b^\ast_\alpha(p',q')=0$, per quasi tutti i
$p',q'$ sugli stessi iperboloidi di massa dei $p,q$,
rispettivamente, e che soddisfano $p'+q'=p+q$. Ma ora abbiamo a
che fare con matrici a traccia nulla, e ci\`{o} permette di
concludere che
\begin{equation}\label{parziale}
\sum_{\alpha}c^\alpha b^\ast_\alpha(p')=\sum_{\alpha}c^\alpha
b^\ast_\alpha(q')=0 \ ,
\end{equation}
almeno per $p'$ e $q'=p+q-p'$ sugli iperboloidi di massa di $p$ e
$q$. Per generalizzare tale risultato a tutti i quadrimpulsi $k$
sui suddetti iperboloidi, si consideri l'insieme $K(p,q)$ di tutti
gli elementi dell'algebra $\sum_{\alpha}c^\alpha B_\alpha$, tali
che
\begin{equation}
\sum_{\alpha}c^\alpha b^\ast_\alpha(p,q)=0 \ .
\end{equation}
Dall'eq. (\ref{def2pt1pt}) segue evidentemente che
\begin{equation}\label{essential}
K(p,q)=K(p)\cap K(q) \ ,
\end{equation}
ed abbiamo gi\`{a} ricavato che, per quasi tutti i $p',q'$ sugli
stessi iperboloidi di massa dei $p,q$, e che soddisfano
$p'+q'=p+q$,
\begin{equation}
K(p,q)=K(p',q') \ .
\end{equation}
Ma dalla (\ref{essential}) segue
\begin{equation}
K(p)\supset K(p,q) \ ,
\end{equation}
e
\begin{equation}
K(p')\supset K(p',q')=K(p,q) \ ,
\end{equation}
sicch\'{e}
\begin{equation}
K(p,p')\supset K(p,q) \ ,
\end{equation}
dove $p+p'\neq p+q$! Iterando questo procedimento si deve quindi
ammettere che
\begin{equation}
K(k)\supset K(p,q) \ ,
\end{equation}
per ogni $k$ soddisfacente le condizioni di \emph{mass shell}. Di
conseguenza, se $\sum_{\alpha}c^\alpha b^\ast_\alpha(p,q)=0$ per
alcuni $p$ e $q$, allora $\sum_{\alpha}c^\alpha
b^\ast_\alpha(k)=0$, per ogni $k$ sugli stessi iperboloidi. Questo
\`{e} ci\`{o} che si voleva dimostrare, e si pu\`{o} infine
concludere che la mappa $I$ definisce effettivamente un
isomorfismo tra i generatori $B_\alpha$ ed i loro rappresentativi
finito-dimensionali sugli stati a due particelle
$b^\ast_\alpha(p,q)$.

Questo risultato ha due conseguenze. Anzitutto, poich\'{e} il
numero di $b^\ast_\alpha(p,q)$ non pu\`{o} superare la loro
dimensione, pari a $N(\sqrt{-p_\mu p^\mu})N(\sqrt{-q_\mu q^\mu})$,
per ipotesi un numero finito, ci pu\`{o} essere al pi\`{u} un
numero finito di generatori di simmetria $B_\alpha$ indipendenti,
dal momento che questo ultimi sono in corrispondenza biunivoca con
le prime tramite $I$. Non \`{e} dunque necessario imporre come
ipotesi indipendente che l'algebra di simmetria generata dai
$B_\alpha$ sia finito-dimensionale.

Inoltre, come preannunciato, possiamo concludere che tale algebra
di simmetria \`{e} la somma diretta di un'algebra di Lie compatta
e semisemplice e di una o pi\`{u} algebre di Lie abeliane, ${\cal
C}\oplus u(1)$, oltre all'algebra generata da combinazioni lineari
dei $P_\mu$ incluse nelle tracce dei $B_\alpha$.

Prendiamo anzitutto in esame il comportamento dei $B^\ast_\alpha$
che generano ${\cal C}$ sotto una trasformazione di Lorentz
$x^{\mu}\rightarrow \Lambda^{\mu}_{\phantom{\mu}\nu}x^{\nu}$.
Quest'ultima agisce sullo spazio di Hilbert della teoria mediante
l'operatore unitario $U(\Lambda)$, e il trasformato
$U(\Lambda)B_\alpha U^{-1}(\Lambda)$, che \`{e} ancora un
generatore hermitiano e commuta con $P_\mu$, deve quindi essere
una combinazione lineare dei $B_\alpha$,
\begin{equation}
U(\Lambda)B_\alpha
U^{-1}(\Lambda)=\sum_{\beta}D^{\beta}_{\phantom{\beta}\alpha}(\Lambda)\,B_\beta
\ ,
\end{equation}
dove i coefficienti reali
$D^{\beta}_{\phantom{\beta}\alpha}(\Lambda)$ definiscono una
rappresentazione del gruppo di Lorentz omogeneo,
\begin{equation}
D(\Lambda_1)D(\Lambda_2)=D(\Lambda_1\Lambda_2) \ .
\end{equation}
In effetti, il gruppo di Lorentz definisce un automorfismo
dell'algebra ${\cal C}$, poich\'{e} i trasformati dei generatori
$U(\Lambda)B_\alpha U^{-1}(\Lambda)$ soddisfano le relazioni di
commutazione (\ref{alg}), e ci\`{o} implica che le costanti di
struttura siano tensori invarianti,
\begin{equation}
C^\gamma_{\alpha\beta}=\sum_{\alpha'\beta'\gamma'}D^{\alpha'}_{\phantom{\alpha}\alpha}(\Lambda)D^{\beta'}_{\phantom{\beta}\beta}(\Lambda)D^{\gamma}_{\phantom{\gamma}\gamma'}(\Lambda^{-1})C^{\gamma\prime}_{\alpha'\beta'}
\ .
\end{equation}
La contrazione di quest'equazione con la corrispondente per
$C^\alpha_{\gamma\delta}$ mostra che la metrica di Cartan-Killing
\begin{equation}
g_{\beta\delta}=\sum_{\alpha\gamma}C^\gamma_{\alpha\beta}C^\alpha_{\gamma\delta}
\end{equation}
\`{e} anch'essa un tensore invariante del gruppo di Lorentz,
\begin{equation}
g_{\beta\delta}=\sum_{\beta'\delta'}D^{\beta'}_{\phantom{\beta}\beta}(\Lambda)D^{\delta'}_{\phantom{\delta}\delta}(\Lambda)g_{\beta'\delta'}
\ .
\end{equation}
Suddividendo i generatori come $B_\alpha=(P_\mu,B_A)$, dove i
$B_A$ corrispondono ai generatori a traccia nulla $B^\ast_\alpha$,
e ricordando che tutti commutano con $P_\mu$, si ottiene che
$C^\alpha_{\gamma\mu}=0=-C^\alpha_{\mu\gamma}$ e
$g_{\mu\alpha}=g_{\alpha\mu}=0$. Ne segue che \`{e} non nulla solo
la metrica di ${\cal C}$, $g_{AB}=\sum_{CD}C^D_{AC}C^C_{BD}$.
Quest'ultima, in quanto metrica di un'algebra di Lie semisemplice
e compatta, \`{e} definita positiva, e pu\`{o} dunque essere
invertita per costruire le matrici $g^{1/2}D(\Lambda)g^{-1/2}$,
che costituiscono una rappresentazione ortogonale (sia $g$ che
$D(\Lambda)$ sono ortogonali, $\det(g^{1/2}D(\Lambda)g^{-1/2})=1$
e
$g^{1/2}D(\Lambda_1)g^{-1/2}g^{1/2}D(\Lambda_2)g^{-1/2}=g^{1/2}D(\Lambda_1\Lambda_2)g^{-1/2}$)
e finito-dimensionale del gruppo di Lorentz. Ma quest'ultimo \`{e}
un gruppo non compatto, e come tale non pu\`{o} ammettere
rappresentazioni unitarie non banali di dimensione finita,
sicch\'{e} $D(\Lambda)=1$, e i generatori $B^\ast_\alpha$
commutano con $U(\Lambda)$ per ogni trasformazione di Lorentz.
Osservando l'azione (\ref{action}), segue che le matrici
$b^\ast_\alpha(p)$, commutando con i \emph{boosts}, devono essere
indipendenti da $p$, e analogamente, commutando con le rotazioni
tridimensionali, devono agire come matrici diagonali sugli indici
di spin. Resta quindi dimostrato che i generatori di ${\cal C}$
danno luogo a simmetrie interne.

Lo stesso si pu\`{o} dimostrare per le algebre $u(1)$. Si noti,
per prima cosa, che per ogni coppia di impulsi $p,q$ esiste un
generatore di Lorentz $J$ che li lascia invariati entrambi: se
sono infatti di tipo luce e paralleli, basta prendere $J$ come il
generatore delle rotazioni attorno all'asse comune a
$\overrightarrow{p}$ e $\overrightarrow{q}$, mentre se sono di
tipo tempo $J$ sar\`{a} il generatore delle rotazioni attorno
all'asse comune di $\overrightarrow{p}$ e $\overrightarrow{q}$ nel
sistema di riferimento del centro di massa, dove
$\overrightarrow{p}=-\overrightarrow{q}$. $J$ si diagonalizza in
tal modo sulla base di stati a due particelle $|p\,m,q\,n\rangle$,
\begin{equation}
J|p\,m,q\,n\rangle=\sigma(m,n)|p\,m,q\,n\rangle \ .
\end{equation}
Denotiamo con $B^\ast_i$ i particolari $B^\ast_\alpha$ che
generano le algebre abeliane. Per ipotesi i $B^\ast_\alpha$
commutano con i $P_\mu$, e poich\'{e} $[J,P_\mu]\sim P_\mu$,
l'identit\`{a} di Jacobi implica che $P_\mu$ commuta con
$[J,B^\ast_\alpha]$, che si pu\`{o} quindi scrivere come
combinazione lineare dei $B_\beta$, e anzi dei $B^\ast_\beta$, dal
momento che il commutatore ha traccia nulla. Ma i $B^\ast_i$,
generando un'algebra abeliana, commuteranno con tutti i
$B^\ast_\alpha$, ed in particolare
\begin{equation}
\left[B^\ast_i,[J,B^\ast_i]\right]=0 \ .
\end{equation}
Prendendone il valore di aspettazione sulla base
$|p\,m,q\,n\rangle$ si ottiene
\begin{equation}\label{EV}
0=\sum_{m',n'}\left(\sigma(m',n')-\sigma(m,n)\right)\left|\left(b^\ast_i(p,q)\right)_{m'n',mn}\right|^2
\ ,
\end{equation}
per ogni $m$ ed $n$. Gli indici $m,n,m',n'$ corrono su un insieme
finito di valori, e se esistesse quindi un $\sigma$ tale che, per
opportuni $m,n$ ed $m',n'$, $\sigma(m,n)=\sigma\neq\sigma(m',n')$
e $\left(b^\ast_i(p,q)\right)_{m'n',mn}\neq 0$, dovrebbe esistere
anche il pi\`{u} piccolo di tali $\sigma$, per il quale il secondo
membro della (\ref{EV}) sarebbe definito positivo. Il solo modo
per garantire che quest'ultima sia sempre soddisfatta consiste
dunque nell'ammettere che
\begin{equation}
\forall m,n,m',n':\quad \sigma(m,n)\neq\sigma(m',n')\Rightarrow
\left(b^\ast_i(p,q)\right)_{m'n',mn}=0 \ ,
\end{equation}
il che significa che $J$ e le $b^\ast_i(p,q)$ sono diagonali sulla
stessa base, ovvero che $[J,B^\ast_i]=0$, in virt\`{u}
dell'isomorfismo $I$. Ma $p+q$ pu\`{o} essere scelto in qualunque
direzione di tipo tempo, sicch\'{e} i generatori abeliani
$B^\ast_i$ devono commutare con tutti i generatori di Lorentz,
\begin{equation}
[J_{\mu\nu},B^\ast_i]=0 \ ,
\end{equation}
e ci\`{o} dimostra che anche i $B^\ast_i$ corrispondono a
simmetrie interne. Il passo A \`{e} cos\`{i} completato.

\vspace{0.5cm}

Estendiamo questi risultati a generatori $A_\alpha$ che non
commutino, per ipotesi, con i generatori delle traslazioni
$P_\mu$, e che agiscano quindi sugli stati ad una particella
``ruotando'' anche l'indice continuo $p$, come nell'eq.
(\ref{actioncompl}). Mostriamo anzitutto che i nuclei $\left({\cal
A}(p,p')\right)_{n'n}$, non nulli soltanto se $p$ e $p'$ sono
entrambi sul \emph{mass shell} (per definizione di operatore di
simmetria), sono anzi non nulli soltanto se $p'=p$. Per farlo, si
osservi preliminarmente che, se $A_\alpha$ \`{e} un operatore di
simmetria, lo \`{e} anche
\begin{equation}
A_\alpha^f\equiv\int d^4 x\, e^{iP\cdot x}A_\alpha\, e^{-iP\cdot
x}f(x) \ ,
\end{equation}
dove $f(x)$ \`{e} una funzione qualsiasi, poich\'{e} la sua azione
sugli stati ad una particella,
\begin{equation}
A_\alpha^f|p\,n\rangle=\sum_{n'}\int d^4
p'\,\tilde{f}(p'-p)\left({\cal
A}_\alpha(p,p')\right)_{n,n'}|p'\,n'\rangle \ ,
\end{equation}
\`{e} identica a quella di $A_\alpha$ a meno del riscalamento dei
nuclei di integrazione tramite la trasformata di Fourier della
funzione $f(x)$,
\begin{equation}
\tilde{f}(p'-p)\equiv\int d^4 x\, e^{i(p'-p)\cdot x}f(x) \ .
\end{equation}
Supponiamo ora che esista un $\Delta\neq 0$ tale che $p$ e
$p+\Delta$ siano sullo stesso \emph{mass shell}, in modo tale che
${\cal A}(p,p+\Delta)\neq 0$. L'ipotesi $2$ restringe ad un numero
finito i possibili iperboloidi di massa su cui vivono stati ad una
particella legati da un processo di scattering, sicch\'{e}, se gli
impulsi $q,p',q'$ sono fissati su alcuni iperboloidi e sono tali
che $p+q=p'+q'$, in generale $q+\Delta,p'+\Delta,q'+\Delta$ non
saranno sugli stessi iperboloidi. Questo significa che, prendendo
$\tilde{f}(k)$ non nulla solo in un piccolo intorno di $\Delta$,
gli $A_\alpha^f$ annullano qualsiasi stato ad una particella con
impulso $q,p',q'$, ma non lo stato con impulso $p$. Ma allora lo
scattering $p\,q\rightarrow p'\,q'$ sarebbe proibito per
simmetria, in contrasto con le ipotesi $3$ e $4$ del teorema. Ne
segue che i nuclei integrali ${\cal A}(p,p')$ devono essere nulli
se $p'\neq p$.

Questo tuttavia non ci riporta automaticamente al caso trattato
nel passo A, poich\'{e} esiste ancora la possibilit\`{a} che i
nuclei integrali contengano non solo $\delta^4(p'-p)$, ma anche
derivate di $\delta^4(p'-p)$. Questo \`{e} il motivo per cui
Coleman e Mandula aggiunsero l'ipotesi $5$, assumendo che tali
nuclei contenessero al pi\`{u} un ordine finito $D_\alpha$ di
derivate di $\delta^4(p'-p)$, ovvero assumendo che l'azione degli
${\cal A}(p,p')$ sugli stati ad una particella fosse quella di un
polinomio di ordine finito $D_\alpha$ nelle derivate
$\partial/\partial p^\mu$,
\begin{equation}
{\cal A}(p,p')=\sum_{n=0}^{D_\alpha}{\cal
A}^{(n)}(p)_{\mu_1...\mu_n}\frac{\partial}{\partial
p_{\mu_1}}...\frac{\partial}{\partial p_{\mu_n}} \ ,
\end{equation}
con coefficienti matriciali dipendenti da $p$. Per utilizzare i
risultati del passo A, \`{e} conveniente costruire l'operatore
\begin{equation}\label{construction}
B_\alpha^{\mu_1...\mu_{D_\alpha}}\equiv[P^{\mu_1},[P^{\mu_2},...[P^{\mu_{D_\alpha}},A_\alpha]]...]
\ ,
\end{equation}
che commuta con $P_\mu$, come si pu\`{o} verificare calcolando
l'elemento di matrice di
$[B_\alpha^{\mu_1...\mu_{D_\alpha}},P_\mu]$ tra stati di impulso
$p'$ e $p$, che \`{e} infatti proporzionale a $D_\alpha+1$ fattori
di $(p'-p)$ su cui agiscono $D_\alpha$ derivate rispetto
all'impulso: il tutto viene calcolato a $p'=p$, e d\`{a} quindi un
risultato nullo. Ma allora, in base a quanto ottenuto in
precedenza, i generatori $B_\alpha^{\mu_1...\mu_{D_\alpha}}$
devono agire sugli stati ad una particella con matrici della forma
\begin{equation}\label{dapassoA}
b_\alpha^{\mu_1...\mu_{D_\alpha}}(p)=b_\alpha^{\ast\mu_1...\mu_{D_\alpha}}+a_\alpha^{\mu\mu_1...\mu_{D_\alpha}}p_\mu\,
1 \ ,
\end{equation}
dove gli $a_\alpha^{\mu\mu_1...\mu_{D_\alpha}}$ sono coefficienti
costanti, mentre le matrici
$b_\alpha^{\ast\,\mu_1...\mu_{D_\alpha}}$, indipendenti da $p$ e a
traccia nulla, generano una simmetria interna, ed entrambi sono
simmetrici, per costruzione, negli indici
$\mu_1...\mu_{D_\alpha}$. La richiesta che i nuclei non portino
gli stati su cui agiscono fuori dal \emph{mass shell} si traduce
nella condizione
\begin{equation}
[A_\alpha,-P_\mu P^\mu]=0 \ ,
\end{equation}
la quale implica che, per $D_\alpha\geq 1$,
\begin{equation}
0=[P^{\mu_1}P_{\mu_1},[P^{\mu_2},...[P^{\mu_{D_\alpha}},A_\alpha]]...]=2P_{\mu_1}B_\alpha^{\mu_1...\mu_{D_\alpha}}
\ ,
\end{equation}
ovvero, in termini delle matrici agenti sugli stati ad una
particella,
\begin{equation}\label{precrucial}
p_{\mu_1}b_\alpha^{\mu_1...\mu_{D_\alpha}}(p)=0 \ ,
\end{equation}
per ogni $p$ di tipo tempo sul \emph{mass shell}. Ma questo
significa che, per $D_\alpha\geq 1$,
\begin{equation}
b_\alpha^{\ast\,\mu_1...\mu_{D_\alpha}}=0 \ ,
\end{equation}
e
\begin{equation}\label{crucial}
a_\alpha^{\mu\mu_1...\mu_{D_\alpha}}=-a_\alpha^{\mu_1\mu...\mu_{D_\alpha}}
\ ,
\end{equation}
dove \`{e} stata usata la relazione $[P_\mu,P_\nu]=0$ \footnote
{Si noti che tale relazione, valida nello spazio piatto, non lo
\`{e} pi\`{u} su uno spazio a curvatura costante non banale come
$AdS$ (si veda il cap. 3).}. Per $D_\alpha\geq 2$, l'ultima
equazione \`{e} inconsistente con la simmetria negli indici
$\mu_1...\mu_{D_\alpha}$ (poich\'{e} utilizzando entrambe le
condizioni si otterrebbe
$a_\alpha^{\mu\mu_1...\mu_{D_\alpha}}=-a_\alpha^{\mu\mu_1...\mu_{D_\alpha}}$),
sicch\'{e} $a_\alpha^{\mu\mu_1...\mu_{D_\alpha}}=0$, e restano
soltanto da esaminare i casi $D_\alpha=0,1$. Ma per $D_\alpha=0$
si ritorna nel caso trattato nel passo A, mentre per $D_\alpha=1$
le (\ref{construction}) e (\ref{dapassoA}), forniscono
\begin{equation}
[P^\nu,A_\alpha]=a_\alpha^{\mu\nu}P^\mu \ ,
\end{equation}
dove $a_\alpha^{\mu\nu}$ \`{e} antisimmetrico in $\mu$ e $\nu$.
Poich\'{e}
$[P^\nu,J^{\rho\sigma}]=i\eta^{\nu\sigma}J^\rho-i\eta^{\nu\rho}J^\sigma$,
questo significa che
\begin{equation}\label{end}
A_\alpha=-\frac{i}{2}a_\alpha^{\mu\nu}J_{\mu\nu}+B_\alpha \ ,
\end{equation}
dove $B_\alpha$ commuta con $P_\mu$, ed \`{e} anch'esso un
generatore di simmetria, perch\'{e} sia $A_\alpha$ che
$J_{\mu\nu}$ lo sono. Ma il passo A garantisce che i $B_\alpha$
sono combinazioni lineari di generatori di una simmetria interna e
di $P_\mu$, sicch\'{e} l'equazione (\ref{end}) completa la
dimostrazione del teorema.

\vspace{0.5cm}

Un'ulteriore sottigliezza compare se si considerano particelle a
massa nulla, poich\'{e} in tal caso $p_\mu p^\mu=0$, e la
(\ref{crucial}) non segue necessariamente dalla
(\ref{precrucial}), ma si pu\`{o} anche avere
\begin{equation}
a_\alpha^{\mu\mu_1...\mu_{D_\alpha}}+a_\alpha^{\mu_1\mu...\mu_{D_\alpha}}\propto
\eta^{\mu\mu_1} \ .
\end{equation}
Tuttavia, se questo caso \`{e} verificato, la tesi del teorema si
generalizza ammettendo che l'algebra di simmetria $G$ consista di
un'algebra di simmetria interna pi\`{u} l'algebra \emph{conforme},
che include l'algebra di Poincar\'{e} come sottoalgebra propria,
ed accoglie come ulteriori generatori di simmetria i $K^\mu$, che
danno luogo alle trasformazioni conformi speciali ($\delta
x^\mu=\rho^\mu(x\cdot x)-x^mu\rho\cdot x$), e $D$, generatore
delle dilatazioni ($\delta x^\mu=\lambda x^\mu$), con le relazioni
di commutazione
\begin{equation}
[P^\mu,D]\,=\,i\,P^\mu \ , \quad [K^\mu,D]\,=\,-i\,K^\mu \ ,
\end{equation}
\begin{equation}
[P^\mu,K^\nu]=2i\eta^{\mu\nu}D+2iJ_{\mu\nu} \ , \quad
[K^\mu,K^\nu]\,=\,0 \ ,
\end{equation}
\begin{equation}
[J^{\rho\sigma},K^\mu]=i\eta^{\mu\rho}K^\sigma-i\eta^{\mu\sigma}K^\rho
\ , \quad [J^{\rho\sigma},D]\,=\,0 \ ,
\end{equation}
oltre alle usuali relazioni che definiscono l'algebra di
Poincar\'{e}.

\chapter{Formalismo a due componenti}
\label{appendice A}

Definiamo gli spinori come rappresentazioni irriducibili del
gruppo delle matrici complesse $2\times 2$ unimodulari $SL(2,C)$,
ricoprimento universale del gruppo di Lorentz proprio $SO(3,1)$,
nello stesso modo in cui nella teoria non relativistica gli
spinori entrano in gioco come rappresentazioni irriducibili del
gruppo $SU(2)$, ricoprimento universale del gruppo delle rotazioni
spaziali $SO(3)$.

Come \`{e} ben noto, il gruppo di Lorentz $O(3,1)$ corrisponde al
gruppo delle trasformazioni spazio-temporali $x^{a}\rightarrow
x'^{a}=\Lambda^{a}_{\phantom{a}b}x^{b}$, con $a,b=0,...,3$ indici
di spazio tangente, che lasciano invariante la forma
$x_{a}\eta^{ab}x_{b}$, dove $\eta^{ab}$ \`{e} la metrica
minkowskiana quadridimensionale, che prendiamo con segnatura
$(-,+,+,+)$. Questa richiesta impone la condizione di
pseudo-ortogonalit\`{a}
\begin{equation}
\Lambda^{a}_{\phantom{a}c}\Lambda^{b}_{\phantom{b}d}\eta^{cd}=\eta^{ab}\
,
\end{equation}
che a sua volta comporta $(\det\Lambda)^{2}=1$, ed il gruppo di
Lorentz proprio $SO(3,1)$ corrisponde alla scelta $\det\Lambda=1$.
Quest'ultimo \`{e} un gruppo continuo e connesso, ma non
semplicemente connesso (cfr., ad esempio, \cite{Gilmore}), ed un
teorema generale assicura che, dato un gruppo non semplicemente
connesso, esiste uno e un solo gruppo semplicemente connesso che
possiede la stessa algebra del primo (ovvero con la stessa
struttura infinitesima, ovvero localmente isomorfo), del quale
viene quindi detto \emph{ricoprimento universale}. Il ricoprimento
universale di $SO(3,1)$ \`{e} $SL(2,C)$, e le matrici ${\cal M}\in
SL(2,C)$ agiscono naturalmente in uno spazio complesso
bidimensionale di elementi $\psi_{\alpha},\,\alpha=1,2$, detti
spinori di Weyl, che ne costituisce la rappresentazione
fondamentale. L'idea alla base del formalismo a due componenti
\`{e} quella di costruire tutte le rappresentazioni del gruppo di
Lorentz partendo dalla rappresentazione fondamentale del suo
ricoprimento universale. Uno spinore di Weyl trasforma sotto
l'azione di un elemento ${\cal M}\in SL(2,C)$ come
\begin{equation}
\psi_{\alpha}\rightarrow \psi'_{\alpha}={\cal
M}_{\alpha}^{\phantom{\alpha}\beta}\psi_{\beta}\ .
\end{equation}
La rappresentazione complessa coniugata, inequivalente, viene di
solito denotata con indici puntati,
$\bar{\psi}_{\dot{\alpha}},\,\dot{\alpha}=1,2$, e trasforma come
\begin{equation}
\bar{\psi}_{\dot{\alpha}}\rightarrow
\bar{\psi}'_{\dot{\alpha}}=({\cal
M}^{\ast})_{\dot{\alpha}}^{\phantom{\alpha}\dot{\beta}}\bar{\psi}_{\dot{\beta}}\
,
\end{equation}
sicch\'{e} $\bar{\psi}_{\dot{\alpha}}=(\psi_{\alpha})^{\ast}$.
$\psi_{\alpha}$ corrisponde alla rappresentazione irriducibile
$(1/2,0)$ di $su(2)\times su(2)$ \footnote {L'algebra di Lorentz e
quella di $su(2)\times su(2)$ sono infatti isomorfe, e siamo
quindi in grado di costruire rappresentazioni (non unitarie) di
dimensione finita di $so(3,1)$ tramite quelle di $su(2)\oplus
su(2)$.}, mentre $\bar{\psi}_{\dot{\alpha}}$ corrisponde alla
rappresentazione $(0,1/2)$, irriducibile anch'essa.

Come $\eta_{ab}$ per $SO(3,1)$, esistono tensori invarianti a due
indici anche per $SL(2,C)$, $\varepsilon_{\alpha\beta}$ e
$\varepsilon_{\dot{\alpha}\dot{\beta}}$,
\begin{eqnarray}
\varepsilon_{\alpha \beta} \, = \, \varepsilon_{\dot{\alpha}
\dot{\beta}}\,=\, \left(\begin{array}{cc}
  0 & 1 \\
  -1~ & 0
\end{array} \right ) \ , \label{matriciepsilon}
\end{eqnarray}
antisimmetrici, propriet\`{a} cruciale per l'invarianza,
\begin{equation}
\varepsilon'_{\alpha \beta}={\cal
M}_{\alpha}^{\phantom{\alpha}\rho}{\cal
M}_{\beta}^{\phantom{\beta}\sigma}\varepsilon_{\rho\sigma}=\det{\cal
M}\,\varepsilon_{\alpha \beta}=\varepsilon_{\alpha \beta} \ ,
\end{equation}
e analogamente per la rappresentazione coniugata. Introduciamo
inoltre i corrispondenti tensori con gli indici alti,
$\varepsilon^{\alpha \beta}$ e $\varepsilon^{\dot{\alpha}
\dot{\beta}}$, tramite la definizione $\varepsilon^{\alpha
\beta}\varepsilon_{\gamma\beta}=\delta_{\gamma}^{\alpha}$, che
implica $ \varepsilon^{\alpha \beta}=\varepsilon^{\dot{\alpha}
\dot{\beta}}=\varepsilon_{\alpha
\beta}=\varepsilon_{\dot{\alpha}\dot{\beta}}$. Questi tensori
vengono usati per alzare ed abbassare gli indici spinoriali,
secondo le convenzioni
\begin{equation}
\psi^{\alpha}=\varepsilon^{\alpha\beta}\psi_{\beta}\ , \quad
\psi_{\alpha}=\psi^{\beta}\varepsilon_{\beta\alpha}\ , \quad
\bar{\psi}^{\dot{\alpha}}=\varepsilon^{\dot{\alpha}\dot{\beta}}\bar{\psi}_{\dot{\beta}}\
, \quad
\bar{\psi}_{\dot{\alpha}}=\bar{\psi}_{\dot{\beta}}\varepsilon^{\dot{\beta}\dot{\alpha}}
\ ,
\end{equation}
che \`{e} necessario stabilire dal momento che i tensori
$\varepsilon$ sono antisimmetrici. Le trasformazioni degli spinori
di Weyl con indici alti sono
\begin{equation}
\psi'^{\alpha}=\psi^{\beta}({\cal
M}^{-1})_{\beta}^{\phantom{\beta}\alpha}\ , \qquad
\bar{\psi}'^{\dot{\alpha}}=\bar{\psi}^{\dot{\beta}}({\cal
M}^{\ast\,-1})_{\dot{\beta}}^{\phantom{\beta}\dot{\alpha}}\ .
\end{equation}

Utilizzando la matrice identit\`{a} $2\times 2$ e le matrici di
Pauli
\begin{equation}
    \sigma^{1} = \left( \begin{array}{cc} 0 & 1 \\ 1 & 0 \end{array}
    \right) \, , \quad \sigma^{2} = \left( \begin{array}{cc} 0 & -i \\
    i & 0 \end{array}\right) \, , \quad \sigma^{3} = \left(
    \begin{array}{cc} 1 & 0 \\ 0 & -1 \end{array}\right) \, ,
    \label{pauli}
\end{equation}
possiamo definire i simboli di van der Waerden
$(\sigma^{a})_{\alpha\dot{\beta}}$ ($a=0,1,2,3$) come
\begin{equation}
(\sigma^{a})_{\alpha\dot{\beta}}\,\equiv\, (1, \sigma^{1},
\sigma^{2}, \sigma^{3}) \ ,
\end{equation}
da cui discendono
\begin{equation}
(\bar{\sigma}^{a})^{\dot{\alpha}\beta}\,\equiv\, (1, -\sigma^{1},
-\sigma^{2}, -\sigma^{3})\,=\,
\varepsilon^{\dot{\alpha}\dot{\delta}}\varepsilon^{\beta\gamma}(\sigma^{a})_{\gamma\dot{\delta}}\
,
\end{equation}
con le propriet\`{a} di hermiticit\`{a}
\begin{equation}
\left((\sigma^{a})_{\alpha\dot{\beta}}\right)^{\dag}\,=\,(\bar{\sigma}^{a})_{\dot{\alpha}\beta}\,=\,(\sigma^{a})_{\beta\dot{\alpha}}
\ , \quad
\left((\sigma^{a})^{\alpha\dot{\beta}}\right)^{\dag}\,=\,(\bar{\sigma}^{a})^{\dot{\alpha}\beta}\,=\,(\sigma^{a})^{\beta\dot{\alpha}}
\ .
\end{equation}
I simboli di van der Waerden soddisfano le relazioni
\begin{eqnarray}
(\sigma^{a})_{\alpha\dot{\alpha}}(\bar{\sigma}_{a})^{\dot{\beta}\beta}
& = & -2\delta_{\alpha}^{\beta}\delta_{\dot{\alpha}}^{\dot{\beta}}
\ , \nonumber \\
(\sigma^{a})_{\alpha}^{\phantom{\alpha}\dot{\alpha}}(\bar{\sigma}^{b})_{\dot{\alpha}}^{\phantom{\alpha}\beta}
& = &
\eta^{ab}\delta_{\alpha}^{\beta}+(\sigma^{ab})_{\alpha}^{\phantom{\alpha}\beta}
\ , \nonumber \\
(\bar{\sigma}^{a})_{\dot{\alpha}}^{\phantom{\alpha}\alpha}(\sigma^{b})_{\alpha}^{\phantom{\alpha}\dot{\beta}}
& = &
\eta^{ab}\delta_{\dot{\alpha}}^{\dot{\beta}}+(\bar{\sigma}^{ab})_{\dot{\alpha}}^{\phantom{\alpha}\dot{\beta}}
\ , \label{completeness}
\end{eqnarray}
dove
$(\sigma^{ab})_{\alpha\beta}=-(\sigma^{ba})_{\alpha\beta}=(\sigma^{ab})_{\beta\alpha}$
e analogamente per le
$(\bar{\sigma}^{ab})_{\dot{\alpha}\dot{\beta}}$. Simmetrizzando la
seconda delle (\ref{completeness}) in $a\leftrightarrow b$ si vede
che i simboli di van der Waerden soddisfano l'algebra di Clifford,
\begin{equation}
(\sigma^{a}\bar{\sigma}^{b}+\sigma^{b}\bar{\sigma}^{a})_{\alpha}^{\beta}=2\eta^{ab}\delta_{\alpha}^{\beta}
\end{equation}
(e un analogo risultato si ottiene simmetrizzando la terza
relazione in (\ref{completeness}), con i ruoli di $\sigma$ e
$\bar{\sigma}$ scambiati), mentre l'antisimmetrizzazione negli
stessi indici fornisce le definizioni
\begin{eqnarray}
(\sigma^{ab})_{\alpha}^{\phantom{\alpha}\beta} & = &
\frac{1}{2}\left[(\sigma^{a})_{\alpha}^{\phantom{\alpha}\dot{\alpha}}(\bar{\sigma}^{b})_{\dot{\alpha}}^{\phantom{\alpha}\beta}-(a\leftrightarrow
b)\right] \ , \nonumber \\
(\bar{\sigma}^{ab})_{\dot{\alpha}}^{\phantom{\alpha}\dot{\beta}} &
= &
\frac{1}{2}\left[(\bar{\sigma}^{a})_{\dot{\alpha}}^{\phantom{\alpha}\alpha}(\sigma^{b})_{\alpha}^{\phantom{\alpha}\dot{\beta}}-(a\leftrightarrow
b)\right] \ .
\end{eqnarray}
Ulteriori utili identit\`{a} sono
\begin{equation}\label{utilexspin3}
(\sigma^{a})_{\alpha\dot{\alpha}}(\sigma^{b})_{\beta\dot{\beta}}-(a\leftrightarrow
b)\,=\,-(\sigma^{ab})_{\alpha\beta}\varepsilon_{\dot{\alpha}\dot{\beta}}-(\sigma^{ab})_{\dot{\alpha}\dot{\beta}}\varepsilon_{\alpha\beta}
\end{equation}
e le propriet\`{a} di dualit\`{a} delle matrici $\sigma^{ab}$ e
$\bar{\sigma}^{ab}$
\begin{equation}\label{duality}
\frac{1}{2}\varepsilon_{abcd}(\sigma^{cd})_{\alpha\beta}\,=\,i(\sigma^{ab})_{\alpha\beta}
\ , \quad
\frac{1}{2}\varepsilon_{abcd}(\bar{\sigma}^{cd})_{\dot{\alpha}\dot{\beta}}\,=\,-i(\bar{\sigma}^{ab})_{\dot{\alpha}\dot{\beta}}
\ ,
\end{equation}
dove $\varepsilon_{abcd}$ \`{e} il tensore totalmente
antisimmetrico quadridimensionale, definito con la convenzione
$\varepsilon^{0123}=-\varepsilon_{0123}=1$.

La connessione tra il gruppo di Lorentz e il suo ricoprimento
universale \`{e} resa manifesta rappresentando, mediante i simboli
di van der Waerden, un quadrivettore $V_{a}$ arbitrario di
$SO(3,1)$ con la matrice hermitiana $2\times 2$ $V_{a}\sigma^{a}$,
\begin{eqnarray}
V_{a}\rightarrow V_{a}\sigma^{a}\,=\,\left(\begin{array}{cc}
  -V_{0}+V_{3}& V_{1}-iV_{2} \\
  V_{1}+iV_{2}~ & -V_{0}-V_{3}
\end{array} \right )\,\equiv\,V \ .
\end{eqnarray}
il cui determinante corrisponde, a parte un segno, alla norma
minkowskiana del quadrivettore,
\begin{equation}\label{corrisp}
\det V=-V_{a}\eta^{ab}V_{b} \ .
\end{equation}
Ma $V$ trasforma come $V'={\cal M}\,V\,{\cal M}^{-1}$, con ${\cal
M}\in SL(2,C)$, e per definizione le trasformazioni di $SL(2,C)$
preservano il determinante di $V$, $\det V'=\det V$. Ci\`{o}
implica, in virt\`{u} della (\ref{corrisp}), che la trasformazione
$V'={\cal M}\,V\,{\cal M}^{-1}$ induce sulle componenti $V_{a}$
una trasformazione di Lorentz propria, la cui forma esplicita in
componenti spinoriali \`{e}
\begin{equation}\label{corrisp2}
V'_{\alpha\dot{\beta}}={\cal
M}_{\alpha}^{\phantom{\alpha}\gamma}({\cal
M}^{\ast})_{\alpha}^{\phantom{\alpha}\gamma}V_{\gamma\dot{\delta}}\
,
\end{equation}
che mostra quindi che ad ogni quadrivettore $V_{a}$ che trasforma
sotto $SO(3,1)$ come $V'^{a}=\Lambda^{a}_{\phantom{a}b}V^{b}$
resta associata la matrice hermitiana
$V_{a}(\sigma^{a})_{\alpha\dot{\beta}}=V_{\alpha\dot{\beta}}$ che
trasforma sotto $SL(2,C)$. La (\ref{corrisp2}) d\`{a} anche
ragione del fatto che i simboli di van der Waerden siano
naturalmente definiti con un indice non puntato e uno puntato,
come \`{e} anche chiaro dal fatto che, per semplice composizione
dei momenti angolari, $V_{\alpha\dot{\beta}}$, che trasforma come
la rappresentazione irriducibile $(1/2,1/2)$, ha esattamente lo
stesso numero di componenti di un quadrivettore. Osserviamo
tuttavia che la corrispondenza tra trasformazioni di $SO(3,1)$ e
del suo ricoprimento universale $SL(2,C)$, che si esprime come
\begin{equation}
V'=V_{a}{\cal M}\sigma^{a}{\cal
M}^{\dag}=\Lambda^{a}_{\phantom{a}b}V^{b}\sigma^{a}\ ,
\end{equation}
ovvero
\begin{equation}\label{corrisp3}
{\cal M}\sigma^{a}{\cal
M}^{\dag}=\Lambda^{b}_{\phantom{b}a}\sigma^{b}\ ,
\end{equation}
non \`{e} una corrispondenza biunivoca, poich\'{e} trasformazioni
indotte da ${\cal M}$ e da $-{\cal M}$ danno luogo alla stessa
$\Lambda^{a}_{\phantom{a}b}$, sicch\'{e}, come \`{e} ben noto, gli
spinori sono rappresentazioni a due valori di $SO(3,1)$.

Dal momento che ad ogni indice di $SO(3,1)$ resta associata, nel
formalismo a due componenti, una coppia di indici
$\alpha\dot{\alpha}$, secondo le relazioni
\begin{equation}
V_{\alpha\dot{\alpha}}=(\sigma^{a})_{\alpha\dot{\alpha}}V_{a}\
,\qquad
V_{a}=-\frac{1}{2}(\sigma^{a})^{\alpha\dot{\alpha}}V_{\alpha\dot{\alpha}}\
,
\end{equation}
possiamo associare ad un tensore di Lorentz di rango arbitrario
$V_{a_{1}...a_{n}}$ il multispinore
$V_{\alpha_{1}...\alpha_{n},\dot{\alpha}_{1}...\dot{\alpha}_{n}}$
tenendo per\`{o} conto della fondamentale propriet\`{a} che
qualsiasi oggetto antisimmetrico in due indici spinoriali dello
stesso tipo \`{e} proporzionale al tensore $\varepsilon$
appropriato. Ad esempio,
\begin{equation}
A_{\alpha\beta}-A_{\beta\alpha}=-\varepsilon_{\alpha\beta}A^{\gamma}_{\gamma}\
,
\end{equation}
il cui coefficiente \`{e} in accordo con la $\varepsilon^{\alpha
\beta}\varepsilon_{\alpha\beta}=2$. Un tensore antisimmetrico
$J_{ab}$ ammette quindi la decomposizione
\begin{equation}
J_{ab}(\sigma^{a})_{\alpha\dot{\alpha}}(\sigma^{b})_{\beta\dot{\beta}}=\varepsilon_{\alpha
\beta}\bar{J}_{\dot{\alpha}\dot{\beta}}+\varepsilon_{\dot{\alpha}\dot{\beta}}J_{\alpha\beta}\
,
\end{equation}
dove
\begin{equation}
J_{\alpha\beta}=-\frac{1}{2}J_{ab}(\sigma^{ab})_{\alpha\beta}\ ,
\quad
\bar{J}_{\dot{\alpha}\dot{\beta}}=-\frac{1}{2}J_{ab}(\bar{\sigma}^{ab})_{\dot{\alpha}\dot{\beta}}\
,
\end{equation}
entrambi simmetrici negli indici spinoriali, mentre
\begin{equation}\label{utilexspin3b}
J_{ab}=-\frac{1}{4}(\sigma_{ab})^{\alpha\beta}J_{\alpha\beta}-\frac{1}{4}(\bar{\sigma}_{ab})^{\dot{\alpha}\dot{\beta}}\bar{J}_{\dot{\alpha}\dot{\beta}}\
.
\end{equation}

\chapter{$\star$-prodotto e integrazione simplettica}
\label{appendice B}

Riportiamo anzitutto alcune formule utili:
\begin{equation}\label{yderivs}
\partial_{\alpha}y_{\beta}=\varepsilon_{\alpha\beta}\ , \qquad \partial^{\alpha}y^{\beta}=\varepsilon^{\alpha\beta}\ ,
\end{equation}
\begin{equation}
\partial_{\alpha}y^{\beta}=\delta_{\alpha}^{\beta}\ , \qquad \partial^{\alpha}y_{\beta}=-\delta^{\alpha}_{\beta}\
,
\end{equation}
dove si intende $\partial_{\alpha}=\frac{\partial}{\partial
y^{\alpha}}$, e analogamente per le variabili spinoriali $z$.
Utilizzando la definizione generale (\ref{defdiffext}) o
(\ref{defintext}) si possono provare le (\ref{contrulez}), da cui
notiamo, in particolare, che
$[z_{\alpha},y_{\beta}]_{\star}=[\bar{z}_{\dot{\alpha}},\bar{y}_{\dot{\beta}}]_{\star}=0$.
Inoltre, dalle regole di contrazione sopra definite si deriva la
formula generale
\begin{equation}\label{compactcontrule}
y_{\alpha(n)}\star
y_{\beta(m)}=\sum_{k=0}^{\textrm{min}(n,m)}i^{k}\,k!{n \choose
k}{m \choose
k}y_{\alpha(n-k)}y_{\beta(m-k)}\varepsilon_{\alpha(k)\beta(k)}\ ,
\end{equation}
che abbiamo scritto in forma estesa in (\ref{contrule}), con la
definizione
\begin{equation}
\varepsilon_{\alpha(k)\beta(k)}=\frac{1}{k!}\sum_{P}\varepsilon_{\alpha_{1}\beta_{P(1)}}...\varepsilon_{\alpha_{k}\beta_{P(k)}}
\ .
\end{equation}
Consideriamo ad esempio il prodotto di due bilineari negli
oscillatori $Y$,
\begin{eqnarray}
y_{\alpha(2)}\star y_{\beta(2)} & = &
y_{\alpha_{1}}y_{\alpha_{2}}y_{\beta_{1}}y_{\beta_{2}}+i(y_{\alpha_{1}}y_{\beta_{2}}\varepsilon_{\alpha_{2}\beta_{1}}+y_{\alpha_{1}}y_{\beta_{1}}\varepsilon_{\alpha_{2}\beta_{2}}
\nonumber \\
 & + &
 y_{\alpha_{2}}y_{\beta_{2}}\varepsilon_{\alpha_{1}\beta_{1}}+y_{\alpha_{2}}y_{\beta_{1}}\varepsilon_{\alpha_{1}\beta_{2}}) \nonumber \\
 & - &
 (\varepsilon_{\alpha_{1}\beta_{1}}\varepsilon_{\alpha_{2}\beta_{2}}+\varepsilon_{\alpha_{1}\beta_{2}}\varepsilon_{\alpha_{2}\beta_{1}})\
 ,
\end{eqnarray}
da cui risulta evidente che il commutatore cancella i termini con
un numero pari di contrazioni e raddoppia quelli contenenti un
numero dispari di contrazioni, e viceversa l'anticommutatore
cancella i termini con un numero dispari e raddoppia quelli con un
numero pari di contrazioni. In particolare, si dimostra in questo
modo che i generatori di spin $s\leq 1$ formano la sottoalgebra
finita massimale dell'algebra infinito-dimensionale di HS, secondo
la (\ref{infdim}).

Tenendo conto delle (\ref{yderivs}) e (\ref{contrulez}) si pu\`{o}
scrivere, ad esempio,
\begin{equation}
[y_{\alpha},y_{\beta}]_{\star}=2i\frac{\partial}{\partial
y_{\alpha}}y_{\beta} \ ,
\end{equation}
e le seguenti regole di contrazione pi\`{u} generali:
\begin{eqnarray}
y_{\alpha}\star F(Y,Z) & = &
y_{\alpha}F(Y,Z)+\left[-i\frac{\partial}{\partial
z^{\alpha}}+i\frac{\partial}{\partial y^{\alpha}}\right]F(Y,Z) \ ,
\nonumber \\
z_{\alpha}\star F(Y,Z) & = &
z_{\alpha}F(Y,Z)+\left[-i\frac{\partial}{\partial
z^{\alpha}}+i\frac{\partial}{\partial y^{\alpha}}\right]F(Y,Z) \ ,
\nonumber \\
F(Y,Z)\star y_{\alpha} & = &
y_{\alpha}F(Y,Z)+\left[-i\frac{\partial}{\partial
z^{\alpha}}-i\frac{\partial}{\partial y^{\alpha}}\right]F(Y,Z) \ ,
\nonumber \\
F(Y,Z)\star z_{\alpha} & = &
z_{\alpha}F(Y,Z)+\left[i\frac{\partial}{\partial
z^{\alpha}}+i\frac{\partial}{\partial y^{\alpha}}\right]F(Y,Z) \ ,
\nonumber \\
\bar{y}_{\dot{\alpha}}\star F(Y,Z) & = &
\bar{y}_{\dot{\alpha}}F(Y,Z)+\left[i\frac{\partial}{\partial
\bar{z}^{\dot{\alpha}}}+i\frac{\partial}{\partial
\bar{y}^{\dot{\alpha}}}\right]F(Y,Z) \ ,
\nonumber \\
\bar{z}_{\dot{\alpha}}\star F(Y,Z) & = &
\bar{z}_{\dot{\alpha}}F(Y,Z)+\left[-i\frac{\partial}{\partial
\bar{z}^{\dot{\alpha}}}-i\frac{\partial}{\partial
\bar{y}^{\dot{\alpha}}}\right]F(Y,Z) \ ,
\nonumber \\
F(Y,Z)\star\bar{y}_{\dot{\alpha}} & = &
\bar{y}_{\dot{\alpha}}F(Y,Z)+\left[i\frac{\partial}{\partial
\bar{z}^{\dot{\alpha}}}-i\frac{\partial}{\partial
\bar{y}^{\dot{\alpha}}}\right]F(Y,Z) \ , \nonumber \\
F(Y,Z)\star\bar{z}_{\dot{\alpha}} & = &
\bar{z}_{\dot{\alpha}}F(Y,Z)+\left[i\frac{\partial}{\partial
\bar{z}^{\dot{\alpha}}}-i\frac{\partial}{\partial
\bar{y}^{\dot{\alpha}}}\right]F(Y,Z) \ , \label{sympldiff}
\end{eqnarray}
dove $F(Y,Z)$ \`{e} un'arbitraria funzione a valori in $\hat{{\cal
A}}^{\star}$.

Nella sezione \ref{esppert} abbiamo incontrato alcune equazioni
differenziali lineari nelle variabili $Z$, del tipo
\begin{equation}\label{sympleqdiff1}
d_{Z}f\,=\,g\,=\,dz^{\alpha}g_{\alpha}+d\bar{z}^{\dot{\alpha}}\bar{g}_{\dot{\alpha}}
\ ,
\end{equation}
\begin{equation}\label{sympleqdiff2}
d_{Z}(dz^{\alpha}f_{\alpha}+d\bar{z}^{\dot{\alpha}}\bar{f}_{\dot{\alpha}})\,=\,h\,=\,\frac{1}{2}dz^{\alpha}\wedge
dz_{\alpha}\,h+\frac{1}{2}d\bar{z}^{\dot{\alpha}}\wedge
d\bar{z}_{\dot{\alpha}}\,\bar{h}+dz^{\alpha}\wedge
d\bar{z}^{\dot{\alpha}}\,h_{\alpha\dot{\alpha}} \ .
\end{equation}
Supponendo che $d_{Z}g=0$, la prima ammette la soluzione generale
\cite{Vasiliev:1992av, Sezgin:1998gg}
\begin{equation}\label{symplint1}
f(z,\bar{z})=f(0,0)+\int_{0}^{1}dt[z^{\alpha}
g_{\alpha}(tz,t\bar{z})+\bar{z}^{\dot{\alpha}}\bar{g}_{\dot{\alpha}}(tz,t\bar{z})]
\ ,
\end{equation}
in cui compare il parametro reale $t\in (0,1)$, ed $f(0,0)$ \`{e}
una costante di integrazione, mentre la seconda si risolve come
\begin{eqnarray}
f_{\alpha}(z,\bar{z}) & = & \frac{\partial}{\partial
z^{\alpha}}k(z,\bar{z})-\int_{0}^{1}dt\,t[z_{\alpha} h(t
z,t\bar{z})+\bar{z}^{\dot{\alpha}}h_{\alpha\dot{\alpha}}(t
z,t\bar{z})] \ , \nonumber \\
f_{\dot{\alpha}}(z,\bar{z}) & = & \frac{\partial}{\partial
\bar{z}^{\dot{\alpha}}}k(z,\bar{z})-\int_{0}^{1}dt\,t[\bar{z}_{\dot{\alpha}}
\bar{h}(t z,t\bar{z})-z^{\alpha}h_{\alpha\dot{\alpha}}(t
z,t\bar{z})] \ , \label{symplint2}
\end{eqnarray}
dove $k(z,\bar{z})$ \`{e} una funzione arbitraria. Per ottenere
queste formule di integrazione si utilizza, in entrambi i casi,
l'identit\`{a}
\begin{equation}\label{chainrule}
t\frac{d}{dt}h(tz)=z^{\alpha}\frac{\partial}{\partial
z^{\alpha}}h(tz) \ .
\end{equation}
\`{E} quindi essenziale, nel risolvere le equazioni (\ref{4ordn}),
(\ref{5ordn}) e (\ref{3ordn}) secondo le (\ref{symplint1}) e
(\ref{symplint2}), calcolare gli $\star$-prodotti prima di mandare
$z,\bar{z}$ in $tz,t\bar{z}$, poich\'{e} le contrazioni modificano
la dipendenza da $z$ e $\bar{z}$, e dunque fa differenza operare
con $t\frac{d}{dt}$ su $A(tZ)\star B(tZ)$ o su $\left(A(Z)\star
B(Z)\right)_{Z\rightarrow tZ}$. Nel primo caso, in particolare, la
(\ref{chainrule}) non vale, e si ha invece
\begin{eqnarray}
&&t\frac{d}{dt}A(t z,t\bar{z})\star B(t
z,t\bar{z})=\left[z^{\alpha}\frac{\partial}{\partial
z^{\alpha}}+\bar{z}^{\dot{\alpha}}\frac{\partial}{\partial
\bar{z}^{\dot{\alpha}}}\right](A\star B) \nonumber \\
&&-2i\varepsilon^{\alpha\beta}\frac{\partial}{\partial
z^{\alpha}}A\star\frac{\partial}{\partial
z^{\beta}}B-2i\varepsilon^{\dot{\alpha}\dot{\beta}}\frac{\partial}{\partial
\bar{z}^{\dot{\alpha}}}A\star\frac{\partial}{\partial
\bar{z}^{\dot{\beta}}}B \ .
\end{eqnarray}

 \end{document}